\pdfoutput=1
\documentclass[lettersize,journal]{IEEEtran}
\usepackage{amsmath,amsfonts}
\usepackage{algorithmic}
\usepackage{algorithm}
\usepackage{array}
\usepackage[caption=false,font=footnotesize,labelfont=rm,textfont=sf]{subfig}
\usepackage{textcomp}
\usepackage{stfloats}
\usepackage{url}
\usepackage{verbatim}
\usepackage{graphicx}
\usepackage{cite}
\usepackage{color}
\usepackage{booktabs} %提供命令\toprule、\midrule、\bottomrule
\usepackage{bbding} % checkmark and xsolid

\begin{document}

\title{Pansharpening via Frequency-Aware Fusion Network with Explicit Similarity Constraints}

\author{Yinghui Xing,~\IEEEmembership{Member,~IEEE,} Yan Zhang, Houjun He,  Xiuwei Zhang and Yanning Zhang,~\IEEEmembership{Senior Member,~IEEE}
        % <-this % stops a space
\thanks{This work was supported in part by the National Natural Science Foundation of China (NFSC) under Grant 62201467, Grant 62101453 and Grant U19B2037; in part by the Guangdong Basic and Applied Basic Research Foundation under Grant 2021A1515110544; in part by the Natural Science Basic Research Program of Shaanxi under Grant 2022JQ-686, and in part by the Project funded by China Postdoctoral Science Foundation under Grant 2022TQ0260. \emph{(Corresponding author: Xiuwei Zhang.)}}

\thanks{Yinghui Xing, Yan Zhang, Xiuwei Zhang and Yanning Zhang are with the National Engineering Laboratory for Integrated Aerospace-Ground-Ocean Big Data Application Technology, School of Computer Science, Northwestern Polytechnical University, Xi’an 710072, China. Yinghui Xing is also with the Research Development Institute of Northwestern Polytechnical University in Shenzhen, Shenzhen 518057, China. (e-mail: xyh\underline{~}7491@nwpu.edu.cn).

Houjun He is with the Information Center of Yellow River Conservancy Commission, Zhengzhou, Henan, China.
}}

%\markboth{IEEE TRANSACTIONS ON GEOSCIENCE AND REMOTE SENSING,~Vol.~XX, No.~X, XX~2022}%
%{Shell \MakeLowercase{\textit{XING et al.}}: FAN}


\maketitle

\begin{abstract}
The process of fusing a  high spatial resolution (HR) panchromatic (PAN) image and a low spatial resolution (LR) multispectral (MS) image to obtain an HRMS image is known as pansharpening. With the development of convolutional neural networks, the performance of pansharpening methods has been improved, however, the blurry effects and the spectral distortion still exist in their fusion results due to the insufficiency in details learning and the frequency mismatch between MS and PAN. Therefore, the improvement of spatial details at the premise of reducing spectral distortion is still a challenge. In this paper, we propose a frequency-aware fusion network (FAFNet) together with a novel high-frequency feature similarity loss to address above mentioned problems. FAFNet is mainly composed of two kinds of blocks, where the frequency aware blocks aim to extract features in the frequency domain with the help of discrete wavelet transform (DWT) layers, and the frequency fusion blocks reconstruct and transform the features from frequency domain to spatial domain with the assistance of inverse DWT (IDWT) layers. Finally, the fusion results are obtained through a convolutional block. In order to learn the correspondence, we also propose a high-frequency feature similarity loss to constrain the HF features derived from PAN and MS branches, so that HF features of PAN can reasonably be used to supplement that of MS. Experimental results on three datasets at both reduced- and full-resolution demonstrate the superiority of the proposed method compared with several state-of-the-art pansharpening models. The codes are available at https://github.com/YinghuiXing/FAFNet.
\end{abstract}


\section{Introduction}
\IEEEPARstart{W}{ith} the increasing demand of earth observation and monitoring, many optical satellites were launched, such as GaoFen-2, QuickBird and WorldView-2, whose sensors can produce bundled multispectral (MS) and panchromatic (PAN) images of the same scene. MS images have high spectral resolution but relatively low spatial resolution, while the single-band PAN images have reverse characteristics. Due to physical limitations of the satellite sensors \cite{9462797}, neither MS nor PAN images have the high resolution both in spatial and spectral domains. However, such a high-quality image is urgently required in practical to effectively facilitate the visual interpretation and other applications, e.g., object detection \cite{7326158}, land-cover classification \cite{5451172} and change detection \cite{7839934}, etc. Therefore, researchers resorted to pansharpening technology, a feasible solution to obtain high-resolution (HR) MS images via the fusion of low-resolution (LR) MS and corresponding HR PAN images. 

Pansharpening has been developed for nearly 40 years, and it can be mainly divided into four categories \cite{9082183}, i.e., component substitution (CS)-based, multiresolution analysis (MRA)-based, variational optimization (VO)-based and deep learning (DL)-based methods. 

CS-based methods, also known as spectral methods, are based on the transformation of MS images to project them into a space, where the spatial and  the spectral components are assumed to be separated to each other, then the spatial components of MS images are substituted by histogram-matched PAN images. Finally, the inverse projection is applied to obtain fusion results \cite{4305344,5523973,4389066,8693555}. Generally, the fusion results of CS-based methods have abundant spatial details, but they are easily affected by spectral distortions.

MRA-based methods rely on the extraction and injection of spatial information, thus are also referred to spatial methods. They extract high-frequency components from PAN images by multiresolution analysis tools \cite{1105917, 4305345, 1512408, vivone2019fast}, and inject them into the interpolated MS images to get HRMS images. In contrast to CS-based methods, these methods preserve spectral information well, but tend to suffer from some spatial degradations.

VO-based methods explore the relationships among PAN, LRMS and HRMS images, and an energy function is designed with various regularization terms like total-variation \cite{6542015}, sparse representation \cite{6849483}, and low-rank \cite{8167324} to obtain the HRMS images. Most of them design spatial and spectral correlation terms based on the similarity of spatial information between HRMS and PAN, and the consistence of spectral information between HRMS and MS \cite{ballester2006variational, 8953524}. In general, VO-based methods can obtain high-quality fusion images, but they have high computational complexity.

\begin{figure*}[h]
\footnotesize
  \centering
  \scriptsize
  \begin{tabular}{{c}{c}{c}}
  \hspace{-4mm}
  \includegraphics[width=2.2in,trim=0 150 0 150,clip]{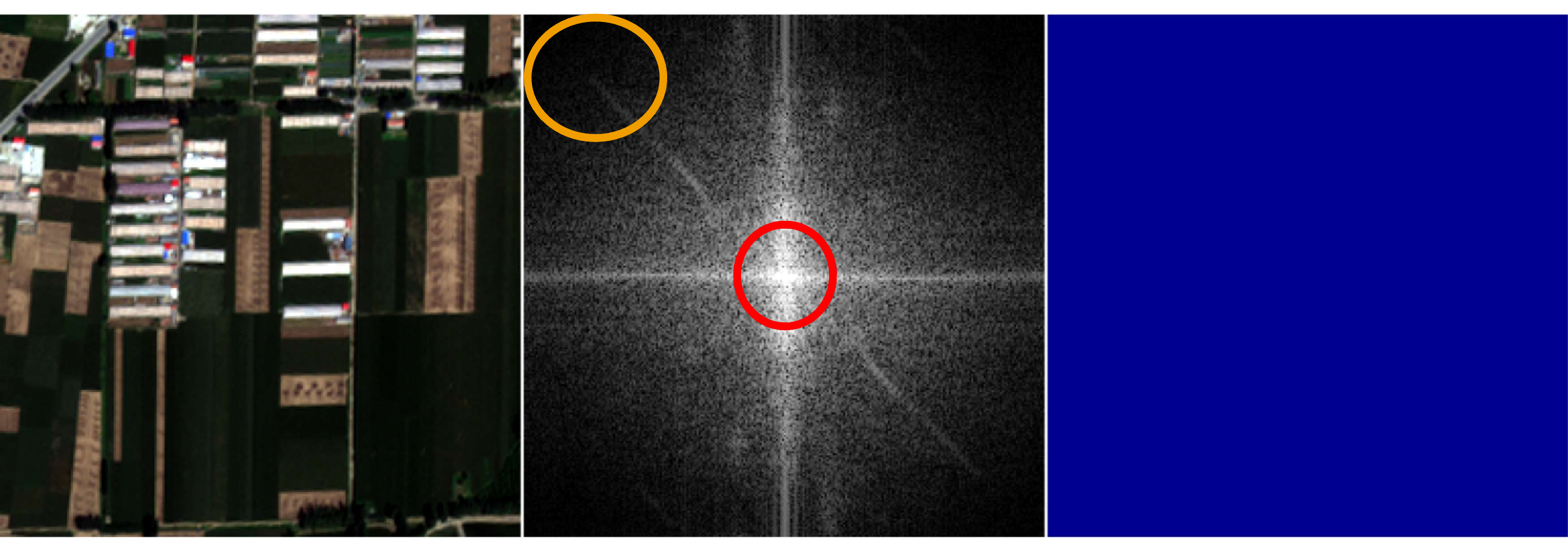} &
  \hspace{-4mm}
  \includegraphics[width=2.2in,trim=0 150 0 150,clip]{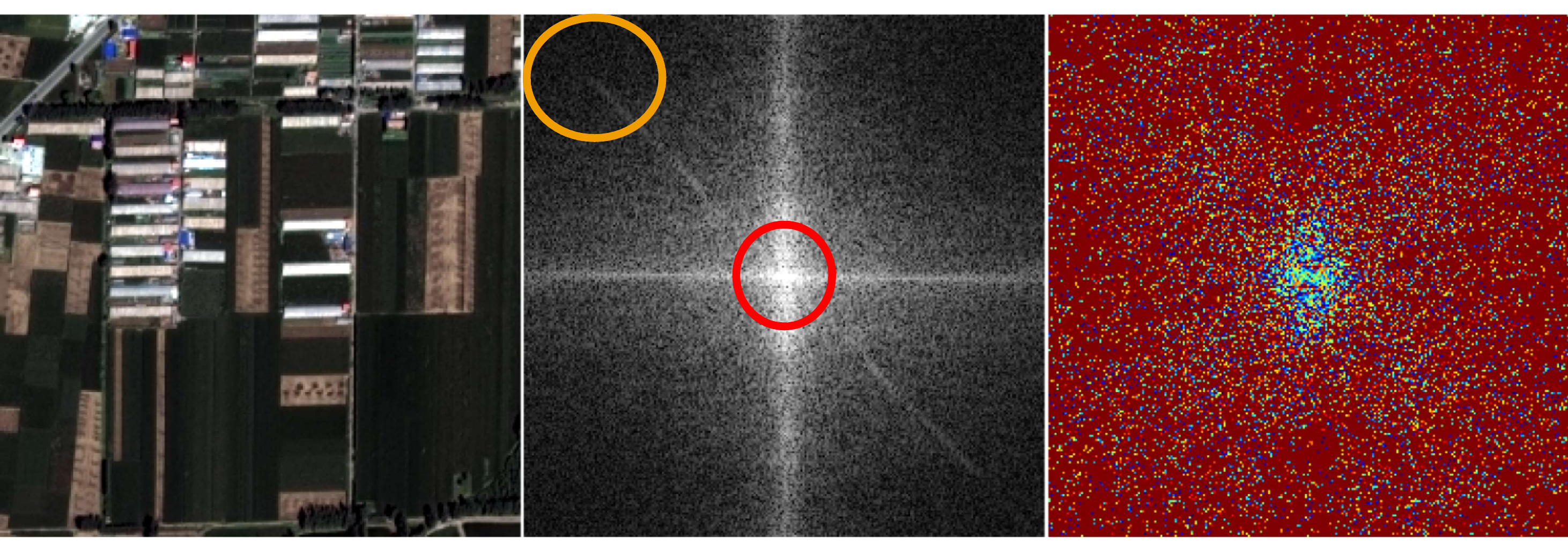} &
  \hspace{-4mm}
  \includegraphics[width=2.2in,trim=0 150 0 150]{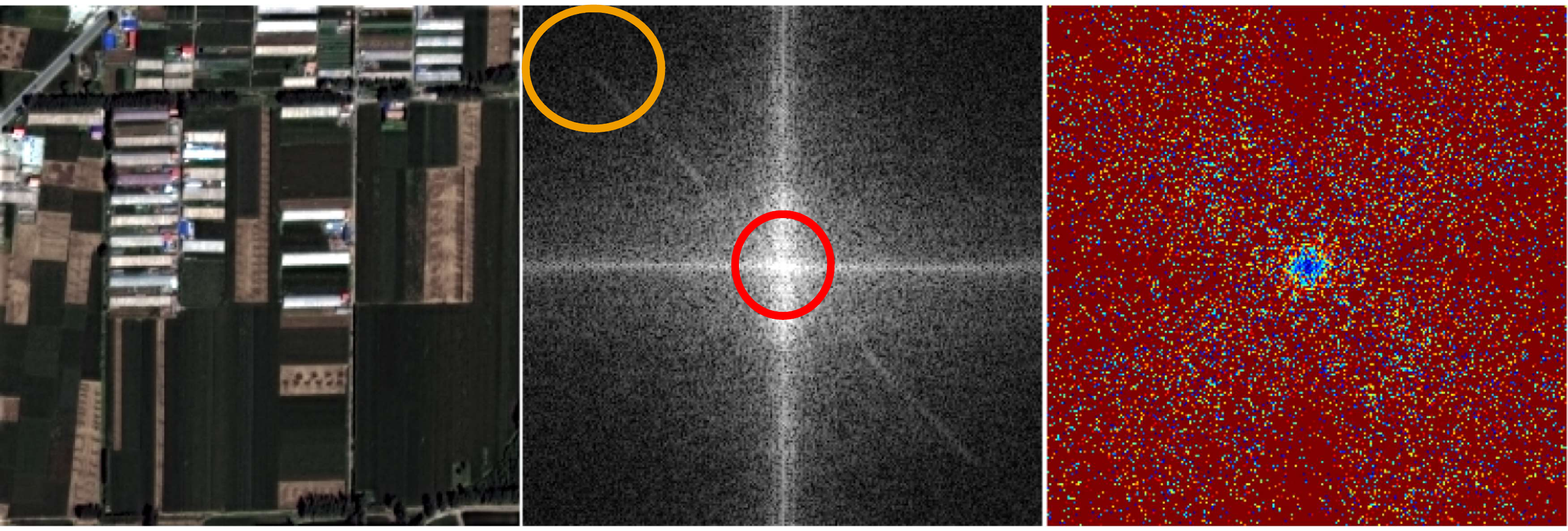} \\

     \hspace{-4mm}Reference &   \hspace{-4mm}GSA &   \hspace{-4mm}MTF-GLP-CBD \\
    
\hspace{-4mm}
  \includegraphics[width=2.2in,trim=0 150 0 150,clip]{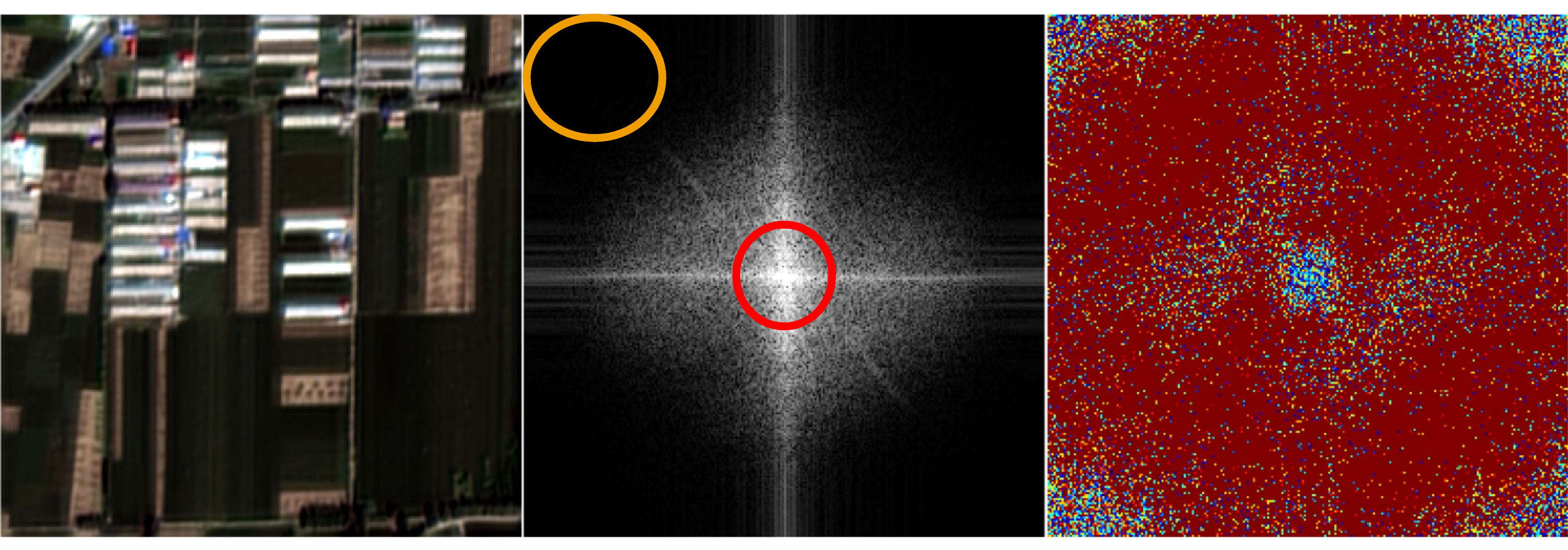} &
  \hspace{-4mm}
  \includegraphics[width=2.2in,trim=0 150 0 150,clip]{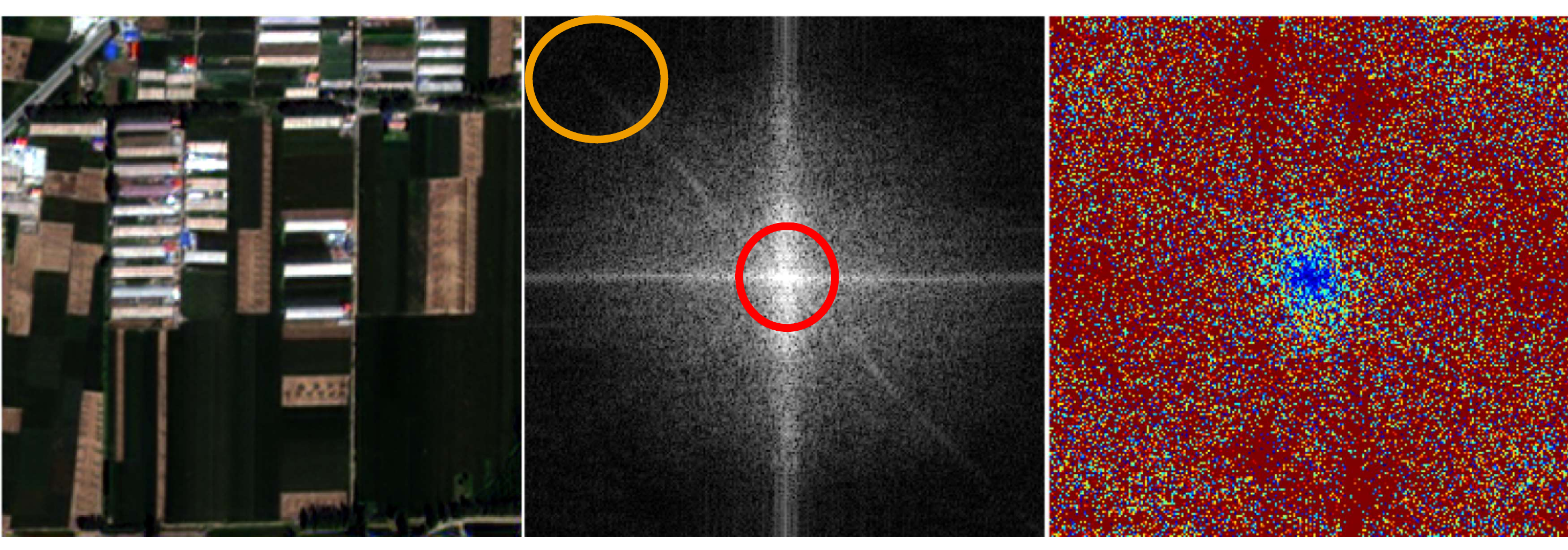} &
  \hspace{-4mm}
  \includegraphics[width=2.2in,trim=0 150 0 150,clip]{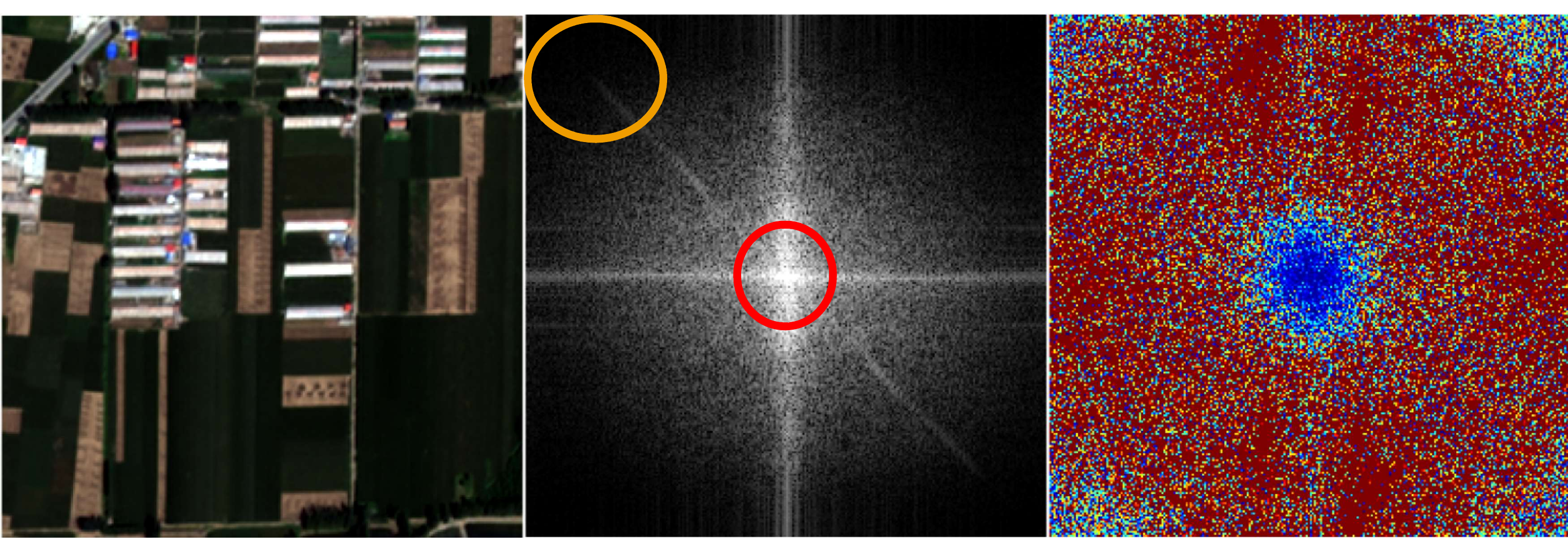} \\

      \hspace{-4mm}GTP-PNet &   \hspace{-4mm}FusionNet &   \hspace{-4mm}FAFNet \\
      
  ~ &\includegraphics[angle=-90,width=0.3\linewidth,trim=220 0 220 0,clip]{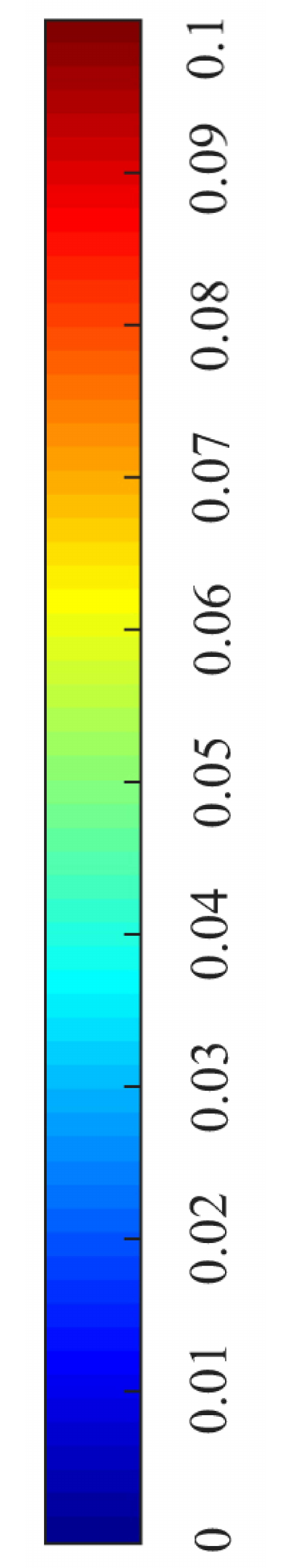}& ~  \\ 
  \end{tabular} \normalsize
  \caption{Visual comparisons between reference and fusion images of different pansharpening methods as well as their spectrum. Note that in order to clearly show the spectrum differences of them, we also calculate their absolute errors, and the absolute error maps are shown in the third column of each method. (One can zoom in for more details)}
  \label{fig_fft}
\end{figure*}

Recently, DL-based methods show superior performance due to the powerful ability of automatic nonlinear representation learning. Among them, convolutional neural networks (CNNs) based pansharpening methods have attracted more attention. The first CNN-based pansharpening neural network (PNN) \cite{masi2016pansharpening} adopted a simple three-layer convolutional structure \cite{7115171}, and obtained desirable fusion results. After that, many CNN-based pansharpening methods were proposed by incorporating some advanced feature extraction structures and fusion strategies. However, their results are easily affected by blurry effects because of the insufficient details learning. Actually, neural networks tend to fit low-frequency components with a high priority \cite{rahaman2019spectral, tancik2020fourier}, which results in the loss of some high-frequency edges and details. In addition, commonly used loss functions, such as mean squared error (MSE) and mean absolute error (MAE), are insensitive to small perturbations, which further leads to the loss of spatial details \cite{gal2021swagan}. Although many works utilized residual learning strategy to enforce networks to learn high-frequency components \cite{8237455, 8334206, ZHANG2021223, 9240949}, they still suffered from spatial and spectral distortions. As can be seen in Fig. \ref{fig_fft}, the spectrum of GSA \cite{4305344} and MTF-GLP-CBD \cite{4305345} results have some unexpected frequency components, so that they suffer from obvious spectral distortions, while that of GTP-PNet \cite{ZHANG2021223} result loses some specific frequency components, resulting in blurry effects and the loss of important boundaries of objects. The spectrum of FusionNet \cite{9240949} seems to be closer to the reference, while the frequency discrepancy also exists. We think the missing or redundant frequency components attribute to insufficient details learning and the frequency inconsistency between MS and PAN.

To deal with the problem of insufficient details learning, we propose a Frequency-Aware Fusion Network (FAFNet) together with a novel High-frequency Feature Similarity (HFS) loss for pansharpening. FAFNet can preserve the high-frequency details by explicitly learning the frequency-aware features with the assistance of DWT/IDWT layers, and also improve the frequency consistency by considering the HFS loss to align the features in the frequency domain. From the frequency spectrum maps and the absolute error maps shown in Fig.~\ref{fig_fft}, we can observe that FAFNet can alleviate the problem of insufficient details learning and at the same time preserve more spectral information.  The main contributions of this paper can be summarized as follows:

\begin{itemize}
\item{We propose a frequency-aware fusion network (FAFNet) to emphasize the details learning in pansharpening. In our work, the frequency features are extracted, aligned and fused automatically. It extracts and updates the frequency components explicitly, which greatly improves the performance of pansharpening.}
\item{A novel High-frequency Feature Similarity (HFS) loss is proposed to learn the correspondence between MS and PAN features, and to further restrict the high-frequency features of PAN images as equivalent as possible to the missing high-frequency parts of MS images. }
\item{DWT and IDWT layers are introduced into our network to make FAFNet work in the frequency domain, so that it can adaptively learn the frequency-aware features through the backward propagations.}
\end{itemize}

The remainder of the paper is outlined as follows. Section \uppercase\expandafter{\romannumeral2} introduces the related works. In Section \uppercase\expandafter{\romannumeral3}, we present the architecture of proposed method as well as the designed loss function in detail. The experimental results and analyses are reported in Section \uppercase\expandafter{\romannumeral4}. Finally, conclusions are drawn in Section \uppercase\expandafter{\romannumeral5}.

\section{Related Works}
%In this section, we first review several CNN-based pansharpening methods relevant to ours and then provide the motivations of proposed method.

%\subsection{CNN-Based Pansharpening Methods for details learning}
To solve the problem of insufficient details learning, many CNN-based methods resort to learning the high-frequency residuals through residual learning (RL) strategy. The use of RL strategy aims to make networks focus on the learning of high-frequency details and thus to address spatial degradation problem in fusion results. Pan-sharpening network (PanNet) \cite{8237455} was the first method that adopted RL strategy to learn the high-frequency components, and its fusion results preserved spatial details well. Similarly, a target-adaptive CNN (TA-PNN) model \cite{8334206} was proposed and the effectiveness of RL strategy was verified through experiments on four datasets. In \cite{ZHANG2021223}, the authors introduced the gradient transformation prior (GTP) into RL architecture, called GTP-PNet, which was composed of two networks, i.e., gradient transformation network (TNet) and pan-sharpening network (PNet). TNet tried to explore the nonlinear mapping between the gradients of PAN and HRMS images. Then PNet was optimized under the guidance of the trained TNet to learn the residuals. FusionNet \cite{9240949} took the differences between duplicated PAN and up-sampled MS images as inputs, and learned the spatial details through a structure-preservation network. These methods reconstructed spatial details by calculating the residuals between source images and target HRMS image.

Apart from residual learning (RL) strategy, some other works  adopted the progressive fusion strategy to compensate the spatial details. A Laplacian pyramid pan-sharpening network (LPPN) \cite{JIN2022158} was designed under the Laplacian pyramid framework and utilized the recursive structure to progressively fuse spatial information at different scales. The dynamic cross feature fusion network (DCFNet) \cite{wu2021dynamic} utilized a high-resolution branch to serve as the main-branch, and two parallel low-resolution branches are then used to progressively supplement information to the main-branch. 
%\textcolor{red}{However, because CNNs have inherent bias of tending to fit low-frequency information~\cite{rahaman2019spectral, tancik2020fourier},} they always suffer from the loss of high-frequency information. 

%\textcolor{red}{Aiming at directly training on full-resolution images instead of simulated reduced-resolution data to avoid the inherent information gap existing between them, unsupervised frameworks are one of promising direction to explore and they are often designing spatial loss term to preserve more spatial structure. For instance, unsupervised pansharpening GAN (UPanGAN) \cite{xu2023upangan} is based on not only a coarse-to-fine progressively fusion scheme learning difference image, but also well-designed loss functions where the difference of downsampled PAN and pseudo PAN are measured as the spatial loss. Zero-Reference GAN (ZeRGAN) \cite{9669094} relys on a multiscale generator architecture to enhance the spatial details, where all the discriminators focus on the consistency of spatial information in the pseudo PAN and the real PAN image. }

Recently, unsupervised pansharpening methods were proposed. They train models directly on images at original resolution to avoid the spatial degradation brought by scale variation. Xu \emph{et al.}~\cite{xu2023upangan} proposed an unsupervised pansharpening GAN (UPanGAN) model, which learned the difference image under a coarse-to-fine progressive fusion scheme, and measured the difference of downsampled PAN and pseudo-PAN by a well-designed spatial loss function. Zero-Reference GAN (ZeRGAN) \cite{9669094} utilized a multiscale generator architecture to enhance the spatial details, where all the discriminators focused on the consistency of spatial information between the pseudo-PAN and the real PAN image.

%However, few of them directly learn frequency-aware features, resulting in the loss of high-frequency information because of the inherent bias of CNNs that tend to fit the low-frequency information \cite{rahaman2019spectral, tancik2020fourier}. 
However, because CNNs have inherent bias of tending to fit low-frequency information~\cite{rahaman2019spectral, tancik2020fourier}, they always suffer from the loss of high-frequency information. In our model, we introduce 2D DWT and IDWT layers to explicitly learn the frequency-aware features, so as to preserve more high-frequency details in the fusion results.

%\subsection{Motivations}
%Recently, many computer vision tasks explicitly and adaptively extracted features in frequency domain for efficient feature learning. Wavelet integrated CNNs (WaveCNets) \cite{9156335} designed DWT and IDWT layers and integrated them into commonly used CNNs for noise-robust image classification, but it dealt with high-level vision tasks, and did not take the details preservation into consideration. Style and wavelet based generative adversarial network (SWAGAN) \cite{gal2021swagan} predicted wavelet coefficients at different scales to make the generated images have more HF details. However, it utilized normal discrete wavelet transforms which lacked the ability of backward propagations when dealing with downstream tasks. Inspired by these tasks, we propose a frequency-aware fusion network (FAFNet) that can adaptively learn the frequency components through backward propagations of not only convolutional layers but also DWT and IDWT layers. Furthermore, a novel high-frequency feature similarity loss is applied to high-frequency features of PAN and MS to improve the consistence between the high- and low-frequency components contained in fusion results.

\begin{figure*}[!t]
  \centering
  \includegraphics[angle=90,width=\linewidth, trim=80 0 80 0,clip] {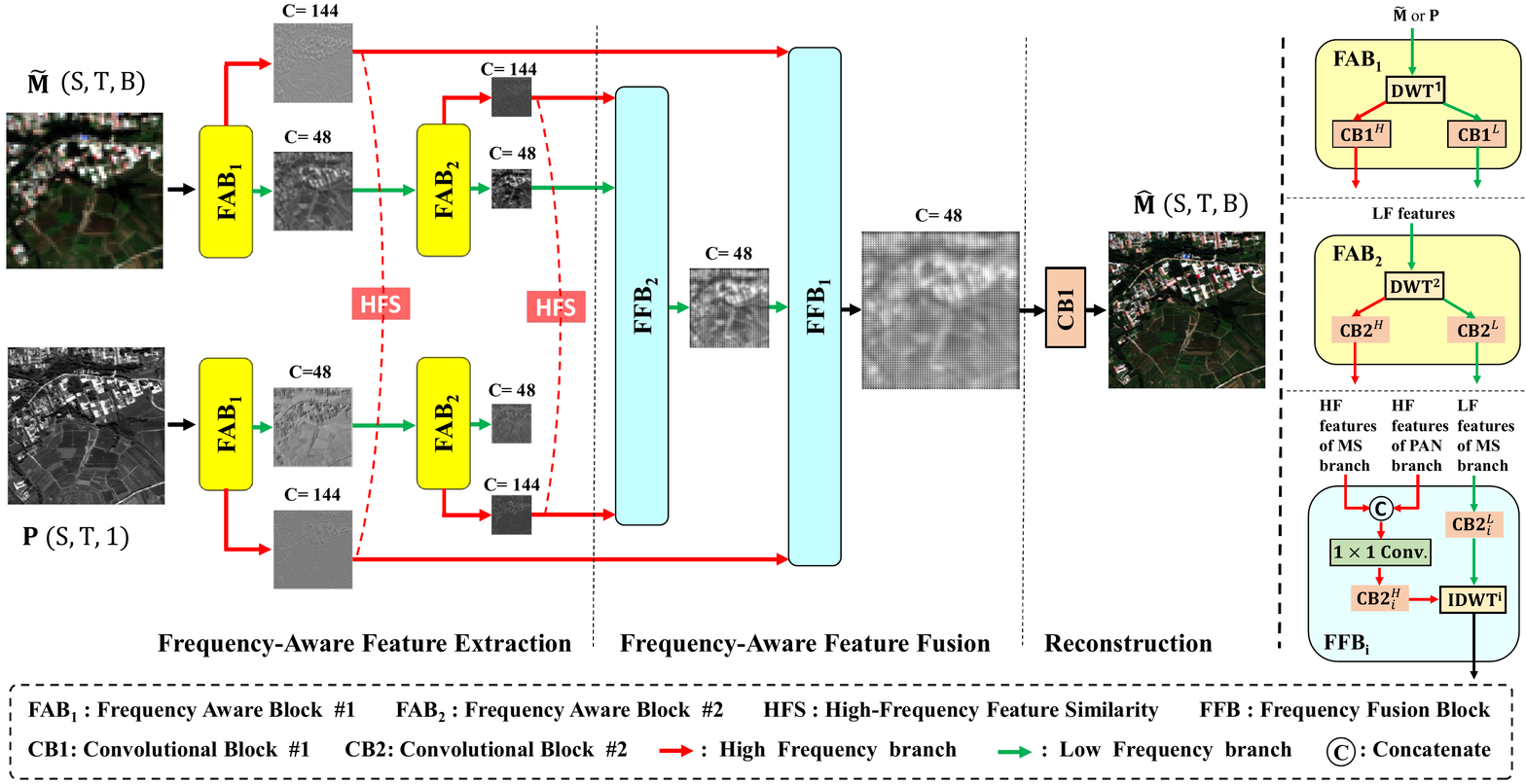}
  \caption{The architecture of proposed FAFNet, where \emph{L} and \emph{H} in the superscript represent low-frequency and high-frequency related modules. S and T denote the height and width of PAN, and B is the number of bands for MS image. C represents the number of feature channels.}
  \label{structure_FAN}
\end{figure*}

\section{Proposed Method}
\subsection{Notations}
MS and PAN images are represented as $\textbf{M}$ $\in$ $\mathbb{R} 
^ {\emph{S / r}\ \times\ \emph{T / r}\ \times\ \emph{B}}$ and
$\textbf{P}$ $\in$ $\mathbb{R} ^ {\emph{S}\ \times\ \emph{T}}$, where $\emph{S}$ and $\emph{T}$ are the height and width of PAN, and $\emph{B}$ denotes the number of MS image bands. $\emph{r}$ is the spatial resolution
ratio of MS to PAN. The pansharpened results and the reference images are represented by $\widehat{\textbf{M}}$
$\in$ $\mathbb{R} ^ {\emph{S}\ \times\ \emph{T}\ \times\ \emph{B}}$
and $\textbf{R}$ $\in$ $\mathbb{R} ^ {\emph{S}\ \times\ \emph{T}\
\times\ \emph{B}}$. In this paper, the up-sampled MS images obtained by bicubic interpolation is denoted by $\widetilde{\textbf{M}}$ $\in$ $\mathbb{R} ^
{\emph{S}\ \times\ \emph{T}\ \times\ \emph{B}}$.

\begin{figure*}[htbp]
\centering
  \begin{tabular}{{c}{c}}
 \hspace{-4mm}
\includegraphics[width=0.5\linewidth,trim=5 90 5 90,clip]{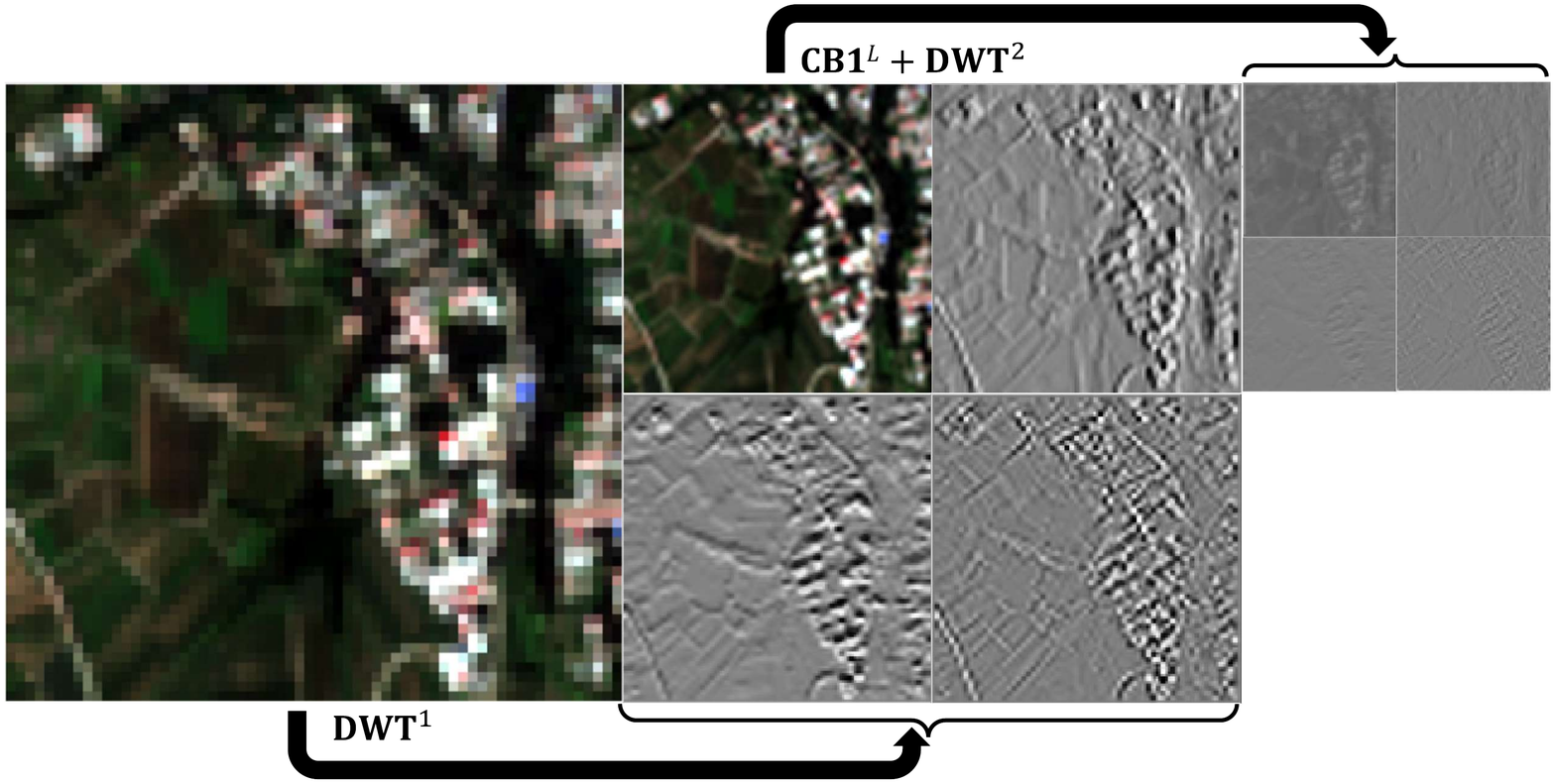}&
\hspace{-4mm}
\includegraphics[width=0.5\linewidth,trim=5 90 5 90,clip]{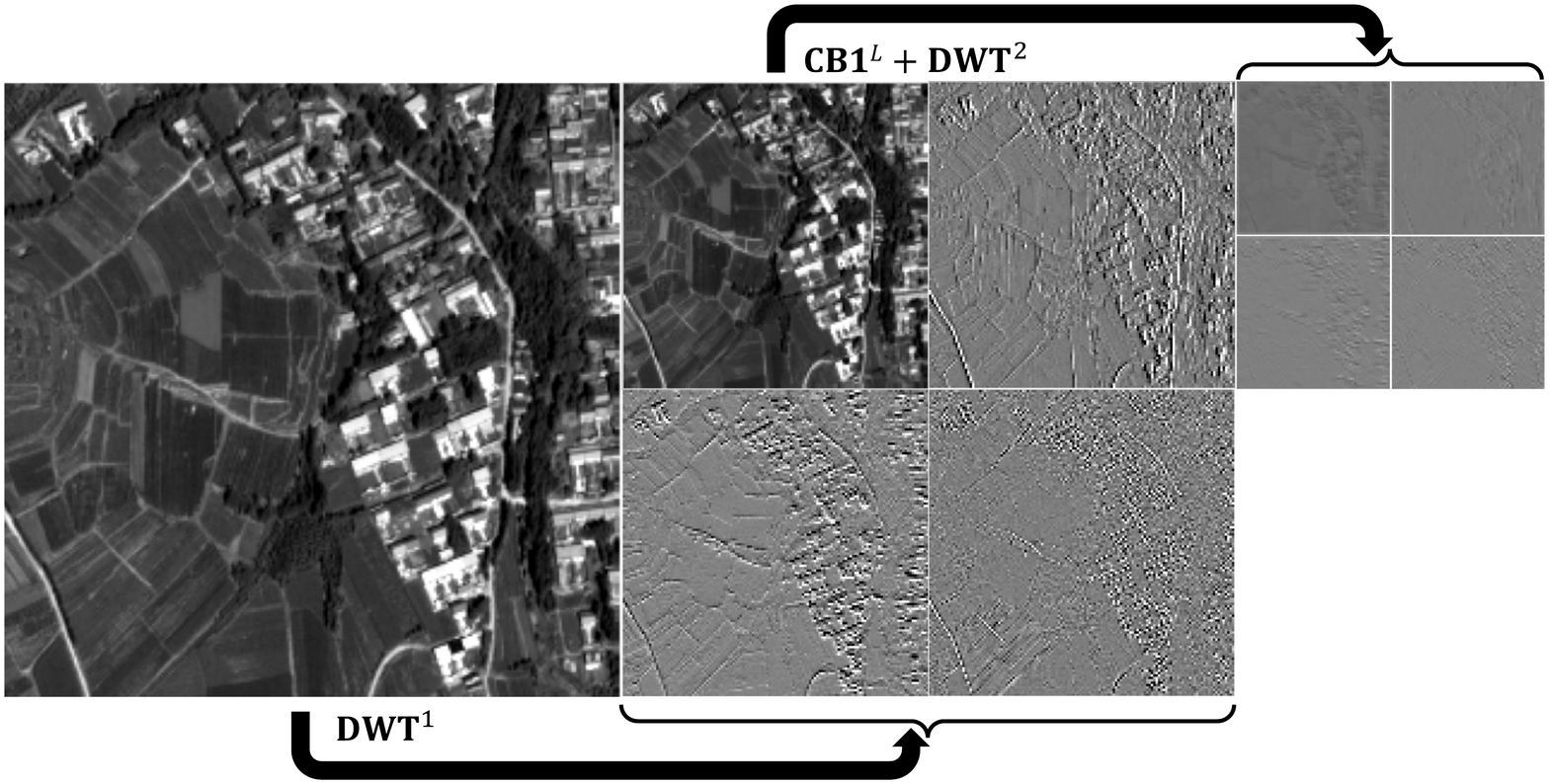}\\
\hspace{-4mm}(a) &   \hspace{-4mm}(b) \\
\end{tabular} \normalsize
\caption{Feature visualizations of (a) MS branch and (b) PAN branch, where first column shows the input image, and the second and the third columns demonstrate the decomposition results of $\textbf{DWT}^1$ and $\textbf{DWT}^2$.}
\label{fig_dwt}
\end{figure*}

\subsection{Network Overview}
The overall architecture of proposed frequency-aware fusion network (FAFNet) is shown in Fig. \ref{structure_FAN}. Following the paradigm of traditional MRA-based methods, FAFNet fuses the MS and PAN images in three steps: feature extraction, feature fusion, and reconstruction. Frequency-aware feature extraction module extracts frequency features of MS and PAN with the assistance of DWT layers, and these features are then fused in the frequency domain by frequency-aware feature fusion module with the corresponding IDWT layers. Finally, the fused features are reconstructed by the reconstruction module to obtain fusion results. To make the frequency features consistent with each other, we also designed a High-frequency Feature Similarity (HFS) loss to learn the correspondence of them. Although follows the similar paradigm as traditional MRA-based methods, the proposed FAFNet has two advantages: 1) FAFNet is based on a nonlinear model, and it can well represent the misalignment in the spectral range between MS and PAN sensors. 2) FAFNet learns the correspondence in the wavelet domain, which allows us to have smooth filters with fewer parameters, and the learnability provides us more flexibility. Next, we will introduce the architecture and operations of each module in detail. In the following, \emph{M} and \emph{P} in the subscript denote the modules in MS and PAN branches, respectively.   

\begin{figure}[htbp]
  \centering
  \includegraphics[width=0.8\linewidth, trim=35 85 25 85,clip] {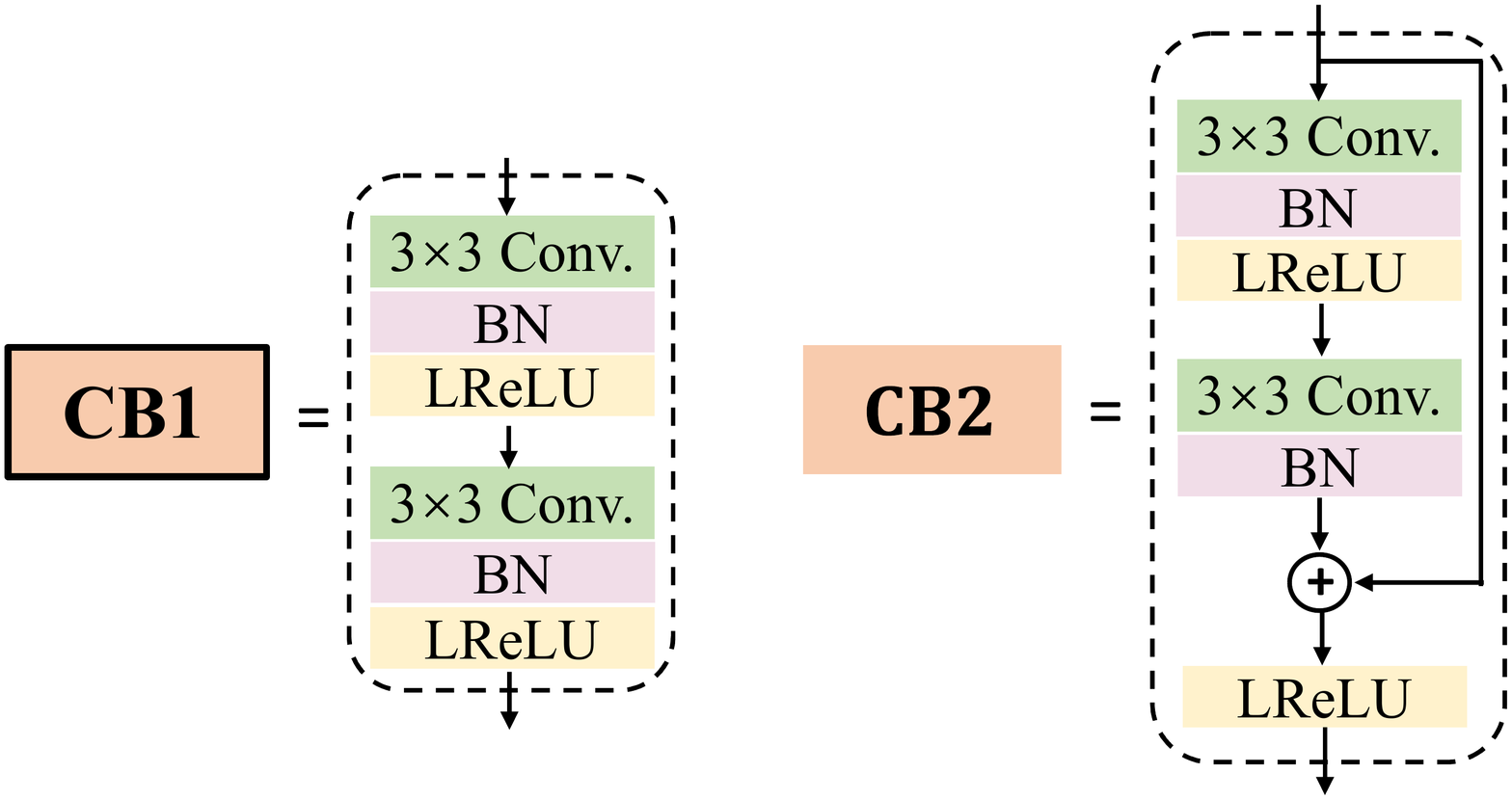}
  \caption{Structure of convolutional blocks (CBs) adopted in the FAFNet, where BN and LReLU stand for batch-normalization and leaky rectified linear unit layers, respectively.}
  \label{structure_CB}
\end{figure}

\subsection{Frequency-Aware Feature Extraction}
We first utilize the discrete wavelet transform (DWT) layer to transform the images into frequency domain. Discrete Wavelet Transform (DWT) is an efficient multi-resolution analysis tool in signal processing \cite{1996Wavelets}, which has the ability of anti-aliasing \cite{9156335}, and can also preserve more structural information in image processing \cite{2020not}. Motivated by the success of \cite{9156335} and the advantages of DWT, we resort to DWT layers to adaptively learn frequency components. Compared with normal DWT, DWT layers are compatible with convolutional layers and their gradients can be back-propagated, which not only make the network learn in the frequency domain, but also make it have the flexibility.

Our proposed frequency-aware feature extraction architecture is composed of a PAN branch and an MS branch, each of which consists of two frequency-aware blocks (FABs), where a DWT layer is coupled with two parallel convolutional blocks (CBs) to transform original images into frequency domain and then to learn frequency-aware features. In order to keep the consistence preliminarily, MS and PAN branches share the same architecture but they have different parameters.

For MS and PAN branches, they take interpolated MS image $\widetilde{\textbf{M}}$ $\in$ $\mathbb{R} ^
{\emph{S}\ \times\ \emph{T}\ \times\ \emph{B}}$ or PAN image $\textbf{P}$ $\in$ $\mathbb{R} ^ {\emph{S}\ \times\ \emph{T}}$ as input, and produce Low-Frequency (LF) and High-Frequency (HF) features via two cascaded FAB blocks.

Due to the fact that MS branch has the same structure as PAN branch, we take PAN branch as an example to introduce the detailed process of frequency-aware feature extraction. As can be seen from Fig. \ref{structure_FAN}, PAN image $\textbf{P}$ $\in$ $\mathbb{R} ^ {\emph{S}\ \times\ \emph{T}}$ is firstly fed into $\text{FAB}_1$, where the image is decomposed by the first DWT layer to obtain the LF components $\textbf{G}_{P}^{L(1)}$ and the HF components $\textbf{G}_{P}^{H(1)}$. Then, two parallel convolutional blocks $\textbf{CB}1^L$ and $\textbf{CB}1^H$, whose architecture can be seen in Fig. \ref{structure_CB}, are used to extract LF-aware features $\textbf{F}_{P}^{L(1)}$ and HF-aware features $\textbf{F}_{P}^{H(1)}$,
% $\textbf{G}_{P}^{L(1)}$  $\textbf{G}_{P}^{H(1)}$ $\textbf{CB}1_P^L$ $\textbf{CB}1_P^H$
\begin{equation}
\label{deqn_ex1}
\begin{aligned}
\textbf{G}_{P}^{L(1)}, \textbf{G}_{P}^{H(1)} &= \textbf{DWT}_{P}^{1}(\textbf{P})\\
\textbf{F}_{P}^{L(1)} &= \textbf{CB1}^{L}(\textbf{G}_{P}^{L(1)})\\
\textbf{F}_{P}^{H(1)} &= \textbf{CB1}^{H}(\textbf{G}_{P}^{H(1)}),
\end{aligned}
\end{equation}
where $\textbf{DWT}_P^1$ refers to the first DWT layer of the PAN branch.

Similar to traditional multiscale wavelet decomposition, another DWT layer $\textbf{DWT}_{P}^{2}$ is  applied to the LF component $\textbf{F}_{P}^{L(1)}$ to further conduct the finer decomposition. Two convolutional blocks $\textbf{CB}2^L$ and $\textbf{CB}2^H$ are adopted in the same way to obtain the second level of LF- and HF-aware features, i.e., $\textbf{F}_{P}^{L(2)}$ and  $\textbf{F}_{P}^{H(2)}$,

\begin{equation}
\label{deqn_ex2}
\textbf{F}_{P}^{x(2)} = \textbf{CB2}^x (\textbf{DWT}_{P}^{2}(\textbf{F}_{P}^{L(1)})),
\end{equation}
where $x \in \{L, H\}$.

Fig. \ref{fig_dwt} shows the visualization of decomposed features through DWT layers. It can be seen that with the assistance of two levels of DWT layers, LF and HF components can be separately processed by convolutional layers. We can also observe from Fig.~\ref{fig_dwt} that PAN features have finer details and they can complement corresponding MS features.

\subsection{Frequency-Aware Feature Fusion} 
After frequency-aware feature extraction, we have four PAN-related features at two scales, i.e., $\textbf{F}_{P}^{L(1)}, \textbf{F}_{P}^{H(1)}, \textbf{F}_{P}^{L(2)}, \textbf{F}_{P}^{H(2)}$, and four corresponding MS-related features $\textbf{F}_{M}^{L(1)}$, $\textbf{F}_{M}^{H(1)}$, $\textbf{F}_{M}^{L(2)}$, $\textbf{F}_{M}^{H(2)}$. Then we should consider the way of combining them to obtain the fused features. In traditional MRA-based methods, LF components are obtained directly from MS image, and HF components are acquired from both PAN and MS images, as shown in Eq. \eqref{deqn_ex3},

\begin{equation}
\label{deqn_ex3}
\begin{aligned}
\widehat{\textbf{M}}_{b} &=\widetilde{\textbf{M}}_{b}+g_{b}\left(\textbf{P}-\textbf{P}^{L}\right) \\
&=\widetilde{\textbf{M}}_{b}^{L}+\widetilde{\textbf{M}}_{b}^{H}+g_{b} \left(\textbf{P}-\textbf{P}^{L}\right)\\
&=\widetilde{\textbf{M}}_{b}^{L}+(\widetilde{\textbf{M}}_{b}^{H}+g_{b} \textbf{P}^{H}),
\end{aligned}
\end{equation}
where $b$ indexes the $b$th bands of MS and $g_{b}$ denotes injection coefficient with respect to $b$th bands.

In order to avoid introducing unexpected frequency components, in this paper, we propose a frequency fusion block (FFB) to fuse above-mentioned features. Note that FFB plays the role of inverse process of FAB, therefore, it mainly consists of two corresponding convolutional blocks and an IDWT layer. Similar to Eq. \eqref{deqn_ex3}, the fused HF components $\textbf{F}_{(i)}^{H}$ of $i$-th level come from both the MS and PAN features. Specifically, the HF features of MS and PAN at the same scale are first concatenated along the channel dimension, then the concatenated features are transformed by a $1\times1$ convolutional layer and a HF-aware convolutional block $\textbf{CB2}_{i}^{H}$, that is,
\begin{equation}
\label{deqn_ex4}
\textbf{F}_{(i)}^{H} = \textbf{CB2}_{i}^{H}(\textbf{Conv}(\textbf{Concat}(\textbf{F}_{P}^{H(i)}, \textbf{F}_{M}^{H(i)}))), i=1, 2,
\end{equation}
where $\textbf{Concat}(\cdot,\cdot)$ denotes the operation that concatenates features along the channel dimension and $\textbf{Conv}(\cdot)$ stands for the $1\times1$ convolutional operation.

The LF components of IDWT layers are obtained only from the MS branch. The LF features of MS firstly pass through a convolutional block $\textbf{CB2}_{i}^{L}$ ($i=1,2$), and then are taken as the LF components of the IDWT layer,
\begin{equation}
\label{deqn_ex5}
\textbf{F}_{(i)}^{L} = \textbf{CB2}_{2}^{L}(\textbf{F}_{M}^{L(i)}).
\end{equation}
%Similarly, the LF components of the second layer is
%\begin{equation}
%\label{deqn_ex6}
%\textbf{F}_{(1)}^{L} = \textbf{CB2}_{1}^{L}(\textbf{F}^{2}).
%\end{equation}

The inverse wavelet transform through the IDWT layers can be formulated as:
\begin{equation}
\label{deqn_ex7}
\textbf{F}^{i} = \textbf{IDWT}^{i}(\textbf{F}_{(i)}^{L}, \textbf{F}_{(i)}^{H}),   i=1,2,
\end{equation}
where $\textbf{F}^{1}$ is the final fused features.

\subsection{Reconstruction Module} 
The reconstruction module takes the fused features as inputs and transforms them from the feature space to the original image space. It is composed of a convolutional block $\textbf{CB1}$, which converts $\textbf{F}^1$ back to the image space to obtain the high-quality fused result $\widehat{\textbf{M}}$,
\begin{equation}
\label{deqn_ex8}
\widehat{\textbf{M}} = \textbf{CB1}(\textbf{F}^1).
\end{equation}

It should be noted that the filter size of each convolutional layer in our model is 3 $\times$ 3 whose padding size is 1, except for the 1 $\times$ 1 convolutional layer with zero-padding. The activation functions used in FAFNet except for the last layer are Leaky Rectified Linear Unit (LReLU) with $\alpha$ = 0.2, and that of the last layer is tanh. 

\subsection{High-Frequency Feature Similarity Loss}
As mentioned before, HF components come from both the MS and PAN branches, while LF components are only obtained from MS branch. Therefore, there is no guarantee that the HF and LF components of fused features can completely be consistent with each other, leading to spectral distortion. To address this problem, we design a High-frequency Feature Similarity (HFS) loss, aiming to constrain the correspondence between HF-aware features of PAN and MS, so as to align the HF-aware features of PAN to that of MS.

\begin{figure}[htbp]
  \centering
\includegraphics[angle=90,width=0.5\linewidth, trim=50 0 50 0,clip] {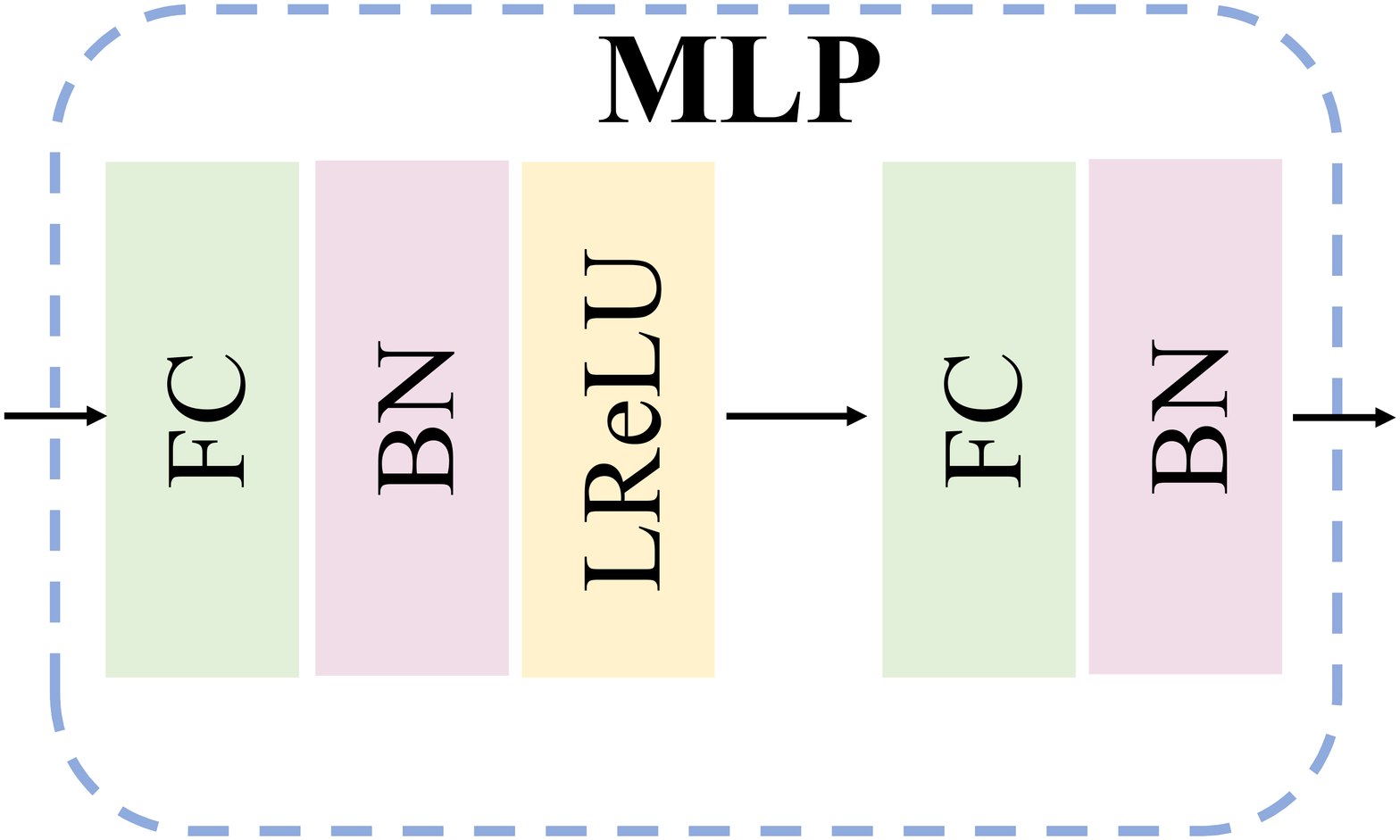}
  \caption{Structure of multi-layer perceptron (MLP) adopted in the high-frequency feature similarity (HFS) loss, where FC indicates the fully-connected layer.}
  \label{structure_MLP}
\end{figure}

In order to compute the HFS loss, the HF-aware features of MS and PAN branches are firstly transformed by a multi-layer perceptron (MLP) as Fig. \ref{structure_MLP} shows. Then we calculate the cross-correlation matrix of them to represent the correlations,
\begin{equation}
\label{deqn_ex9}
 \textbf{C} = CC(\textbf{MLP}(R(\textbf{F}_{1})), \textbf{MLP}(R(\textbf{F}_{2}))),
\end{equation}
where $\textbf{F}_{1}$ and $\textbf{F}_{2}$ represent two fourth-order tensor features that will be aligned, and $R(\cdot)$ is the operator that reshapes a fourth-order tensor feature with the shape of ${\emph{D}_{1}\ \times\ \emph{D}_{2}\ \times\ \emph{D}_{3}\ \times\ \emph{D}_{4}}$ to a matrix with the shape of ${\emph{D}_{1}\ \times\ \emph{D}_{2}\emph{D}_{3}\emph{D}_{4}}$. $\textbf{MLP}(\cdot)$ represents the operations of MLP structure, and $CC(\cdot,\cdot)$ denotes the calculation of cross-correlation matrix $\textbf{C}$. 

There are two reasons why the HF-aware features are transformed by the MLP structure. Firstly, we use MLP structure to reduce the dimension of HF-aware features to facilitate the computation of cross-correlation matrix. The second reason is that MLP structure can further aggregate the non-local information of features, bringing in more discriminative feature representations.

The diagonal elements in $\textbf{C}$ denote the correlations of two representations from corresponding branches, therefore, the HFS loss is formulated as,
\begin{equation}
\label{deqn_ex10}
 \mathcal{L}_\text{HFS}= \frac {1}{\emph{N}}\sum_{i}^{\emph{N}}(1-\text{c}_{i i})^{2}+\lambda\frac {1}{\emph{N(N-1)}} \sum_{i}^{\emph{N}} \sum_{j \neq i}^{\emph{N-1}} \text{c}_{i j}^{2},
\end{equation}
where $\text{c}_{i i}$ and $\text{c}_{i j}$ denote the  diagonal and off-diagonal elements of cross-correlation matrix $\textbf{C} \in \mathbb{R}^{N\times N}$. $\lambda$ is the balance parameter and is set to 5 $\times$ 10$^{-3}$. The diagonal elements of $\textbf{C}$ are constrained to be 1, which means two representations from the same group of PAN and MS should be as close as possible. At the same time, off-diagonal ones of $\textbf{C}$ are constrained to be 0, which aims to decorrelate two representations from different groups to produce instance-specific representations.

Because our FAFNet model has two DWT layers, we can compute two HFS losses $\mathcal{L}_\text{HFS}^{1}$ and $\mathcal{L}_\text{HFS}^{2}$ with $\textbf{F}_{P}^{H(i)}$ and $\textbf{F}_{M}^{H(i)}$.
%replace $\textbf{F}_{1}$ and $\textbf{F}_{2}$ in Eq. \ref{deqn_ex9}, that is
\begin{equation}
\label{deqn_ex11}
 \mathcal{L}_\text{HFS}^{i}= \mathcal{L}_\text{HFS}(\textbf{F}_{P}^{H(i)},\textbf{F}_{M}^{H(i)}), i=1,2.
\end{equation}

\subsection{Overall Loss Function}
The HFS losses are combined with the mean absolute error (MAE) loss to constitute the overall loss function, 
\begin{equation}
 \label{deqn_ex12}
 \mathcal{L}_\text{MAE} = (1/\emph{B})\sum _{b=1}^{\emph{B}}||\textbf{R}_{b} - \widehat{\textbf{M}}_{b} ||_{1}
\end{equation}

\begin{equation}
 \label{deqn_ex13}
 \mathcal{L}= \mathcal{L}_\text{MAE}+\beta( \mathcal{L}_\text{HFS }^{1} + \mathcal{L}_\text{HFS}^{2}).
\end{equation}
where $\mathcal{L}_\text{MAE}$ is the MAE loss and $\beta$ is the balance parameter.

\begin{table*}[htbp]
\caption{Details of Used Datasets.\label{tab_dataset}}
\centering
\renewcommand{\arraystretch}{1.5}
\begin{tabular}{cccccccc}
\toprule
Sensor & Spatial resolution & \multicolumn{2}{c}{Training data} & \multicolumn{2}{c}{Validation data} & \multicolumn{2}{c}{Test data}\\		
\cline{3-8}
~ & ~ & Size & Number & Size & Number & Size & Number\\	
\midrule
QuickBird (QB) & 2.4m MS & 64 $\times$ 64 $\times$ 4 MS & 6943 & 64 $\times$ 64 $\times$ 4 MS & 743 & 256 $\times$ 256 $\times$ 4 MS & 156\\
~  & 0.6m PAN & 256 $\times$ 256 PAN & ~ & 256 $\times$ 256 PAN & ~ & 1024 $\times$ 1024 PAN & ~\\
\midrule
WorldView-4 (WV-4) & 1.2m MS & 64 $\times$ 64 $\times$ 4 MS & 7938 & 64 $\times$ 64 $\times$ 4 MS & 772 & 256 $\times$ 256 $\times$ 4 MS & 271\\
~ & 0.3m PAN & 256 $\times$ 256 PAN & ~ & 256 $\times$ 256 PAN & ~ & 1024 $\times$ 1024 PAN & ~\\
\midrule
WorldView-2 (WV-2) & 2.0m MS & 64 $\times$ 64 $\times$ 8 MS & 9641 & 64 $\times$ 64 $\times$ 8 MS & 945 & 256 $\times$ 256 $\times$ 8 MS & 136\\
~ & 0.5m PAN & 256 $\times$ 256 PAN & ~ & 256 $\times$ 256 PAN & ~ & 1024 $\times$ 1024 PAN & ~\\
\bottomrule
\end{tabular}
\end{table*}

\begin{figure*}[htbp]
\centering
\subfloat[]{\includegraphics[width=1.1in,trim=120 0 120 0,clip]{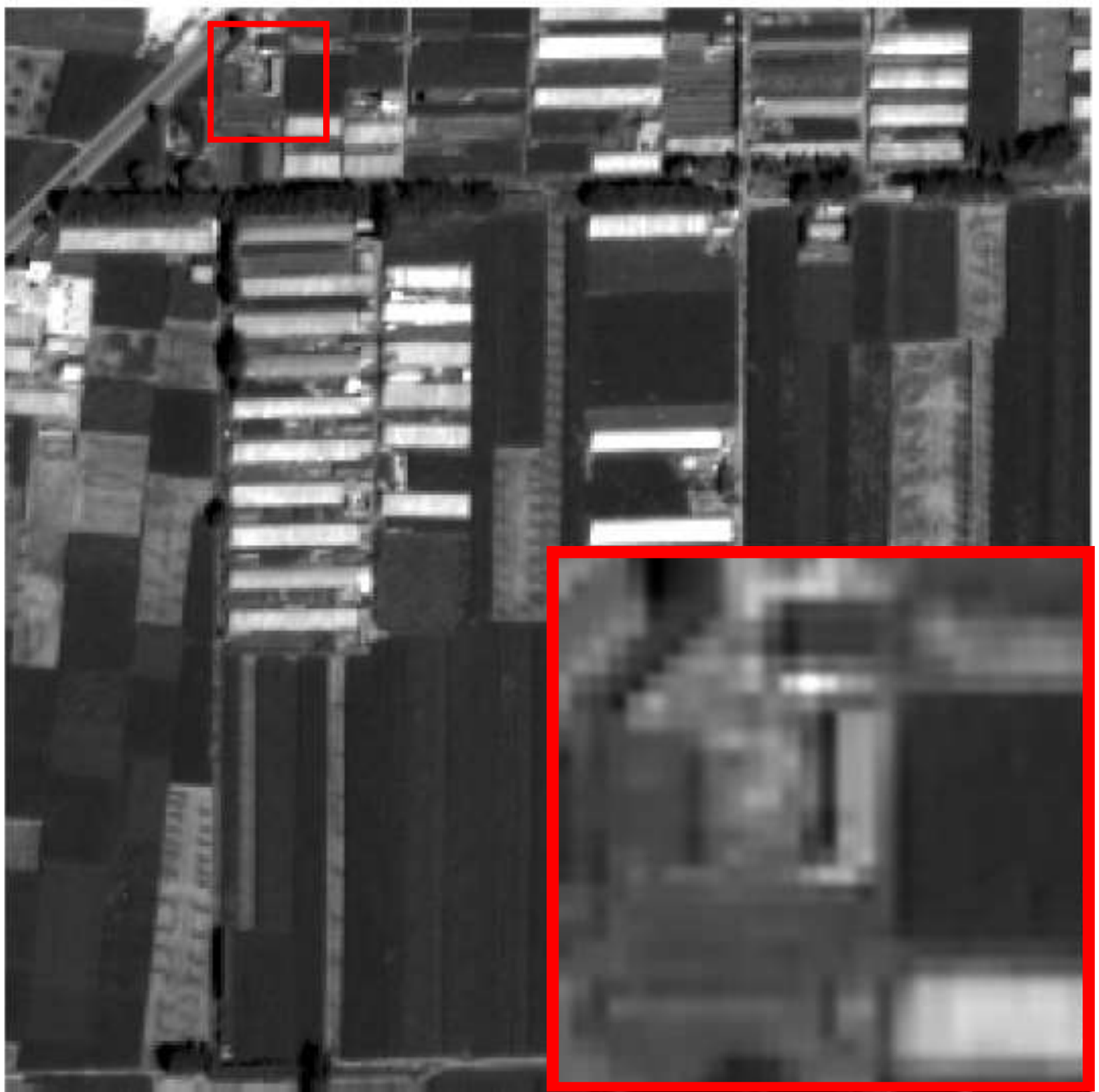}
\label{PAN}}
\subfloat[]{\includegraphics[width=1.1in,trim=120 0 120 0,clip]{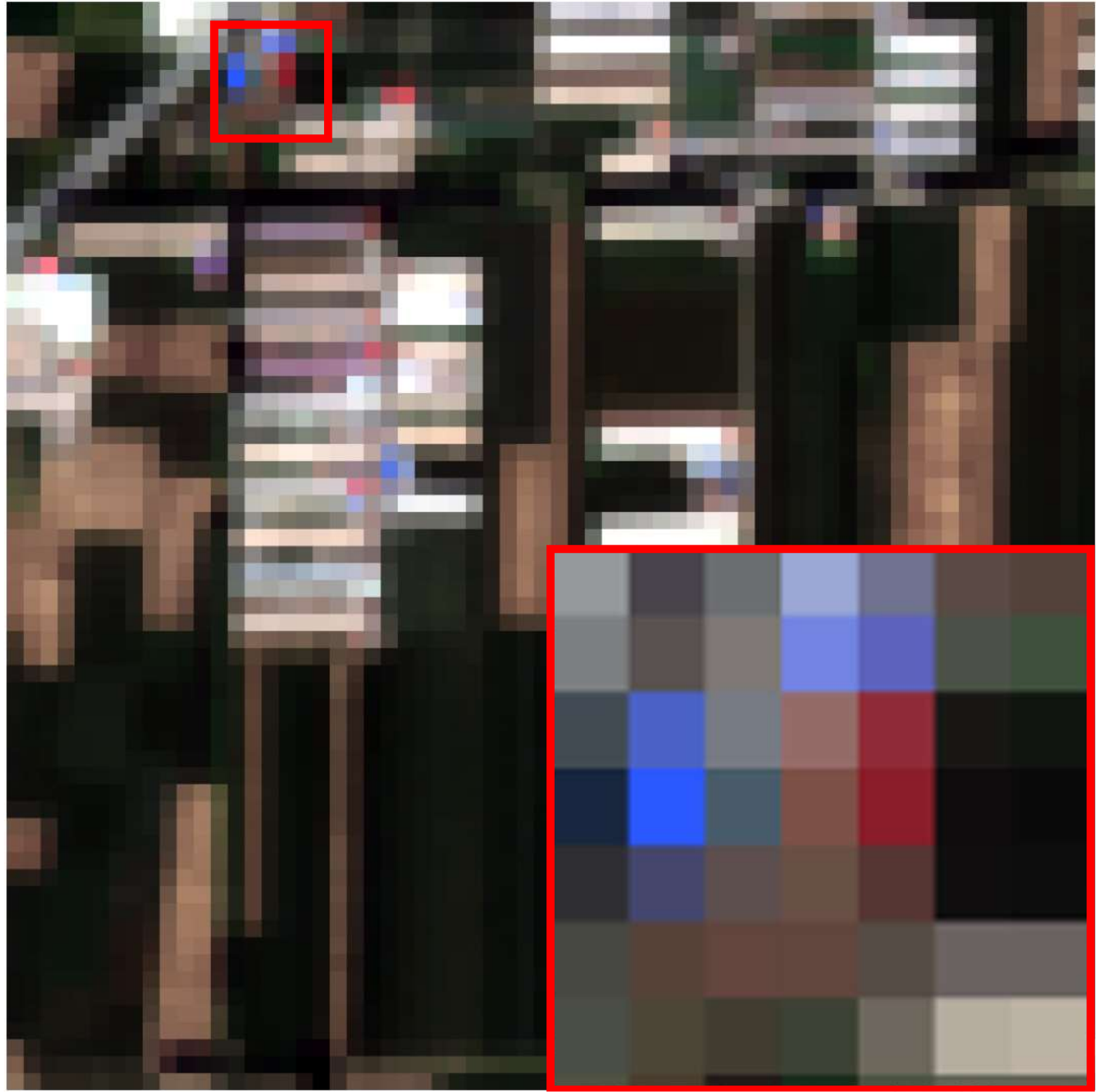}
\label{up-sampled MS}}
\subfloat[]{\includegraphics[width=1.1in,trim=120 0 120 0,clip]{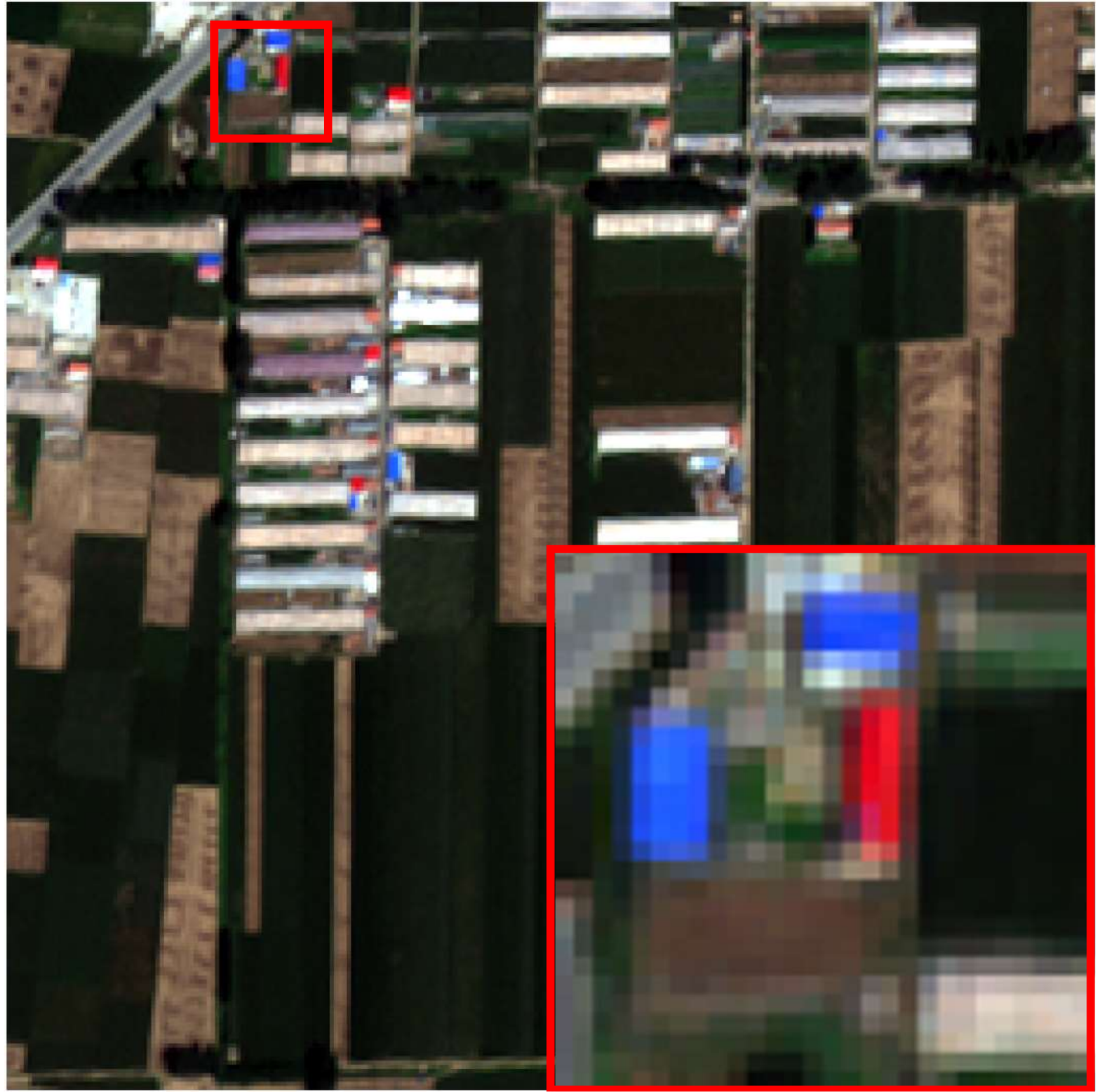}
\label{Reference}}
\subfloat[]{\includegraphics[width=1.1in,trim=120 0 120 0,clip]{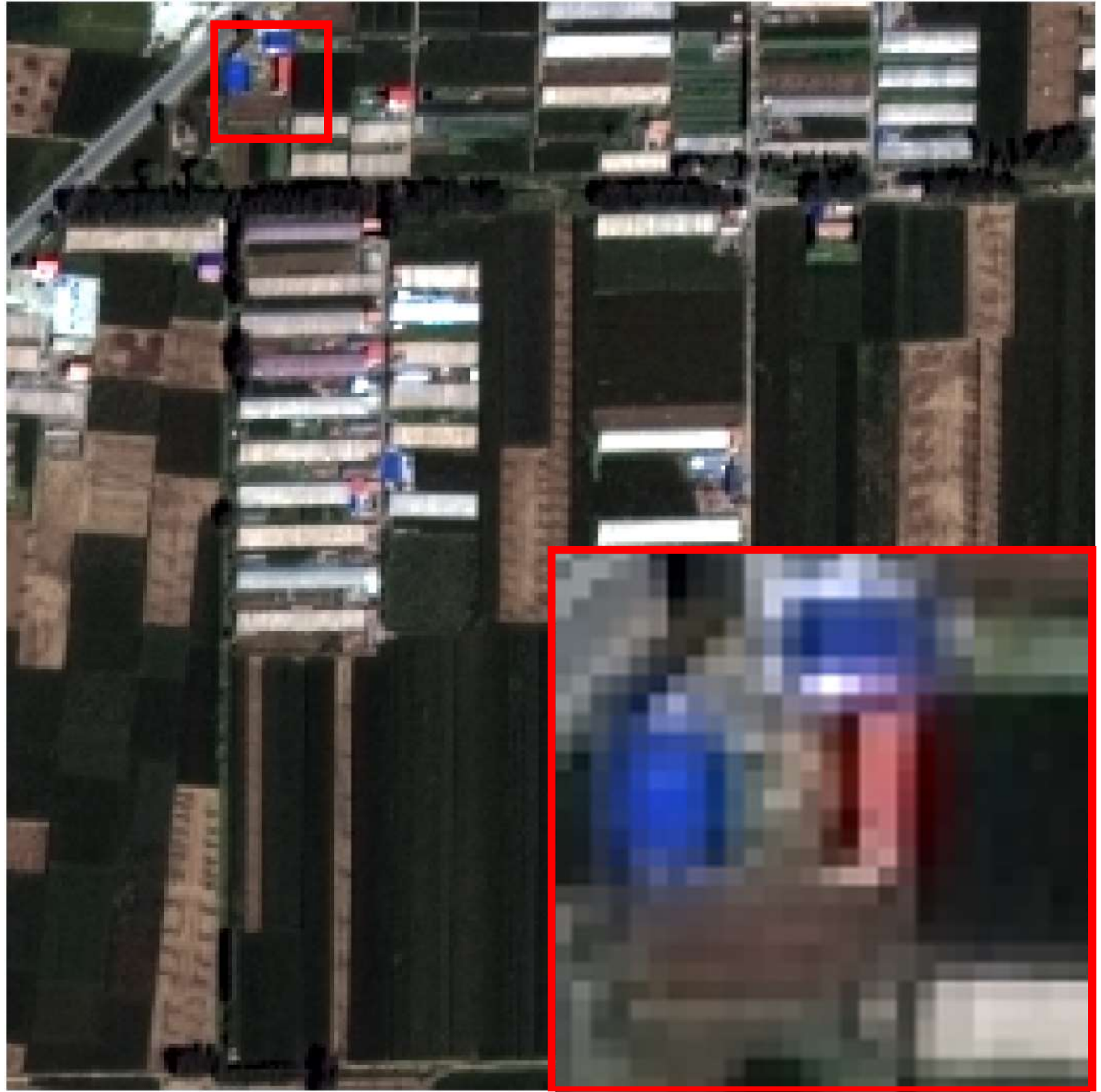}
\label{GSA}}
\subfloat[]{\includegraphics[width=1.1in,trim=120 0 120 0,clip]{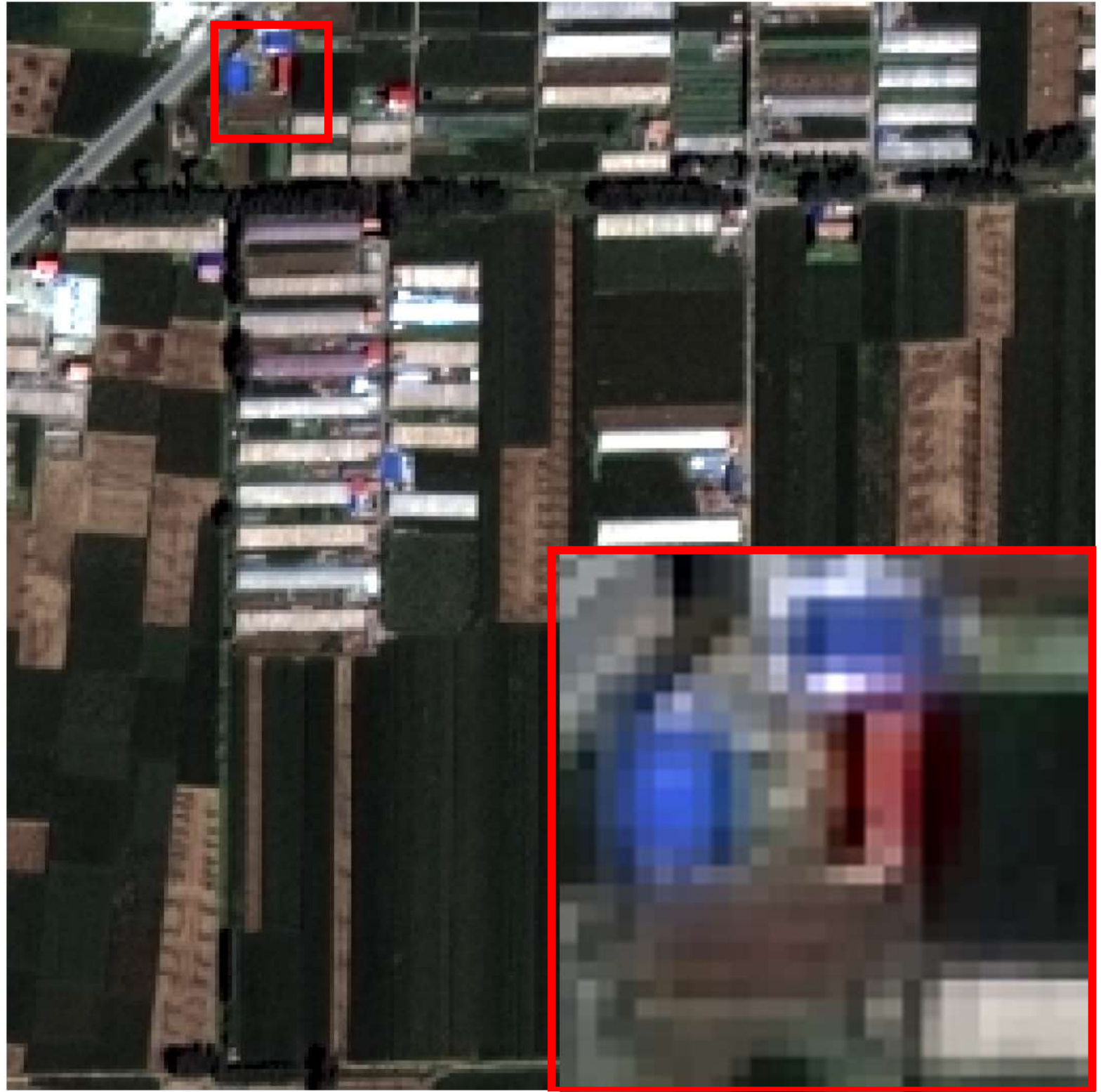}
\label{BDSD-PC}}

\vspace{-0.1in}
\subfloat[]{\includegraphics[width=1.1in,trim=120 0 120 0,clip]{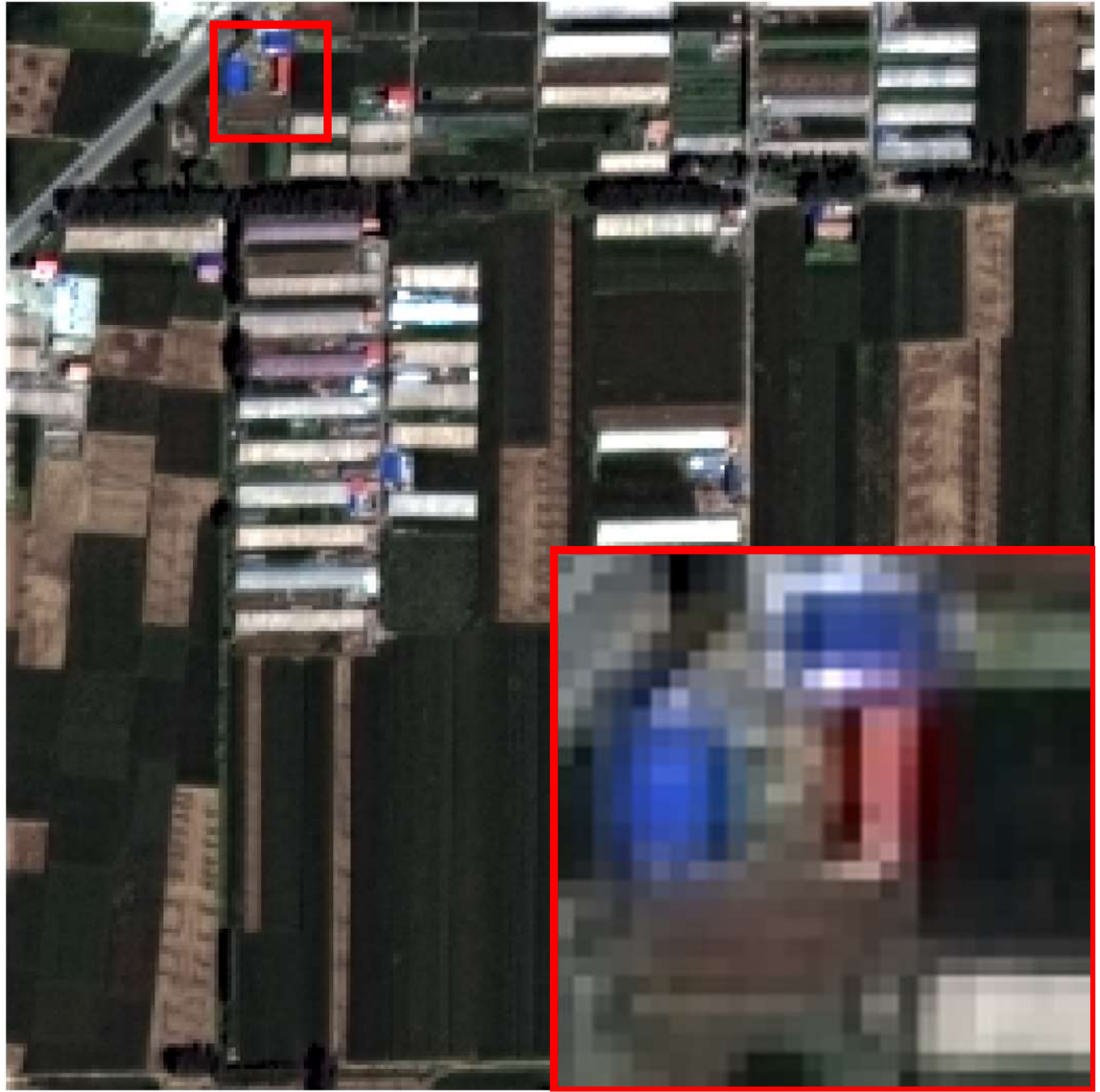}
\label{MTF-GLP-CBD}}
\subfloat[]{\includegraphics[width=1.1in,trim=120 0 120 0,clip]{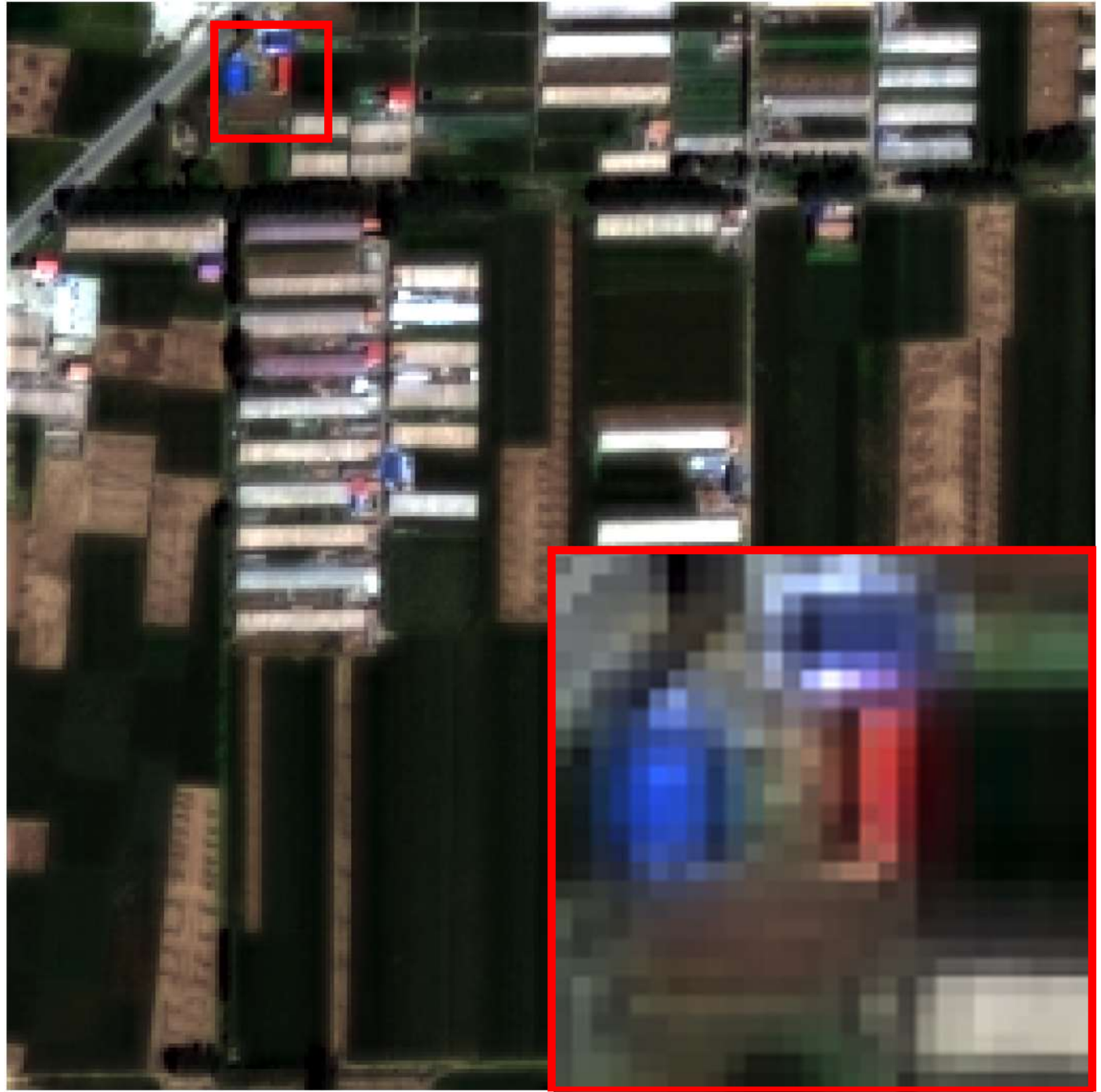}
\label{AWLP-H}}
\subfloat[]{\includegraphics[width=1.1in,trim=120 0 120 0,clip]{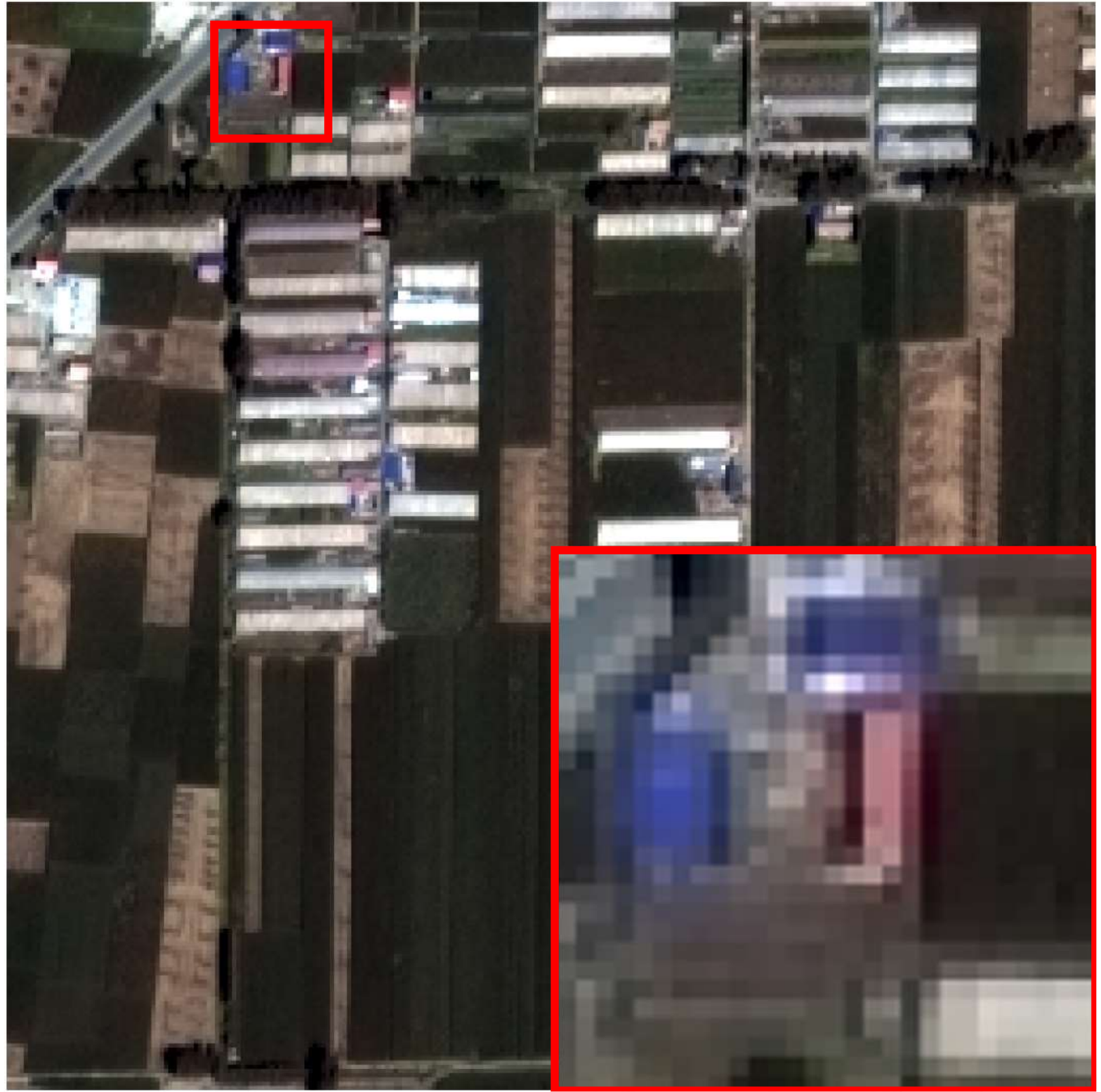}
\label{PWMBF}}
\subfloat[]{\includegraphics[width=1.1in,trim=120 0 120 0,clip]{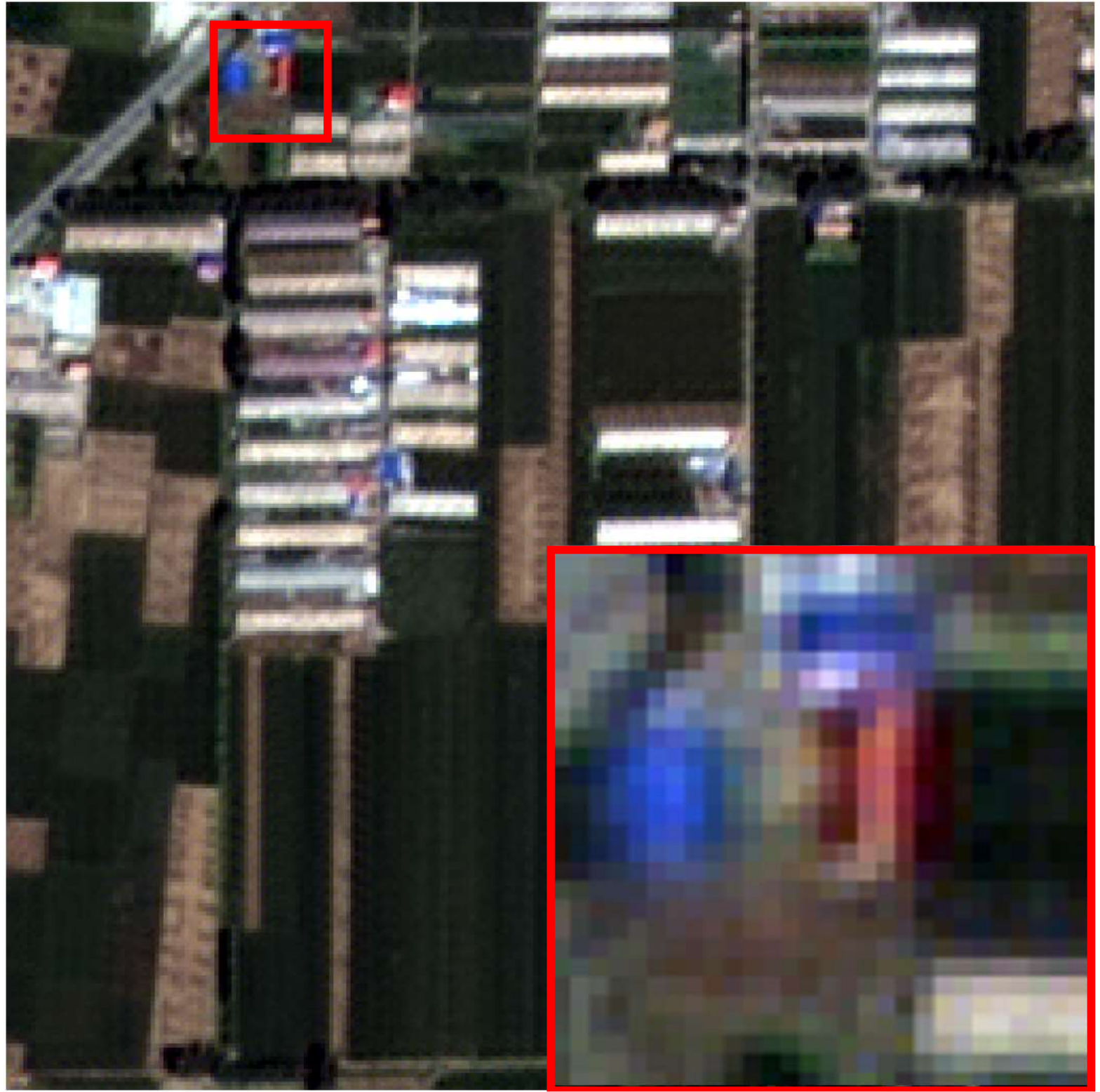}
\label{PanNet}}
\subfloat[]{\includegraphics[width=1.1in,trim=120 0 120 0,clip]{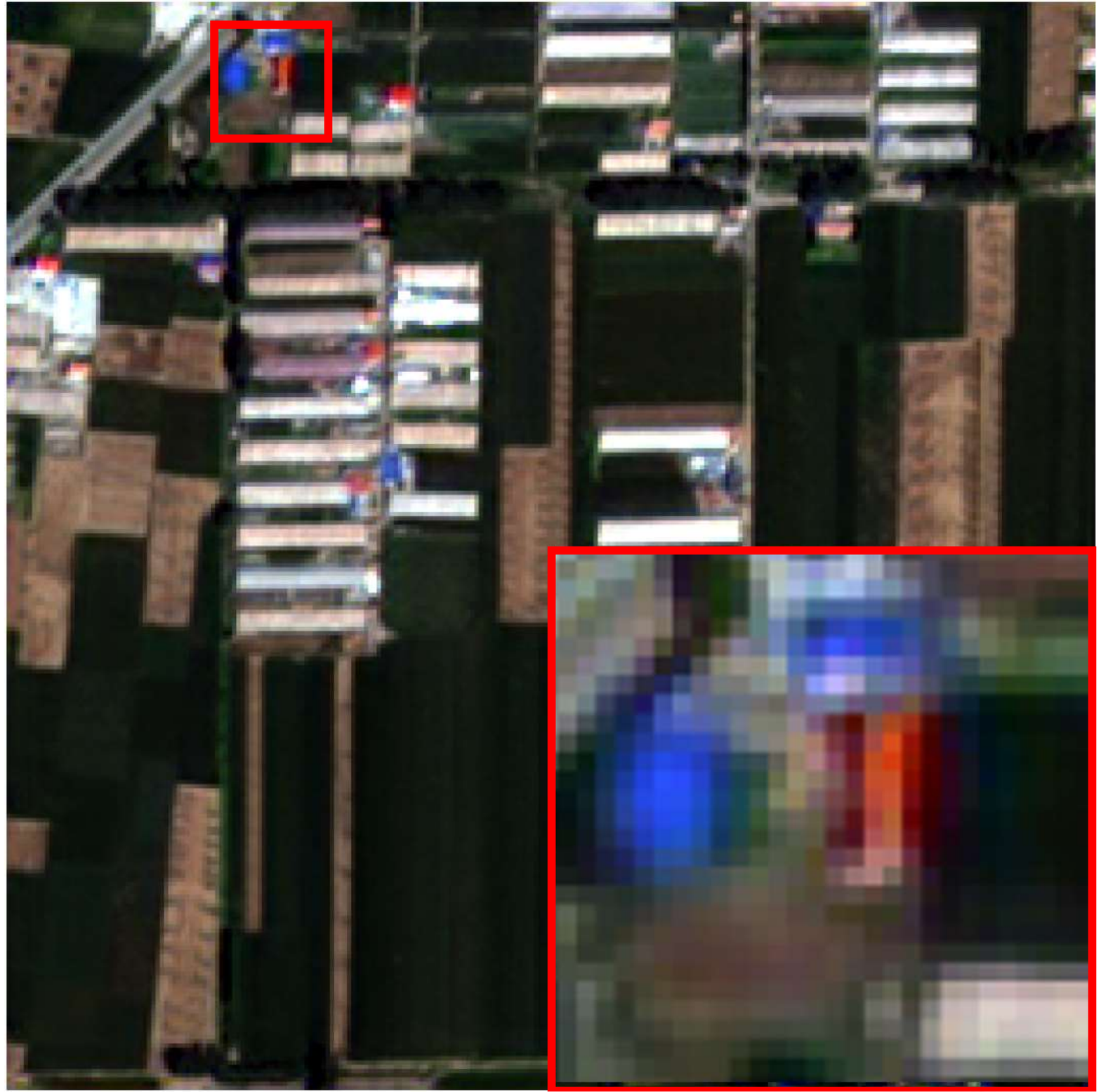}
\label{FusionNet}}

\vspace{-0.1in}
\subfloat[]{\includegraphics[width=1.1in,trim=120 0 120 0,clip]{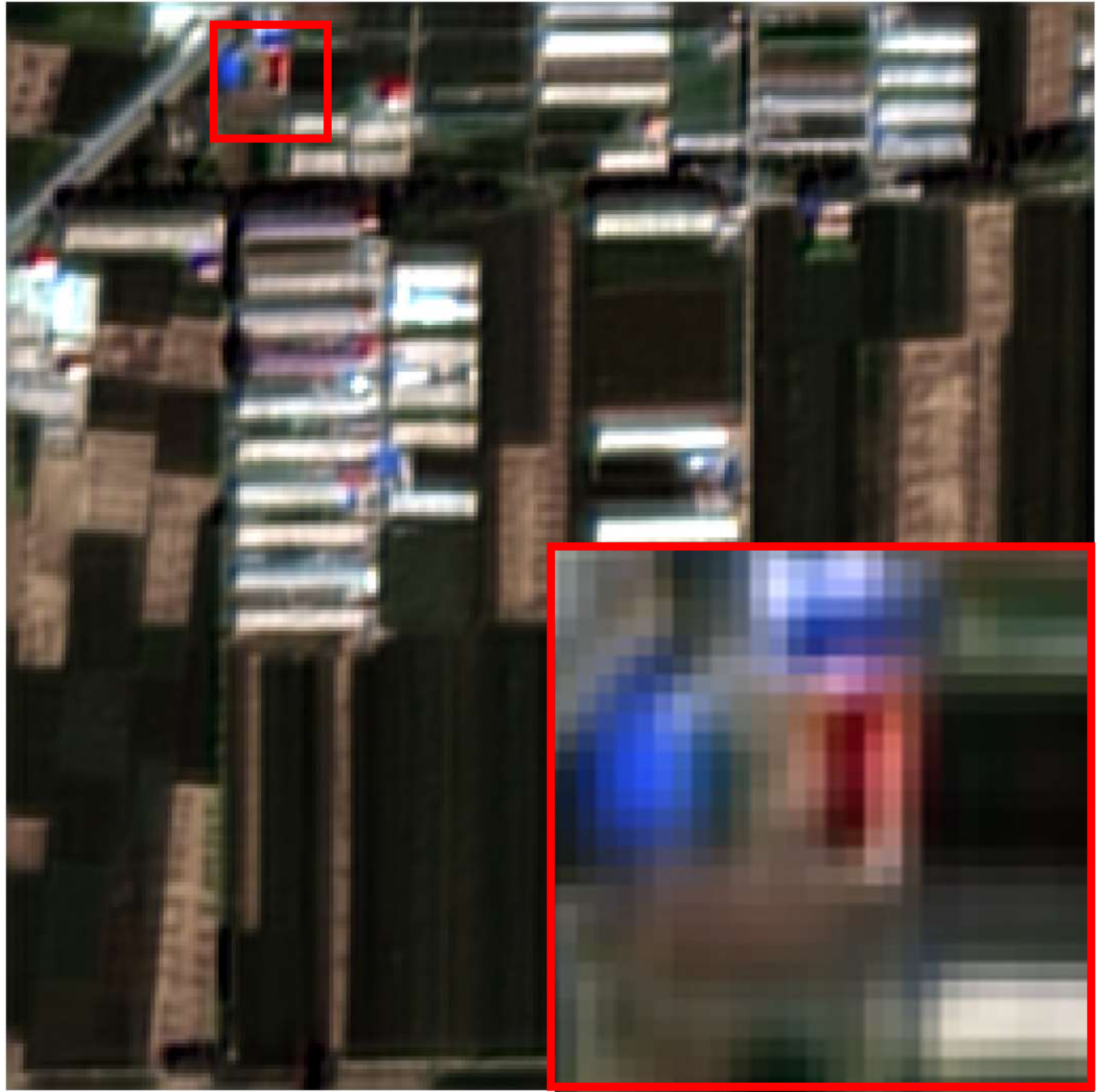}
\label{GTP-PNet}}
\subfloat[]{\includegraphics[width=1.1in,trim=120 0 120 0,clip]{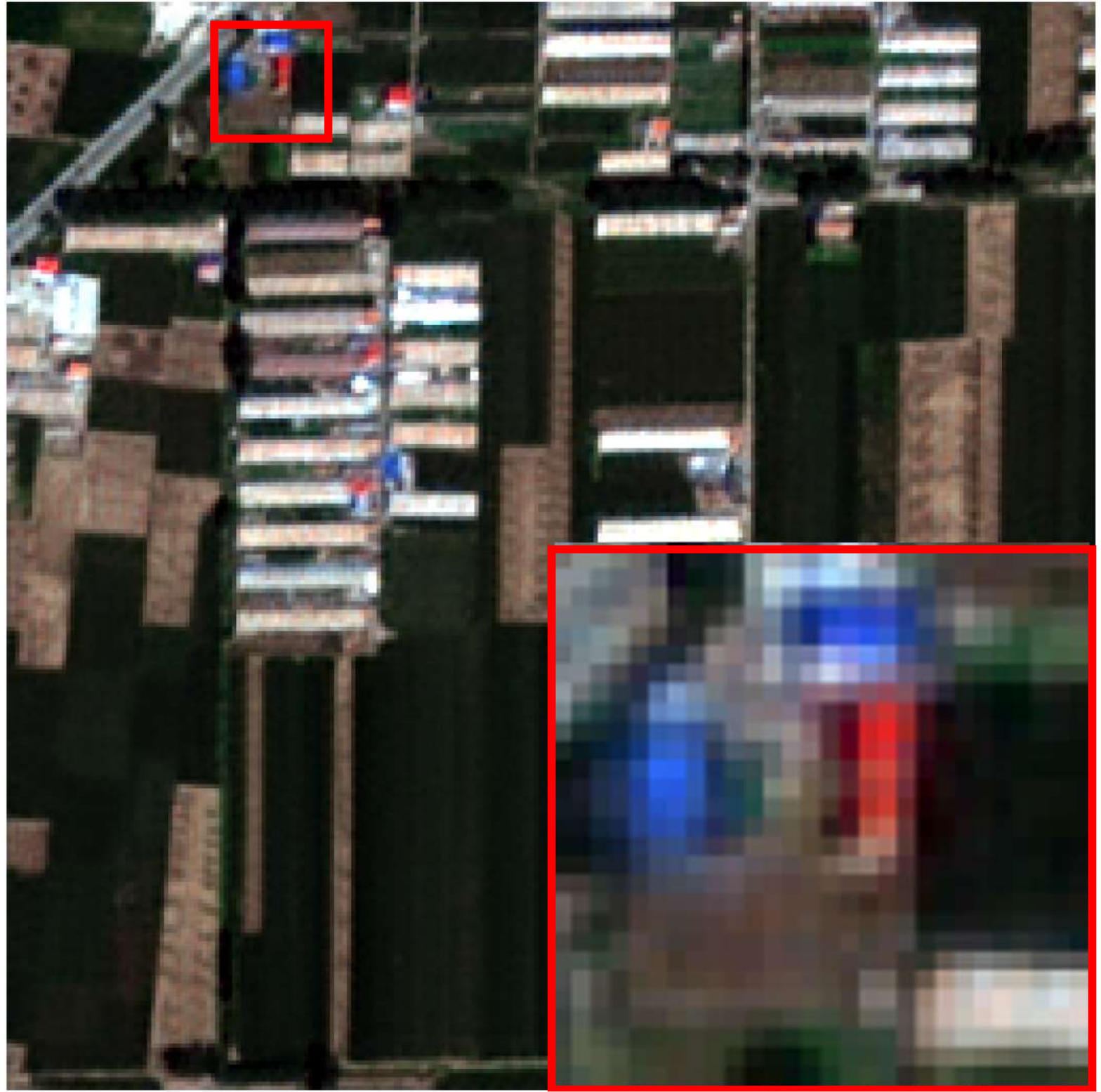}
\label{LPPN}}
\subfloat[]{\includegraphics[width=1.1in,trim=120 0 120 0,clip]{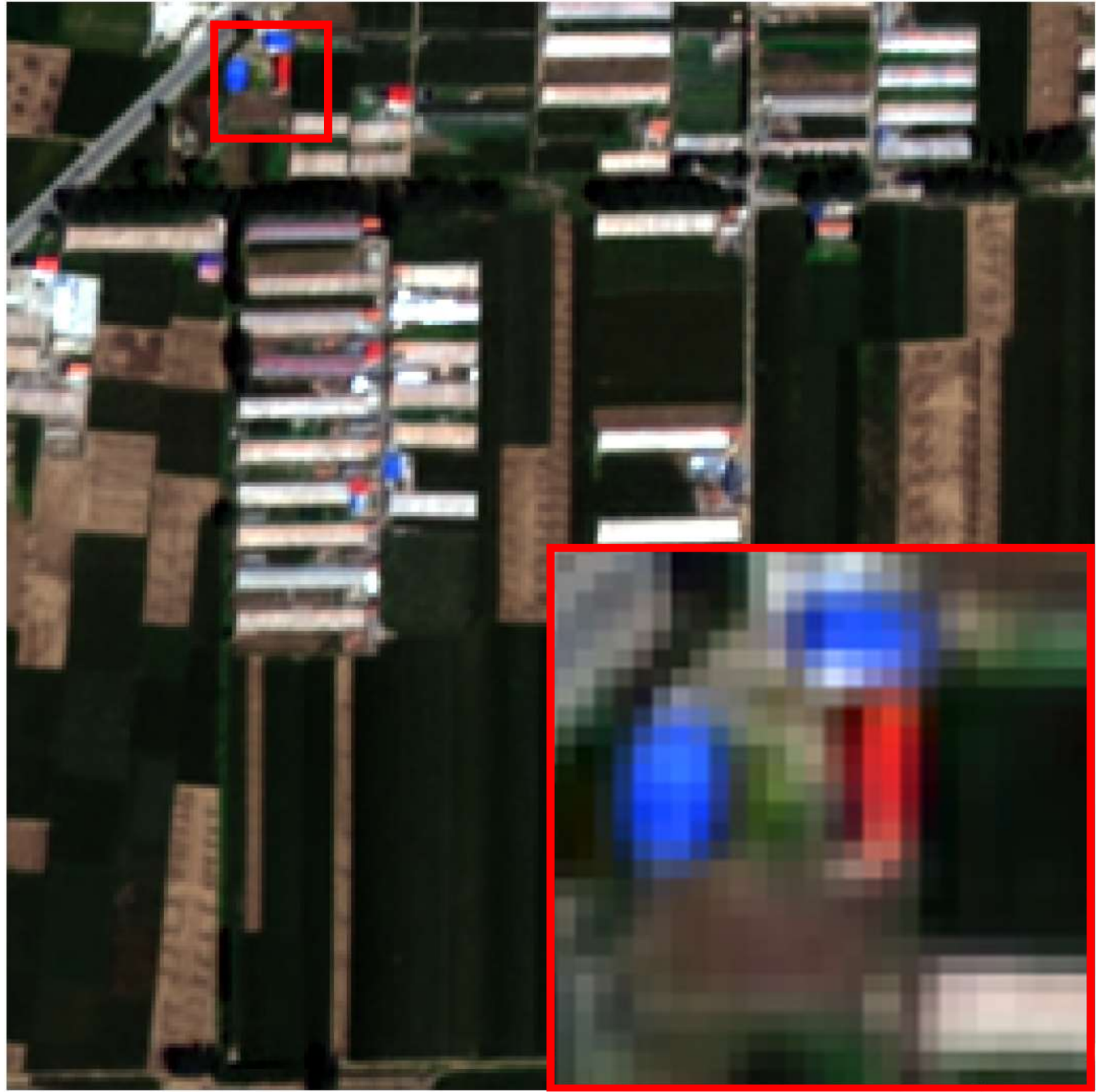}
\label{FAFNet}}
\caption{Visual comparison of different methods on WorldView-4 (WV-4) dataset at reduced resolution. (a) PAN. (b) Up-sampled MS. (c) Reference. (d) GSA \cite{4305344}. (e) BDSD-PC \cite{8693555}. (f) MTF-GLP-CBD \cite{4305345}. (g) AWLP-H \cite{vivone2019fast}. (h) PWMBF \cite{6951484}. (i) PanNet \cite{8237455}. (j) FusionNet \cite{9240949}. (k) GTP-PNet \cite{ZHANG2021223}. (l) LPPN \cite{JIN2022158}. (m) FAFNet.}
\label{fig_wv4_reduce}
\end{figure*}

\begin{figure*}[h]
\centering
\subfloat[]{\includegraphics[width=1in,trim=120 0 120 0,clip]{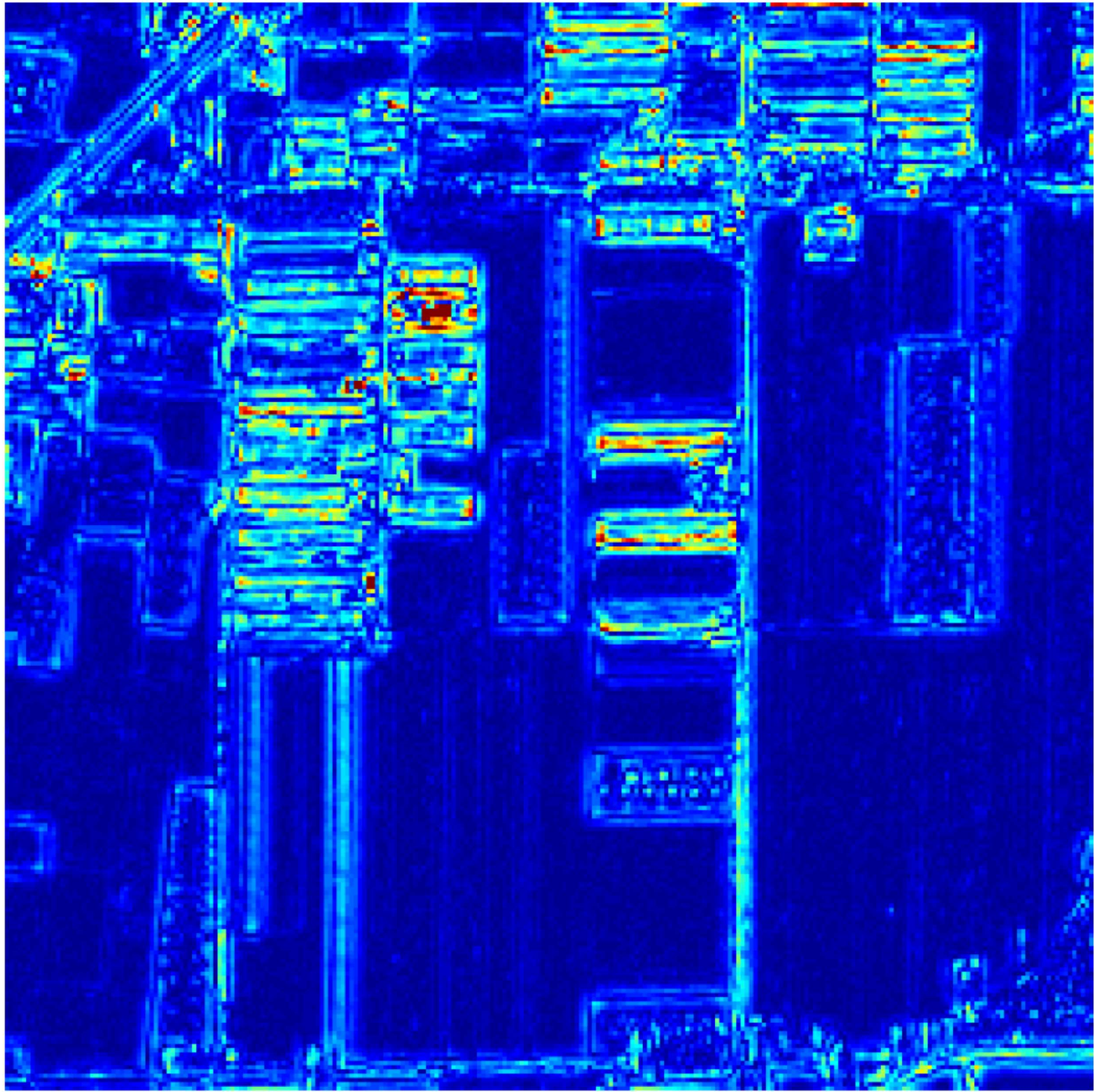}
\label{up-sampled MS-error}}
\subfloat[]{\includegraphics[width=1in,trim=120 0 120 0,clip]{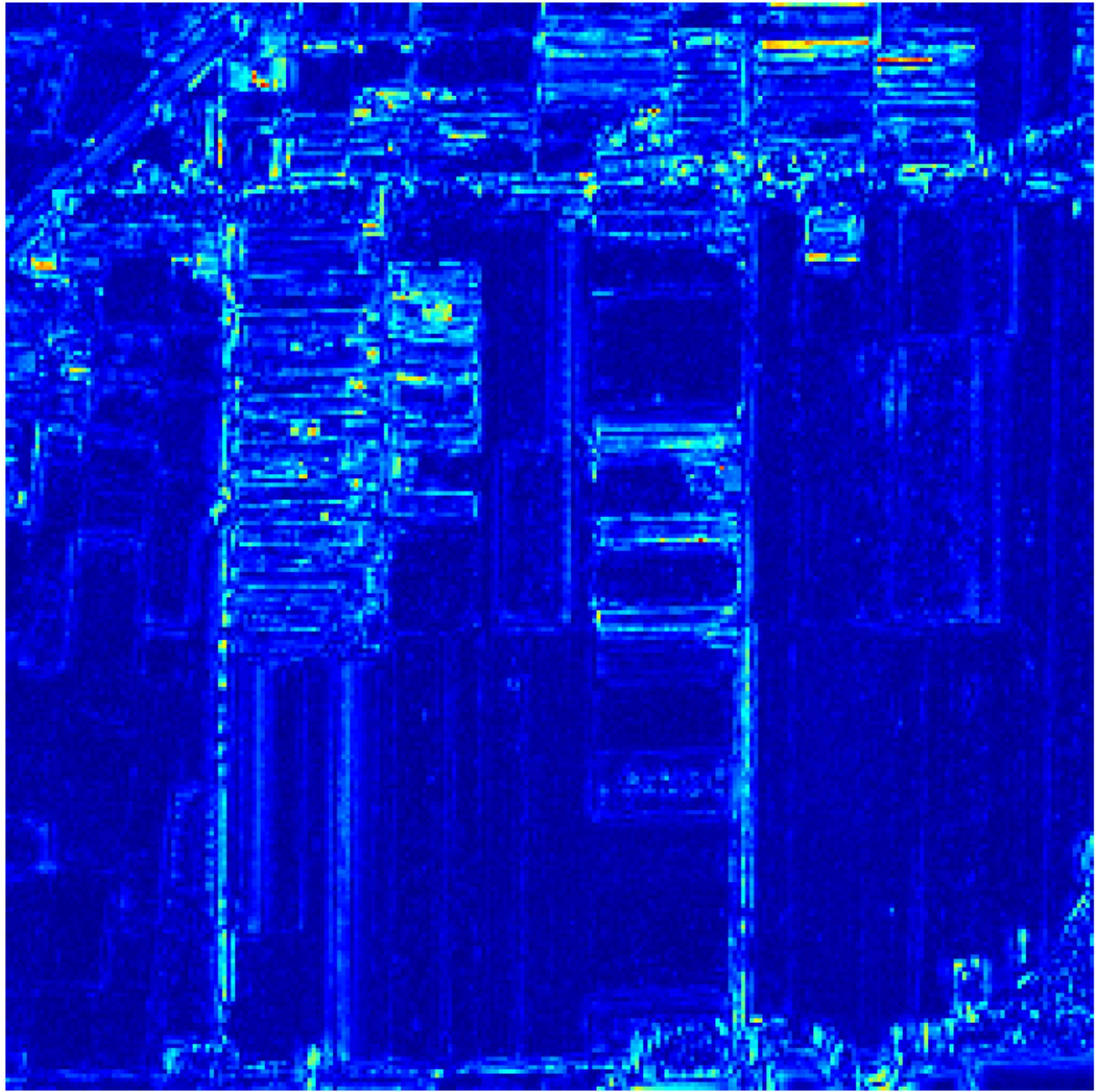}
\label{GSA}}
\subfloat[]{\includegraphics[width=1in,trim=120 0 120 0,clip]{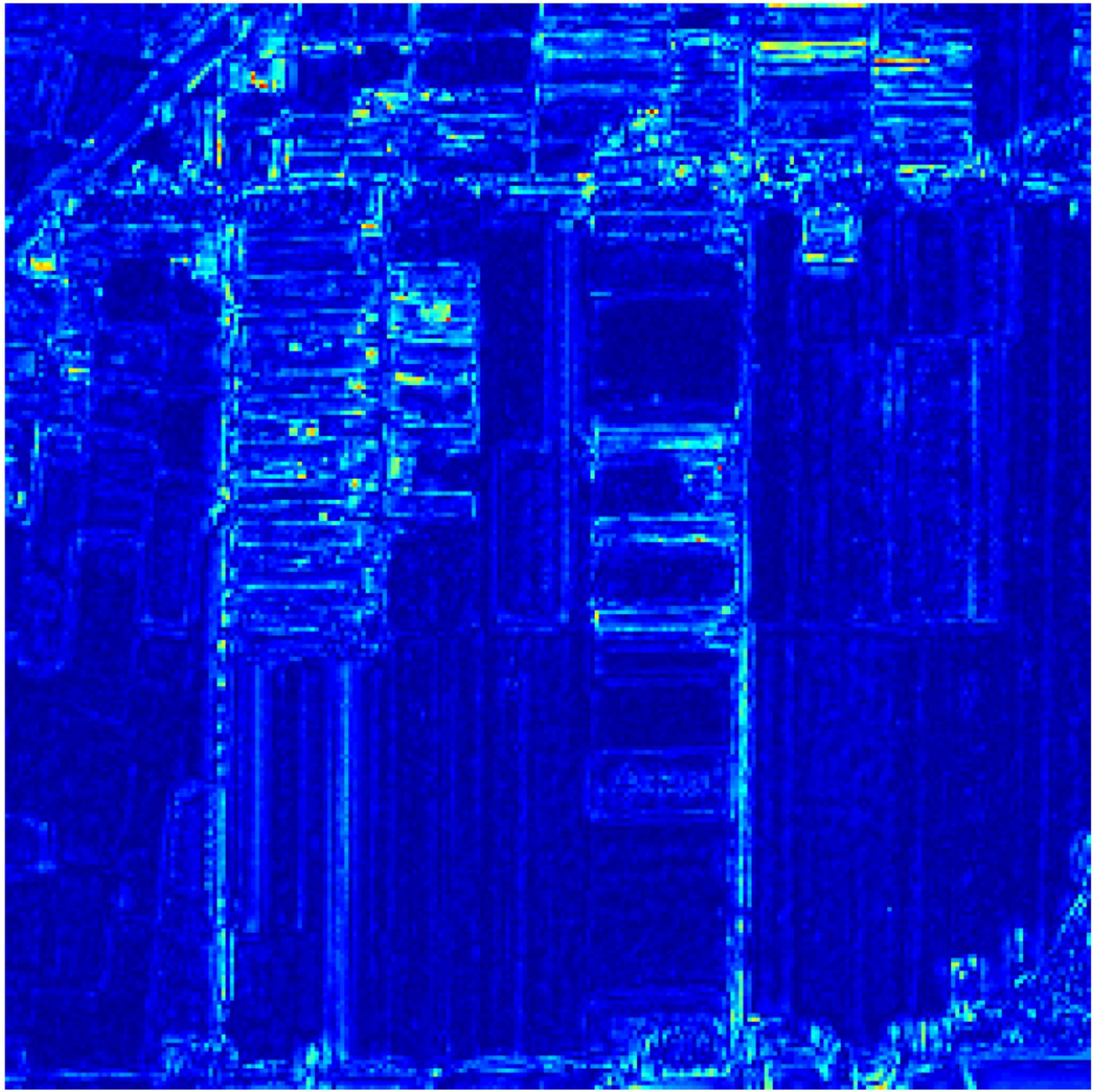}
\label{BDSD-PC}}
\subfloat[]{\includegraphics[width=1in,trim=120 0 120 0,clip]{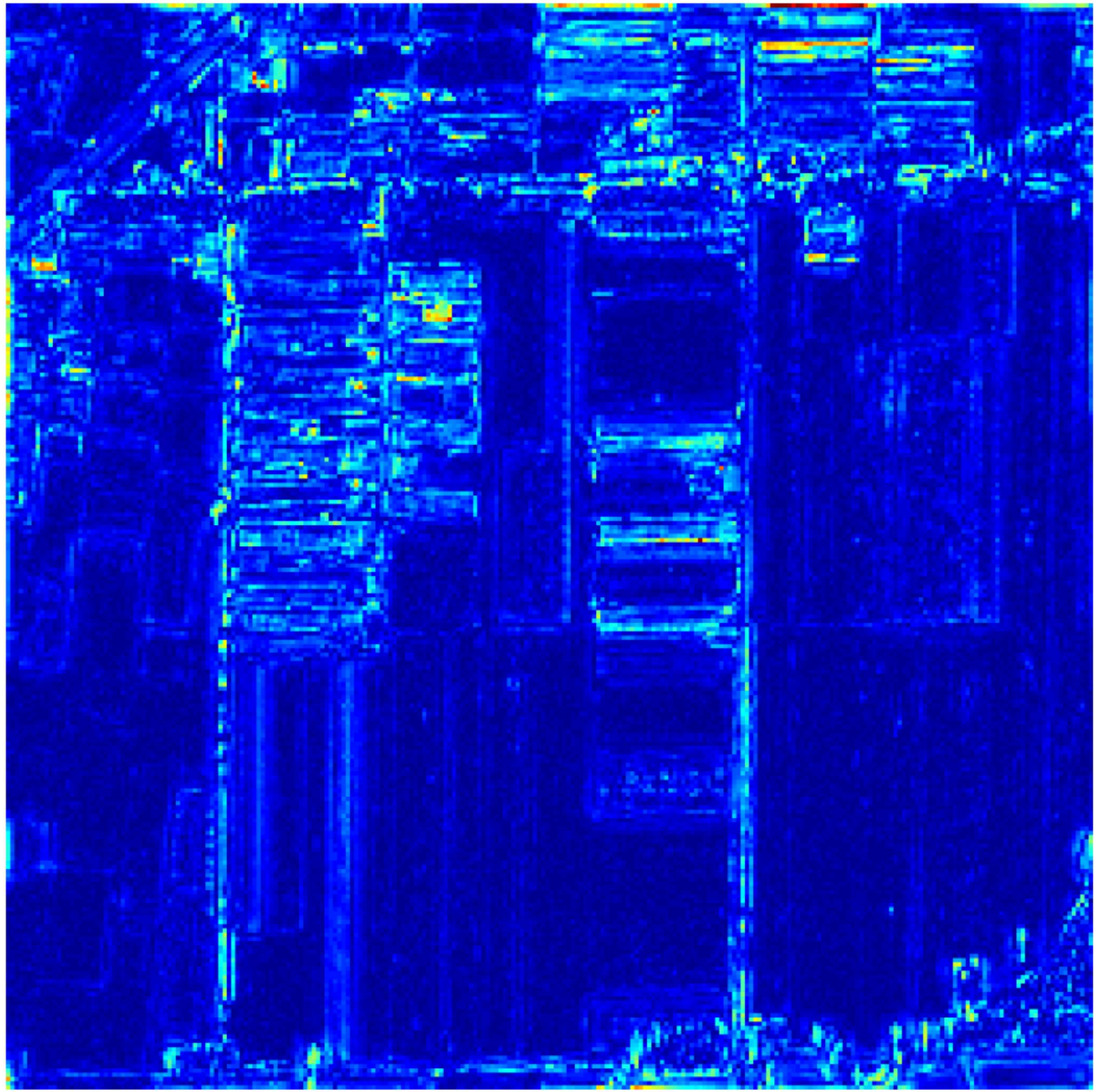}
\label{MTF-GLP-CBD}}
\subfloat[]{\includegraphics[width=1in,trim=120 0 120 0,clip]{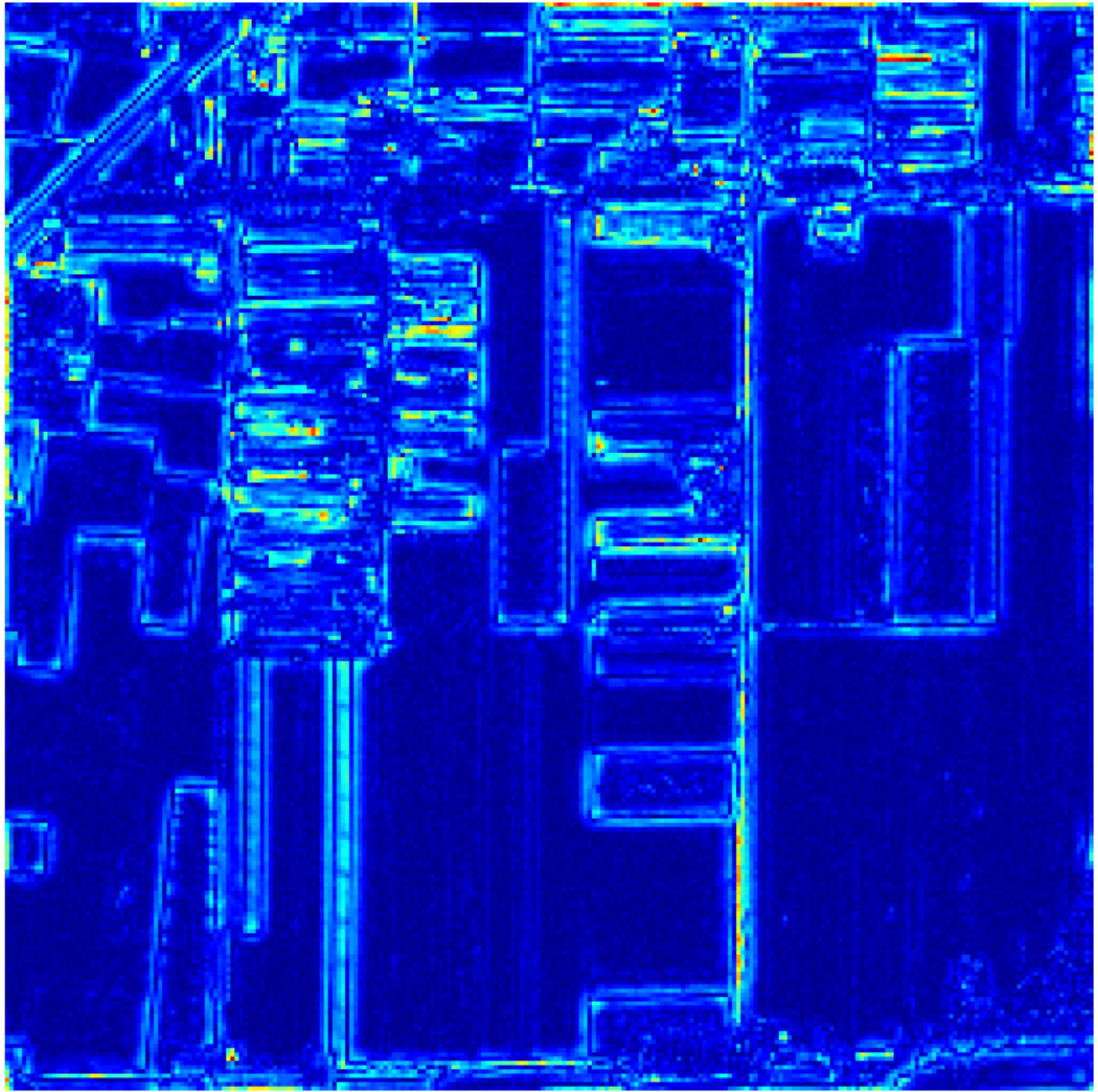}
\label{AWLP-H}}
\subfloat[]{\includegraphics[width=1in,trim=120 0 120 0,clip]{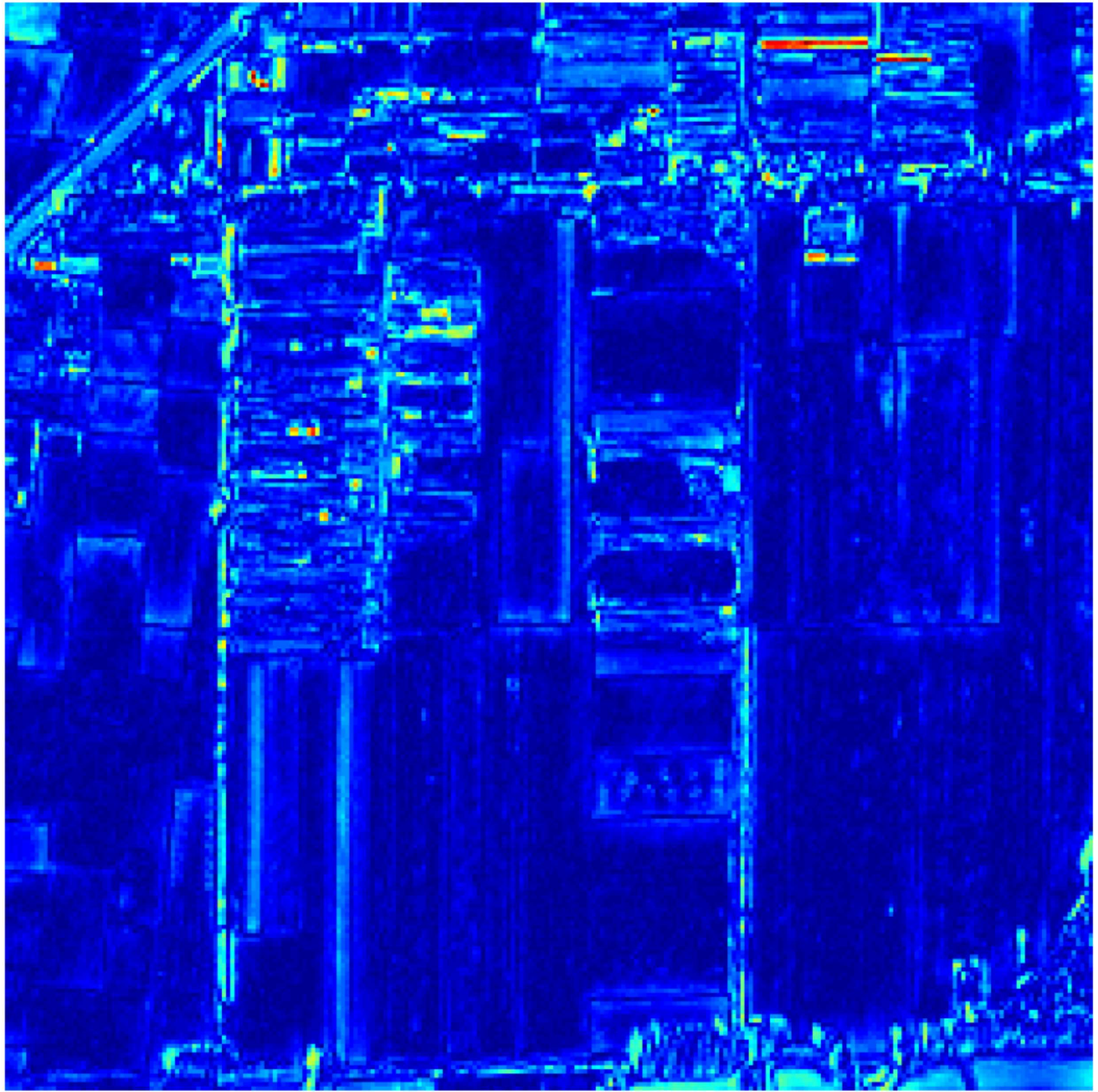}
\label{PWMBF}}

\vspace{-0.1in}
\subfloat[]{\includegraphics[width=1in,trim=120 0 120 0,clip]{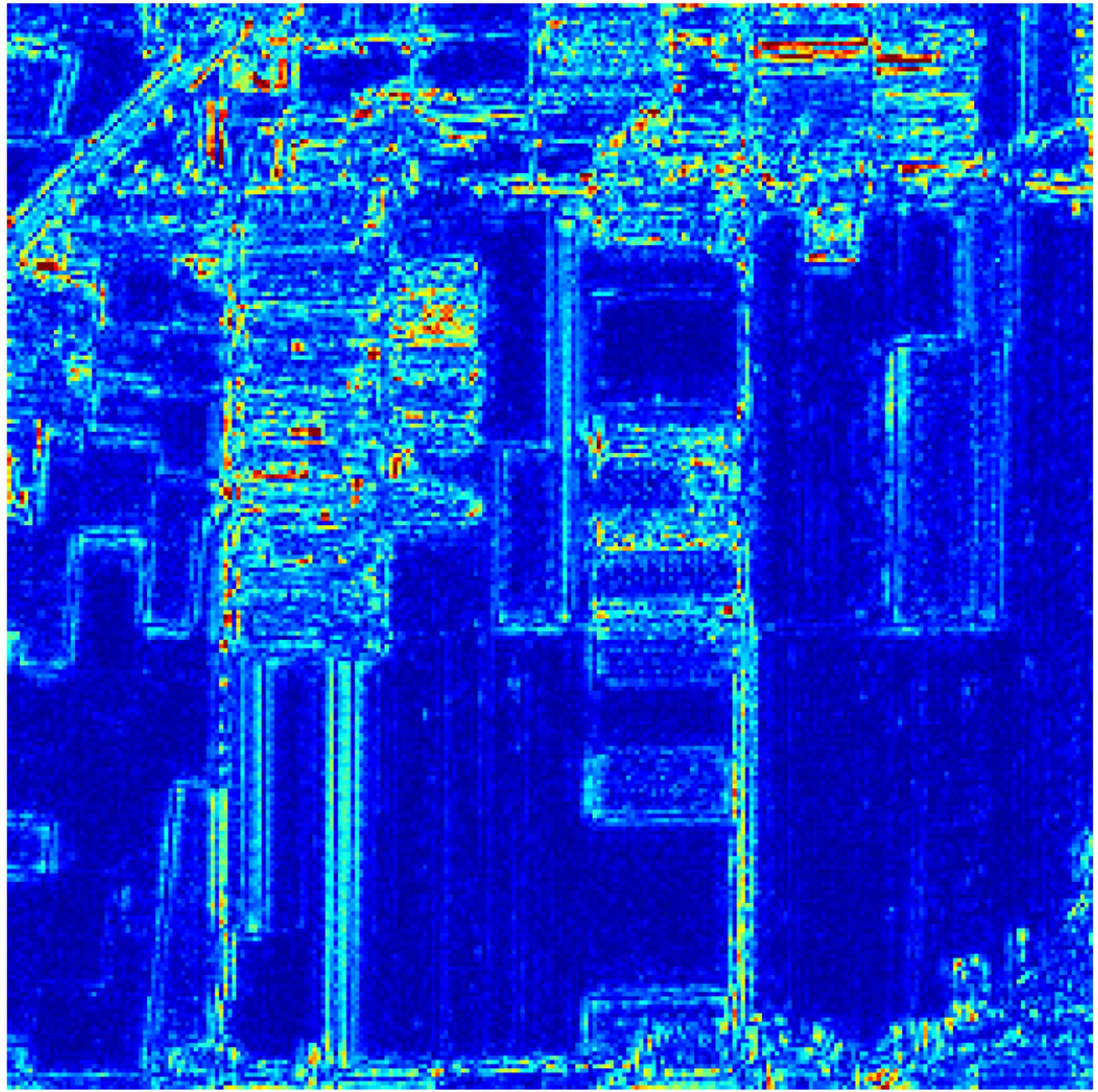}
\label{PanNet}}
\subfloat[]{\includegraphics[width=1in,trim=120 0 120 0,clip]{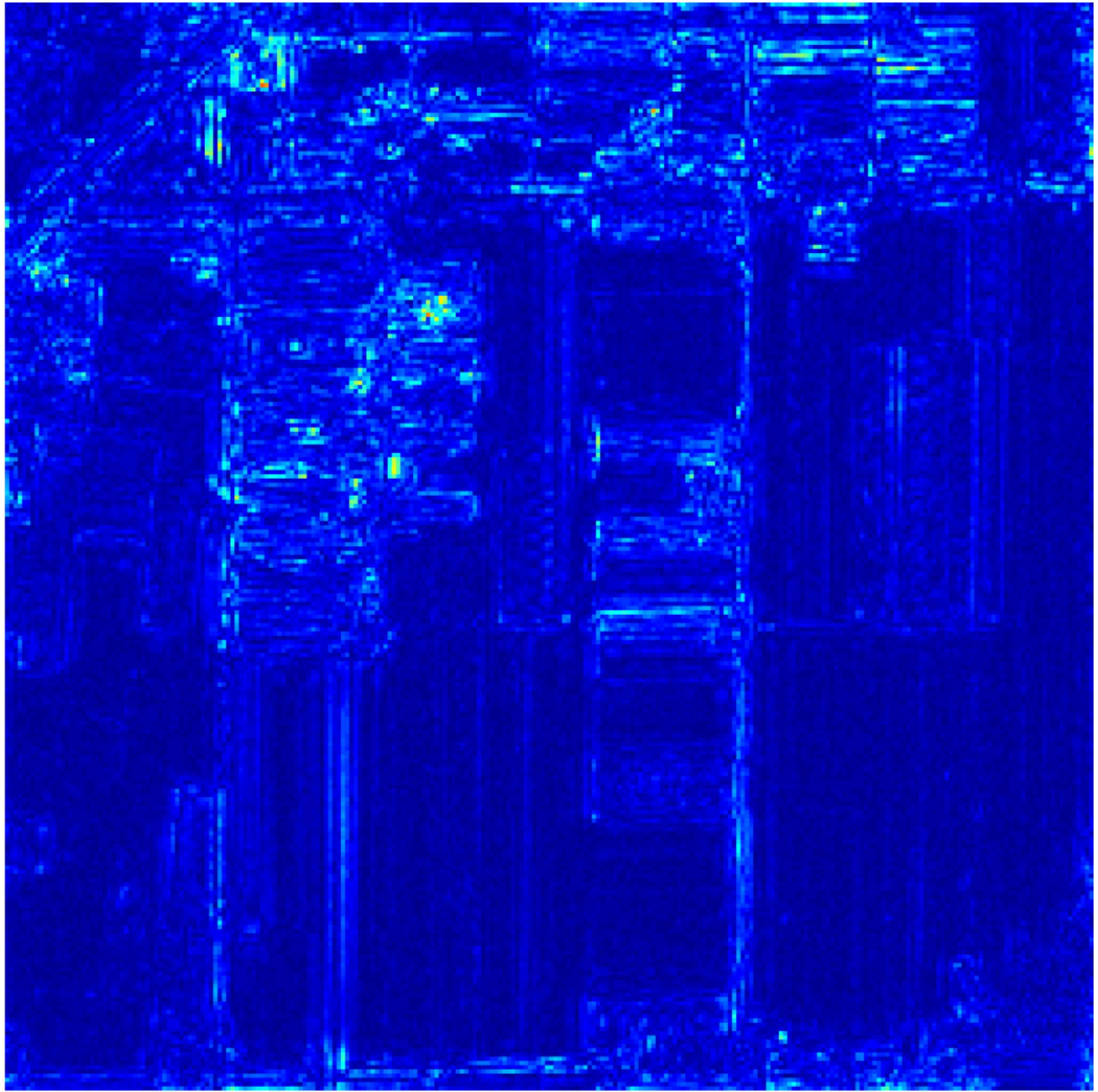}
\label{FusionNet}}
\subfloat[]{\includegraphics[width=1in,trim=120 0 120 0,clip]{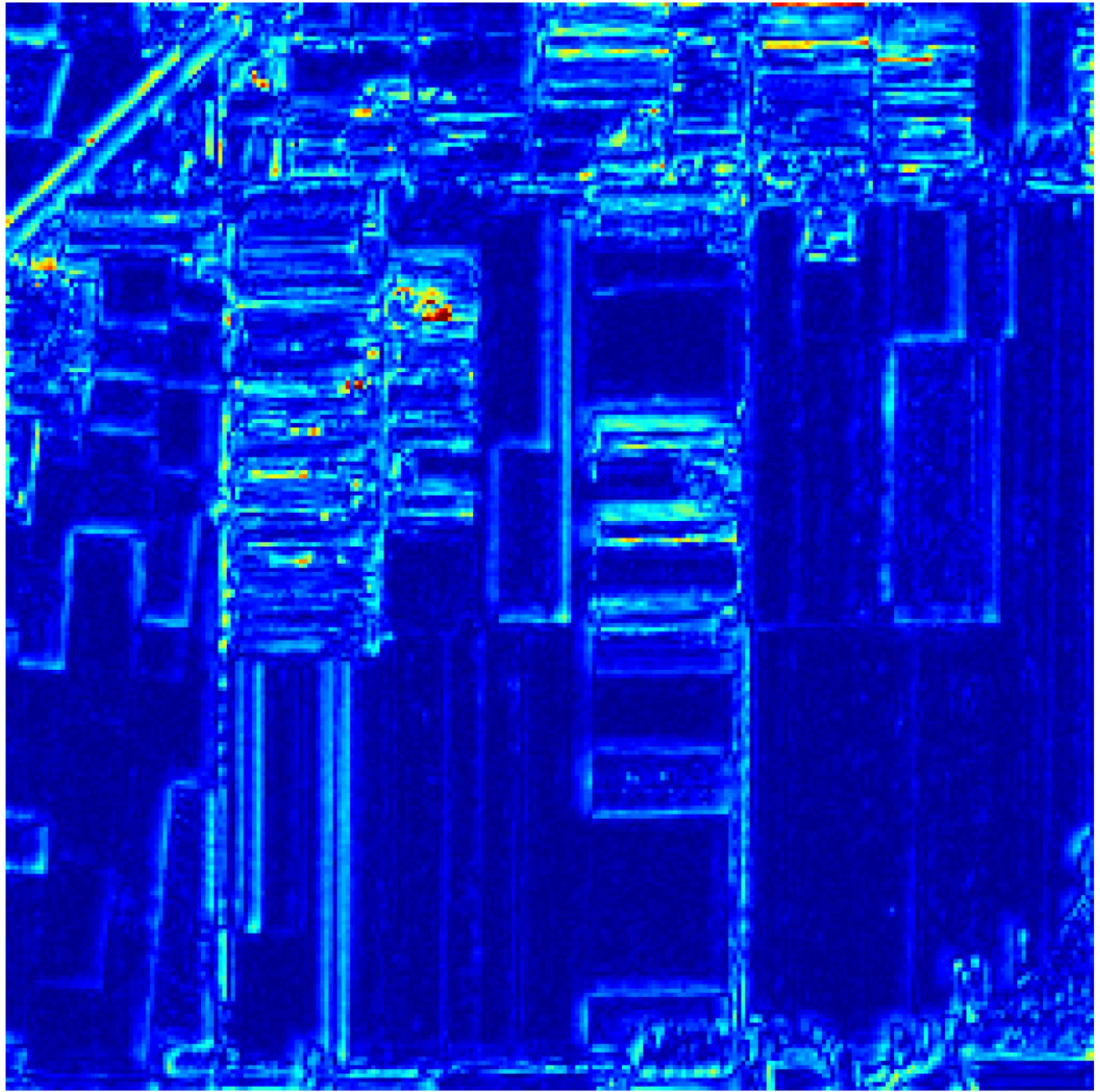}
\label{GTP-PNet}}
\subfloat[]{\includegraphics[width=1in,trim=120 0 120 0,clip]{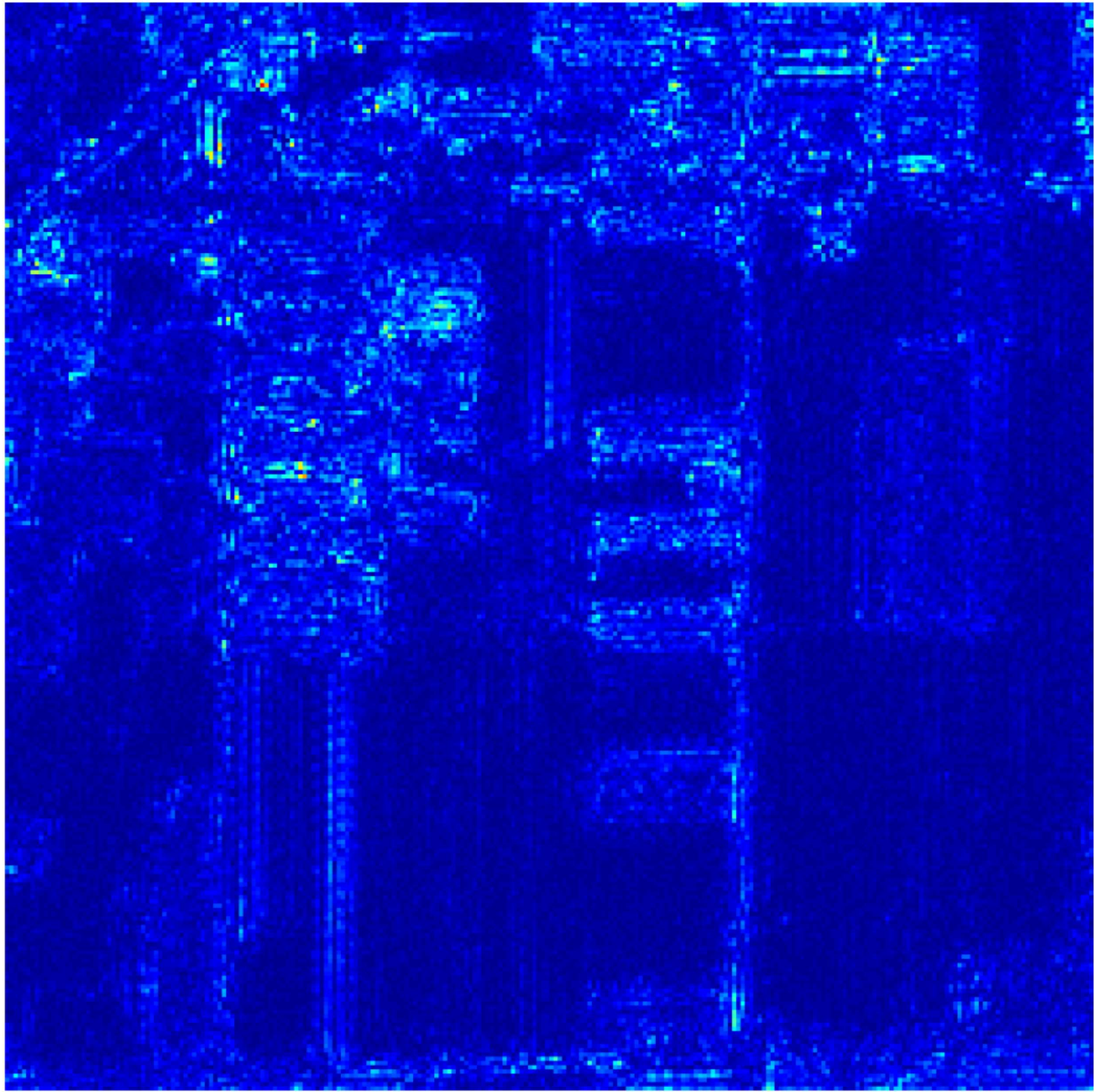}
\label{LPPN}}
\subfloat[]{\includegraphics[width=1in,trim=120 0 120 0,clip]{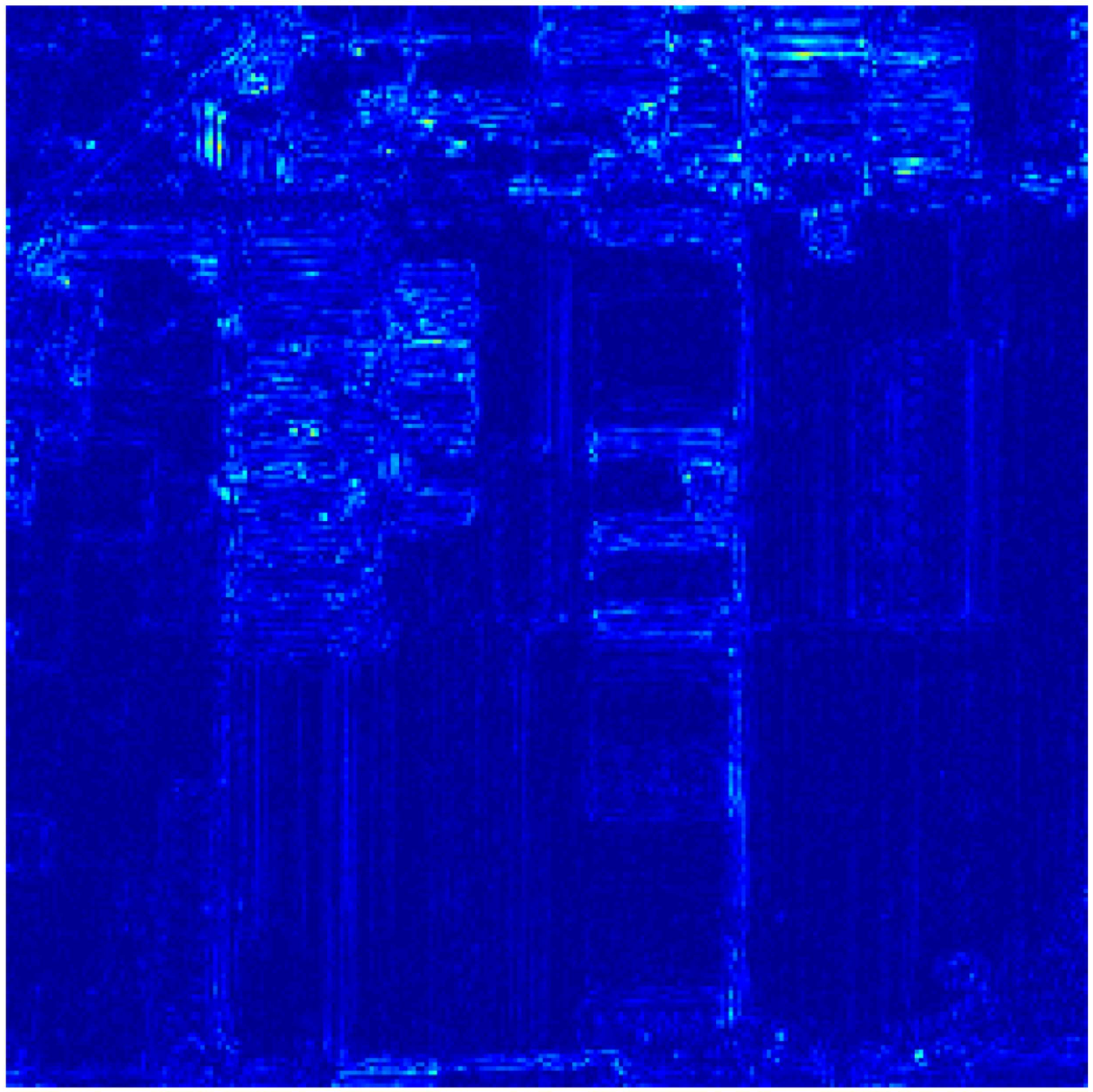}
\label{FAFNet}}
\subfloat[]{\includegraphics[width=1in,trim=120 0 120 0,clip]{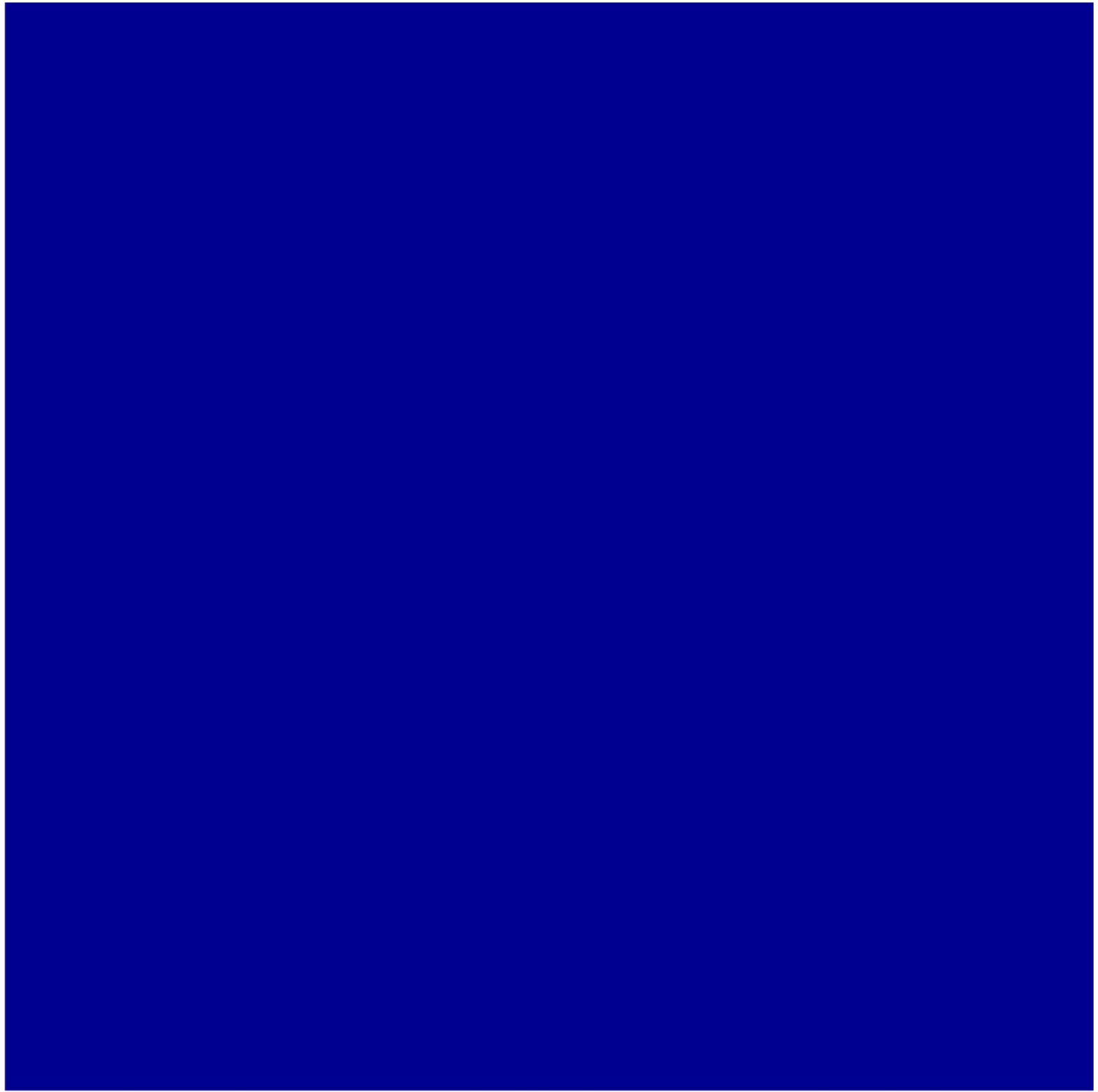}
\label{Reference}}

\subfloat{\includegraphics[angle=-90,width=0.3\linewidth,trim=220 0 220 0,clip]{figures/wv4/error/bar.pdf}
}
\caption{Absolute error maps (AEMs) between reference image and (a)Up-sampled MS, (b) GSA \cite{4305344}, (c) BDSD-PC \cite{8693555}, (d) MTF-GLP-CBD \cite{4305345}, (e) AWLP-H \cite{vivone2019fast}, (f) PWMBF \cite{6951484}, (g) PanNet \cite{8237455}, (h) FusionNet \cite{9240949}, (i) GTP-PNet \cite{ZHANG2021223}, (j) LPPN \cite{JIN2022158}, (k) FAFNet and (l) Reference on WV-4 dataset at reduced resolution.}
\label{fig_wv4_reduce_error}
\end{figure*}

\section{Experimental Results and Analysis}
In order to verify the effectiveness of the proposed method, we compare our model with nine state-of-the-art methods, including Gram-Schmidt adaptive (GSA) \cite{4305344}, band-dependent spatial-detail with physical constraints (BDSD-PC) \cite{8693555}, Modulation Transfer Function-generalized Laplacian pyramid with context-based decision (MTF-GLP-CBD) \cite{4305345}, additive wavelet luminance proportional with haze-correction (AWLP-H) \cite{vivone2019fast}, PCA/Wavelet model-based fusion (PWMBF) \cite{6951484}, PanNet \cite{8237455}, FusionNet \cite{9240949}, GTP-PNet \cite{ZHANG2021223} and LPPN \cite{JIN2022158}. Experiments are conducted on three satellite datasets at both reduced and full resolution. In the reduced-resolution experiments, the widely used evaluaton indexes, spectral angle mapper (SAM) \cite{yuhas1992discrimination}, erreur relative globale adimensionnelle de synth$\grave{e}$se (ERGAS) \cite{wald2002data}, Q$2^n$ \cite{garzelli2009hypercomplex} and the spatial correlation coefficient (SCC) \cite{zhou1998wavelet} are adopted, while in the full-resolution experiments, the spectral distortion index ($D_\lambda$), the spatial distortion index ($D_s$) and the quality with no reference (QNR) \cite{alparone2008multispectral} index are used to evaluate the quality of fusion results. The optimal values for SAM, ERGAS, $D_\lambda$ and $D_s$ are 0, while those for SCC, QNR, and Q$2^n$ are 1.

\subsection{Datasets}
Experiments are conducted on three datasets acquired from WorldView-4 (WV-4), QuickBird (QB) and WorldView-2 (WV-2) satellites. All datasets are composed of paired multi-bands MS and single-band PAN images. The MS images of WV-4 and QB datasets have four spectral bands (i.e., blue, green, red, and near infrared) while those of WV-2 dataset have not only the blue, green, red and near infrared bands, but also the coastal, yellow, red edge, and near-infrared 2 bands. The details of the datasets are listed in Table \ref{tab_dataset}.

In the reduced-resolution experiment, the original MS and PAN images are firstly filtered by a $5\times5$ Gaussian smooth kernel and then downsampled by a factor of 4. The original MS images are treated as references, and these downsampled pairs of MS and PAN images are inputs of all the comparison methods \cite{wald1997fusion}. In the full-resolution experiment, original MS and PAN pairs are directly taken as inputs.

\subsection{Training Details}
The training samples are obtained under the Wald's protocol \cite{wald1997fusion}, where the original MS and PAN patches are filtered by a $5\times5$ Gaussian kernel and downsampled by a factor of 4, then the original MS patches are treated as references. The sizes of training MS, PAN and reference patches are $16\times16\times \emph{B}$, $64 \times 64$, and $64\times 64 \times \emph{B}$ ($\emph{B}=4$ for WV-4 and QB datasets, and $\emph{B}=8$ for WV-2 dataset), respectively.

Our proposed model is achieved under the PyTorch framework \cite{Paszke2017AutomaticDI} and the parameters are updated by the Adam optimizer. The learning rate is initialized to $1\times10^{-4}$. The batch size is empirically set to 32 and the training process terminates after 2000 epochs. Additionally, we train four CNN-based methods, PanNet \cite{8237455}, FusionNet \cite{9240949}, GTP-PNet \cite{ZHANG2021223} and LPPN \cite{JIN2022158}, on our datasets according to the default settings provided by the authors to guarantee the fair comparisons. All experiments are supported by an Nvidia GTX 1080Ti GPU.

\subsection{Experiments at Reduced Resolution}
In this section, experiments at reduced resolution are conducted on the WV-4, QB and WV-2 datasets. Both the visual inspections and the quantitative assessments are presented to fully verify the superiority of our approach. We magnify representative regions to show the details of the fusion results. Furthermore, we also compute the absolute error maps (AEMs) between the fusion results and the reference image, which can provide more insights about the performance of the compared methods. For quantitative assessments, we list the mean values and the standard deviations of the indexes in Tables \ref{tab_wv4}-\ref{tab_wv2}, where the best results are marked in bold and the second-best results are underlined.

The fusion results and AEMs on the WV-4 dataset are shown in Figs. \ref{fig_wv4_reduce}-\ref{fig_wv4_reduce_error}. It can be observed that the fusion results of GSA \cite{4305344}, BDSD-PC \cite{8693555}, MTF-GLP-CBD \cite{4305345}, AWLP-H \cite{vivone2019fast}, PWMBF \cite{6951484}, PanNet \cite{8237455} and GTP-PNet \cite{ZHANG2021223} suffer from varying degrees of spectral distortion. Although the results of FusionNet \cite{9240949} and LPPN \cite{JIN2022158} have better spectral fidelity in terms of the color of soil, roof and vegetations, their boundaries between different objects tend to be blurry. Compared with them, the proposed FAFNet preserves the spatial and spectral information better, which can be seen especially from the zoomed-in regions in Fig. \ref{fig_wv4_reduce}. The AEMs shown in Fig. \ref{fig_wv4_reduce_error} also verify the effectiveness of the proposed method in terms of the preservation of spatial and spectral information. Table \ref{tab_wv4} lists the quantitative assessments on 271 pairs of test data from the WV-4 dataset. From Table \ref{tab_wv4}, it is clear that our proposed method obtains the best result in terms of the ERGAS, Q4, SAM and SCC indexes, which are in consistent with visual inspections. 
% regarding to and ranks second in terms of , 
\begin{table*}[h]
\caption{Average Quantitative Results on 271 Pairs of Test Data from WV-4 Dataset.\label{tab_wv4}}
\centering
\begin{tabular}{c|cccc|ccc}
\toprule
~ & \multicolumn{4}{c|}{\textbf{Reduced resolution}} & \multicolumn{3}{c}{\textbf{Full resolution}} \\		
\midrule
Methods & ERGAS & Q4 & SAM & SCC  & QNR & $D_\lambda$ & $D_s$ \\
\midrule
Ideal value & 0 & 1 & 0 & 1 & 1 & 0 & 0\\	
\midrule
GSA \cite{4305344} & 1.9888±0.2670 & 0.8692±0.0014 & 2.6112±0.9380 & 0.9026±0.0010  &  0.8704±0.0017 & 0.0465±0.0003 & 0.0876±0.0009 \\
BDSD-PC \cite{8693555} & 2.0073±0.2359 & 0.8655±0.0016 & 2.7308±0.8334 & 0.9152±0.0005 & 0.9073±0.0021 & 0.0331±0.0003 & 0.0621±0.0011 \\
MTF-GLP-CBD \cite{4305345} & 2.0775±0.2891 & 0.8653±0.0016 & 2.6658±0.9698 & 0.9007±0.0010  & 0.8694±0.0016 & 0.0548±0.0003 & 0.0806±0.0008\\
AWLP-H \cite{vivone2019fast}  & 1.9218±0.2833 & 0.8866±0.0008 & 2.4787±0.8440 & 0.9269±0.0004 & 0.9011±0.0009  & 0.0469±0.0003 & 0.0548±0.0004\\
PWMBF \cite{6951484} & 2.1375±0.2610 & 0.8482±0.0014 & 3.0205±1.0255 & 0.8951±0.0007 & 0.8309±0.0016  & 0.0754±0.0004  & 0.1018±0.0008 \\
PanNet \cite{8237455} & 1.8679±0.1500 & 0.8346±0.0047 & 2.5500±0.5736 & 0.9201±0.0002 & 0.9092±0.0013  & 0.0363±0.0001 & 0.0568±0.0009\\
FusionNet \cite{9240949} & 1.6754±0.4231 & 0.8672±0.0074 & 2.3900±1.5702 & 0.9542±0.0001  & 0.9419±0.0015 & 0.0263±0.0007 & 0.0330±0.0003 \\
GTP-PNet \cite{ZHANG2021223} & 2.3218±0.2699 & 0.8284±0.0018 & 3.0114±0.8521 & 0.8723±0.0006 & 0.9341±0.0008 & 0.0215±0.0001 & 0.0454±0.0005\\
LPPN \cite{JIN2022158} & \underline{1.5032±0.1560} & \underline{0.8835±0.0049} & \underline{2.0922±0.4901} & \underline{0.9566±0.0001} & \underline{0.9491±0.0009} & \underline{0.0209±0.0002} & \underline{0.0308±0.0004} \\
FAFNet & \textbf{1.1364±0.1101} & \textbf{0.9356±0.0015} & \textbf{1.5026±0.2782} & \textbf{0.9717±0.0001} & \textbf{0.9594±0.0003} & \textbf{0.0168±0.0001} & \textbf{0.0243±0.0001}\\
\bottomrule
\end{tabular}
\end{table*}

\begin{figure*}[h]
\centering
\subfloat[]{ \includegraphics[width=1.1in,trim=120 0 120 0,clip]{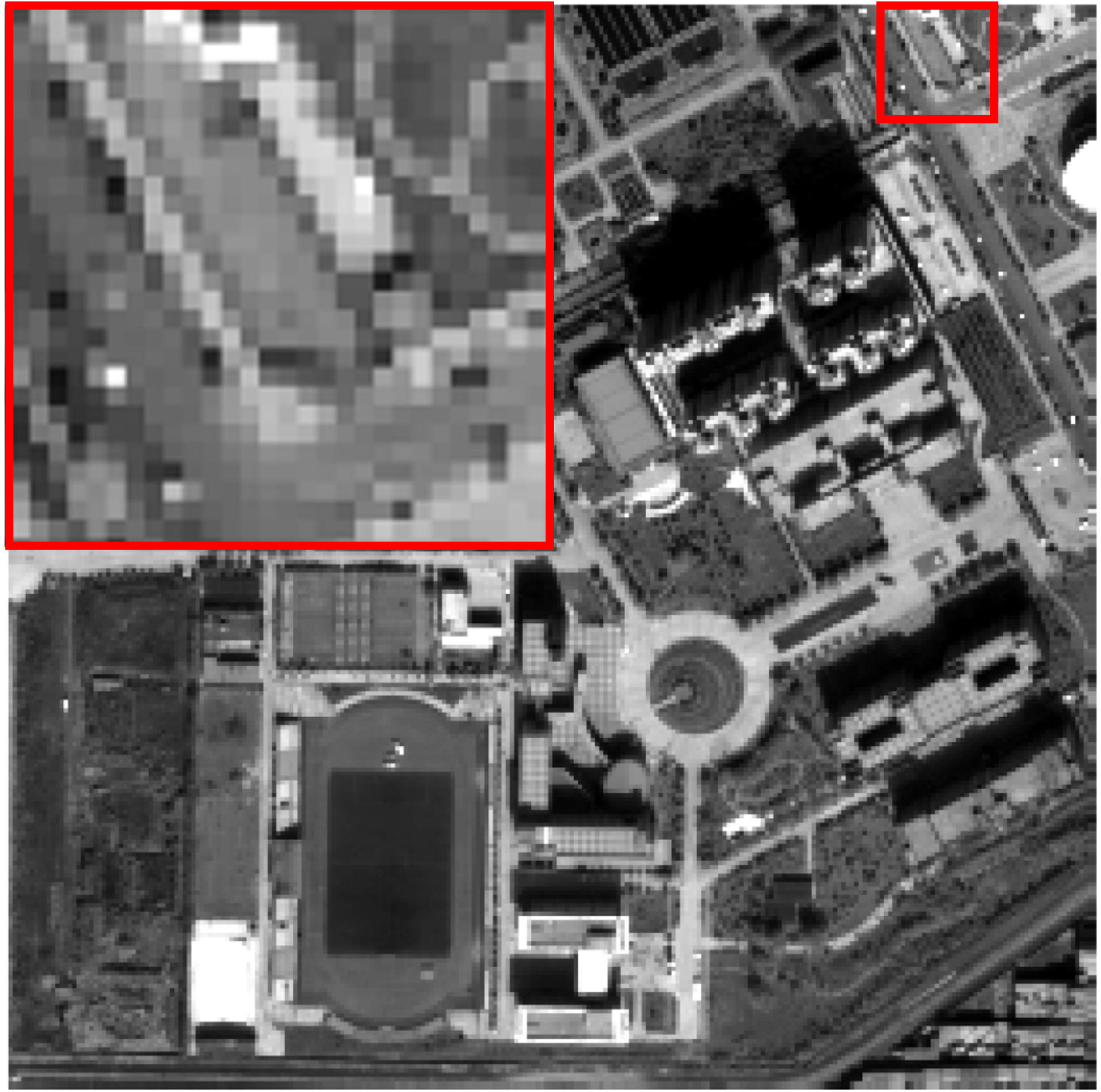}
\label{PAN}}
\subfloat[]{\includegraphics[width=1.1in,trim=120 0 120 0,clip]{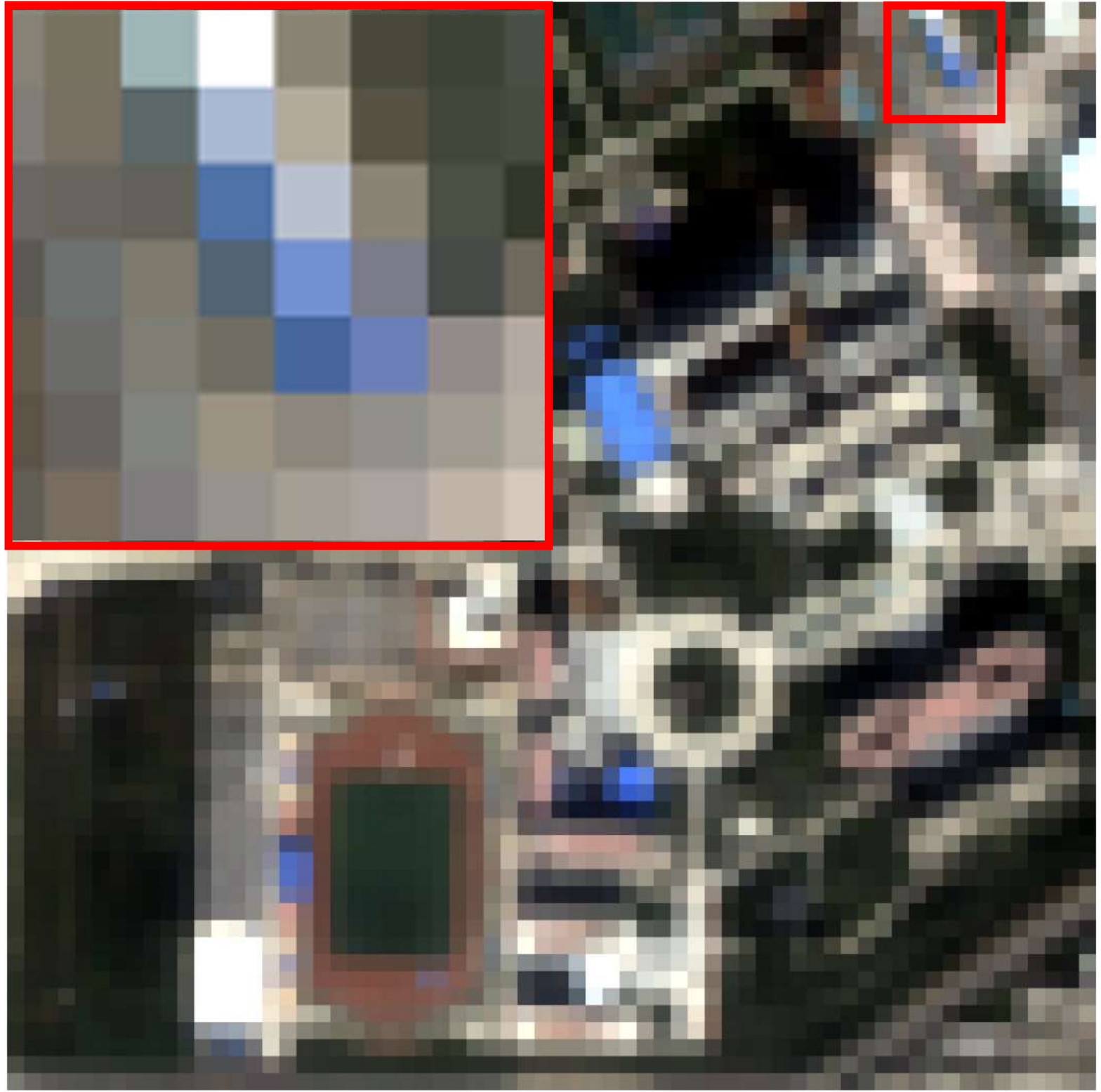}
\label{up-sampled MS}}
\subfloat[]{\includegraphics[width=1.1in,trim=120 0 120 0,clip]{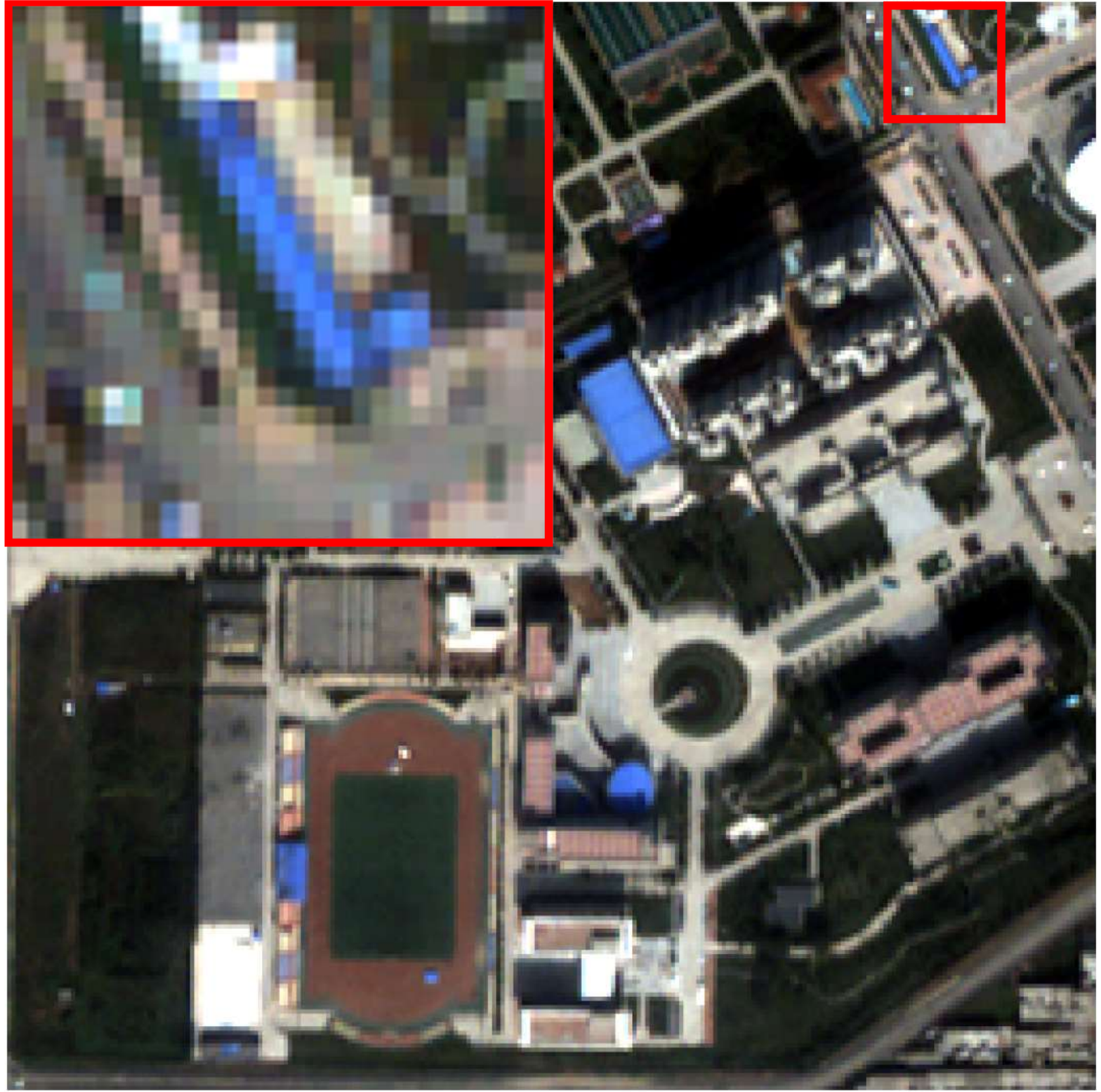}
\label{Reference}}
\subfloat[]{\includegraphics[width=1.1in,trim=120 0 120 0,clip]{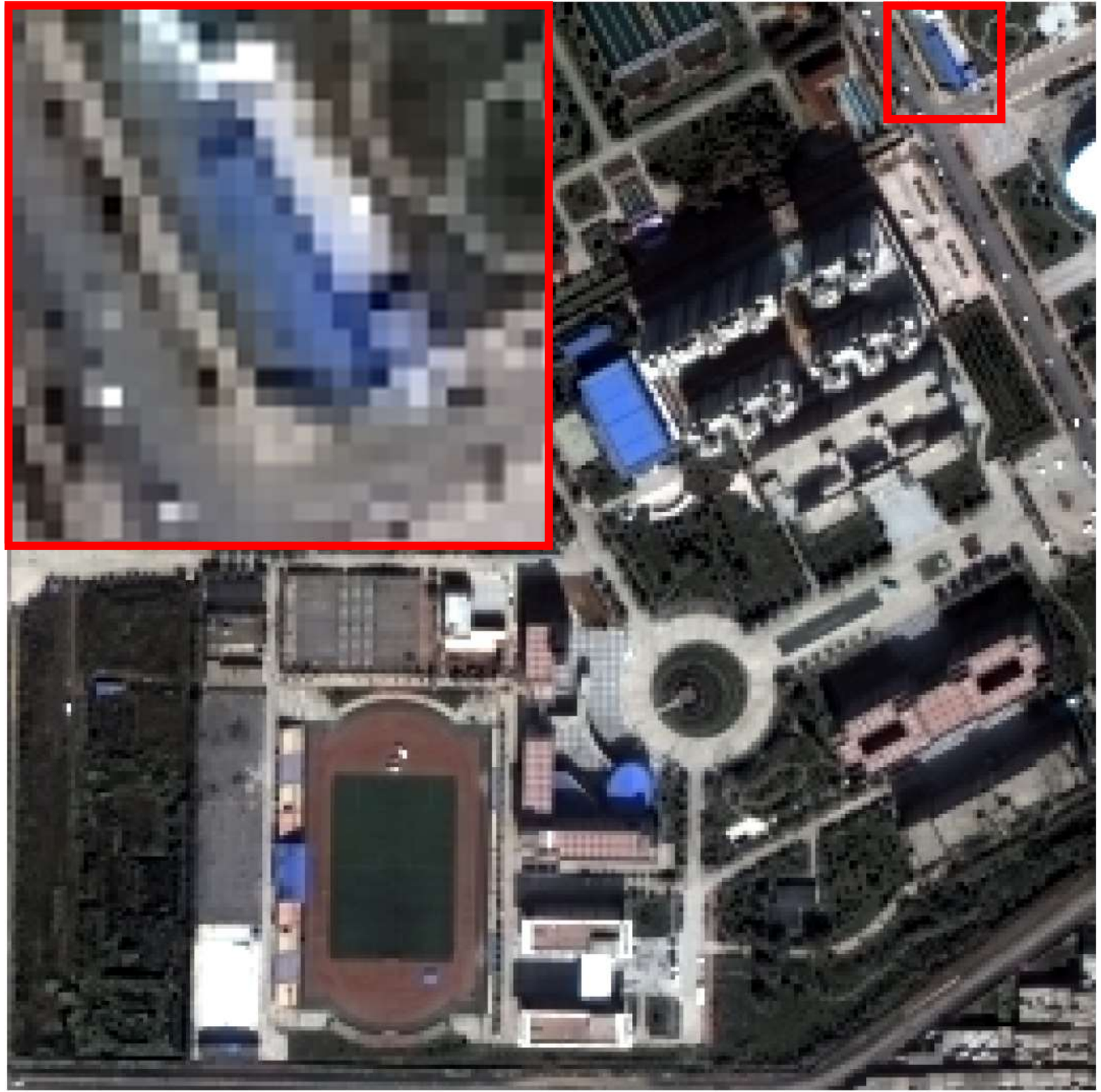}
\label{GSA}}
\subfloat[]{\includegraphics[width=1.1in,trim=120 0 120 0,clip]{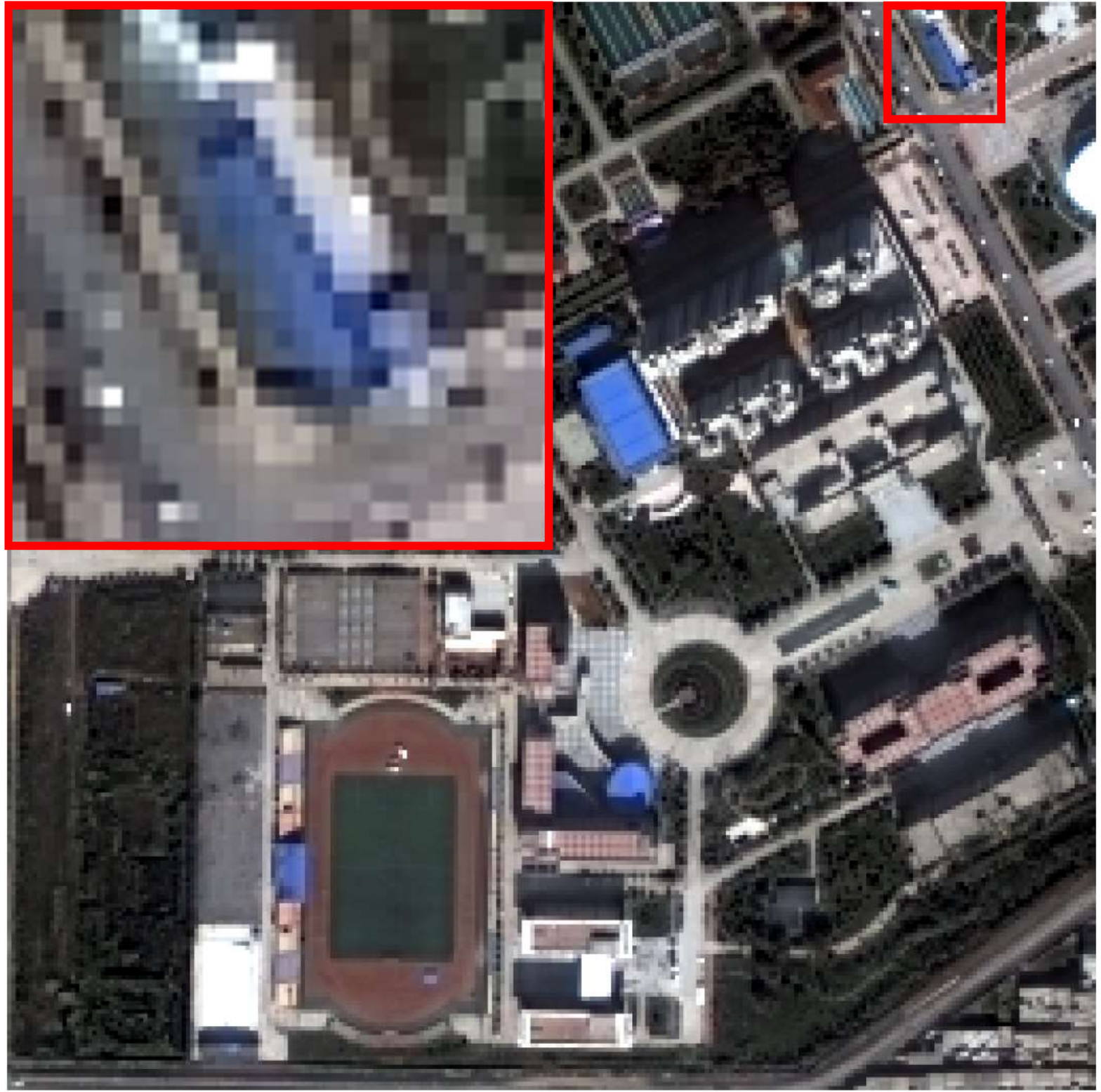}
\label{BDSD-PC}}

\vspace{-0.1in}
\subfloat[]{\includegraphics[width=1.1in,trim=120 0 120 0,clip]{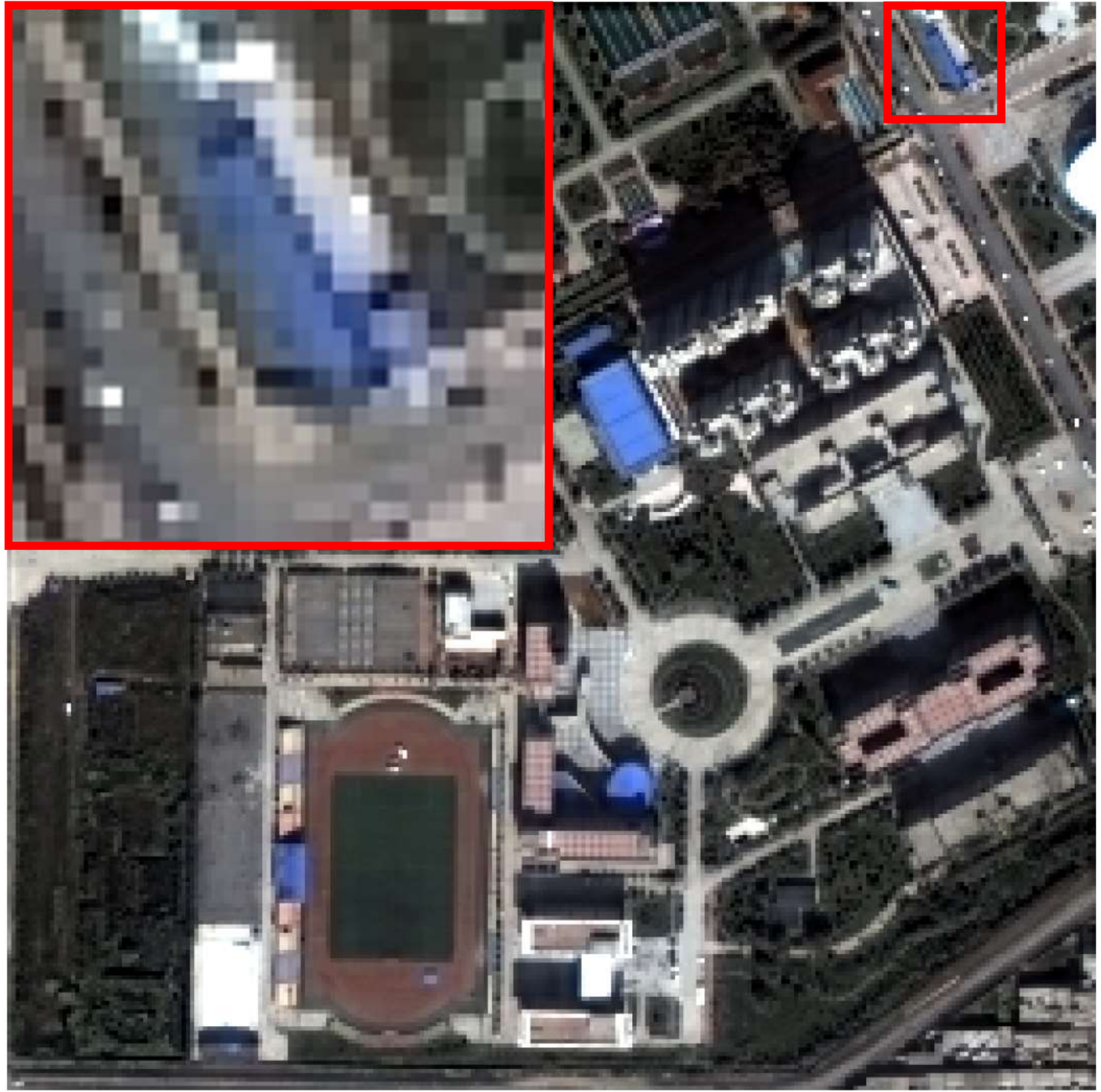}
\label{MTF-GLP-CBD}}
\subfloat[]{\includegraphics[width=1.1in,trim=120 0 120 0,clip]{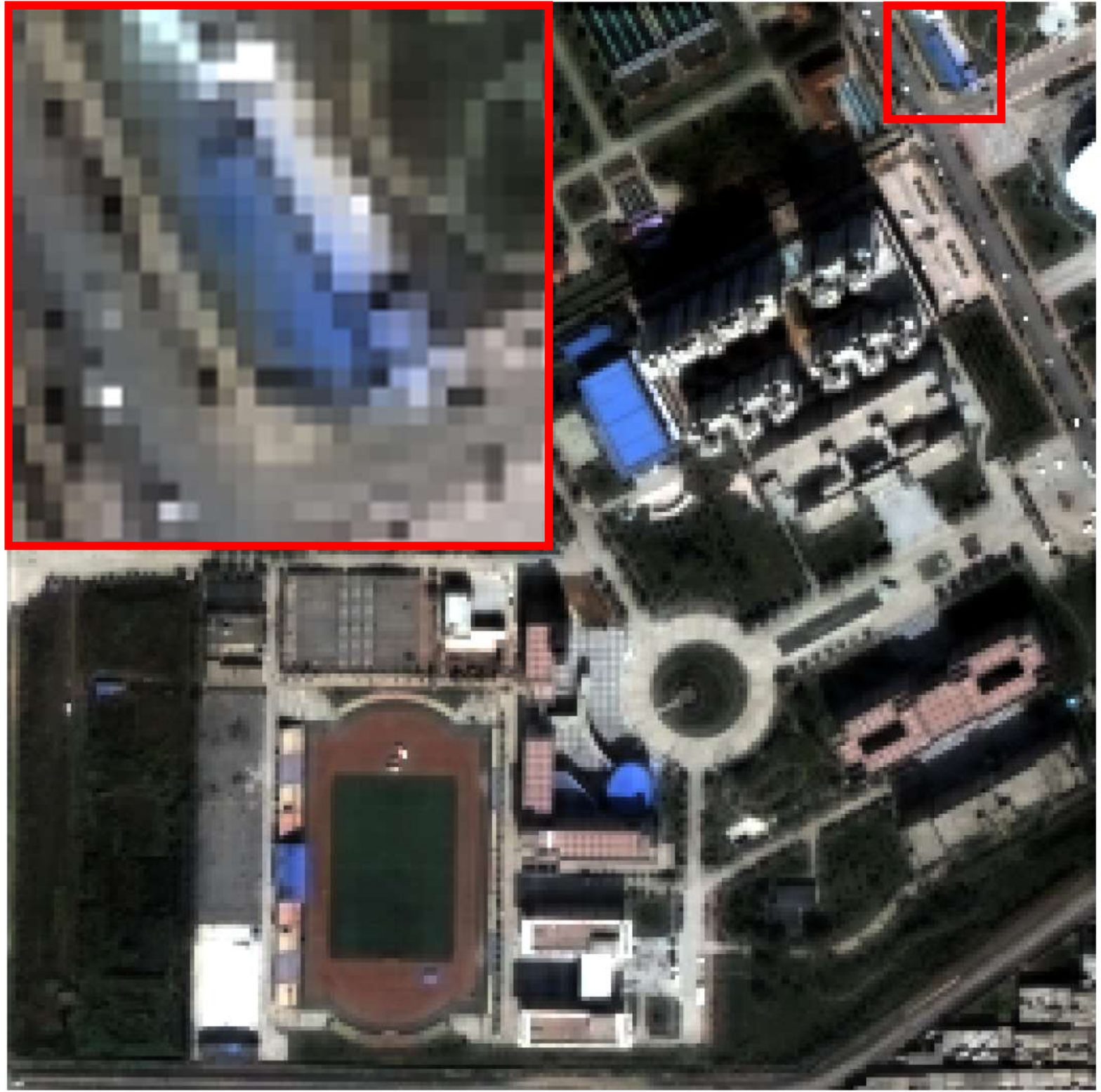}
\label{AWLP-H}}
\subfloat[]{\includegraphics[width=1.1in,trim=120 0 120 0,clip]{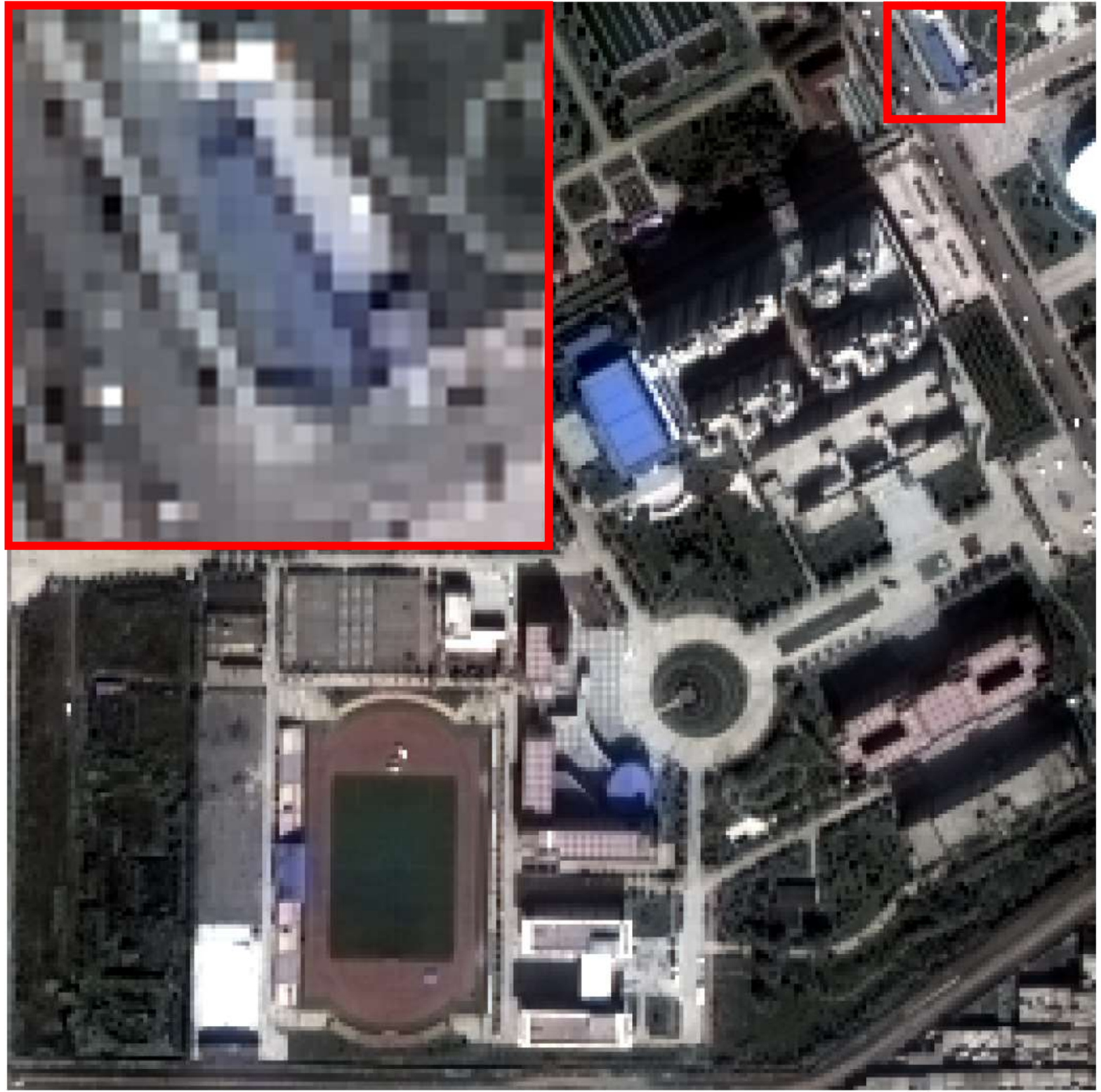}
\label{PWMBF}}
\subfloat[]{\includegraphics[width=1.1in,trim=120 0 120 0,clip]{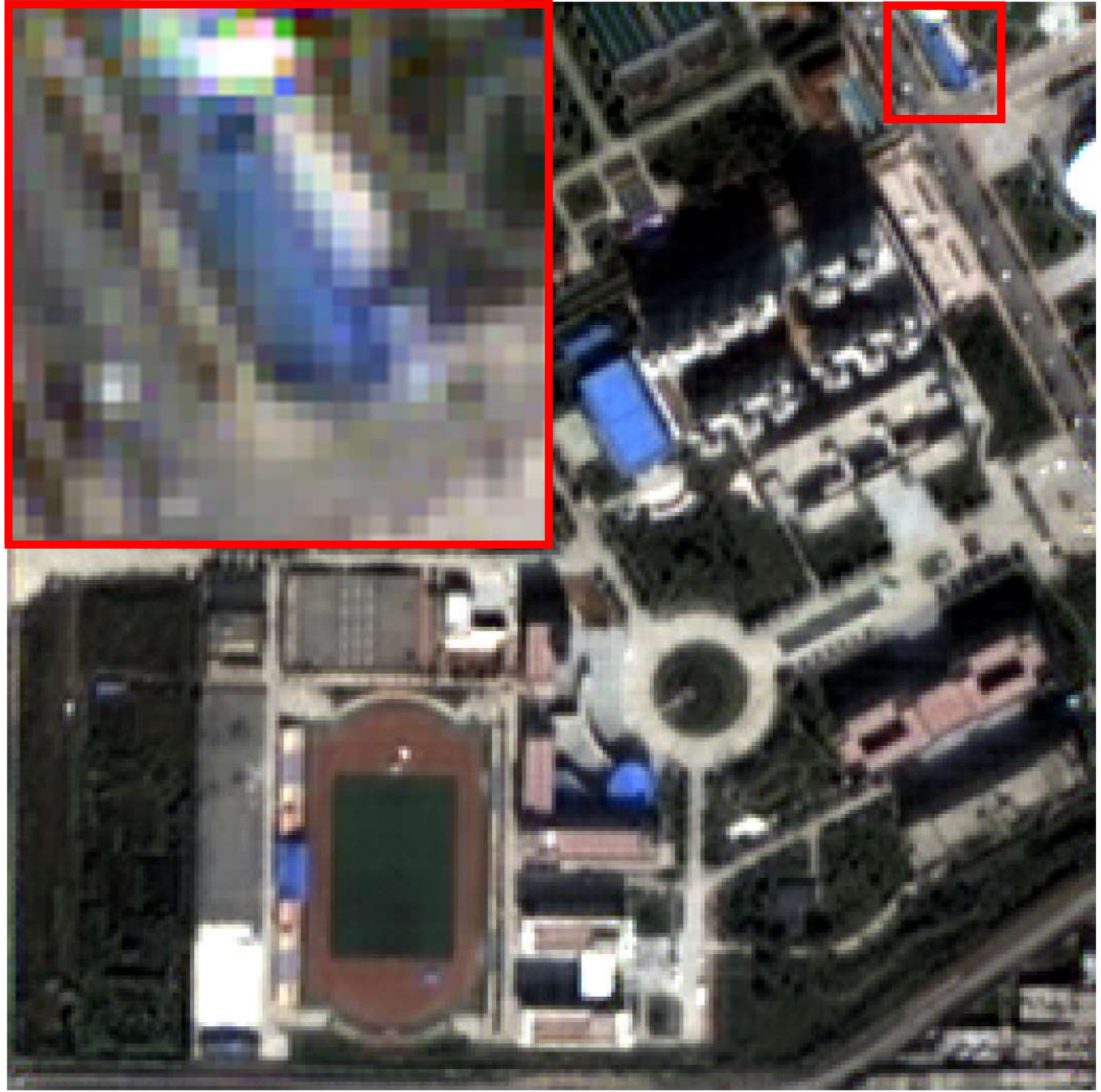}
\label{PanNet}}
\subfloat[]{\includegraphics[width=1.1in,trim=120 0 120 0,clip]{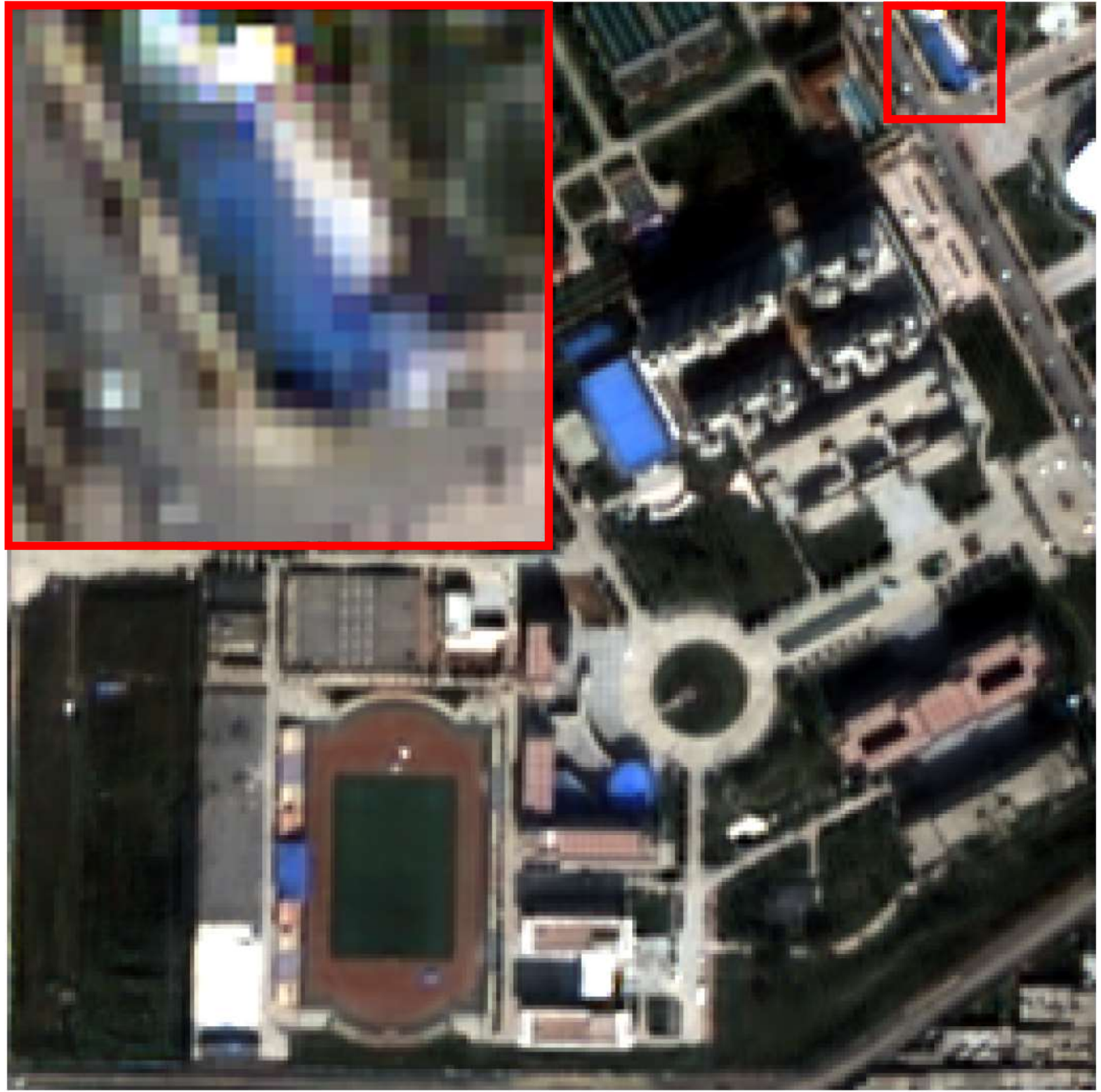}
\label{FusionNet}}

\vspace{-0.1in}
\subfloat[]{\includegraphics[width=1.1in,trim=120 0 120 0,clip]{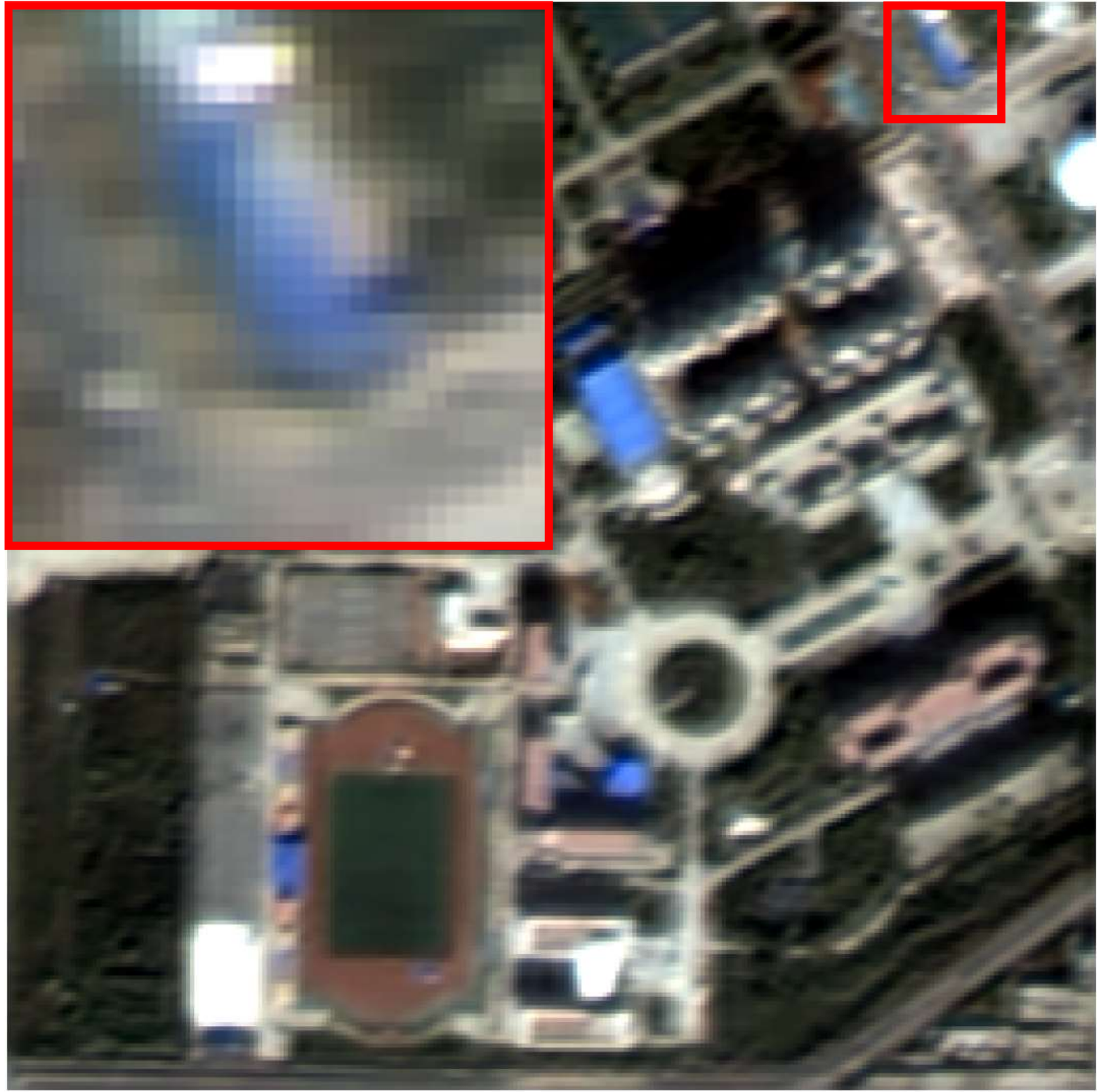}
\label{GTP-PNet}}
\subfloat[]{\includegraphics[width=1.1in,trim=120 0 120 0,clip]{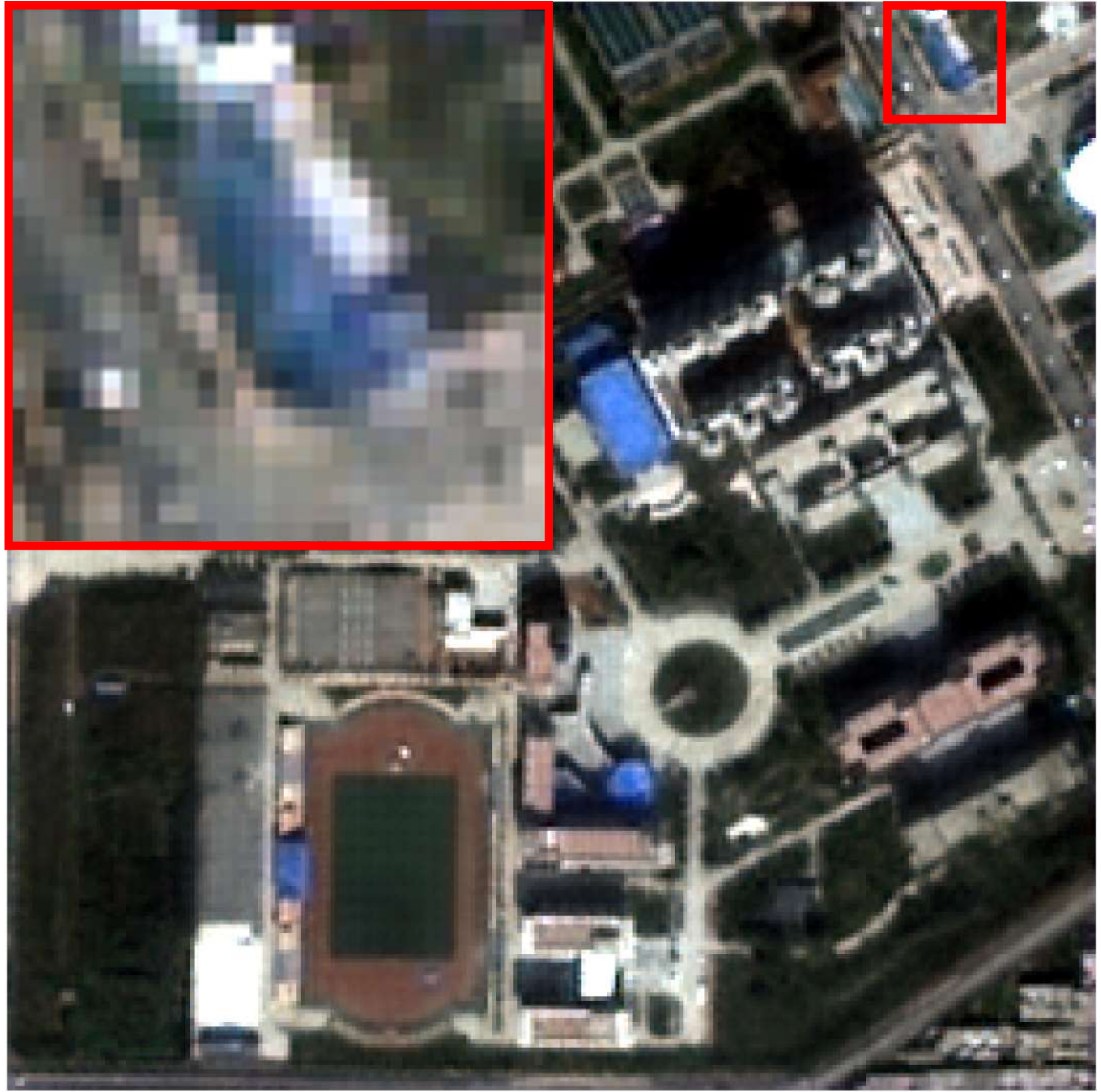}
\label{LPPN}}
\subfloat[]{\includegraphics[width=1.1in,trim=120 0 120 0,clip]{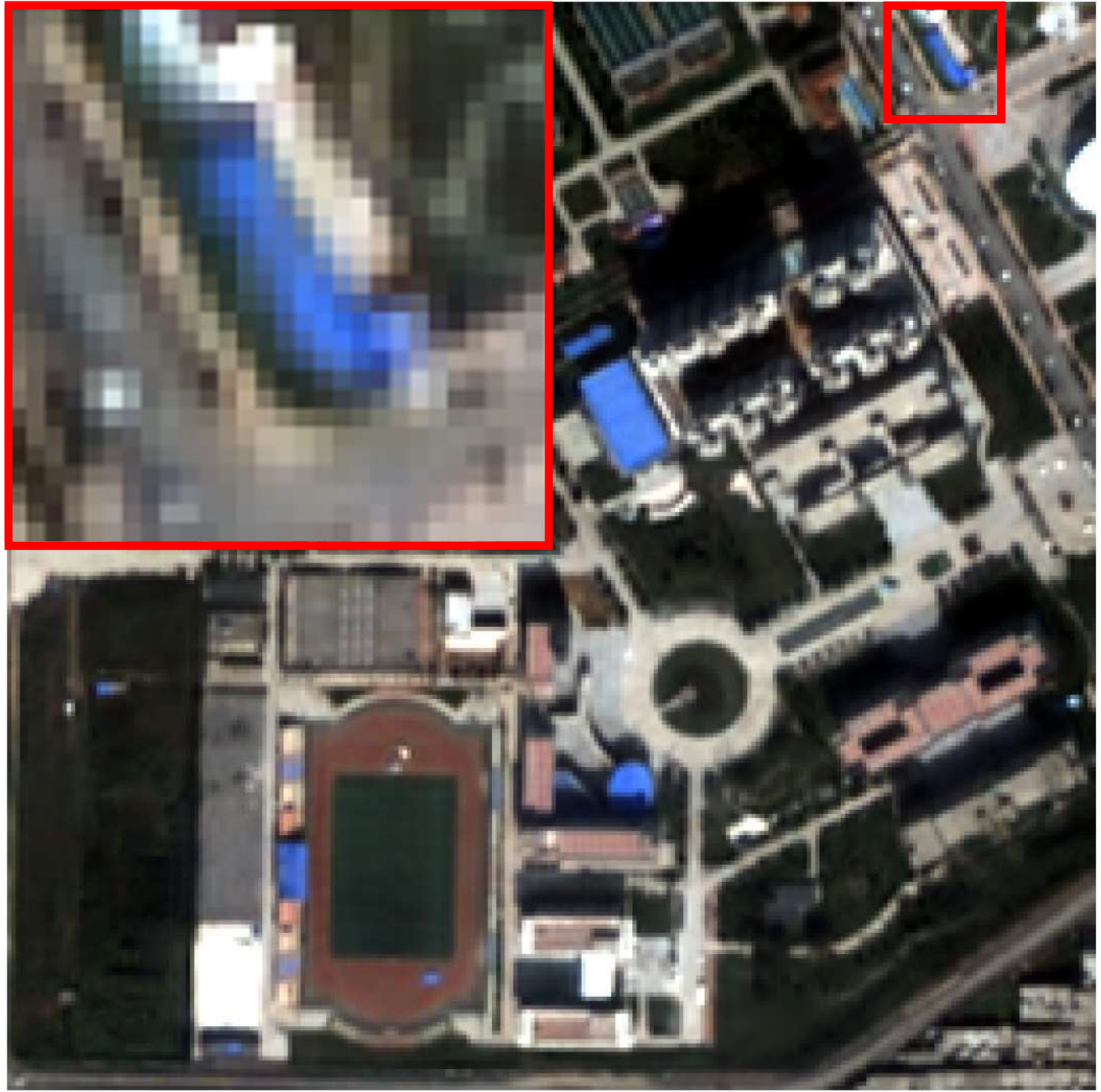}
\label{FAFNet}}
\caption{Visual comparison of different methods on QuickBird (QB) dataset at reduced resolution. (a) PAN. (b) Up-sampled MS. (c) Reference. (d) GSA \cite{4305344}. (e) BDSD-PC \cite{8693555}. (f) MTF-GLP-CBD \cite{4305345}. (g) AWLP-H \cite{vivone2019fast}. (h) PWMBF \cite{6951484}. (i) PanNet \cite{8237455}. (j) FusionNet \cite{9240949}. (k) GTP-PNet \cite{ZHANG2021223}. (l) LPPN \cite{JIN2022158}. (m) FAFNet.}
\label{fig_qb_reduce}
\end{figure*}

\begin{figure*}[h]
\centering
\subfloat[]{\includegraphics[width=1in,trim=120 0 120 0,clip]{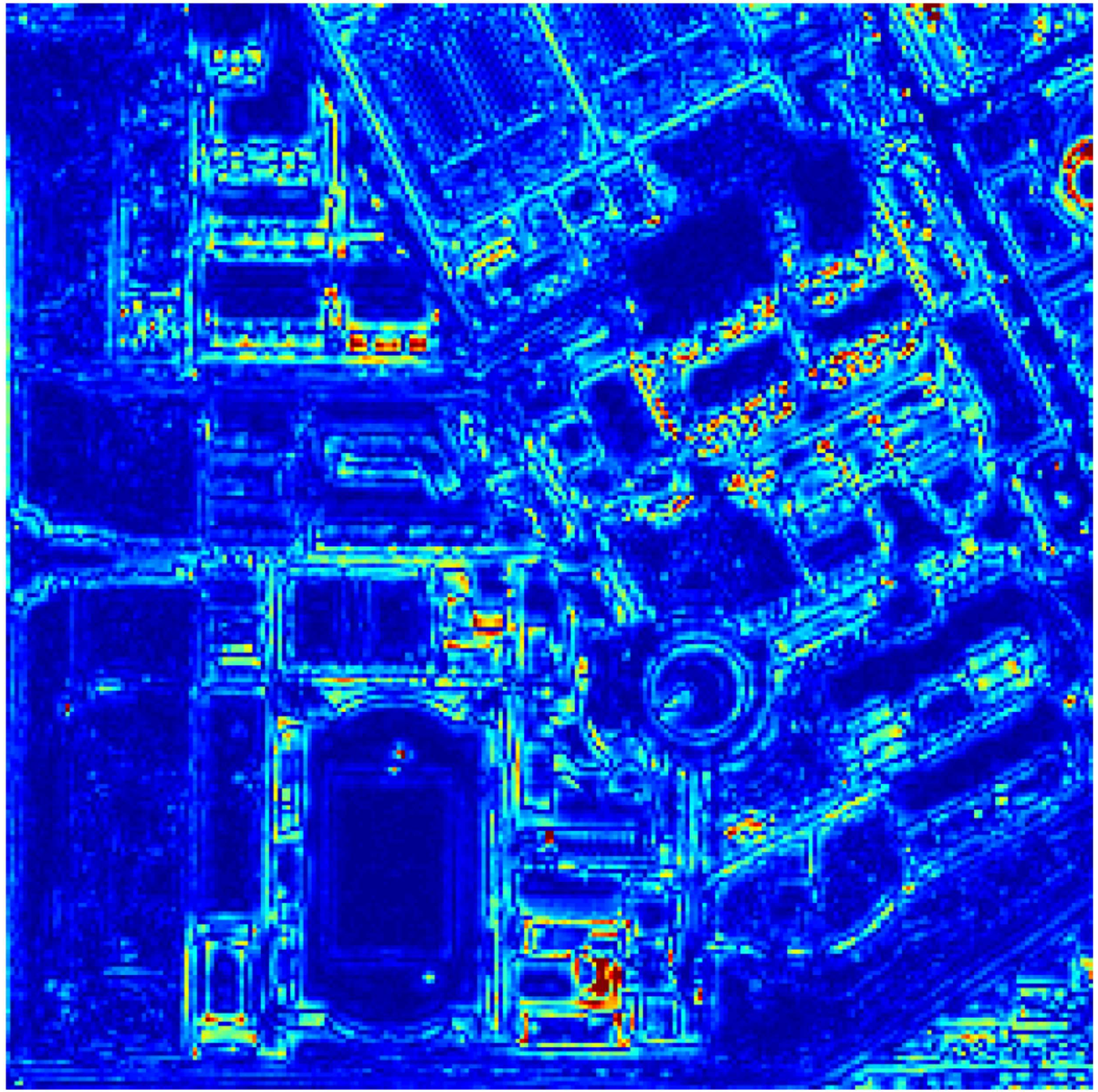}
\label{up-sampled MS-error}}
\subfloat[]{\includegraphics[width=1in,trim=120 0 120 0,clip]{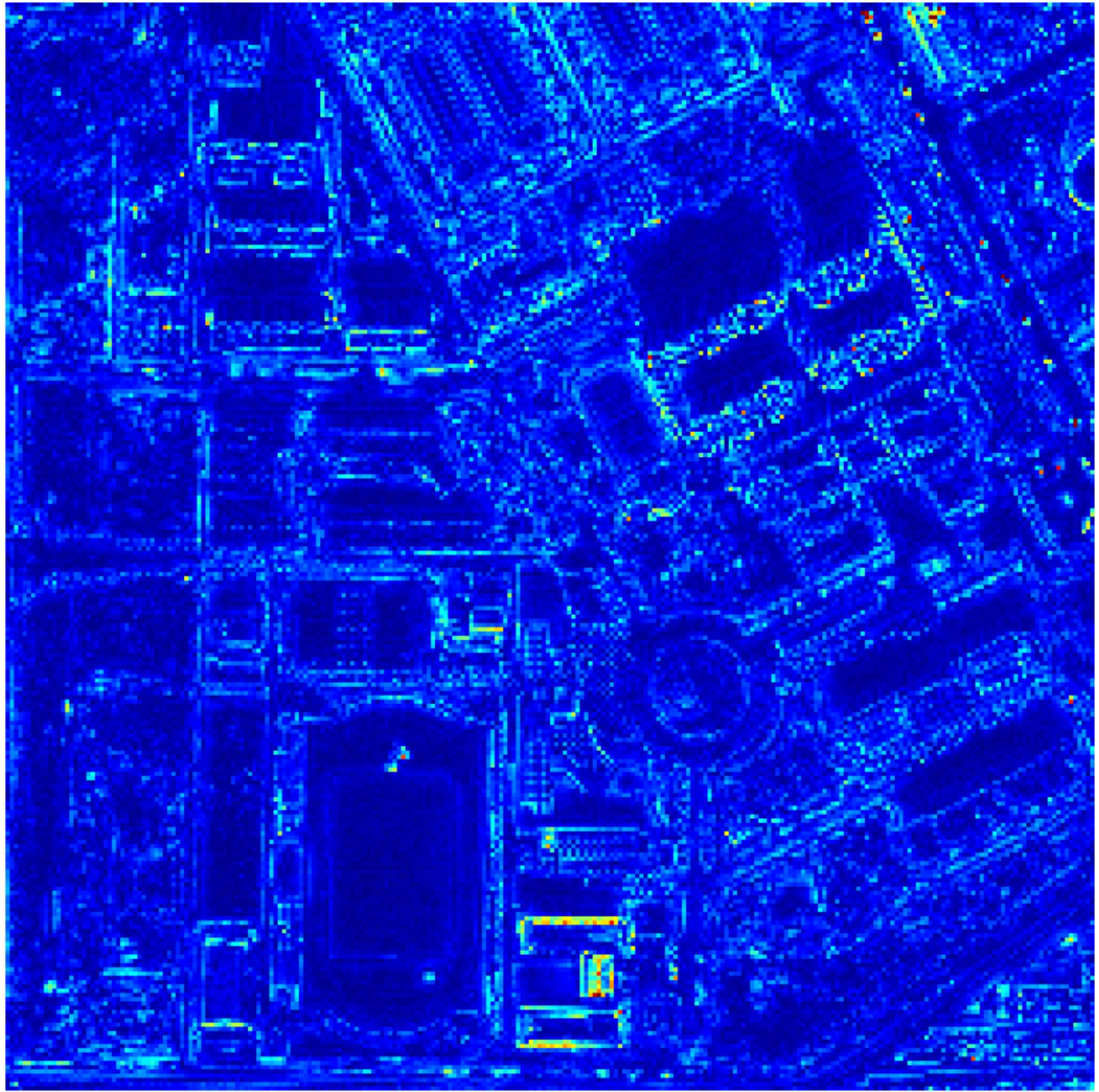}
\label{GSA}}
\subfloat[]{\includegraphics[width=1in,trim=120 0 120 0,clip]{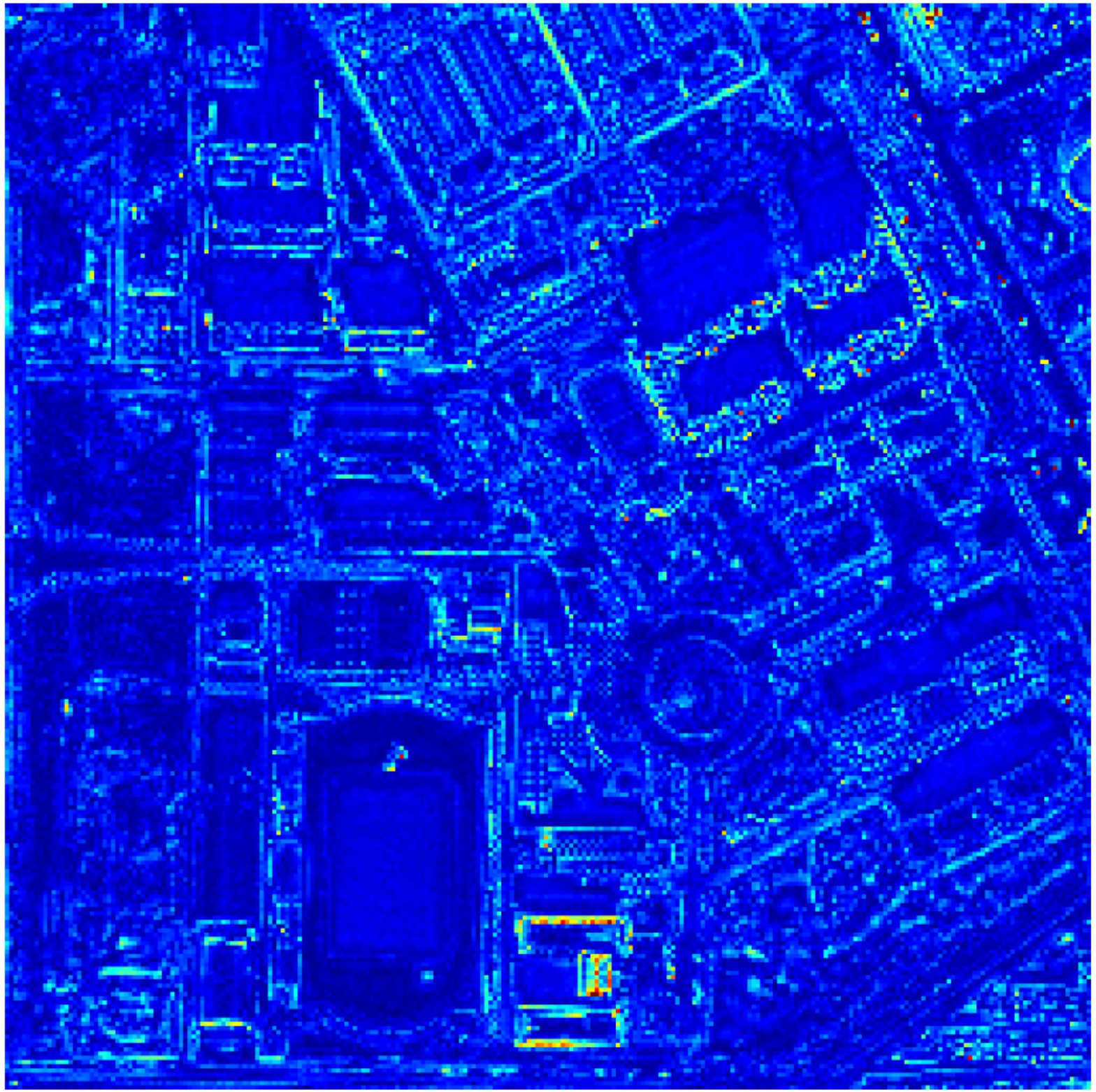}
\label{BDSD-PC}}
\subfloat[]{\includegraphics[width=1in,trim=120 0 120 0,clip]{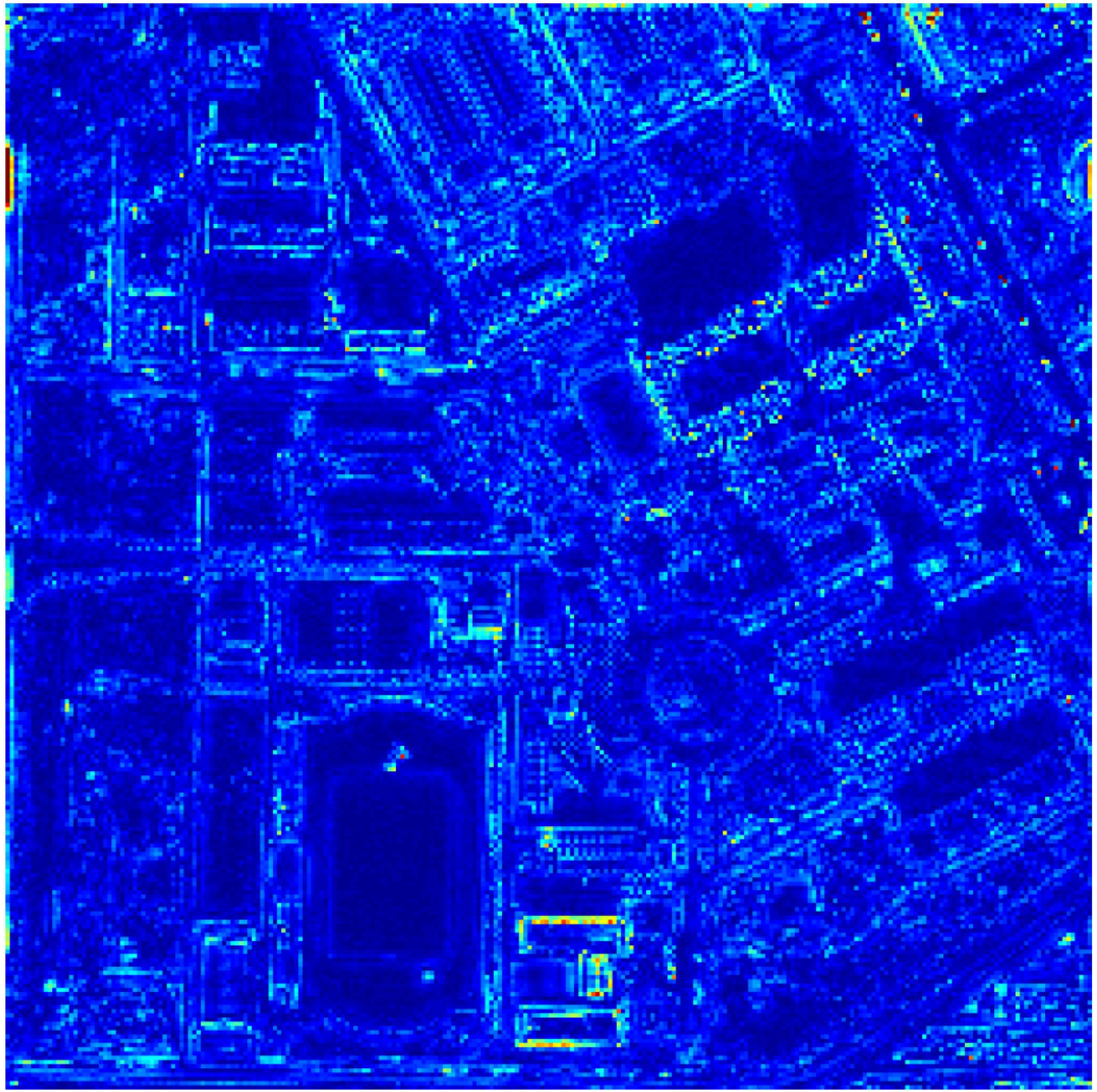}
\label{MTF-GLP-CBD}}
\subfloat[]{\includegraphics[width=1in,trim=120 0 120 0,clip]{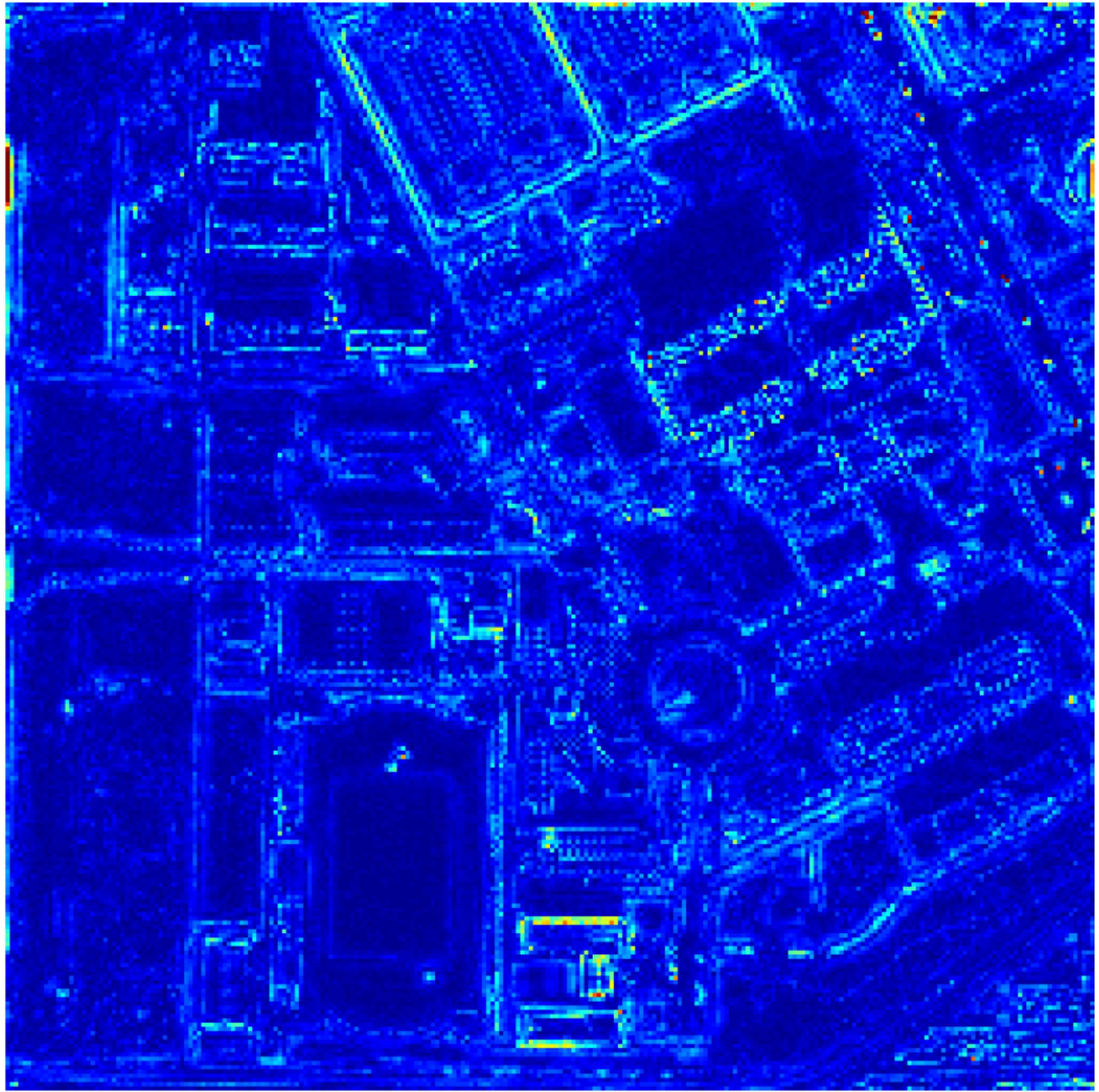}
\label{AWLP-H}}
\subfloat[]{\includegraphics[width=1in,trim=120 0 120 0,clip]{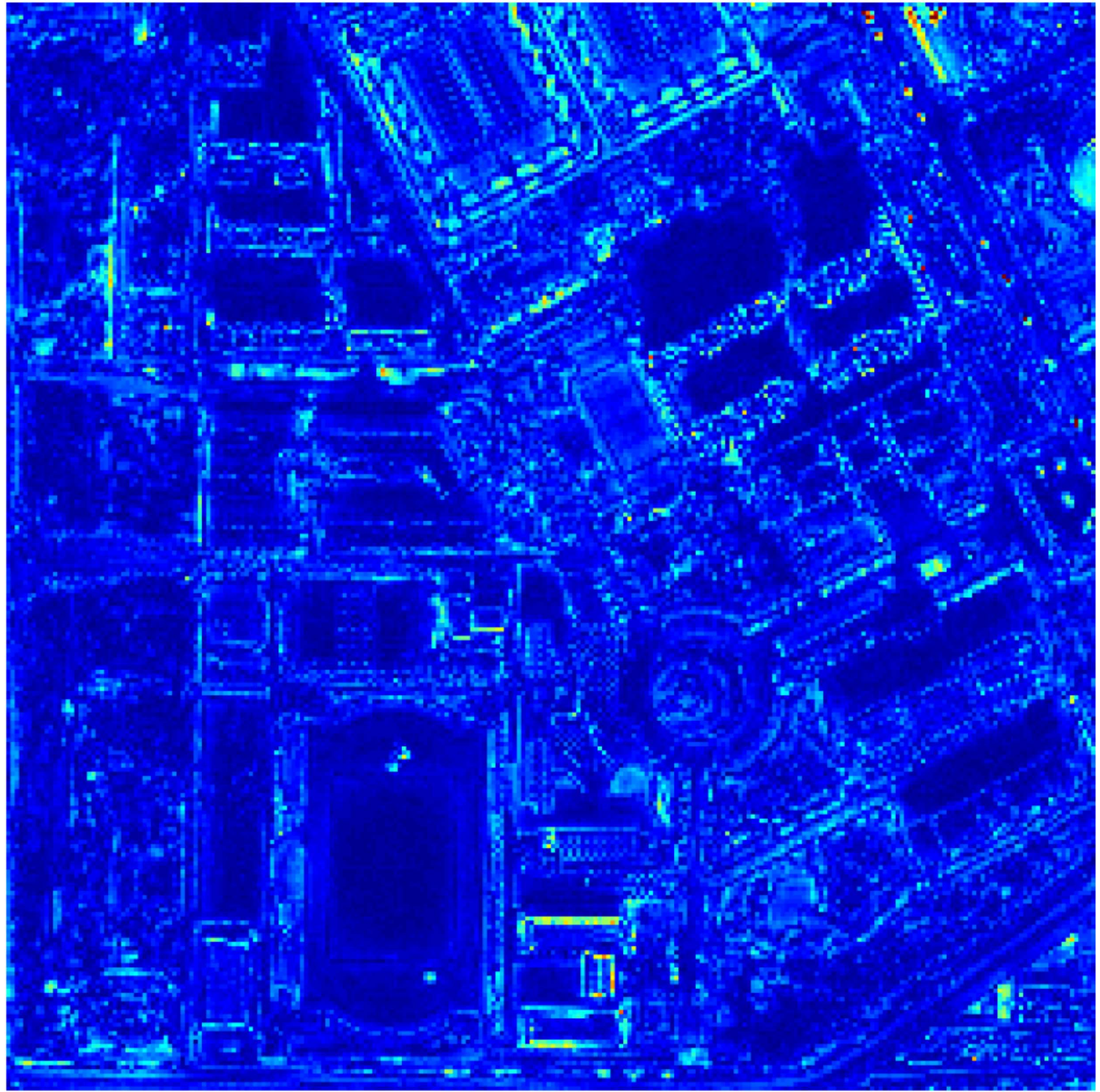}
\label{PWMBF}}

\vspace{-0.1in}
\subfloat[]{\includegraphics[width=1in,trim=120 0 120 0,clip]{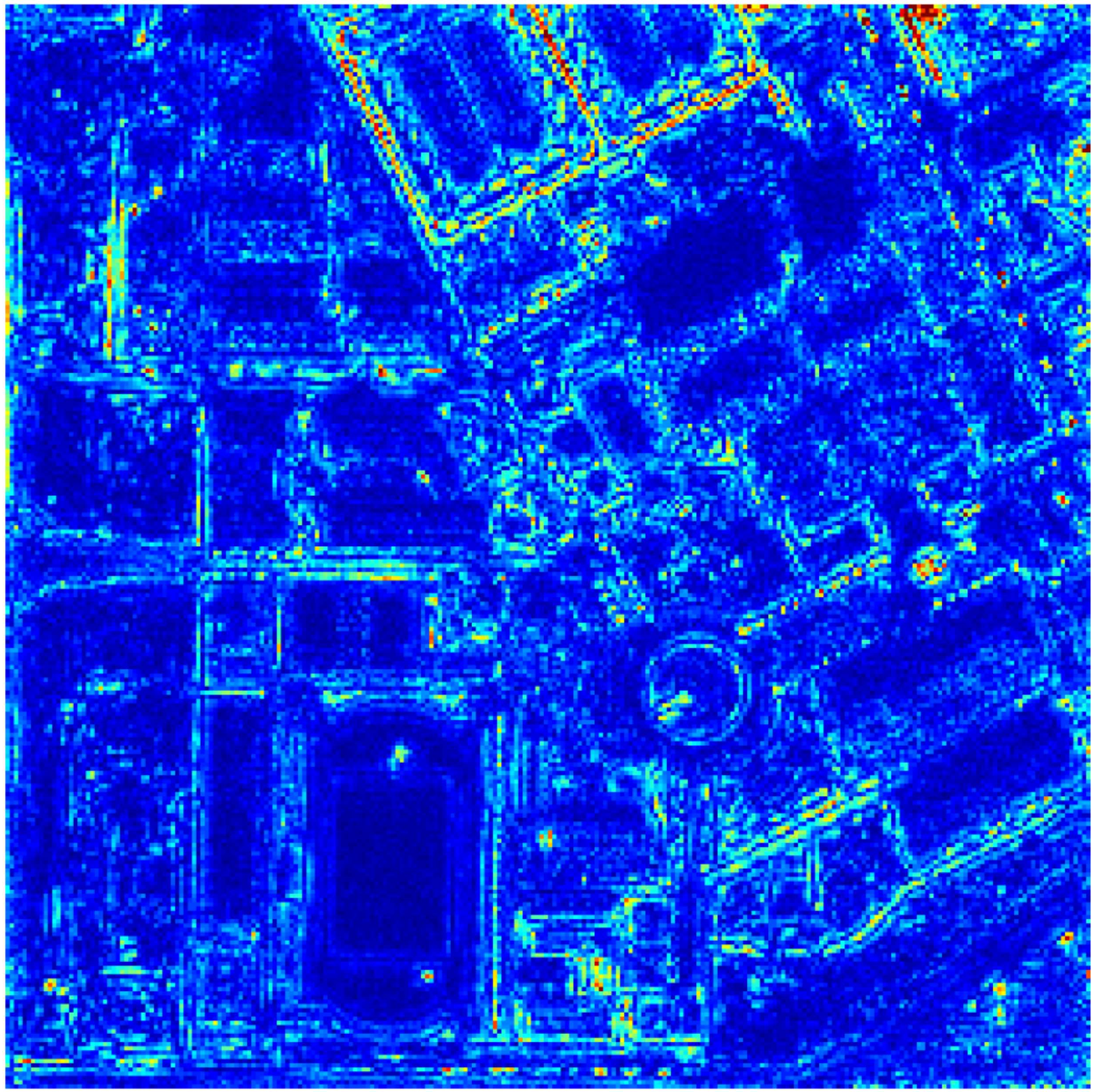}
\label{PanNet}}
\subfloat[]{\includegraphics[width=1in,trim=120 0 120 0,clip]{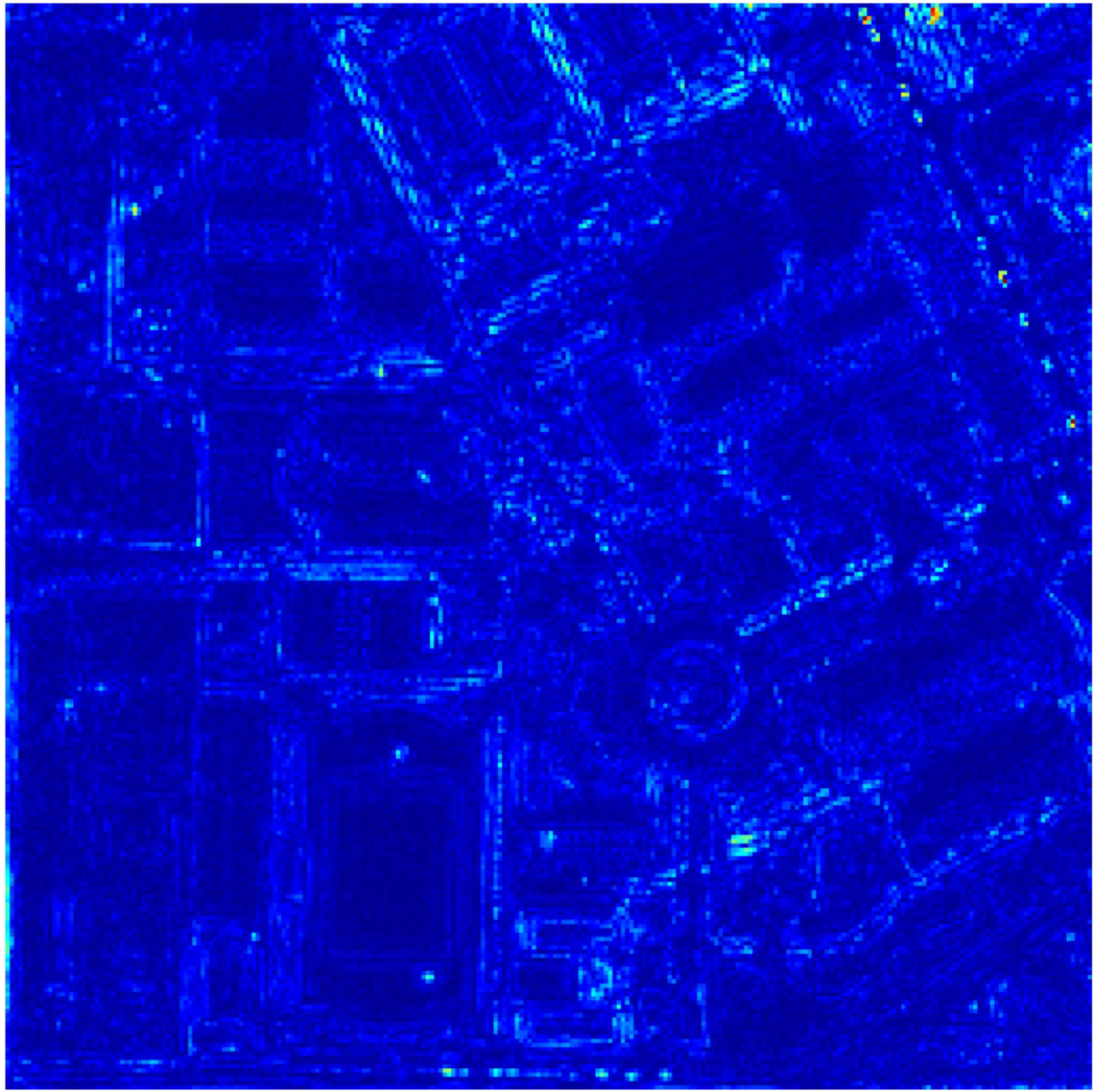}
\label{FusionNet}}
\subfloat[]{\includegraphics[width=1in,trim=120 0 120 0,clip]{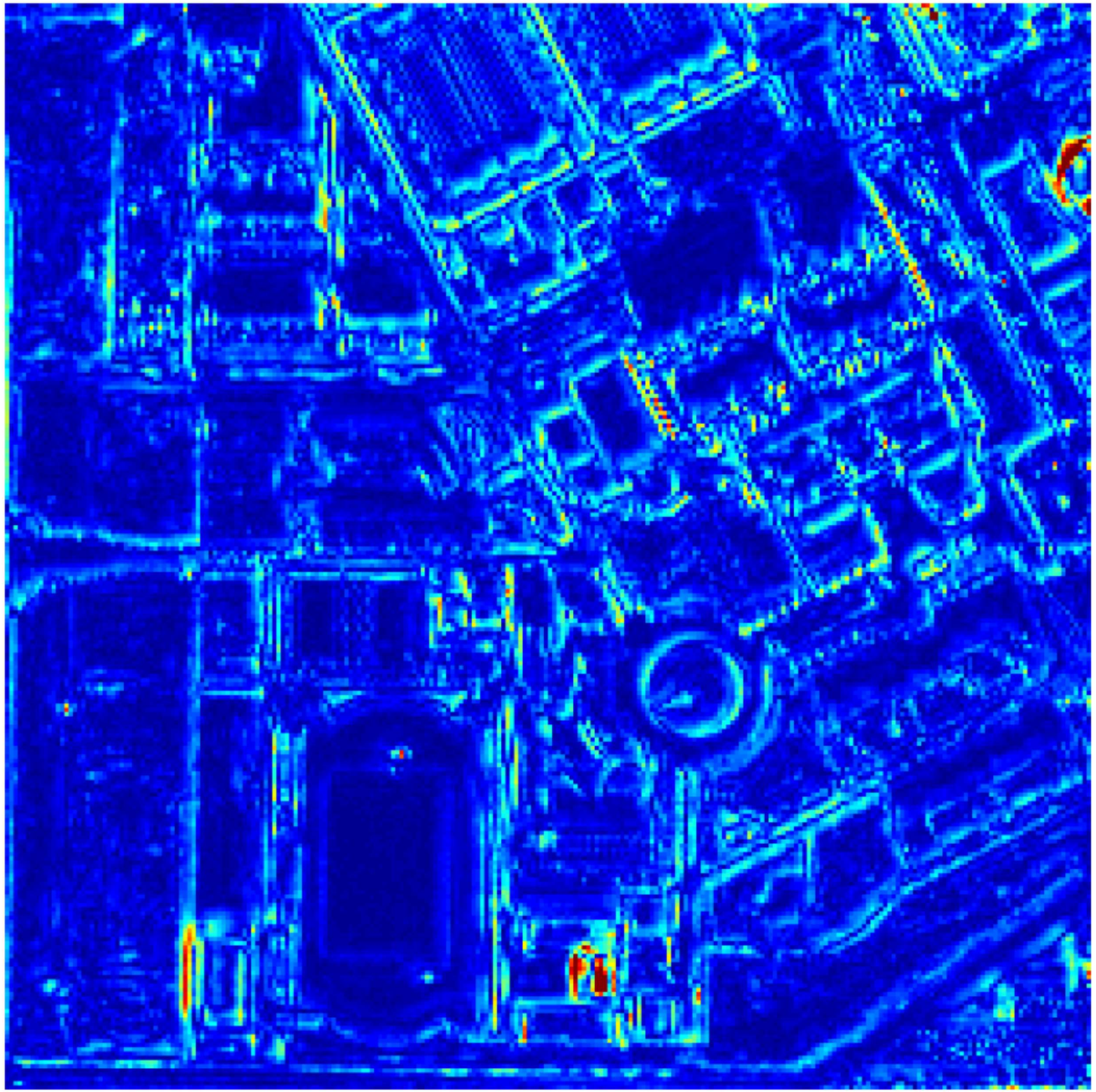}
\label{GTP-PNet}}
\subfloat[]{\includegraphics[width=1in,trim=120 0 120 0,clip]{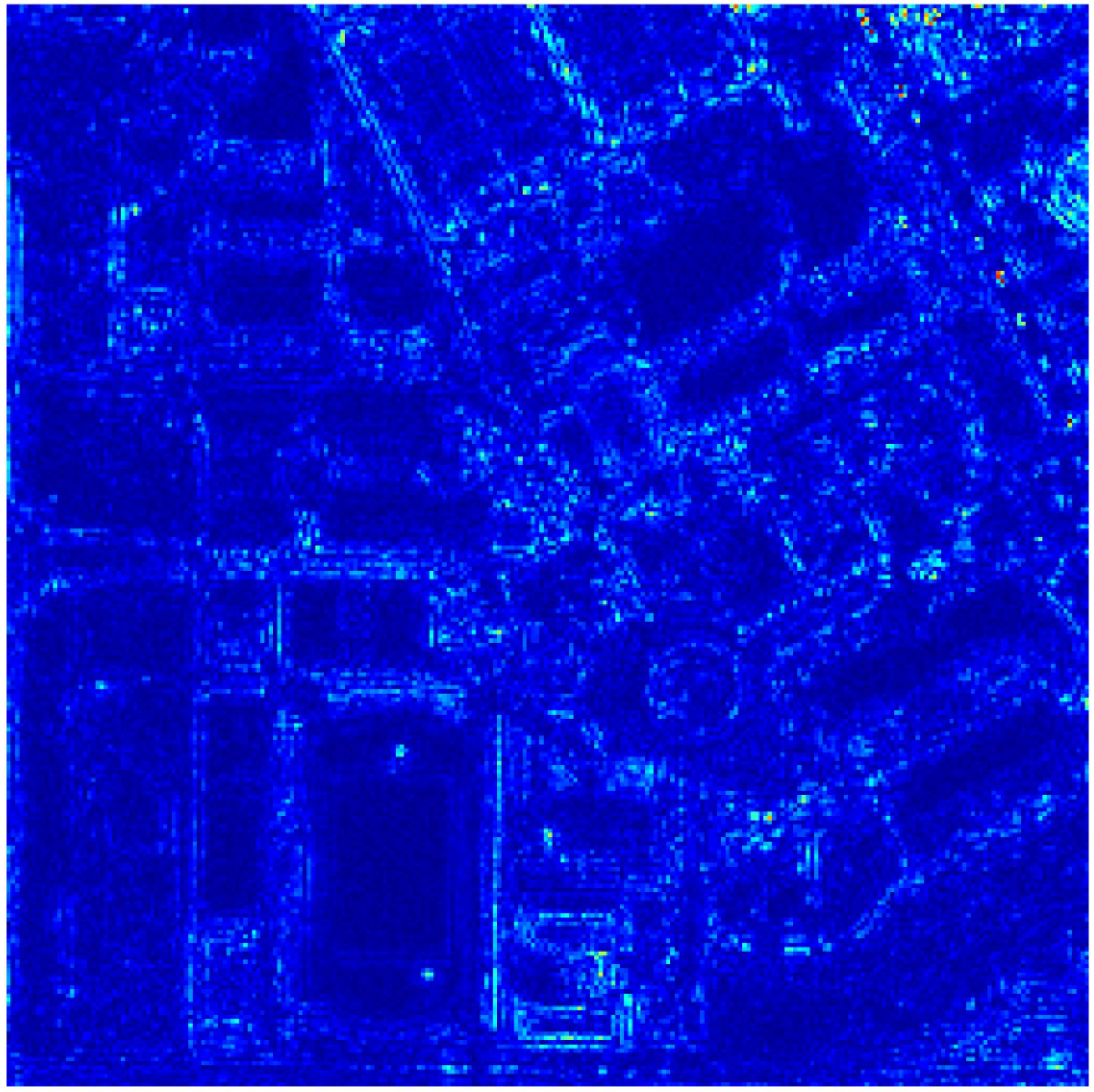}
\label{LPPN}}
\subfloat[]{\includegraphics[width=1in,trim=120 0 120 0,clip]{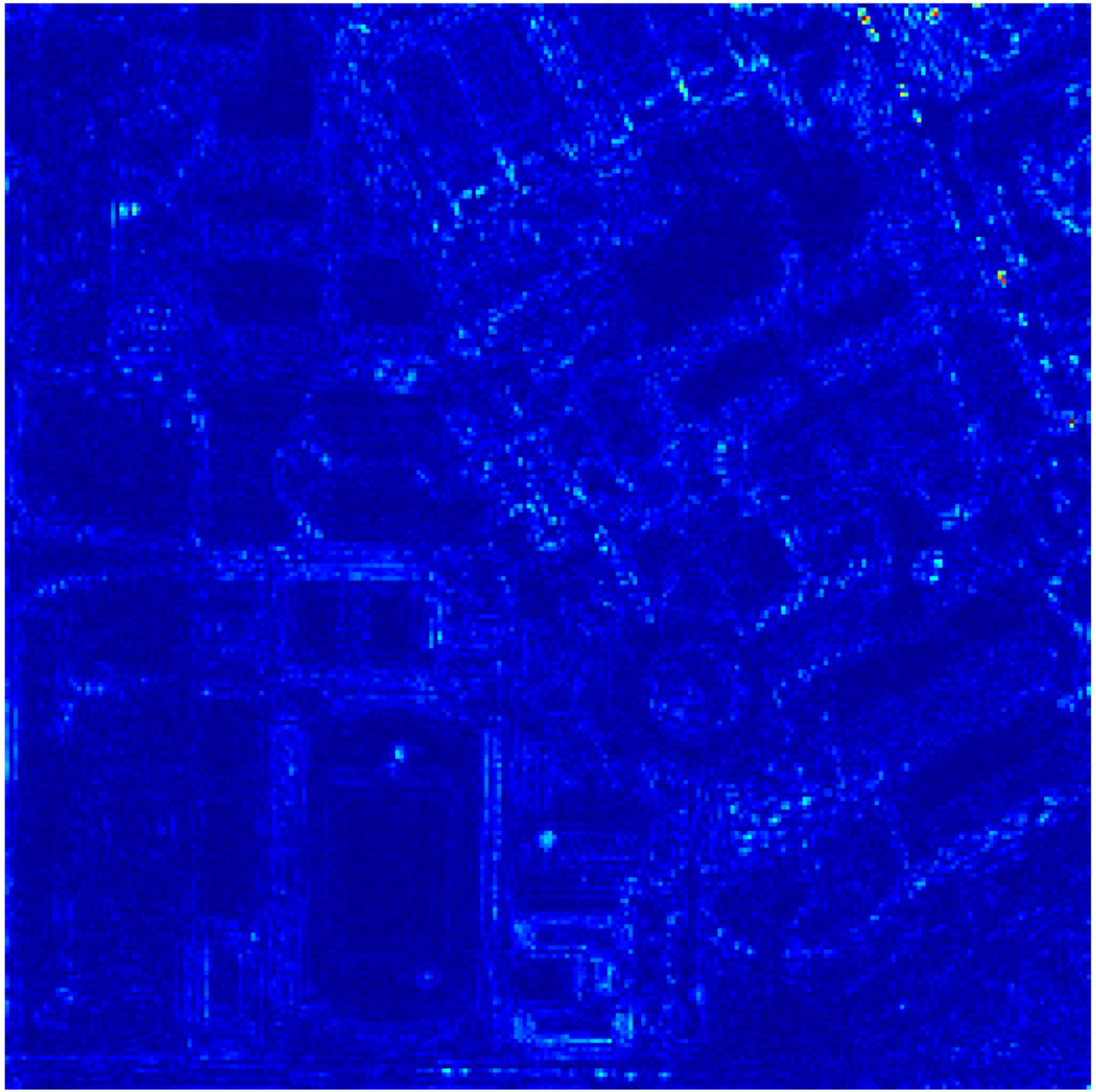}
\label{FAFNet}}
\subfloat[]{\includegraphics[width=1in,trim=120 0 120 0,clip]{figures/wv4/error/ref-error.pdf}
\label{Reference}}

\subfloat{\includegraphics[angle=-90,width=0.3\linewidth,trim=220 0 220 0,clip]{figures/wv4/error/bar.pdf}
}
\caption{Absolute error maps (AEMs) between reference image and (a)Up-sampled MS, (b) GSA \cite{4305344}, (c) BDSD-PC \cite{8693555}, (d) MTF-GLP-CBD \cite{4305345}, (e) AWLP-H \cite{vivone2019fast}, (f) PWMBF \cite{6951484}, (g) PanNet \cite{8237455}, (h) FusionNet \cite{9240949}, (i) GTP-PNet \cite{ZHANG2021223}, (j) LPPN \cite{JIN2022158}, (k) FAFNet and (l) Reference on QB dataset at reduced resolution.}
\label{fig_qb_reduce_error}
\end{figure*}

Figs. \ref{fig_qb_reduce}-\ref{fig_qb_reduce_error} show fusion results and the corresponding AEMs on QB dataset, in which Figs. \ref{fig_qb_reduce}(a)-(c) show the PAN, up-sampled MS and reference images, and the fusion results of comparison methods and the proposed FAFNet are shown in Figs. \ref{fig_qb_reduce}(d)-(m). The pansharpened results of GSA \cite{4305344}, BDSD-PC \cite{8693555}, MTF-GLP-CBD \cite{4305345}, AWLP-H \cite{vivone2019fast}, PWMBF \cite{6951484}, PanNet \cite{8237455}, GTP-PNet \cite{ZHANG2021223}, FusionNet \cite{9240949} and LPPN \cite{JIN2022158} are affected by different levels of spectral distortion and spatial degradation, especially for the color and the shape of buildings in the zoomed-in regions and small objects on the playground. The fusion product of FAFNet is the closest to the reference. Additionally, absolute error maps (AEMs) between fusion results and the reference image are demonstrated in Fig. \ref{fig_qb_reduce_error}. From Fig. \ref{fig_qb_reduce_error}, we can find that FAFNet and FusionNet \cite{9240949} have less information loss. The same conclusion can also be drawn from the quantitative assessments shown in Table \ref{tab_qb} that the proposed FAFNet and FusionNet \cite{9240949} obtain the best and the second-best indexes. All the indexes listed in Table \ref{tab_qb} are calculated on the 156 pairs of test data from the QB dataset, from which we can conclude that FAFNet is superior to other compared methods on QB dataset at the reduced resolution.

\begin{table*}[htbp]
\caption{Average Quantitative Results on 156 Pairs of Test Data from QB Dataset.\label{tab_qb}}
\centering
\begin{tabular}{c|cccc|ccc}
\toprule
~ & \multicolumn{4}{c|}{\textbf{Reduced resolution}} & \multicolumn{3}{c}{\textbf{Full resolution}} \\		
\midrule
Methods & ERGAS & Q4 & SAM & SCC  & QNR & $D_\lambda$ & $D_s$ \\
\midrule
Ideal value & 0 & 1 & 0 & 1 & 1 & 0 & 0 \\	
\midrule
GSA \cite{4305344} & 2.5949±0.2115 & 0.8446±0.0080 & 2.9398±0.5032 & 0.9289±0.0004  & 0.7946±0.0059 & 0.0753±0.0006 & 0.1412±0.0055\\
BDSD-PC \cite{8693555}  & 2.5481±0.4372 & 0.8742±0.0016 & 3.0162±0.6257 & 0.9356±0.0003 & 0.8820±0.0014 & 0.0375±0.0004 & 0.0840±0.0007\\
MTF-GLP-CBD \cite{4305345} & 2.5192±0.3413 & 0.8733±0.0021 & 2.9481±0.5714 & 0.9301±0.0004  & 0.8477±0.0016 & 0.0653±0.0004 & 0.0935±0.0006\\
AWLP-H \cite{vivone2019fast} & 2.3095±0.2934 & 0.9057±0.0008 & 2.4911±0.4131 & 0.9443±0.0005 & 0.8706±0.0011 & 0.0567±0.0003 & 0.0774±0.0004\\
PWMBF \cite{6951484} & 2.5040±0.3275 & 0.8534±0.0028 & 3.2283±0.6586 & 0.9317±0.0003 & 0.8289±0.0014 & 0.0778±0.0004 & 0.1015±0.0006\\
PanNet \cite{8237455} & 1.8434±0.1917 & 0.9100±0.0019 & 2.6711±0.4395 & 0.9533±0.0002 & 0.8911±0.0010 & 0.0448±0.0002 & 0.0674±0.0004\\
FusionNet \cite{9240949}  & \underline{1.5123±0.1034} & \underline{0.9355±0.0009} & \underline{2.0824±0.2044} & \underline{0.9705±0.0001} & 0.9167±0.0006 & 0.0269±0.0002 & 0.0581±0.0003 \\
GTP-PNet \cite{ZHANG2021223} & 2.8528±0.5879 & 0.8398±0.0015 & 3.2133±0.6455 & 0.9009±0.0005 & 0.9231±0.0015 & 0.0318±0.0002 & 0.0469±0.0008\\
LPPN \cite{JIN2022158} & 1.7728±0.1726 & 0.9223±0.0010 & 2.3360±0.2806 & 0.9579±0.0001 & \underline{0.9416±0.0006} & \underline{0.0261±0.0005} & \underline{0.0332±0.0001}\\
FAFNet & \textbf{1.2712±0.0943} & \textbf{0.9580±0.0004} & \textbf{1.7147±0.1559} & \textbf{0.9785±0.0000} & \textbf{0.9695±0.0002} & \textbf{0.0123±0.0001} & \textbf{0.0185±0.0001}\\
\bottomrule
\end{tabular}
\end{table*}

\begin{figure*}[htbp]
\centering
\subfloat[]{\includegraphics[width=1.1in,trim=120 0 120 0,clip]{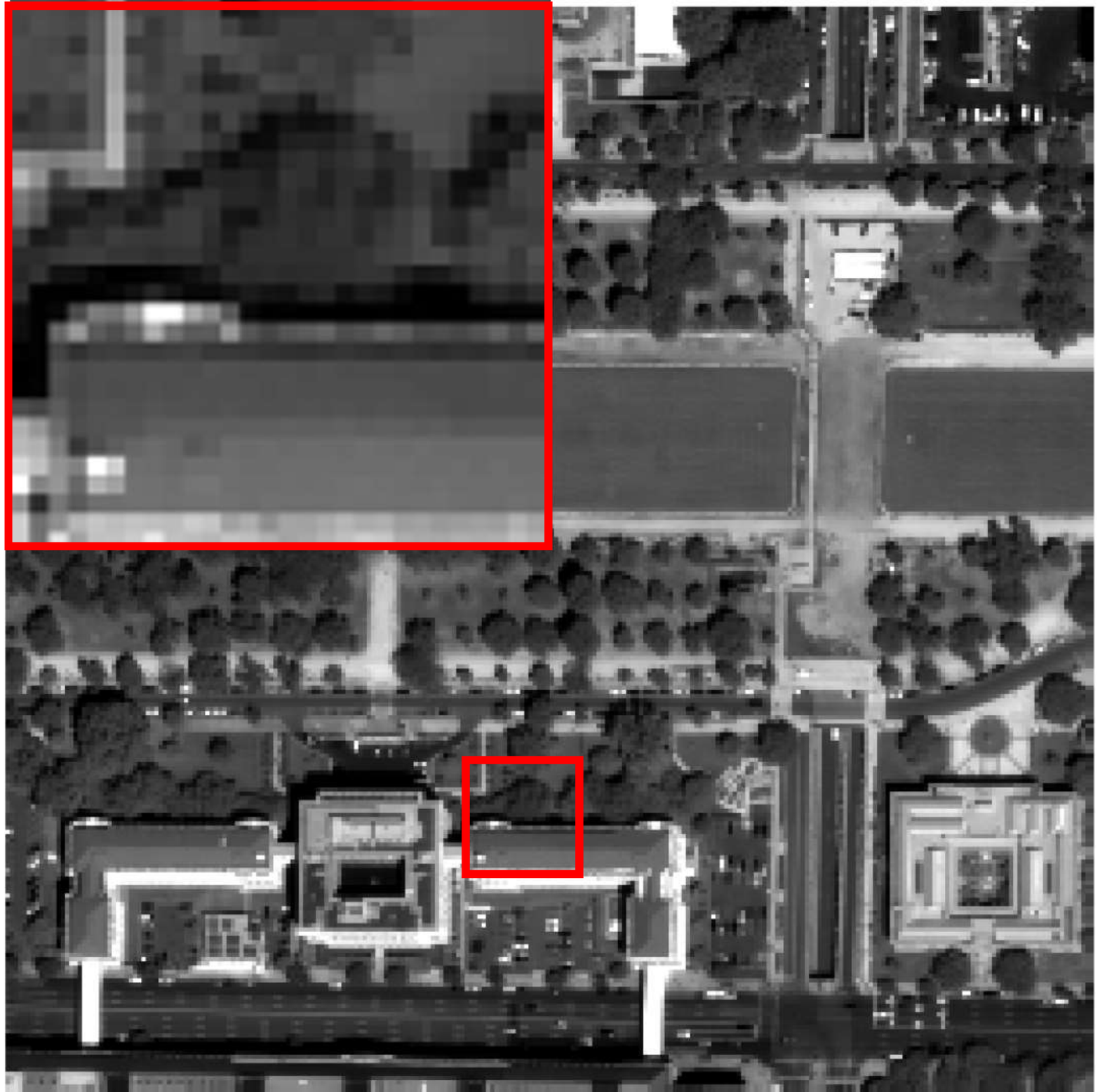}
\label{PAN}}
\subfloat[]{\includegraphics[width=1.1in,trim=120 0 120 0,clip]{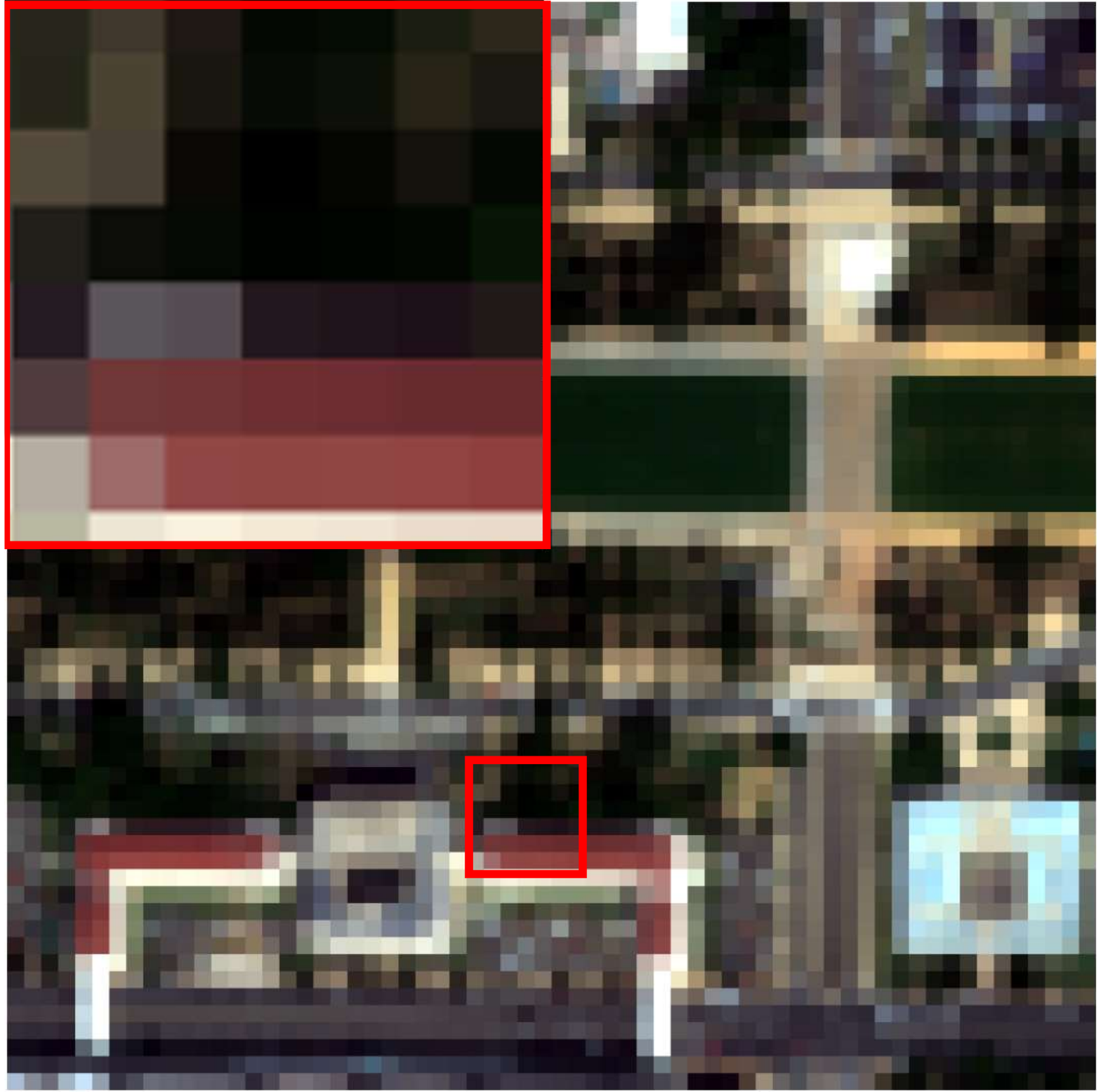}
\label{up-sampled MS}}
\subfloat[]{\includegraphics[width=1.1in,trim=120 0 120 0,clip]{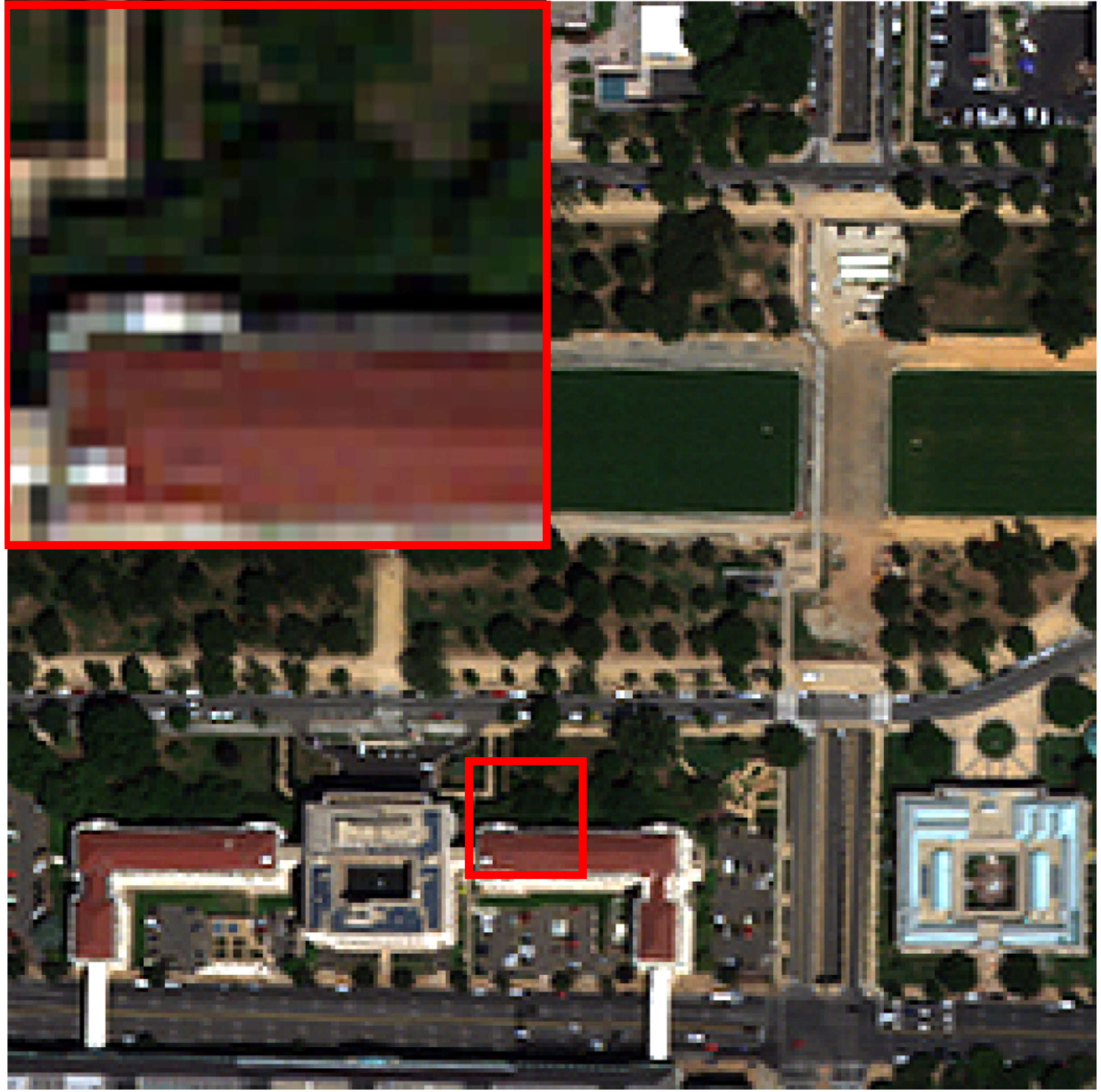}
\label{Reference}}
\subfloat[]{\includegraphics[width=1.1in,trim=120 0 120 0,clip]{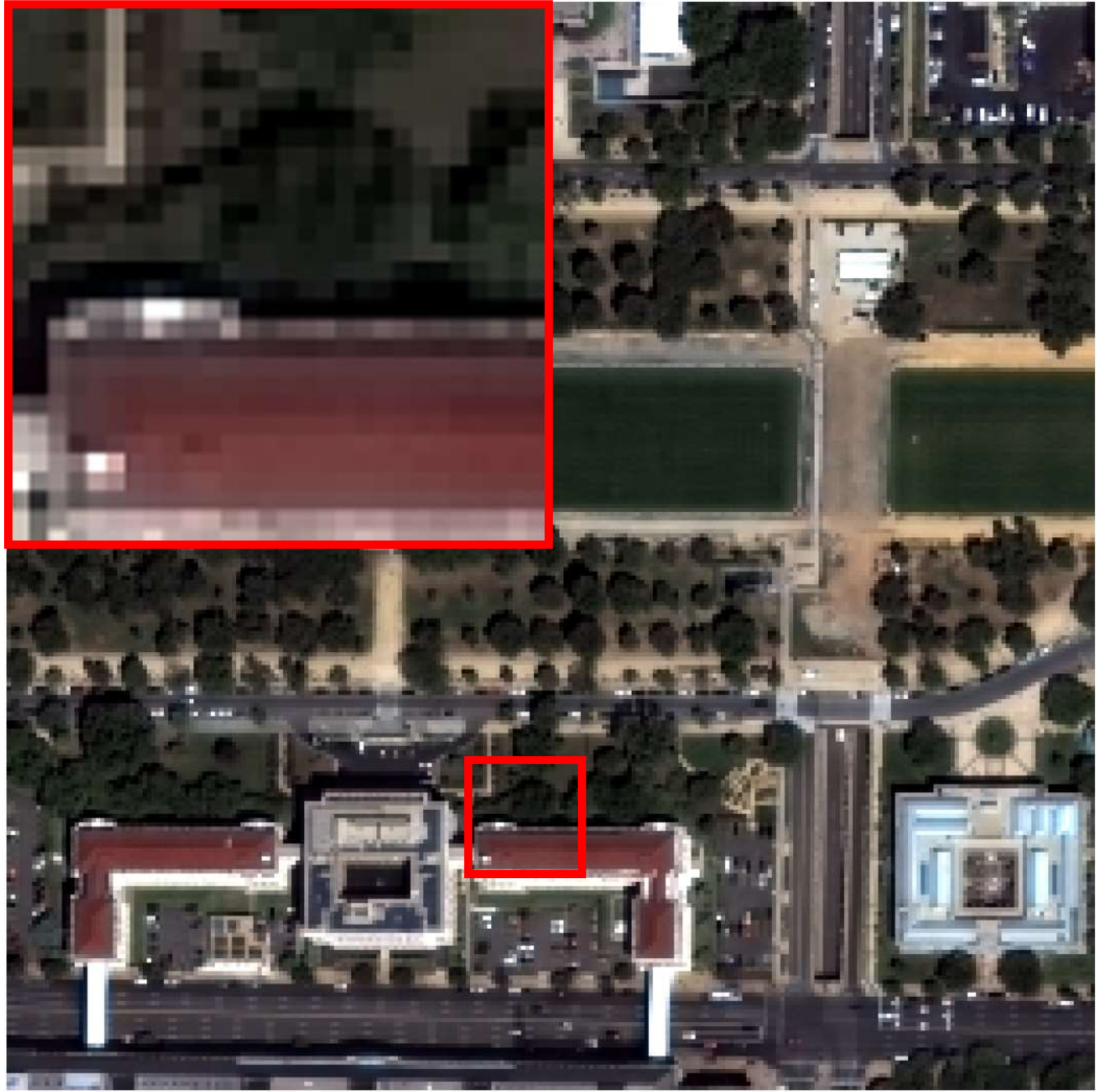}
\label{GSA}}
\subfloat[]{\includegraphics[width=1.1in,trim=120 0 120 0,clip]{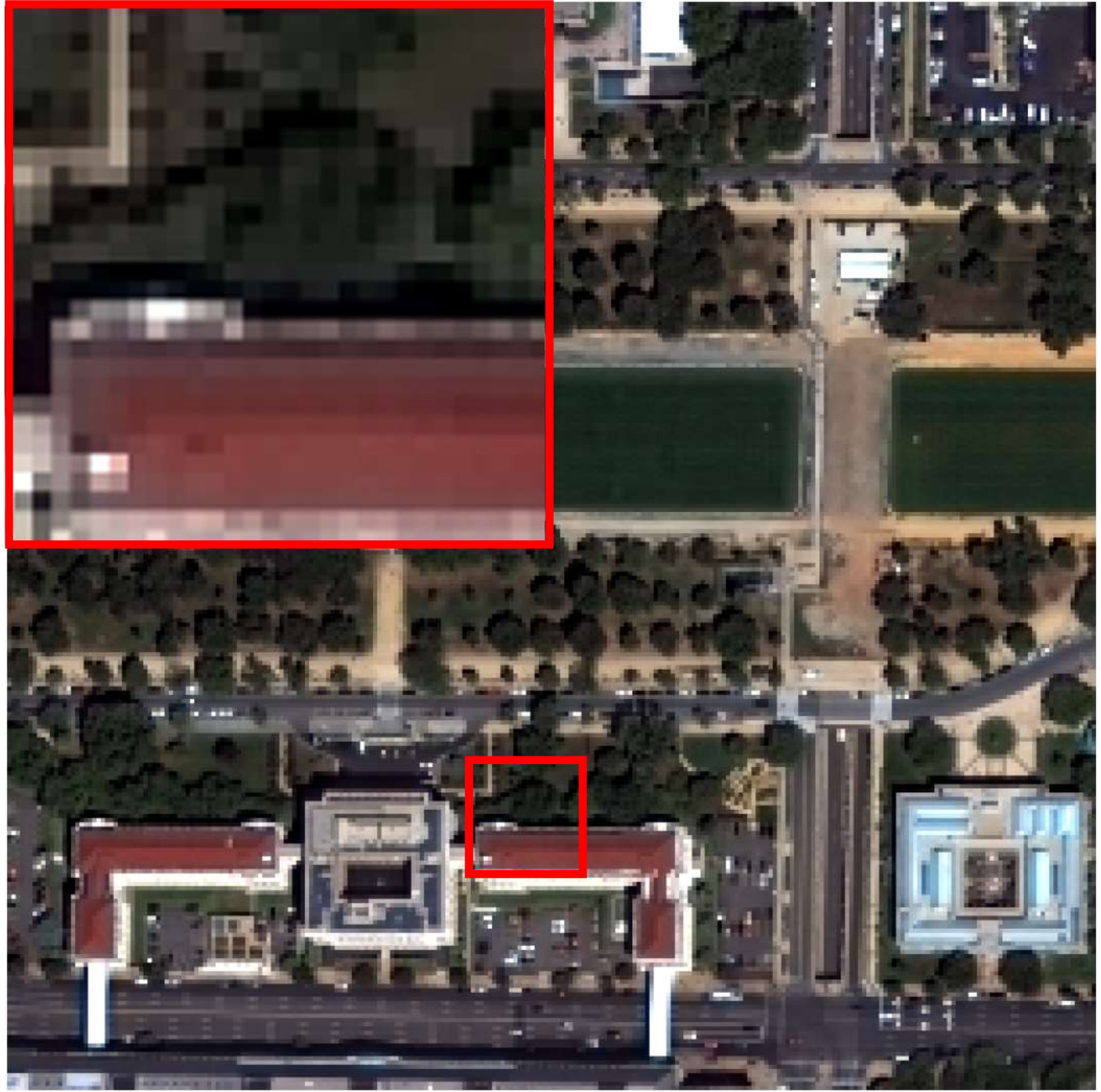}
\label{BDSD-PC}}

\vspace{-0.1in}
\subfloat[]{\includegraphics[width=1.1in,trim=120 0 120 0,clip]{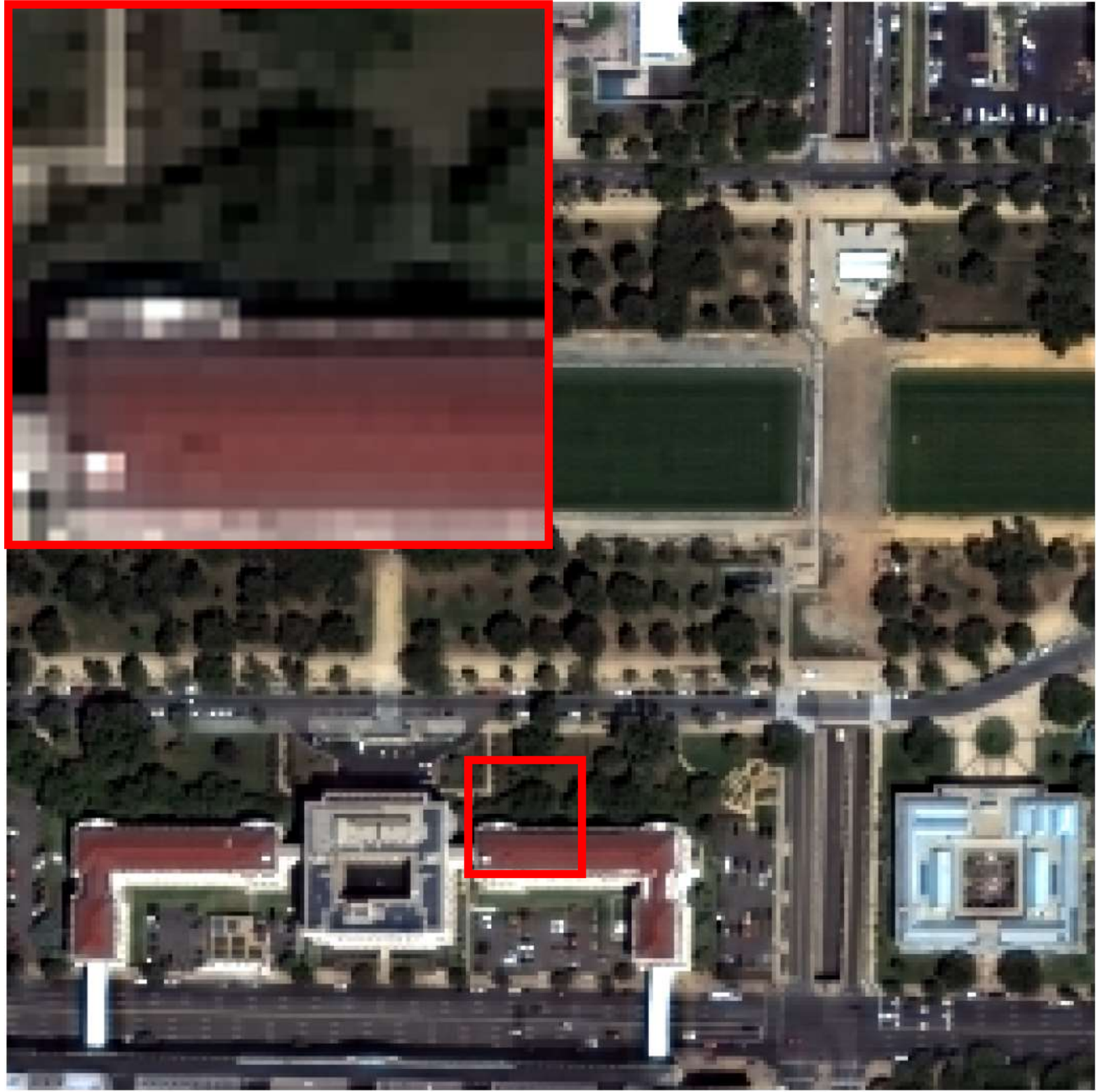}
\label{MTF-GLP-CBD}}
\subfloat[]{\includegraphics[width=1.1in,trim=120 0 120 0,clip]{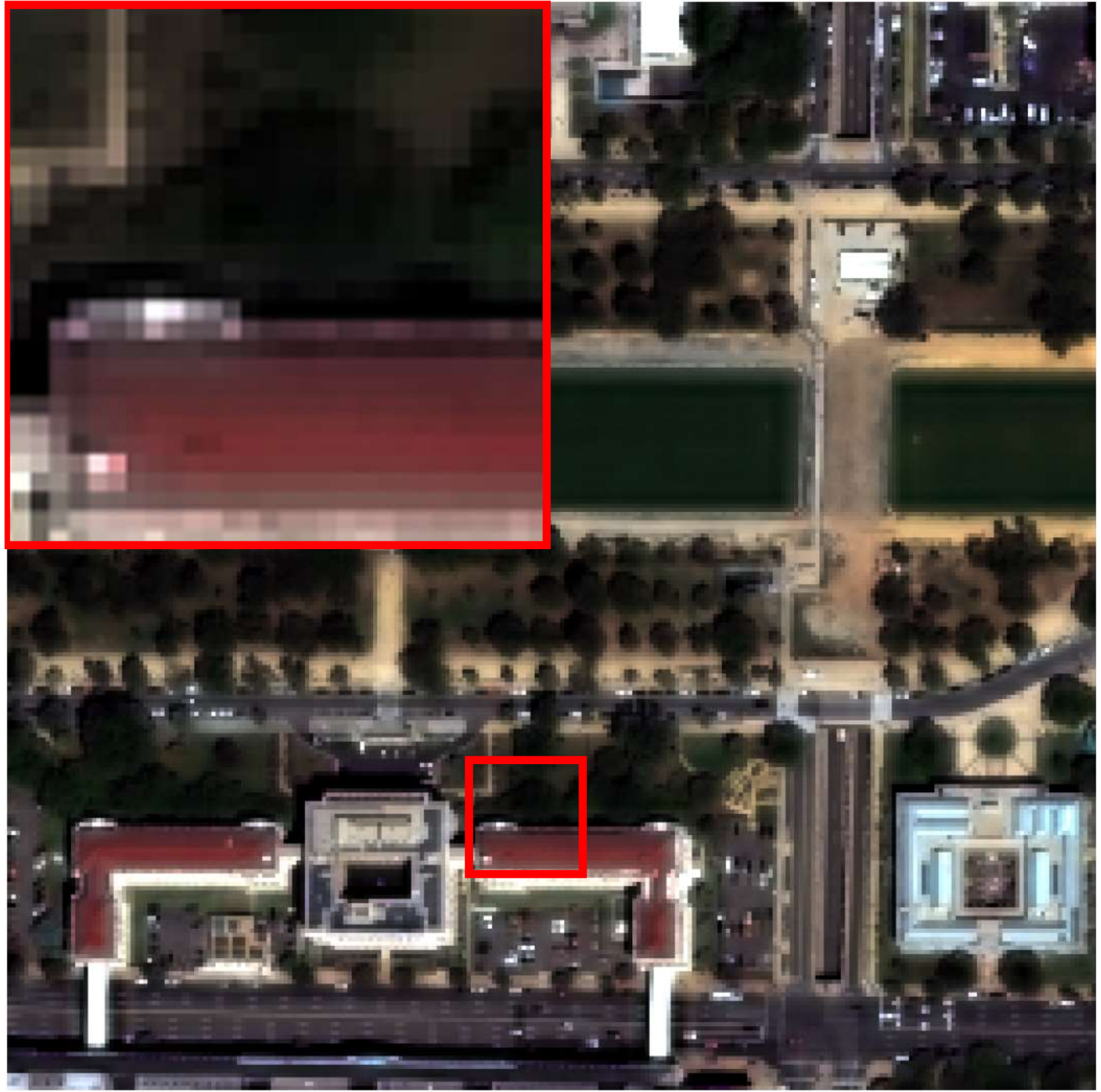}
\label{AWLP-H}}
\subfloat[]{\includegraphics[width=1.1in,trim=120 0 120 0,clip]{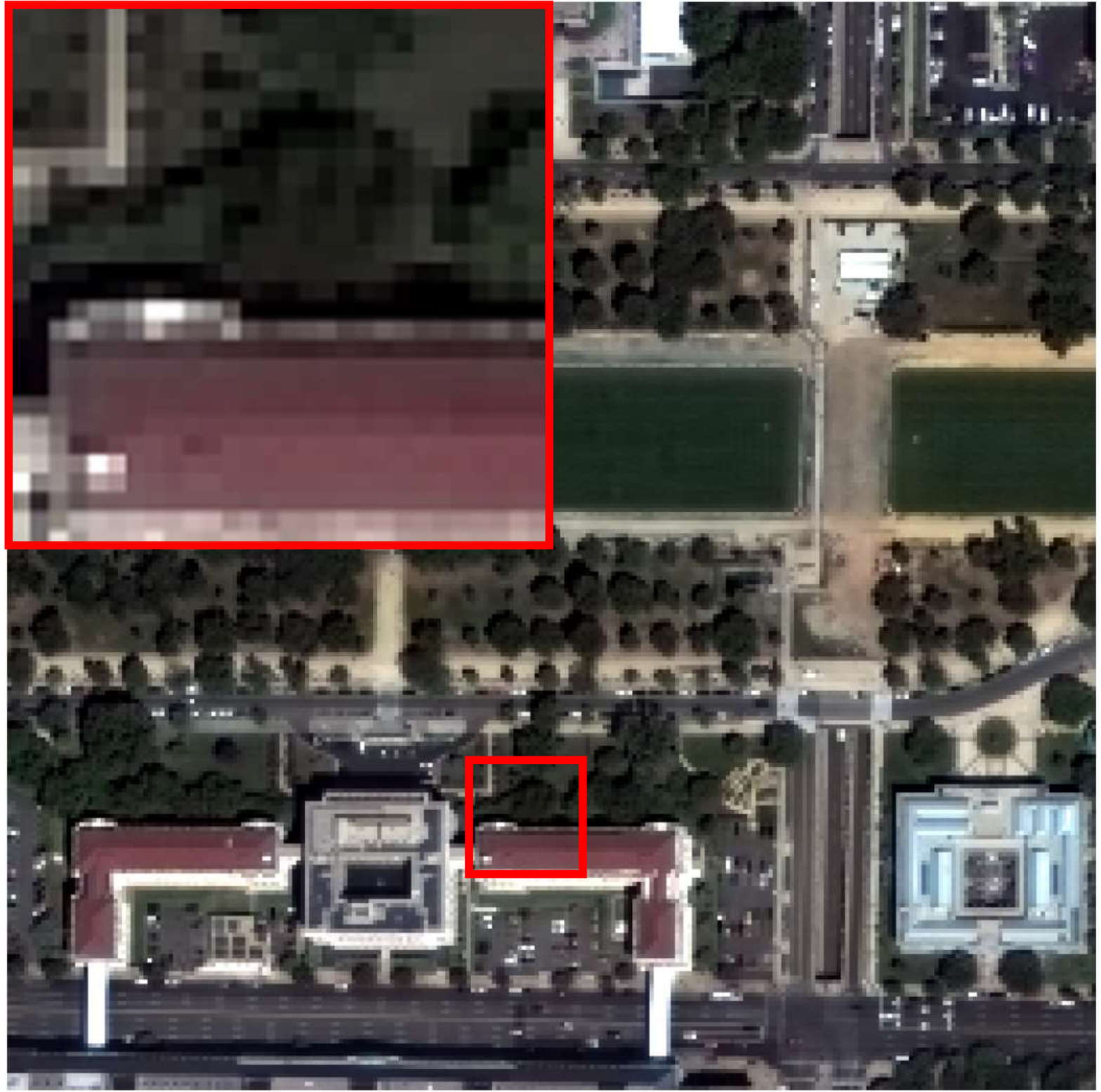}
\label{PWMBF}}
\subfloat[]{\includegraphics[width=1.1in,trim=120 0 120 0,clip]{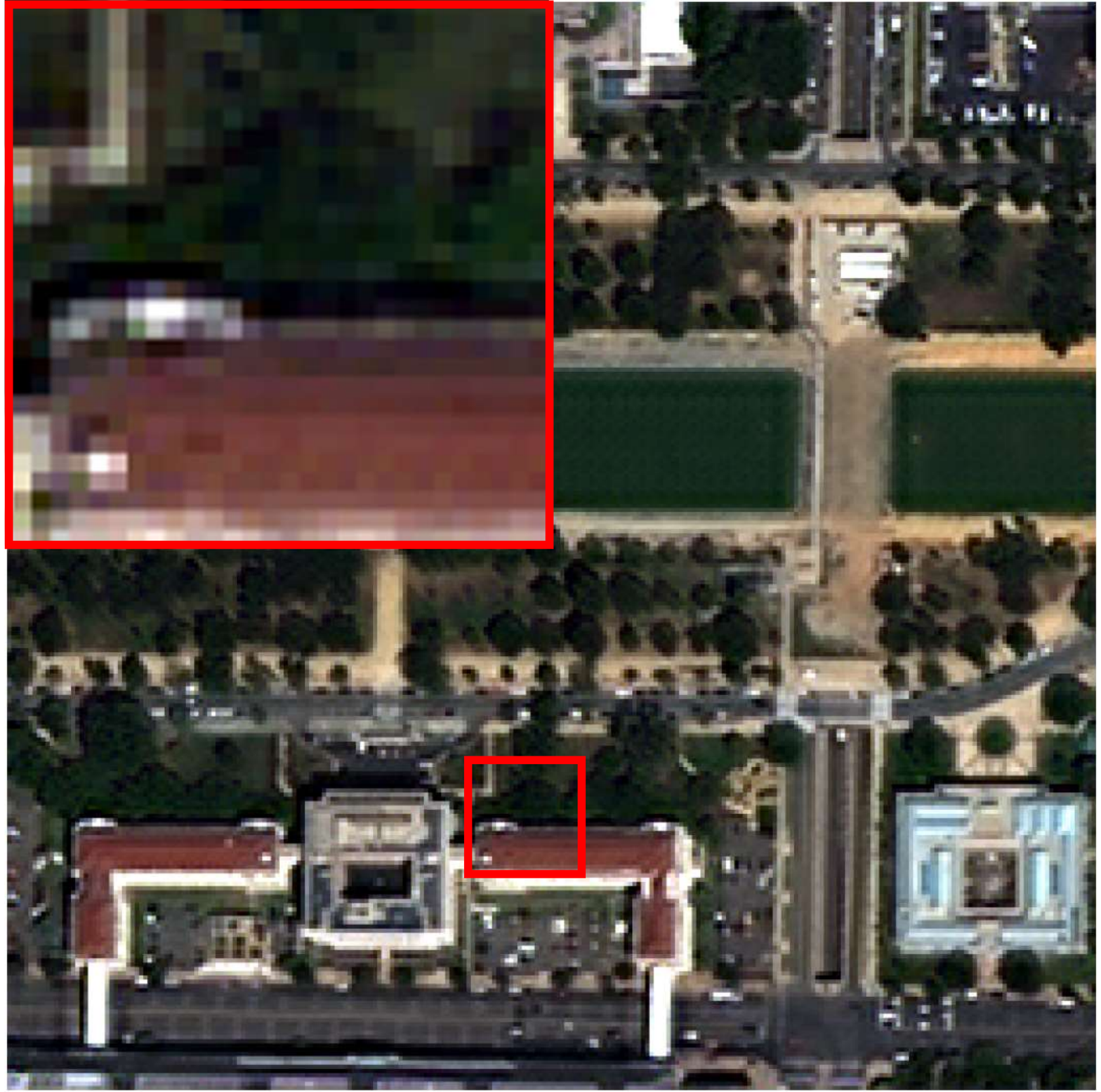}
\label{PanNet}}
\subfloat[]{\includegraphics[width=1.1in,trim=120 0 120 0,clip]{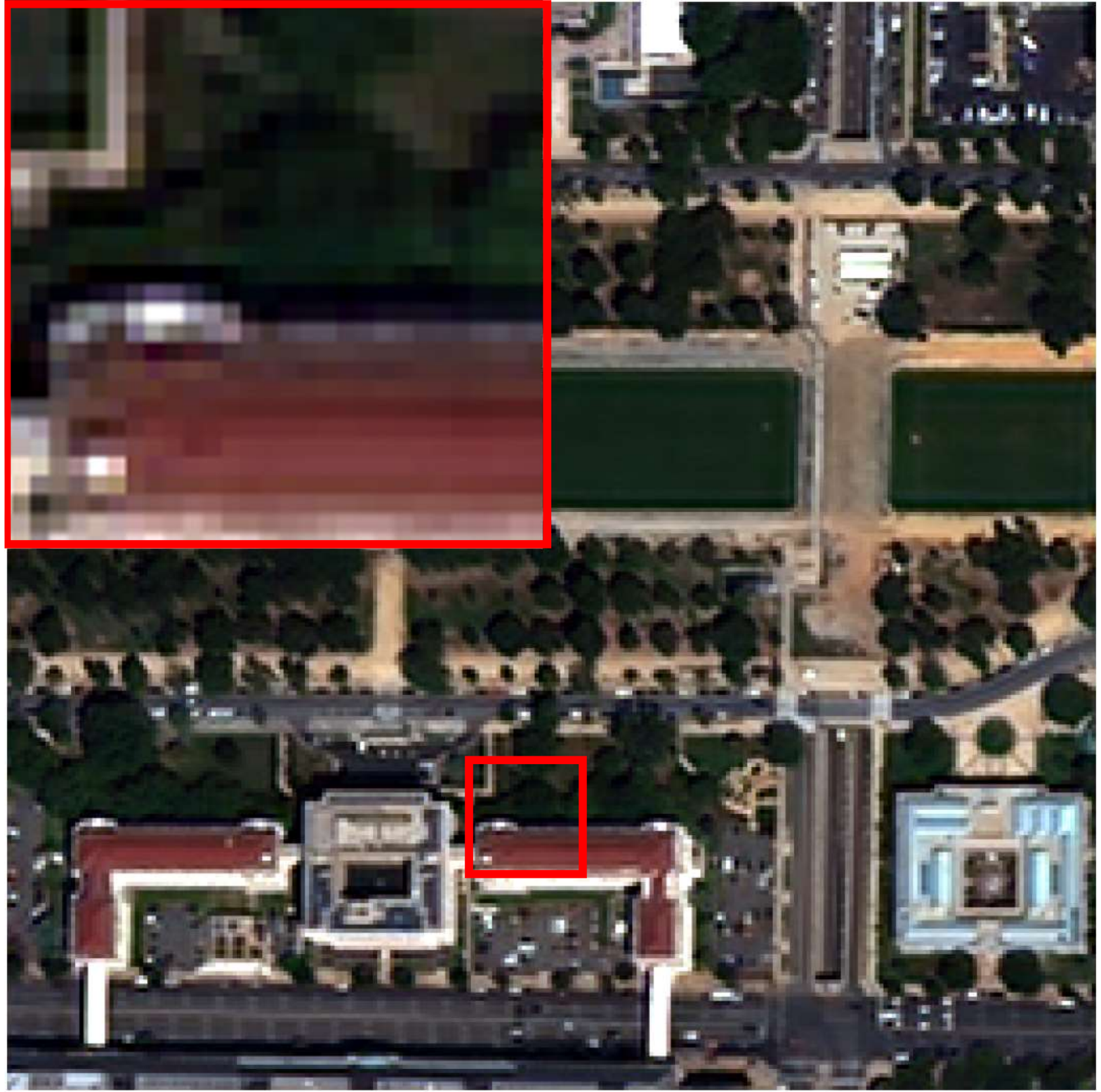}
\label{FusionNet}}

\vspace{-0.1in}
\subfloat[]{\includegraphics[width=1.1in,trim=120 0 120 0,clip]{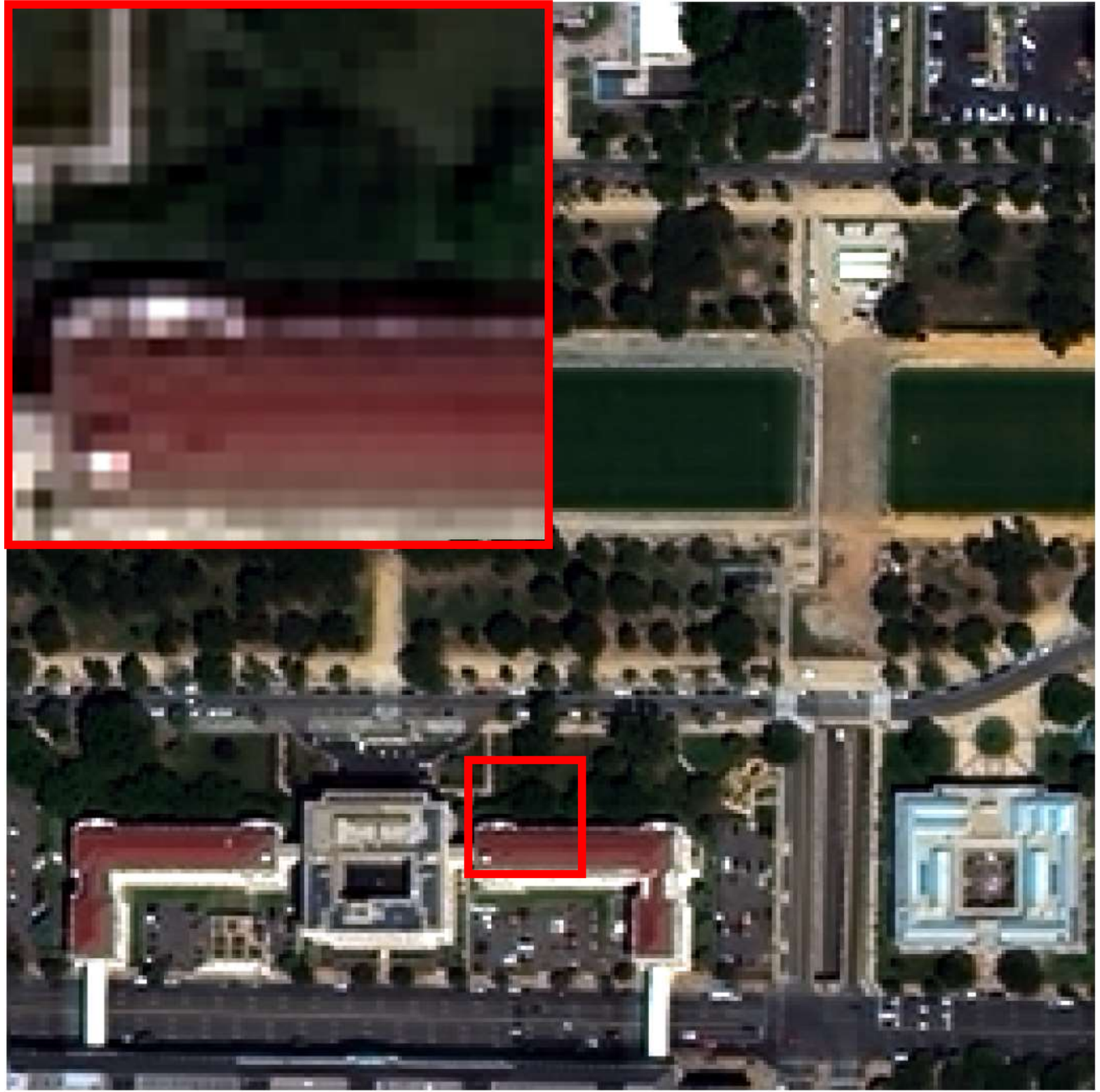}
\label{GTP-PNet}}
\subfloat[]{\includegraphics[width=1.1in,trim=120 0 120 0,clip]{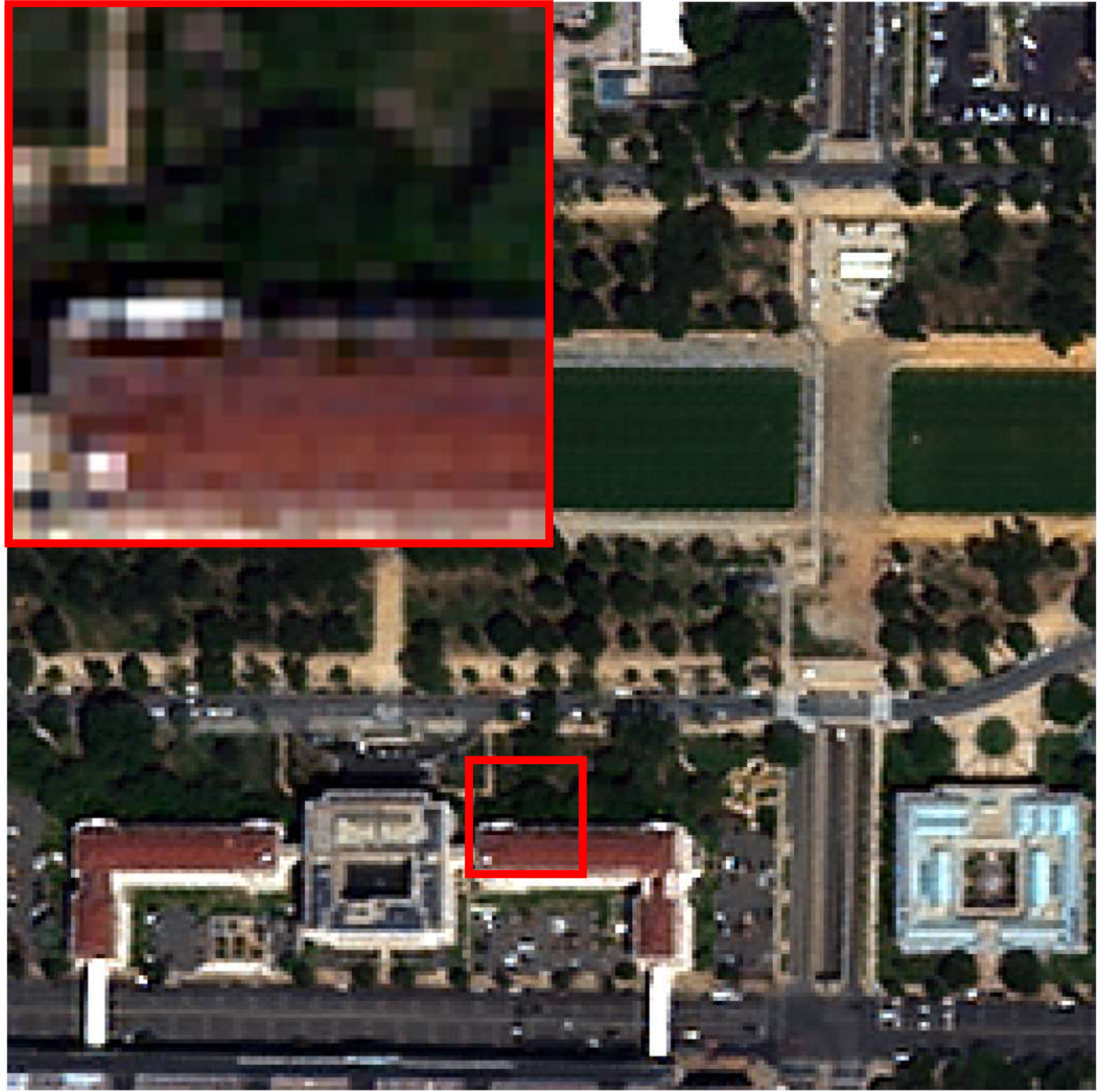}
\label{LPPN}}
\subfloat[]{\includegraphics[width=1.1in,trim=120 0 120 0,clip]{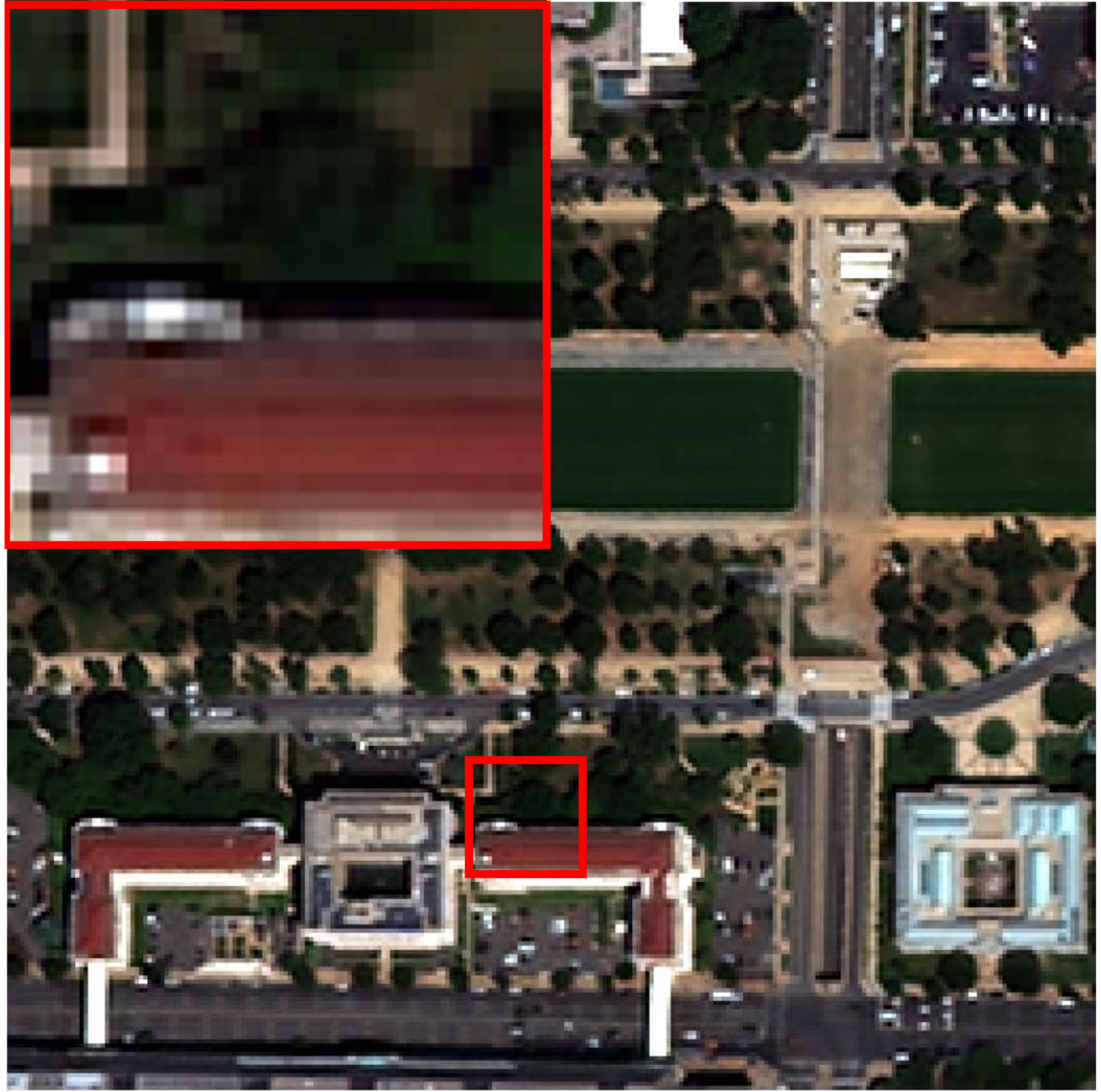}
\label{FAFNet}}
\caption{Vsual comparison of different methods on WorldView-2 (WV-2) dataset at reduced resolution. (a) PAN. (b) Up-sampled MS. (c) Reference. (d) GSA \cite{4305344}. (e) BDSD-PC \cite{8693555}. (f) MTF-GLP-CBD \cite{4305345}. (g) AWLP-H \cite{vivone2019fast}. (h) PWMBF \cite{6951484}. (i) PanNet \cite{8237455}. (j) FusionNet \cite{9240949}. (k) GTP-PNet \cite{ZHANG2021223}. (l) LPPN \cite{JIN2022158}. (m) FAFNet.}
\label{fig_wv2_reduce}
\end{figure*}

\begin{figure*}[htbp]
\centering
\subfloat[]{\includegraphics[width=1in,trim=120 0 120 0,clip]{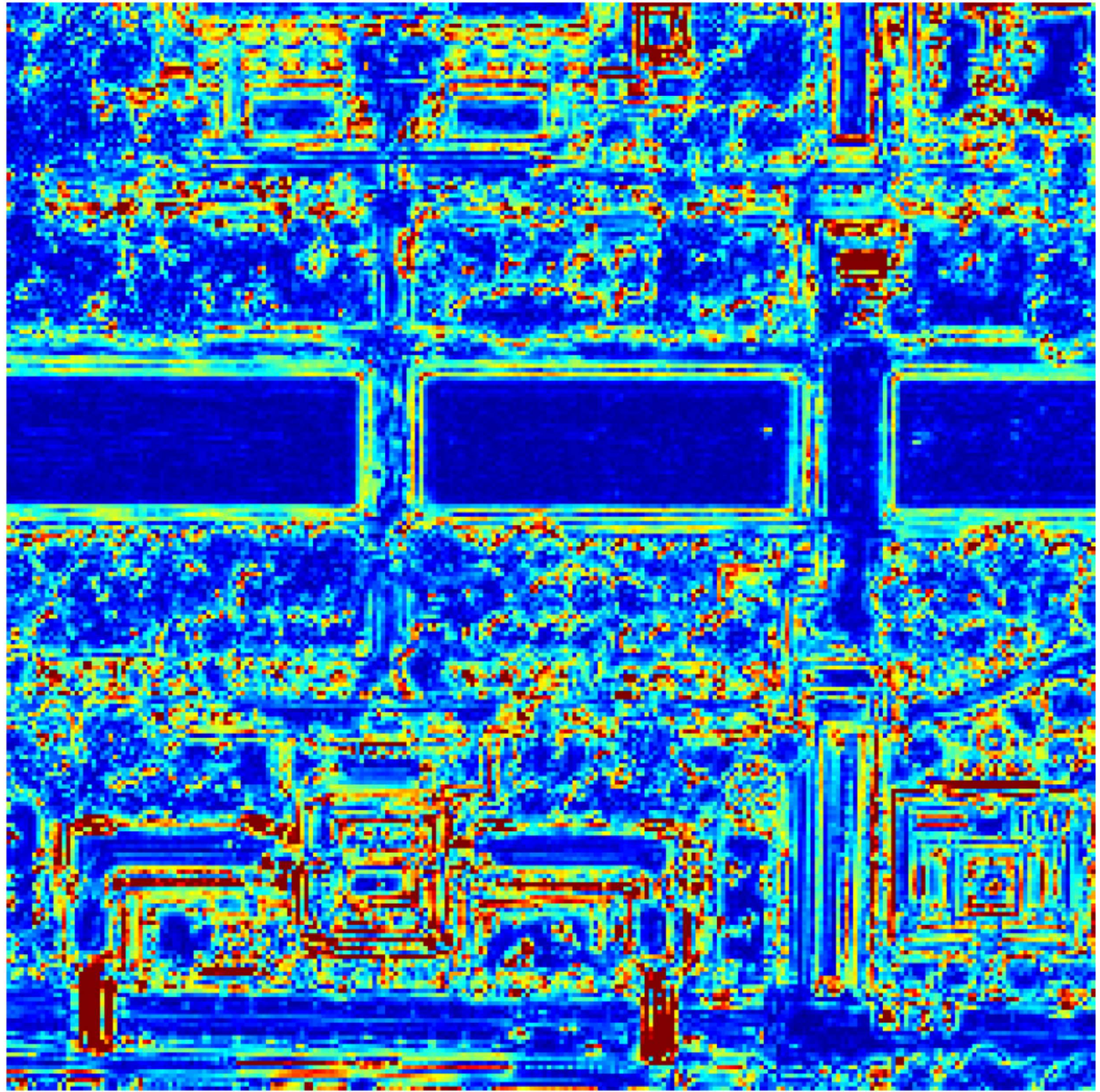}
\label{up-sampled MS-error}}
\subfloat[]{\includegraphics[width=1in,trim=120 0 120 0,clip]{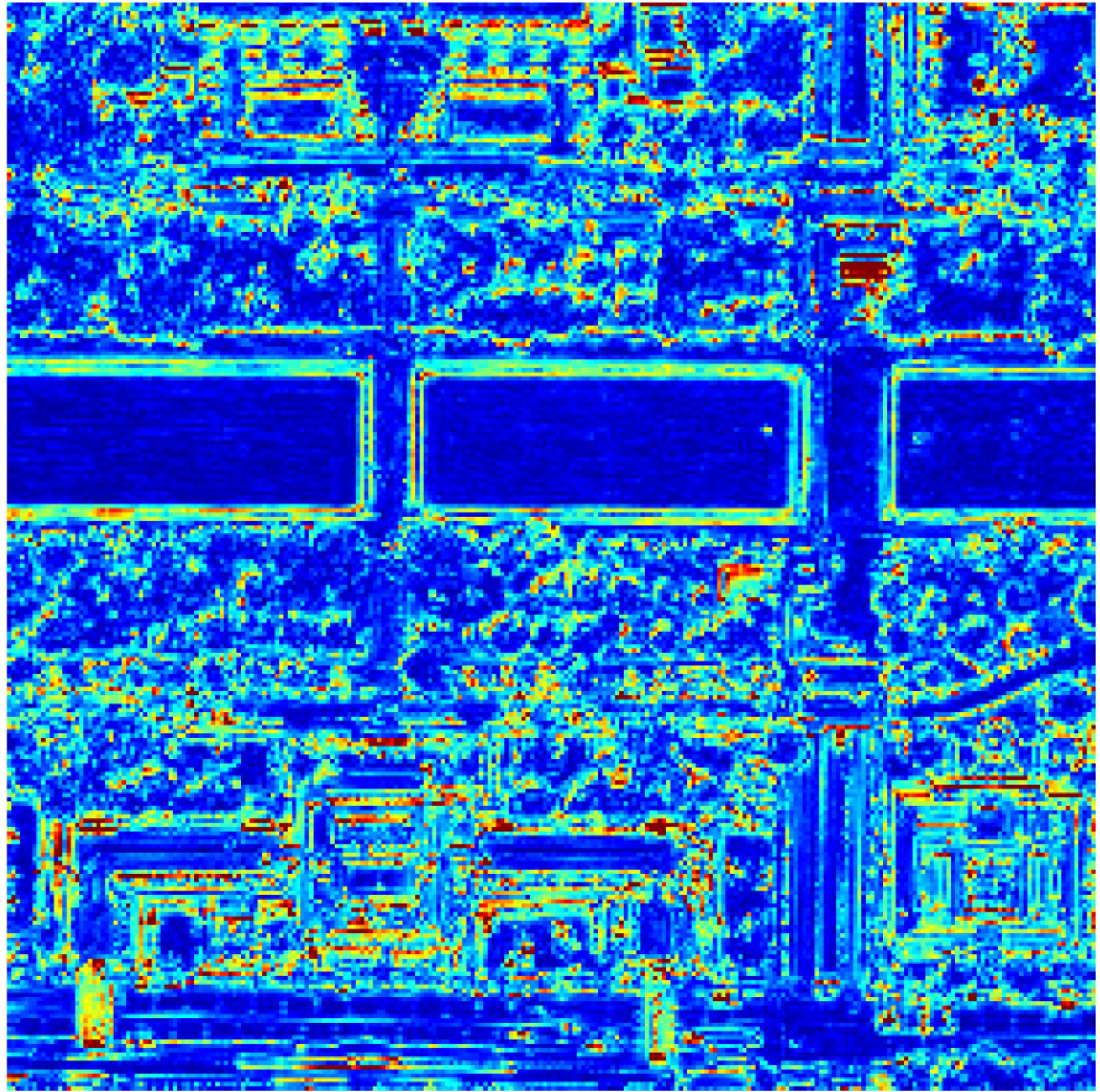}
\label{GSA}}
\subfloat[]{\includegraphics[width=1in,trim=120 0 120 0,clip]{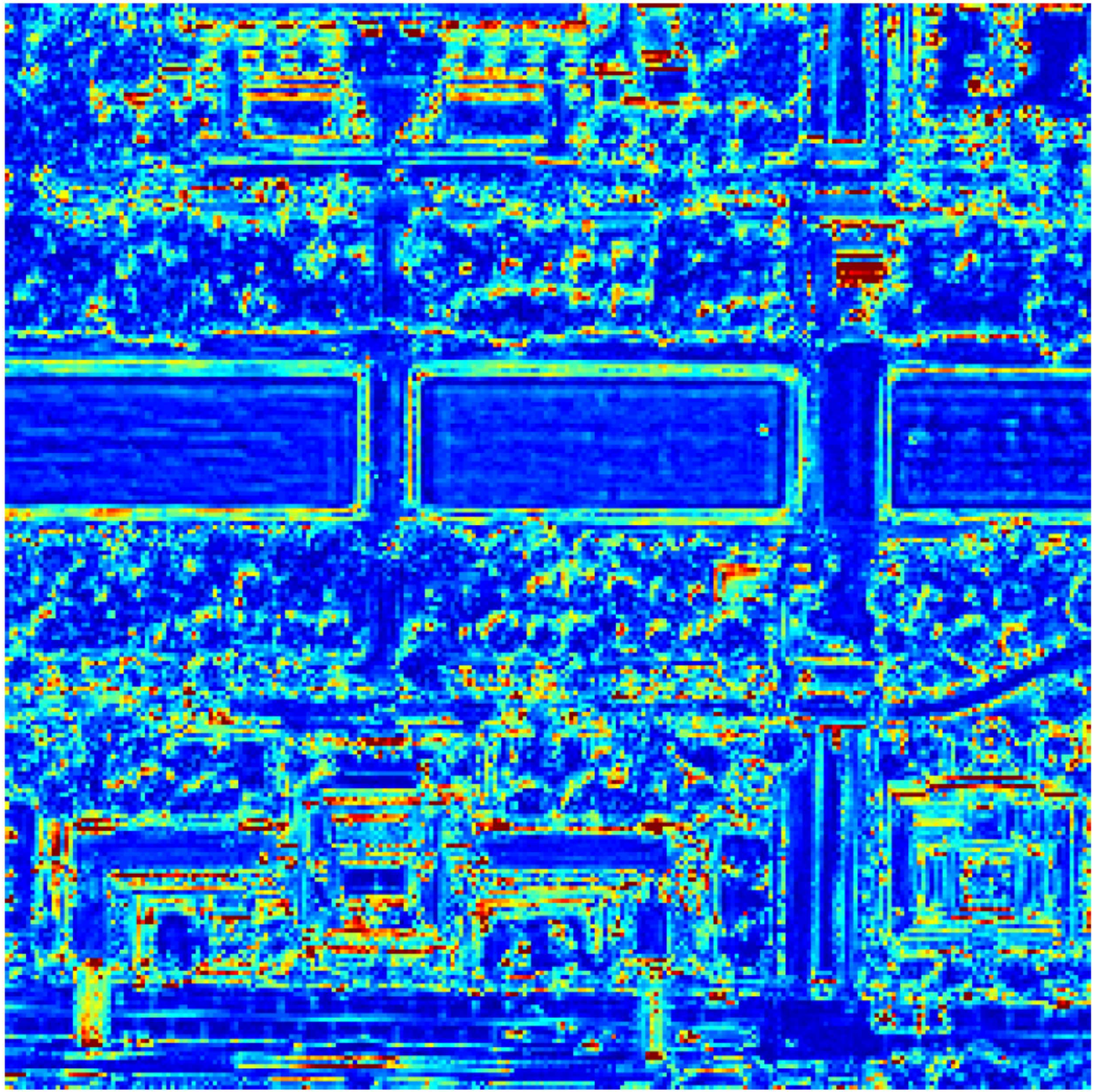}
\label{BDSD-PC}}
\subfloat[]{\includegraphics[width=1in,trim=120 0 120 0,clip]{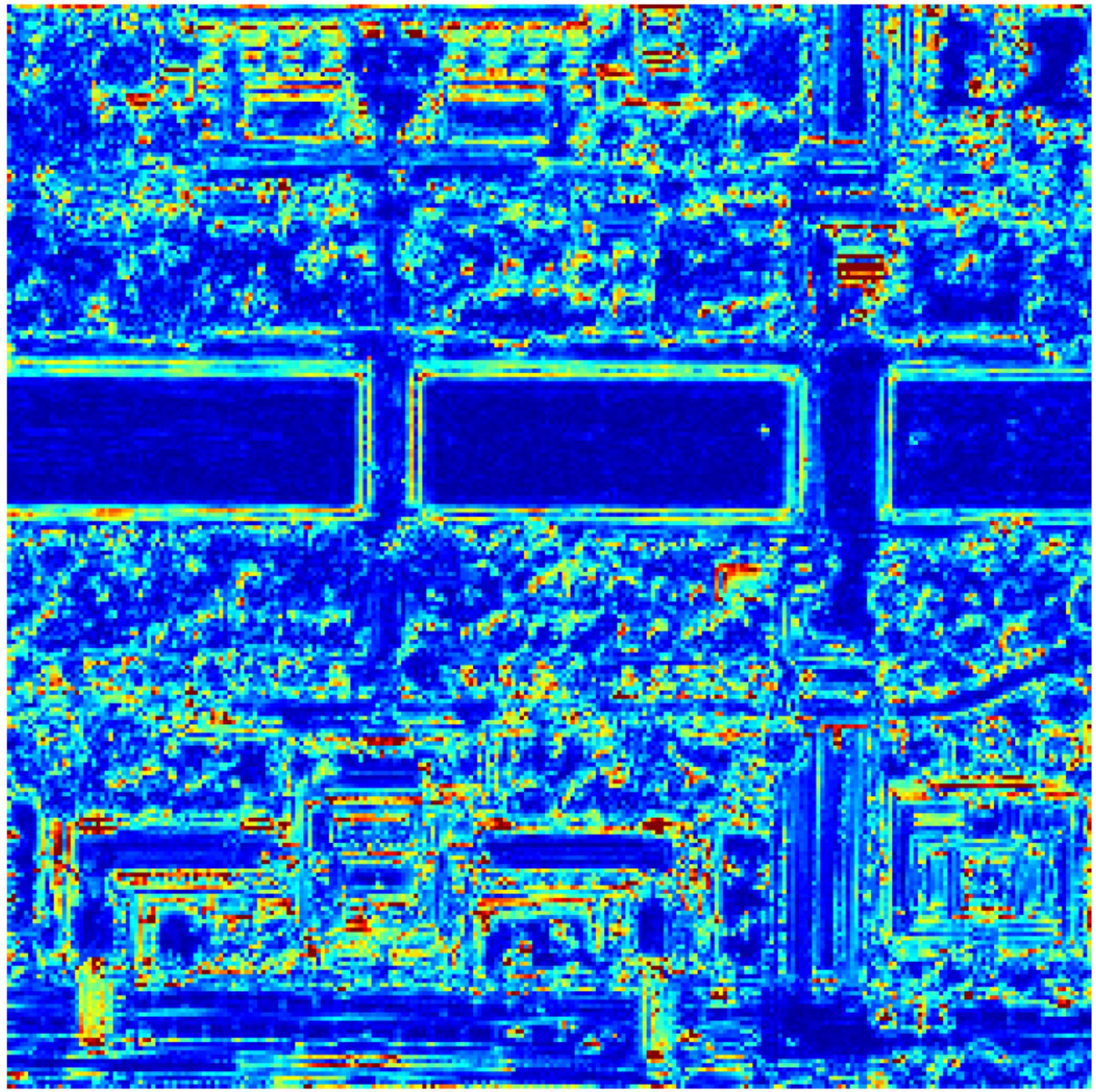}
\label{MTF-GLP-CBD}}
\subfloat[]{\includegraphics[width=1in,trim=120 0 120 0,clip]{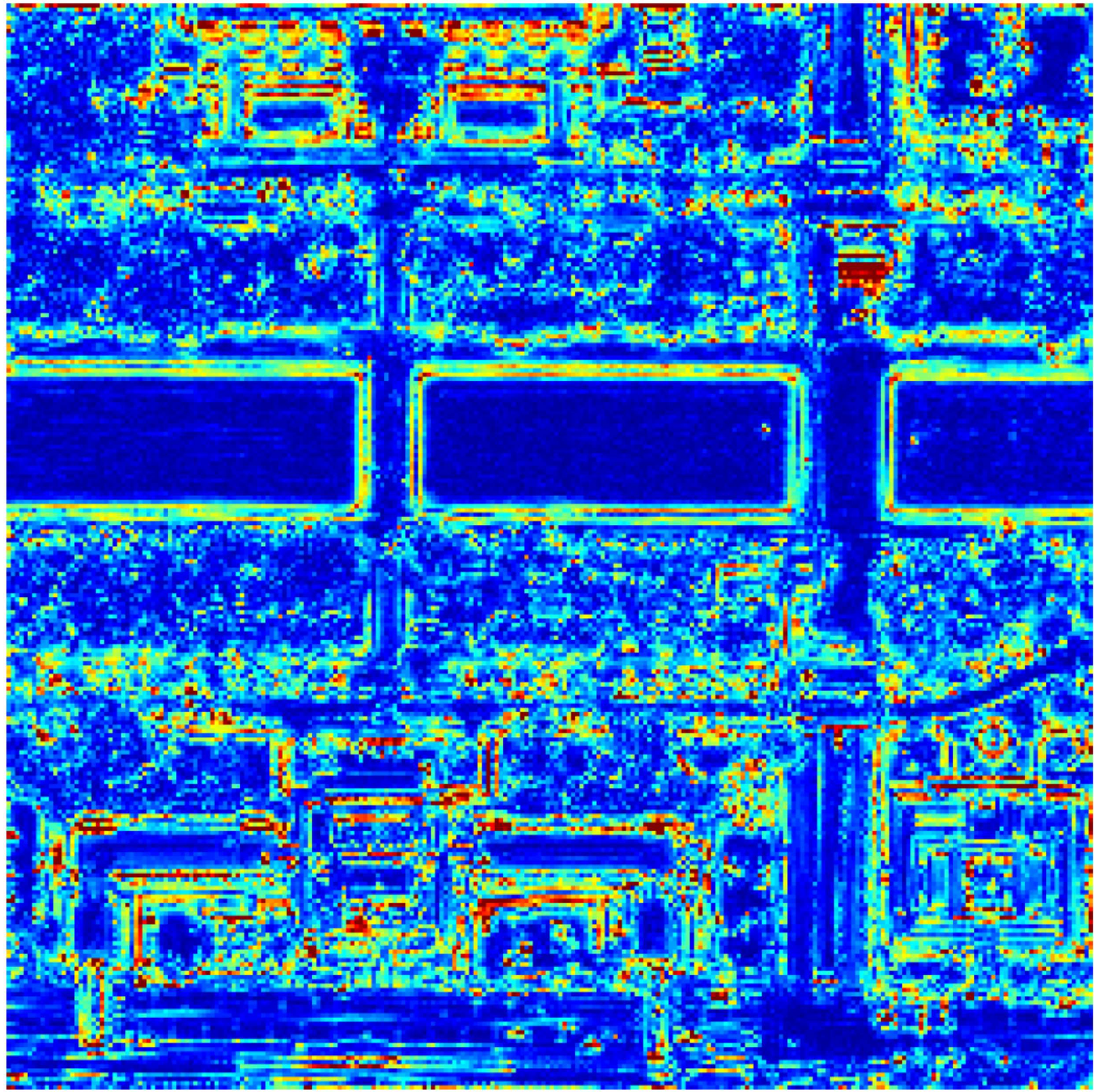}
\label{AWLP-H}}
\subfloat[]{\includegraphics[width=1in,trim=120 0 120 0,clip]{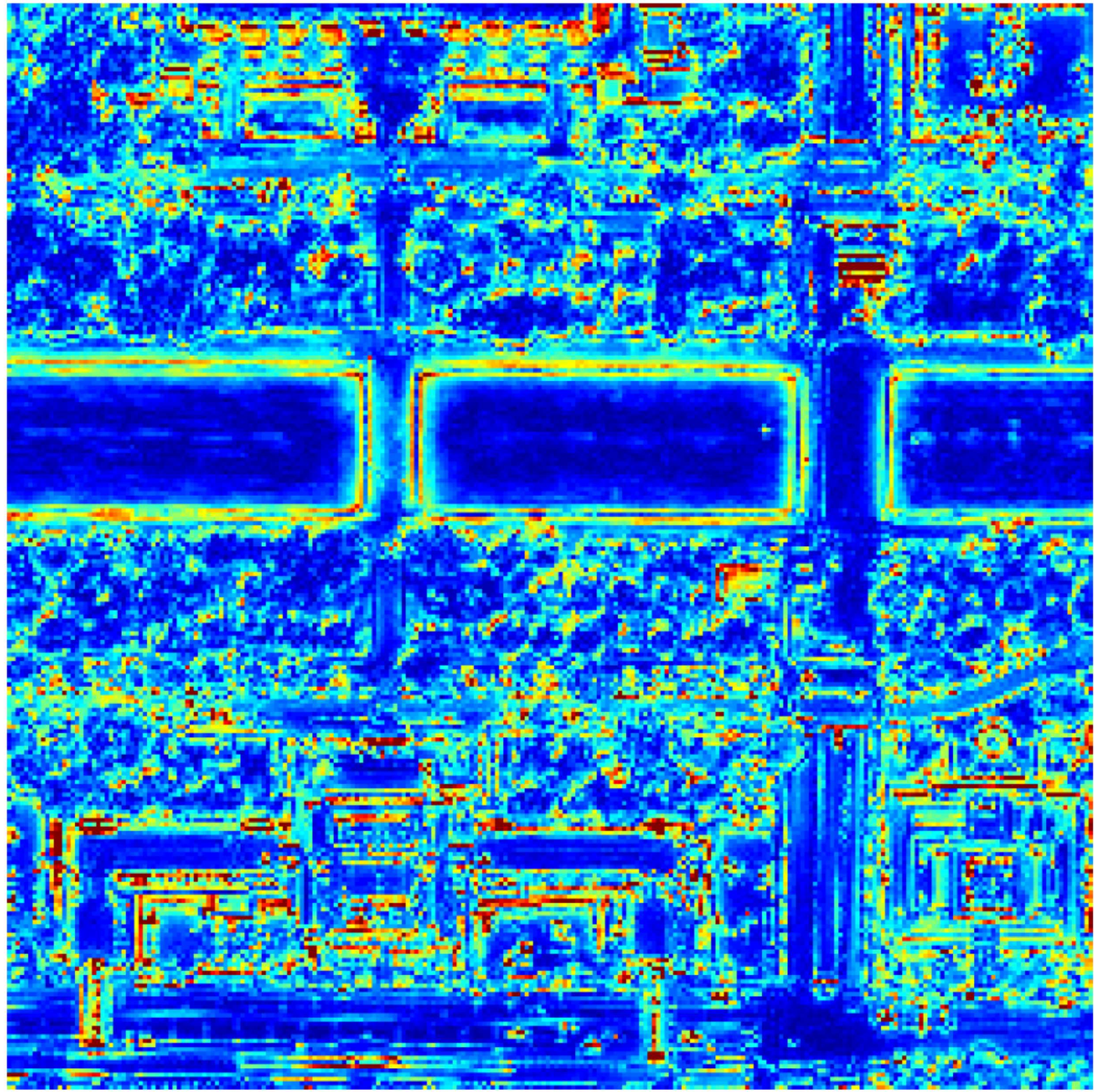}
\label{PWMBF}}

\vspace{-0.1in}
\subfloat[]{\includegraphics[width=1in,trim=120 0 120 0,clip]{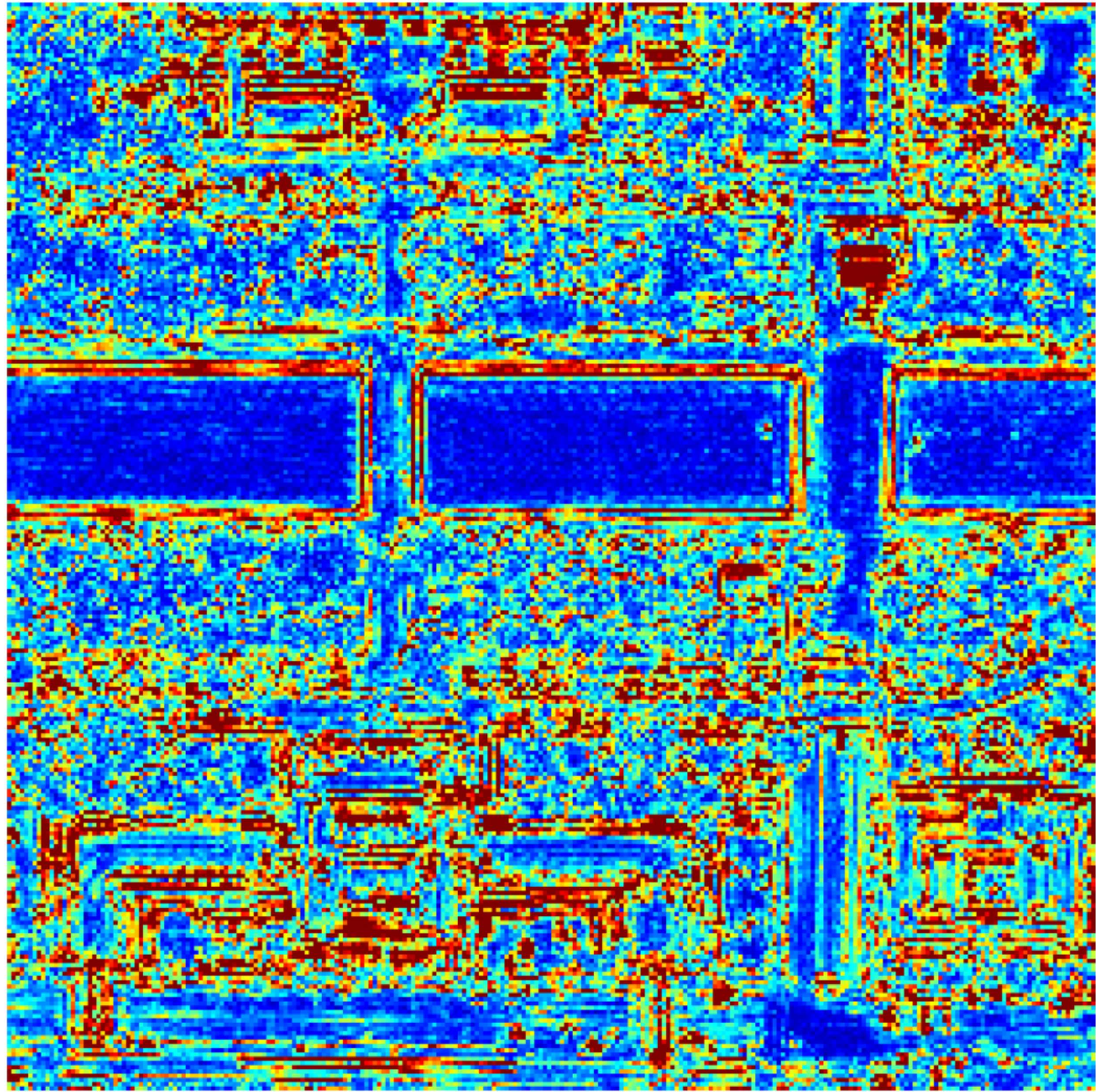}
\label{PanNet}}
\subfloat[]{\includegraphics[width=1in,trim=120 0 120 0,clip]{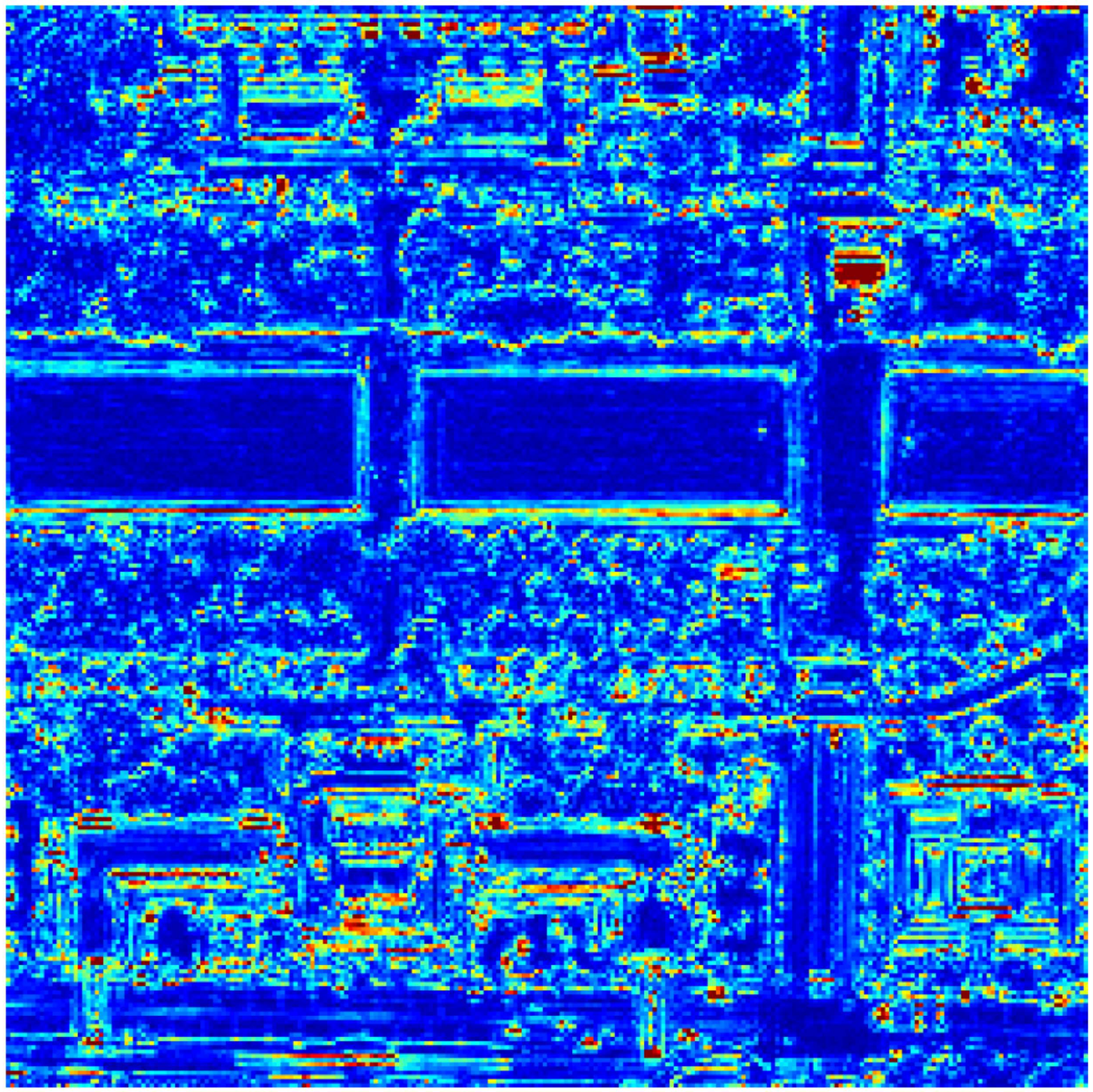}
\label{FusionNet}}
\subfloat[]{\includegraphics[width=1in,trim=120 0 120 0,clip]{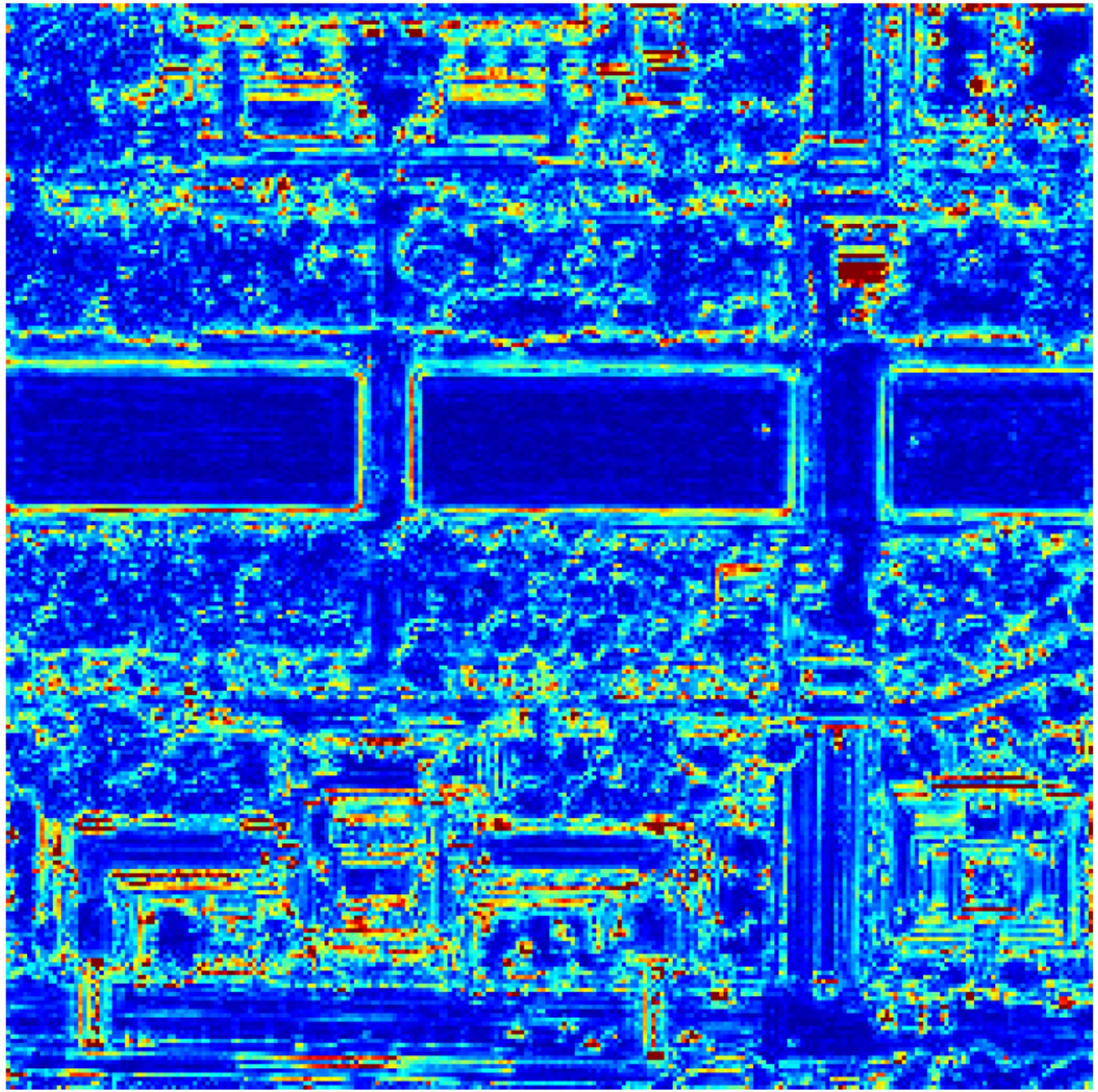}
\label{GTP-PNet}}
\subfloat[]{\includegraphics[width=1in,trim=120 0 120 0,clip]{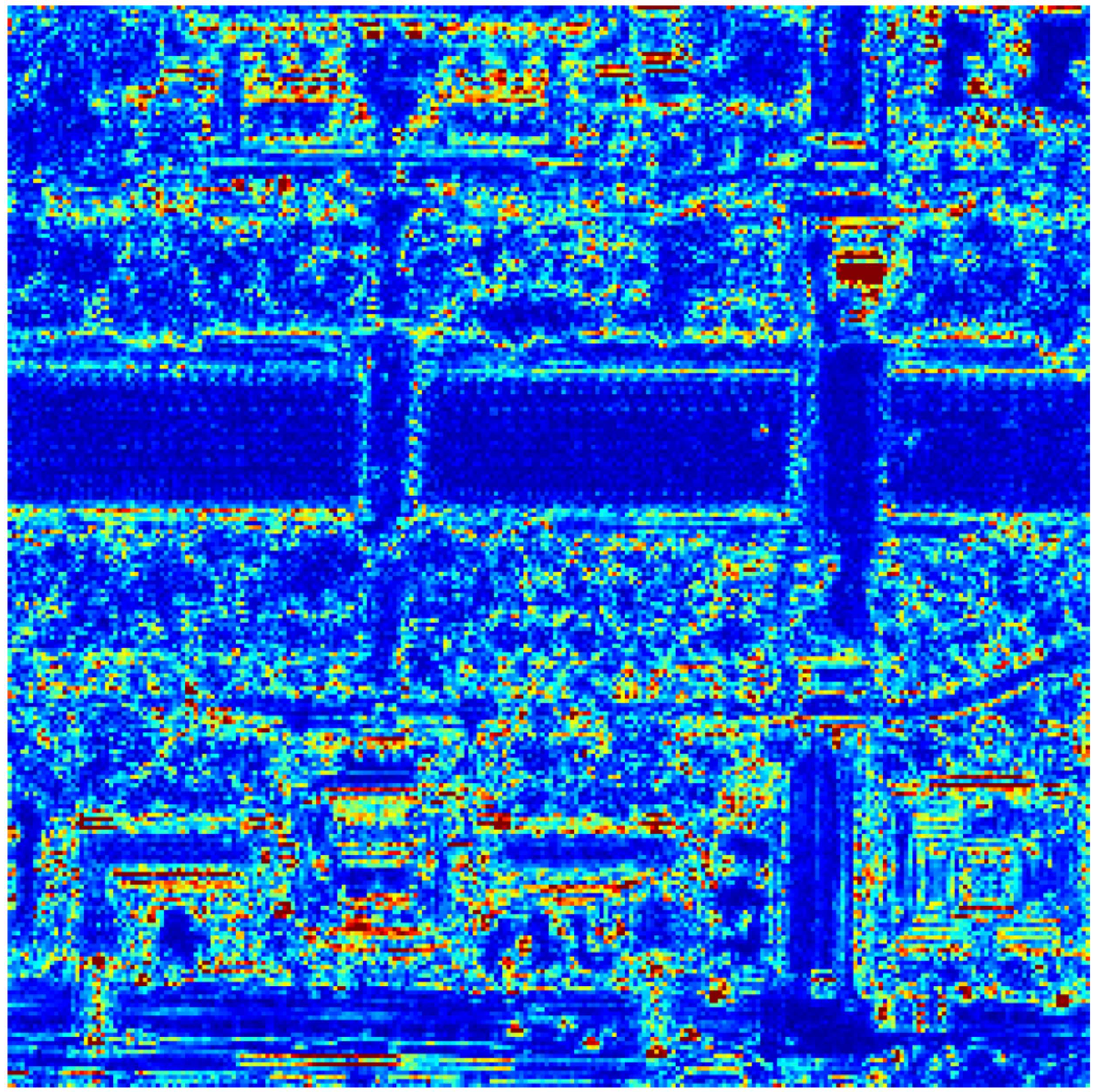}
\label{LPPN}}
\subfloat[]{\includegraphics[width=1in,trim=120 0 120 0,clip]{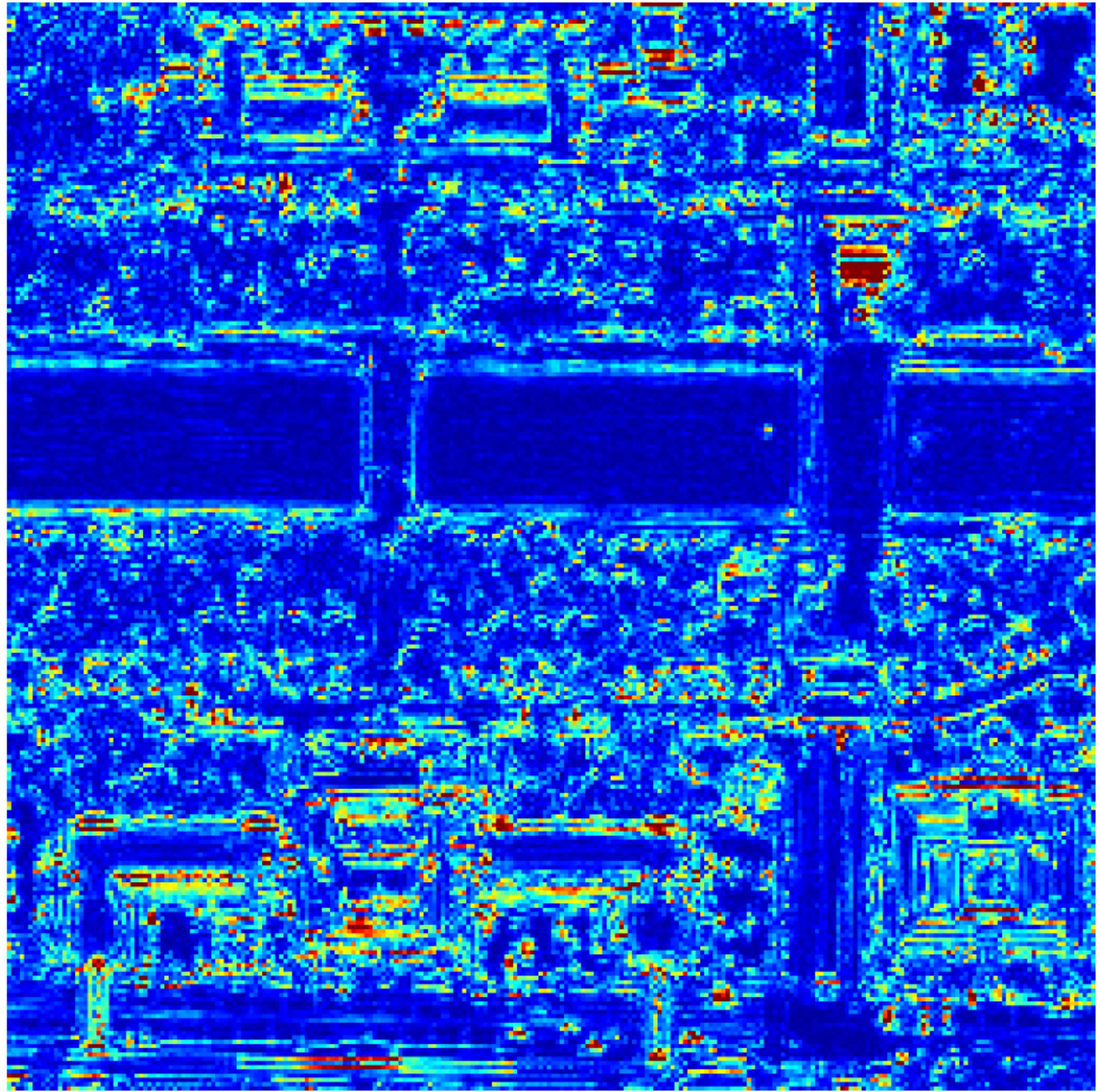}
\label{FAFNet}}
\subfloat[]{\includegraphics[width=1in,trim=120 0 120 0,clip]{figures/wv4/error/ref-error.pdf}
\label{Reference}}

\subfloat{\includegraphics[angle=-90,width=0.3\linewidth,trim=220 0 220 0,clip]{figures/wv4/error/bar.pdf}
\label{bar}}
\caption{Absolute error maps (AEMs) between reference image and (a)Up-sampled MS, (b) GSA \cite{4305344}, (c) BDSD-PC \cite{8693555}, (d) MTF-GLP-CBD \cite{4305345}, (e) AWLP-H \cite{vivone2019fast}, (f) PWMBF \cite{6951484}, (g) PanNet \cite{8237455}, (h) FusionNet \cite{9240949}, (i) GTP-PNet \cite{ZHANG2021223}, (j) LPPN \cite{JIN2022158}, (k) FAFNet and (l) Reference on WV-2 dataset at reduced resolution.}
\label{fig_wv2_reduce_error}
\end{figure*}

To verify the effectiveness of proposed method for eight-band MS image pansharpening, we conduct experiments on the WV-2 dataset. The experimental results are presented in Figs. \ref{fig_wv2_reduce}-\ref{fig_wv2_reduce_error}, where Figs. \ref{fig_wv2_reduce}(d)-(m) show the fusion results of compared methods. The results of GSA \cite{4305344}, BDSD-PC \cite{8693555}, MTF-GLP-CBD \cite{4305345} and PWMBF \cite{6951484} methods have sufficient spatial details, but they are affected by serious spectral distortion, which can be observed from the color of soil. Although AWLP-H \cite{vivone2019fast}, PanNet \cite{8237455}, GTP-PNet \cite{ZHANG2021223} and LPPN \cite{JIN2022158} suffer from less spectral distortion, they are affected by the spatial degradation. The advantages of FAFNet can be observed from the zoomed-in regions of Fig. \ref{fig_wv2_reduce}(m) that the color of road, roof, grasses and the shape of objects (such as the line between the road, the white part of roof) are more reliable. From Fig. \ref{fig_wv2_reduce_error}, we can see that the error maps of FusionNet \cite{9240949} and FAFNet are comparable, but residuals in that of FAFNet is less. In addition, the quality assessments on 136 pairs of test data from WV-2 dataset are shown in Table \ref{tab_wv2}, from which we can also conclude that the proposed FAFNet outperforms the compared state-of-the-art methods with respect to four adopted evaluation metrics.

\subsection{Experiments at Full Resolution }
\begin{figure*}[htbp]
\centering
\subfloat[]{\includegraphics[width=1.1in,trim=120 0 120 0,clip]{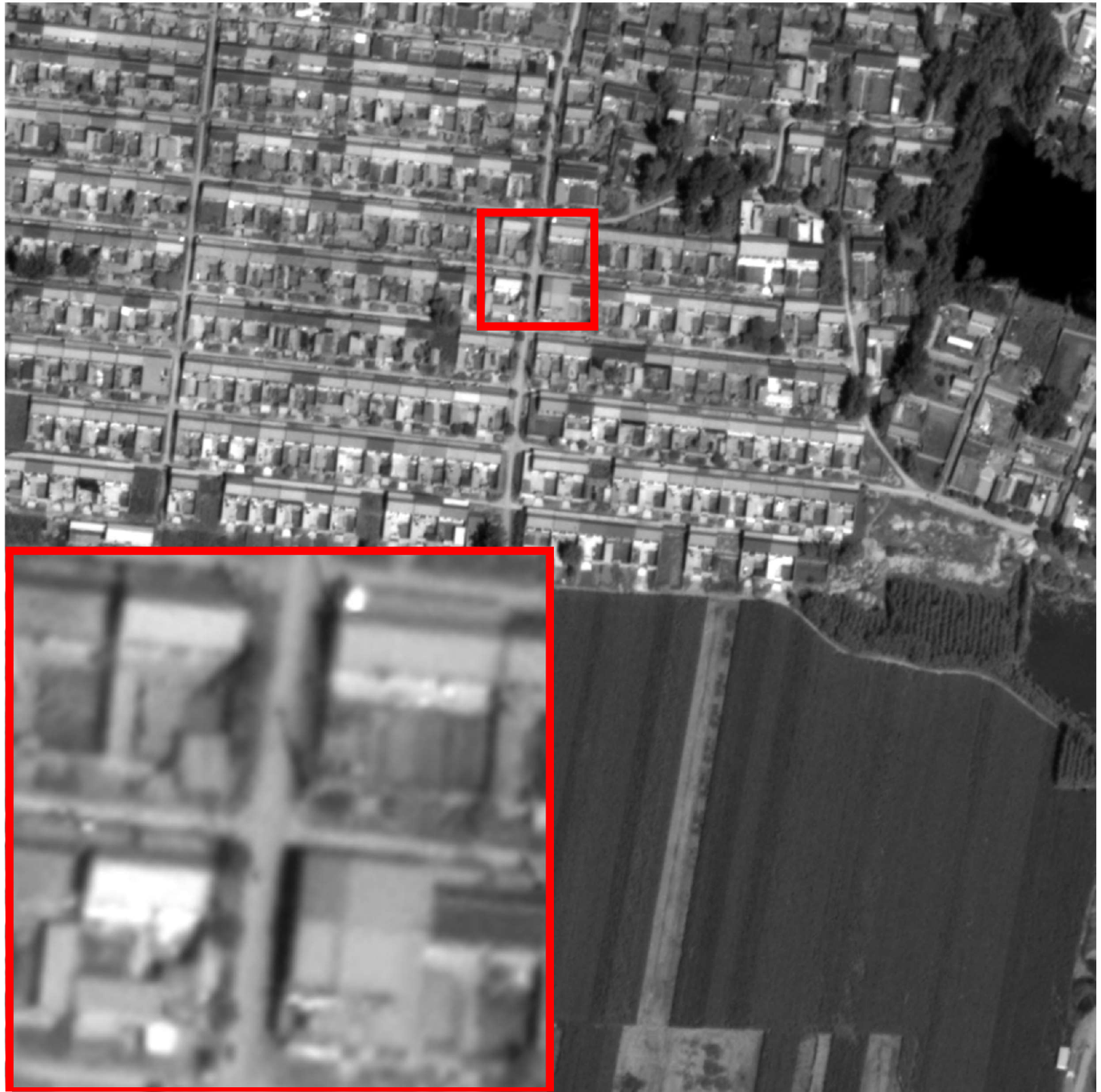}
\label{PAN}}
\subfloat[]{\includegraphics[width=1.1in,trim=120 0 120 0,clip]{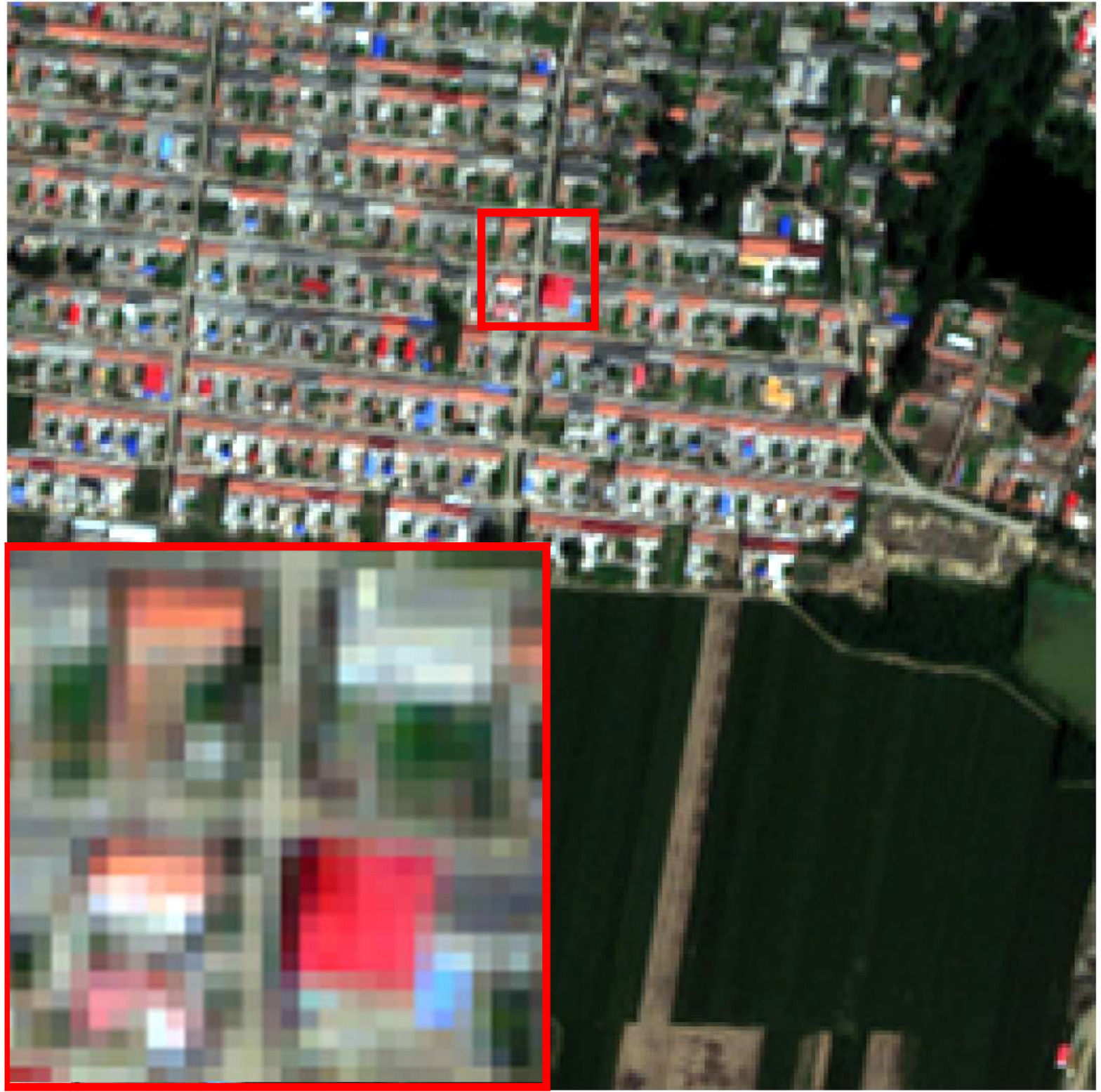}
\label{up-sampled MS}}
\subfloat[]{\includegraphics[width=1.1in,trim=120 0 120 0,clip]{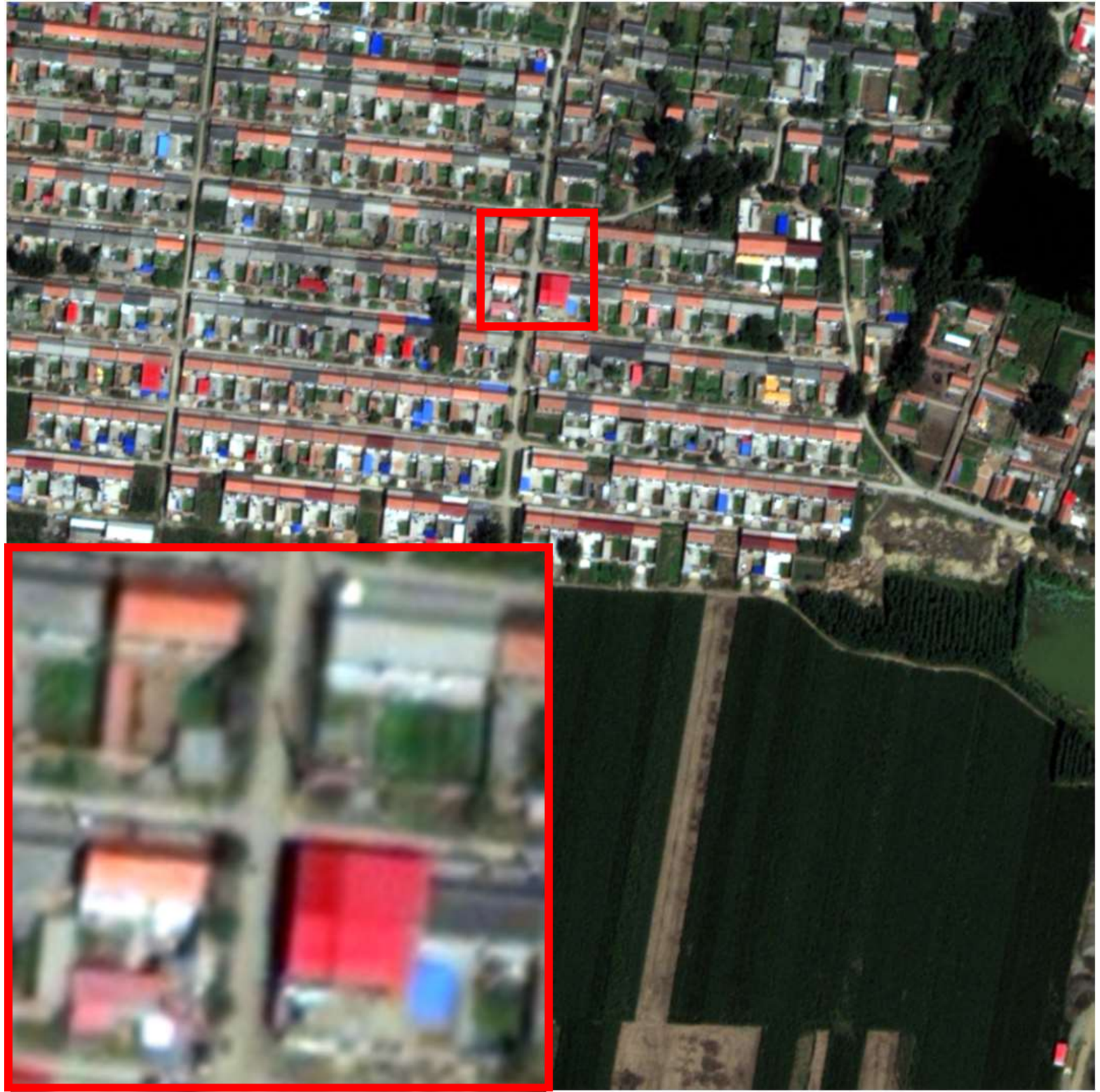}
\label{GSA}}
\subfloat[]{\includegraphics[width=1.1in,trim=120 0 120 0,clip]{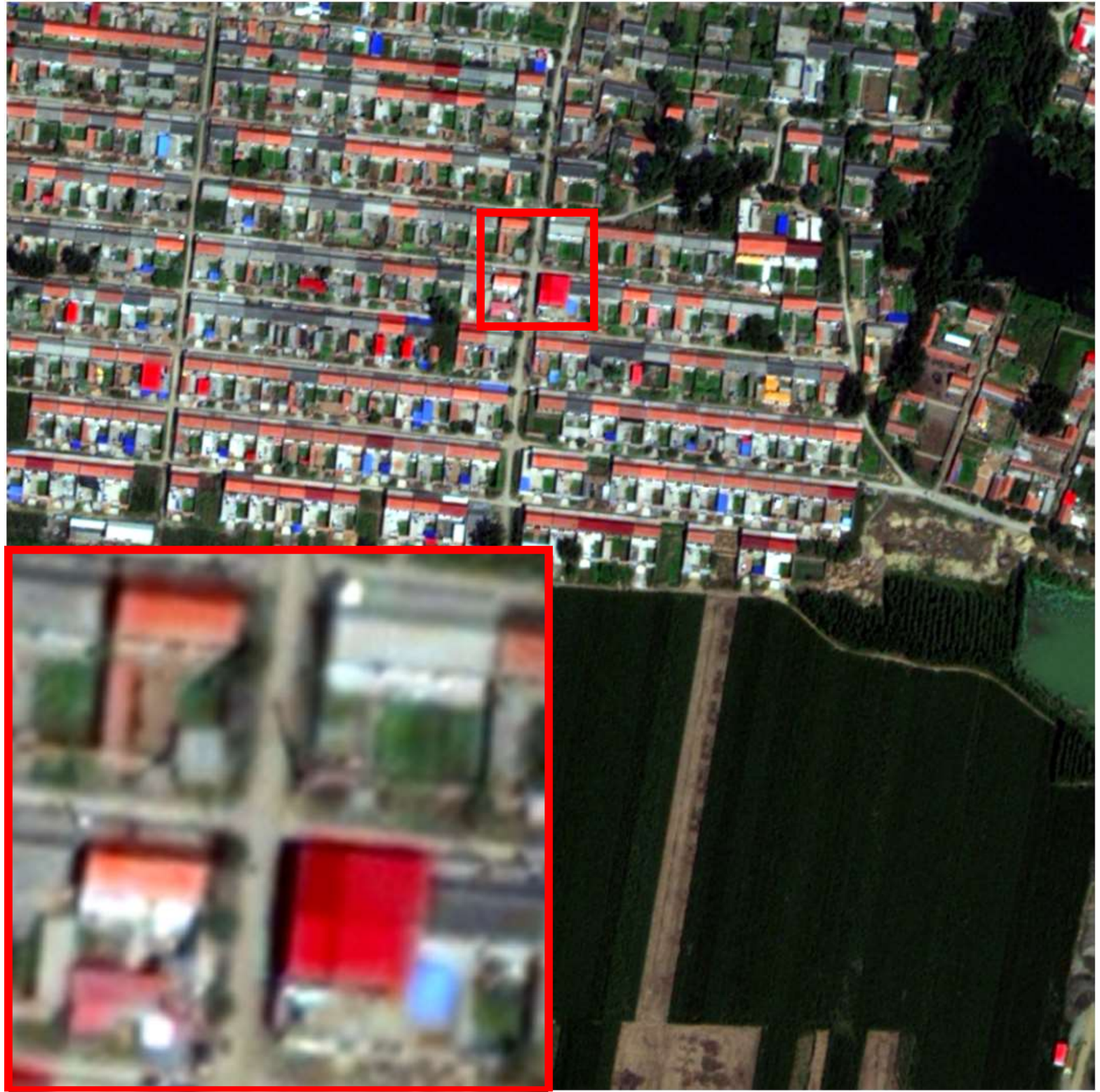}
\label{BDSD-PC}}
\subfloat[]{\includegraphics[width=1.1in,trim=120 0 120 0,clip]{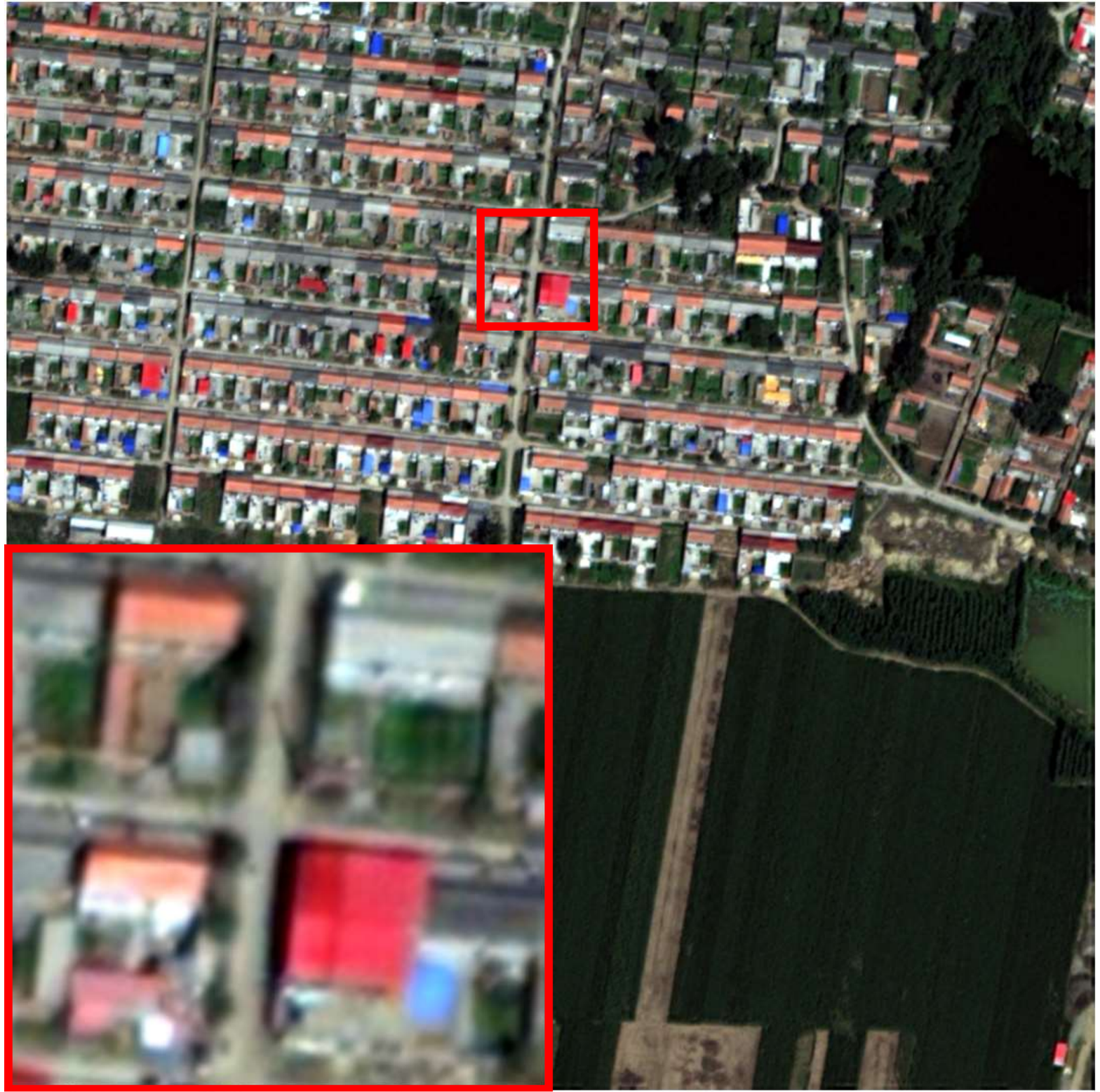}
\label{MTF-GLP-CBD}}
\subfloat[]{\includegraphics[width=1.1in,trim=120 0 120 0,clip]{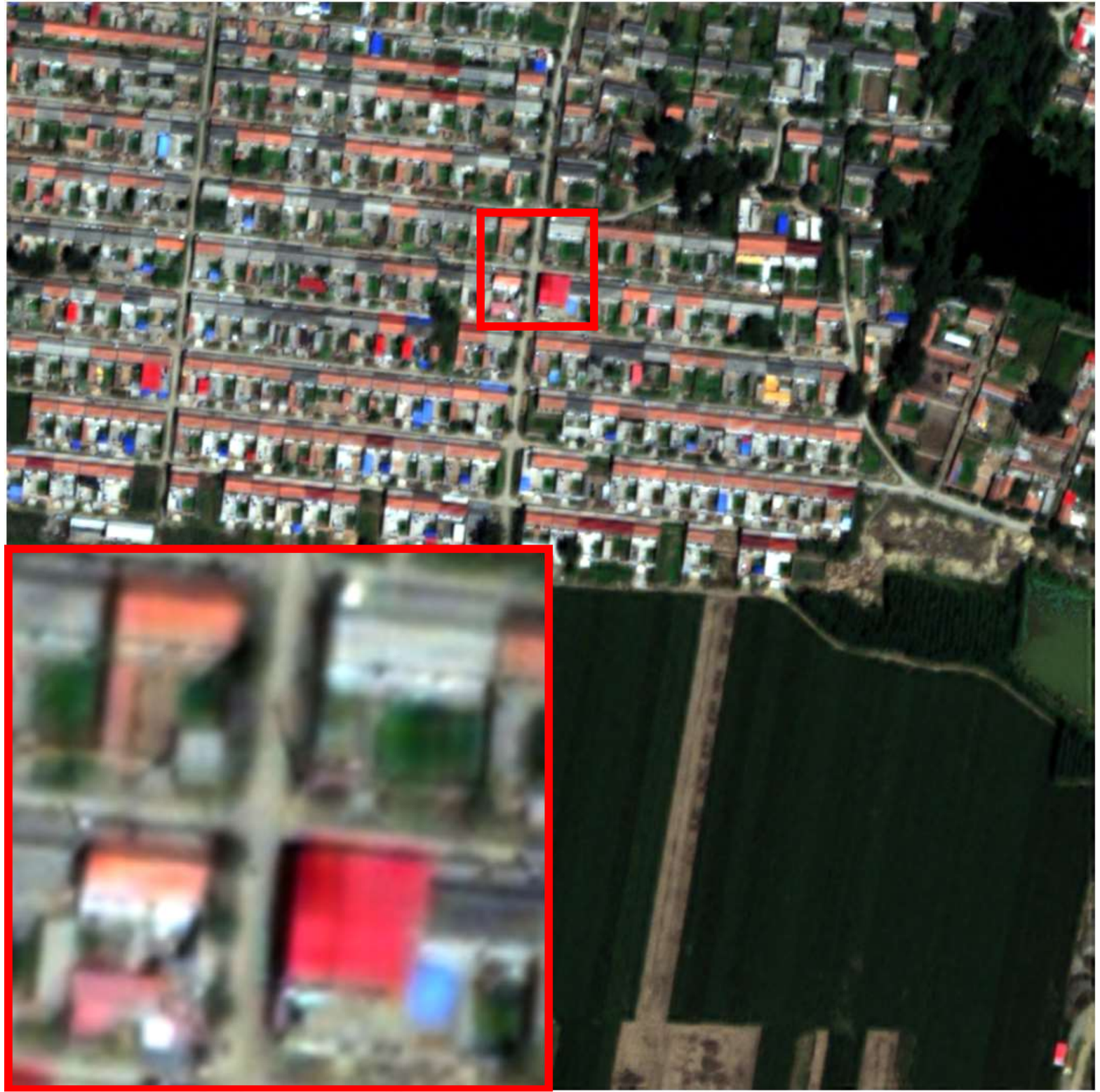}
\label{AWLP-H}}

\vspace{-0.1in}
\subfloat[]{\includegraphics[width=1.1in,trim=120 0 120 0,clip]{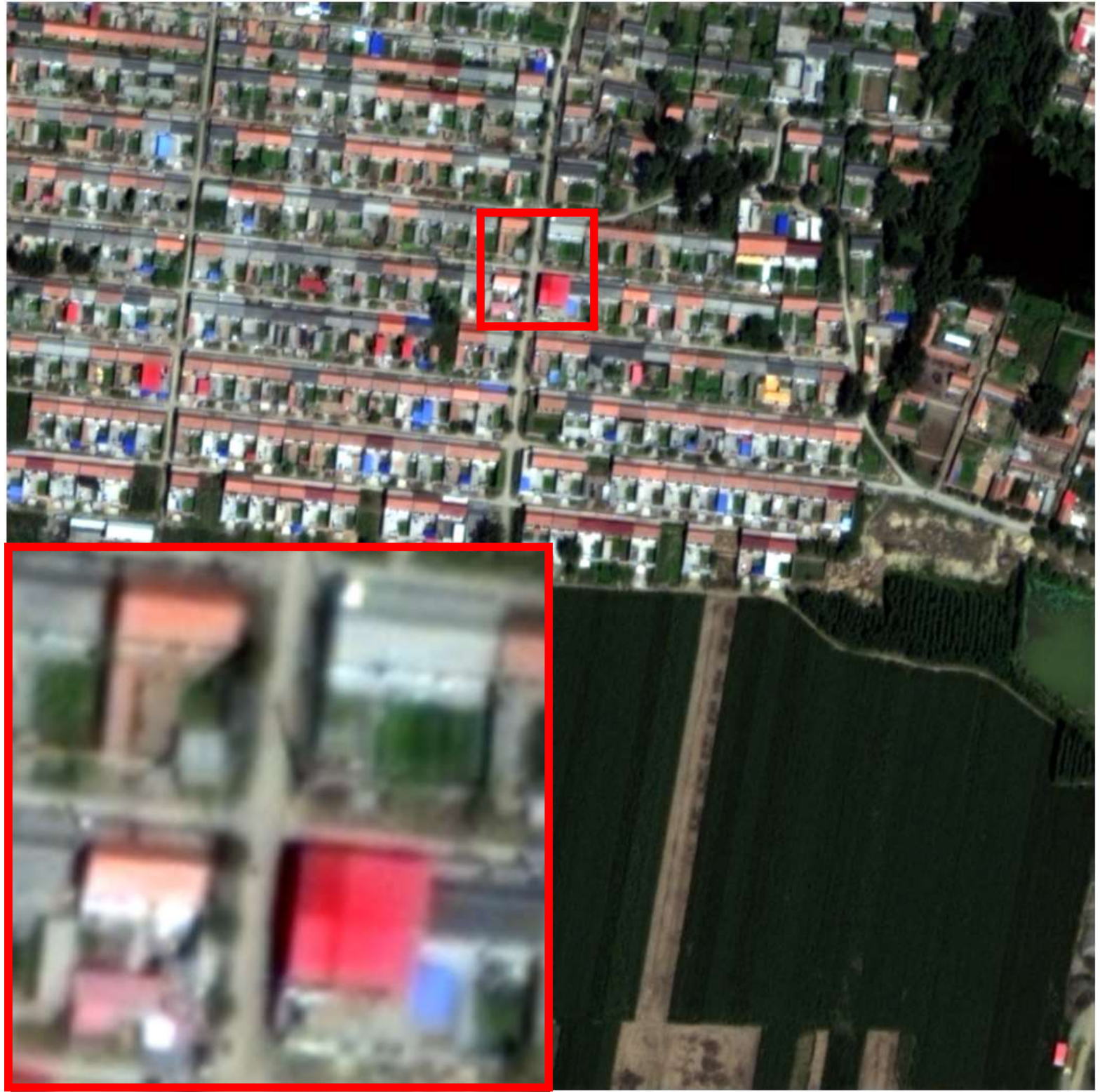}
\label{PWMBF}}
\subfloat[]{\includegraphics[width=1.1in,trim=120 0 120 0,clip]{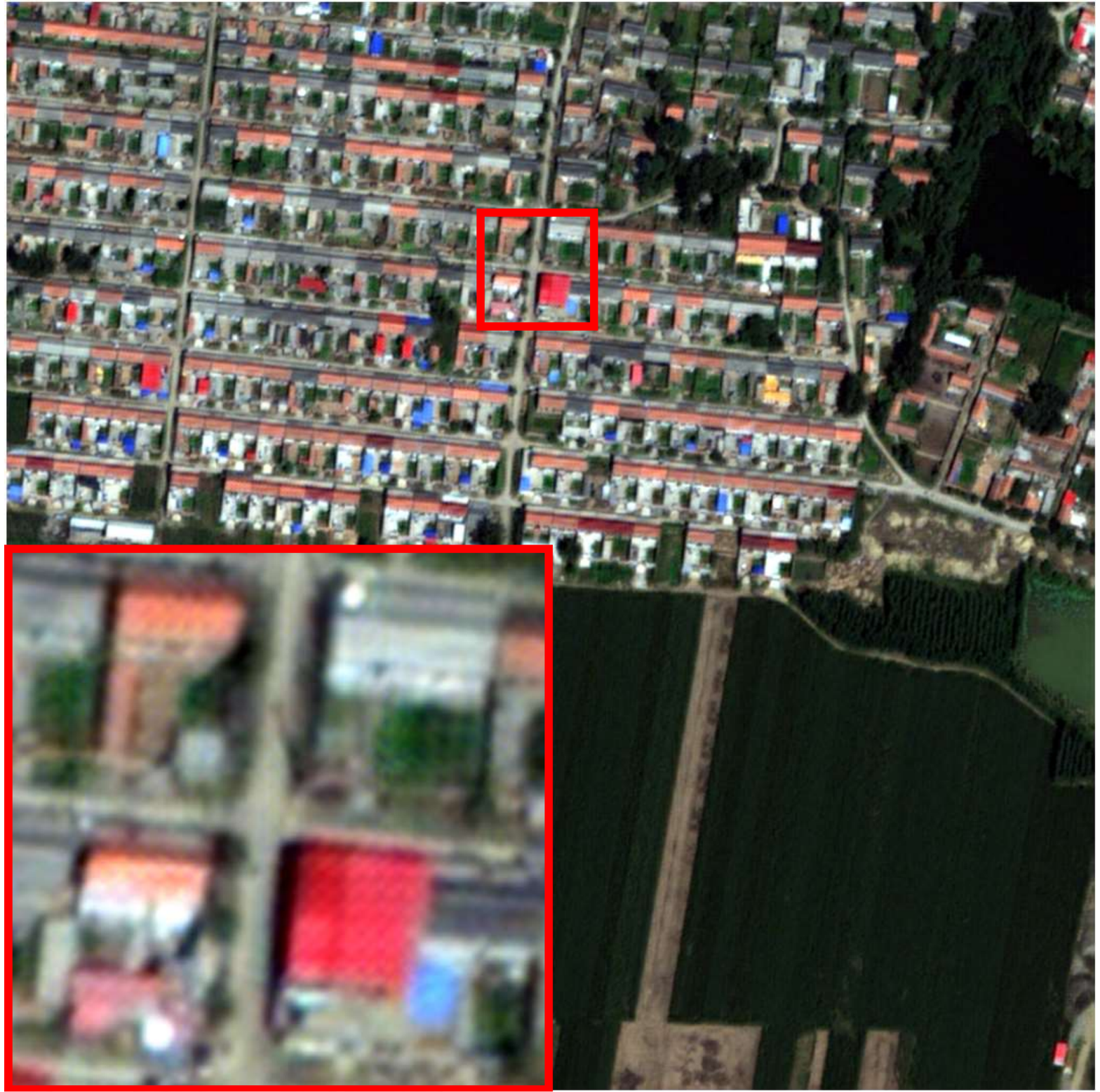}
\label{PanNet}}
\subfloat[]{\includegraphics[width=1.1in,trim=120 0 120 0,clip]{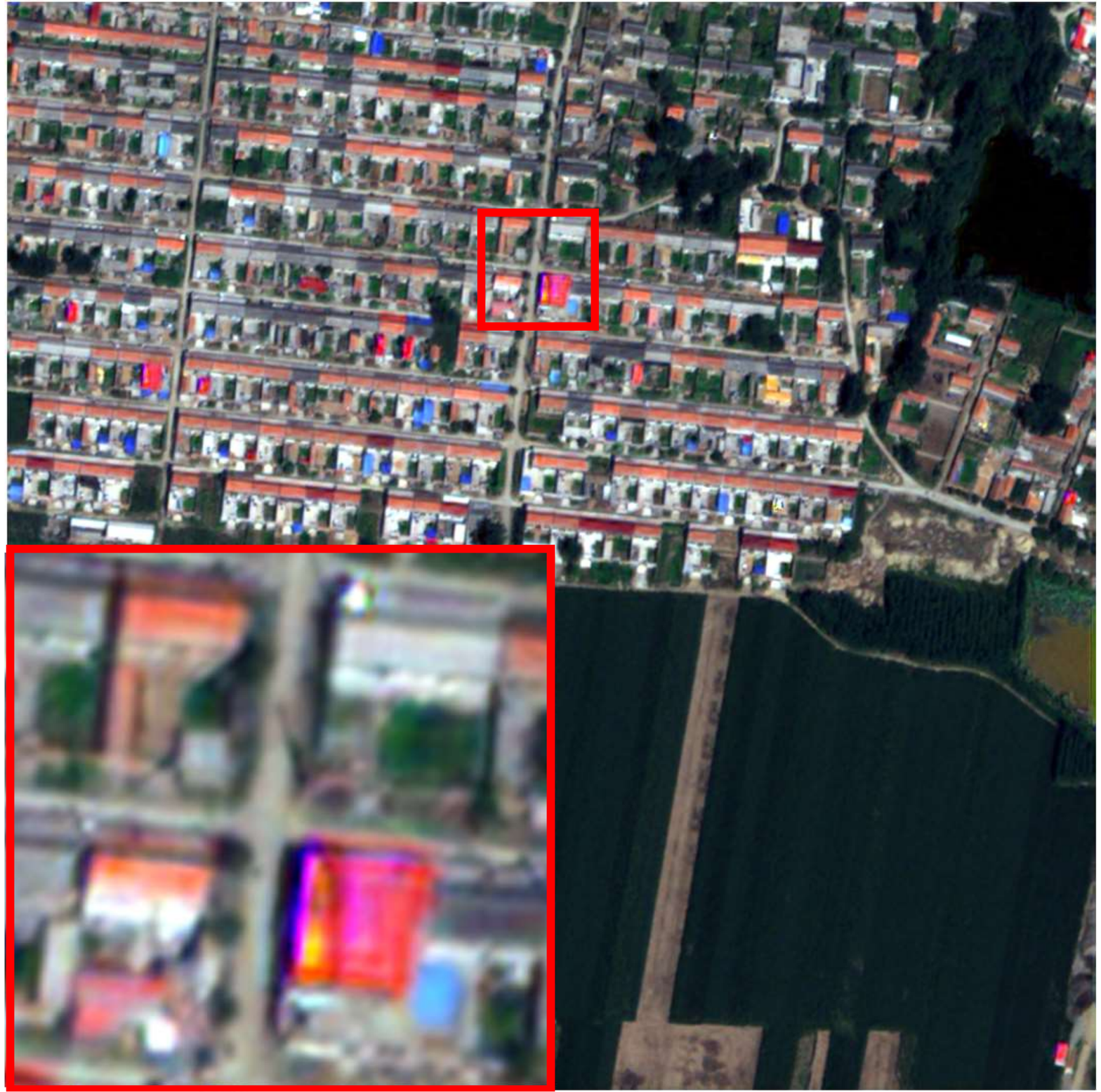}
\label{FusionNet}}
\subfloat[]{\includegraphics[width=1.1in,trim=120 0 120 0,clip]{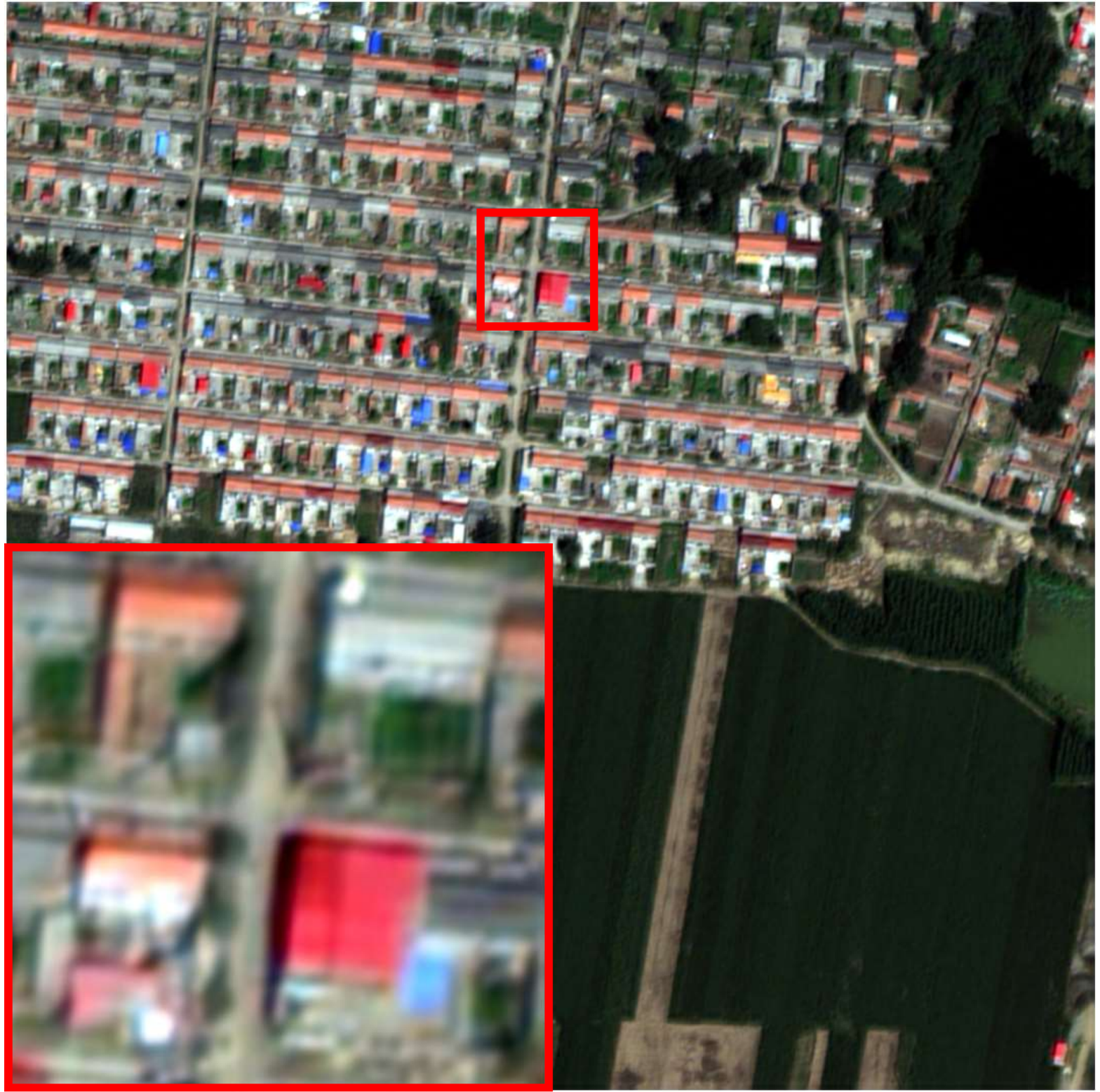}
\label{GTP-PNet}}
\subfloat[]{\includegraphics[width=1.1in,trim=120 0 120 0,clip]{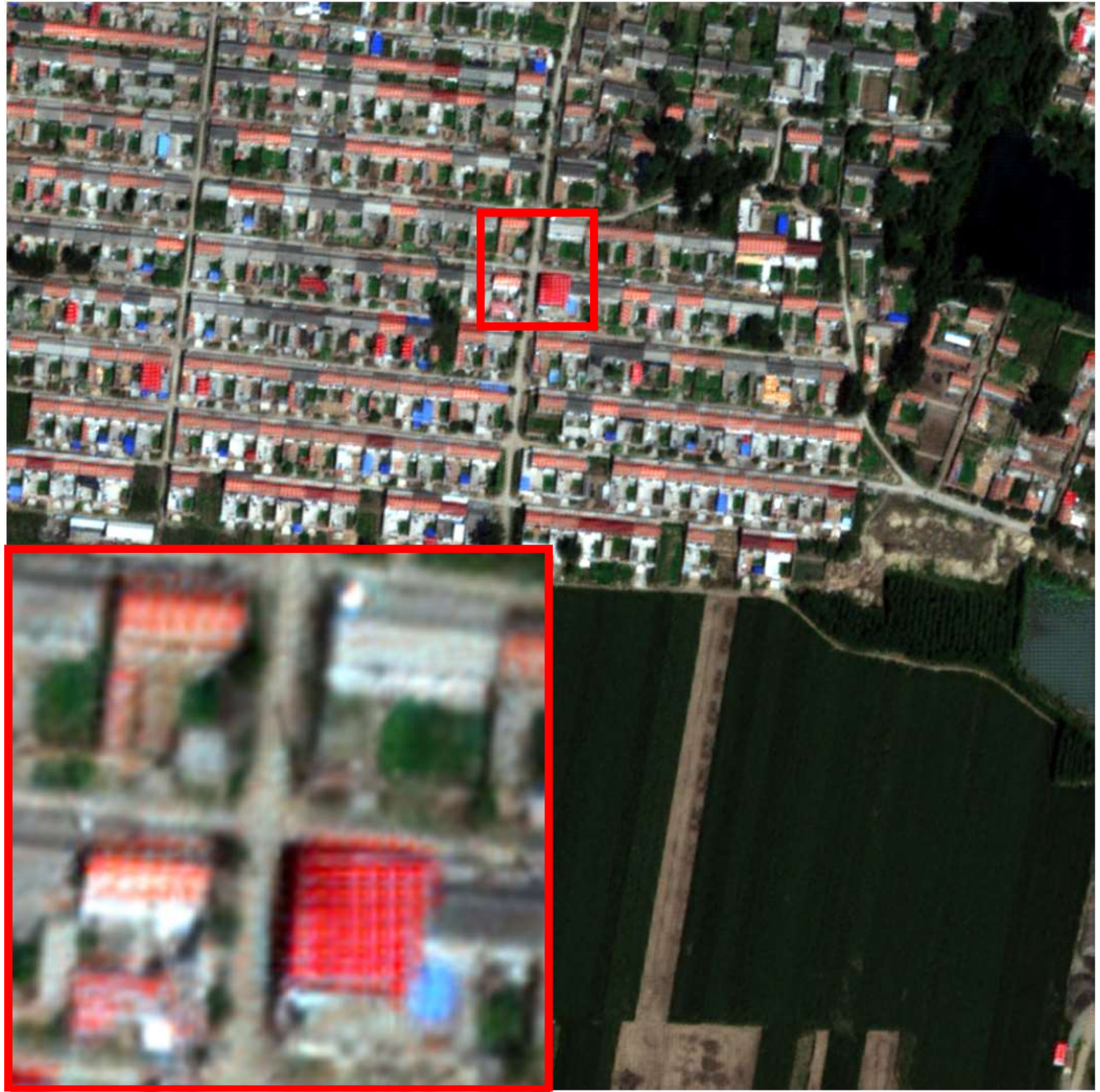}
\label{LPPN}}
\subfloat[]{\includegraphics[width=1.1in,trim=120 0 120 0,clip]{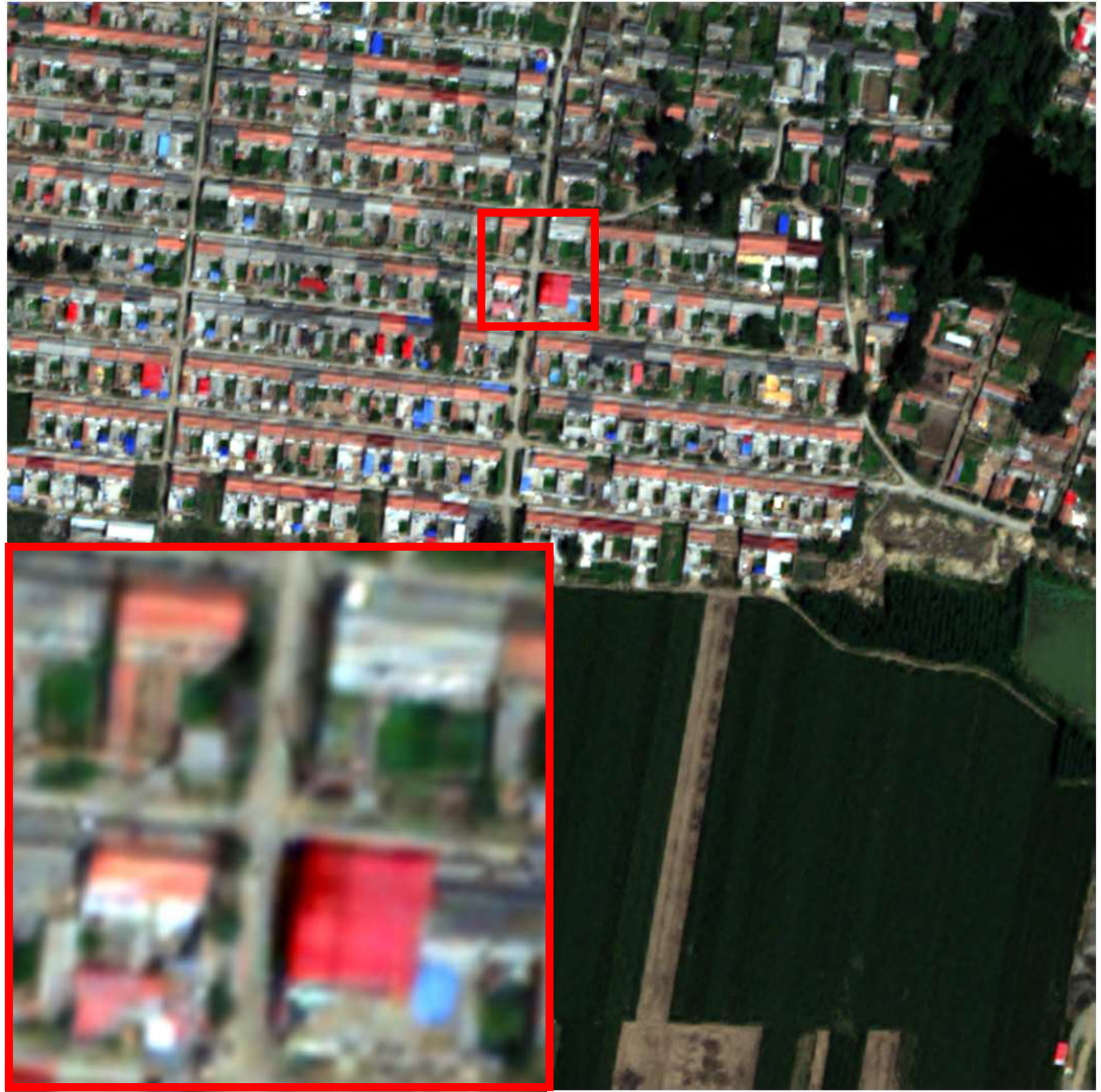}
\label{FAFNet}}
\caption{Visual comparison of different methods on WorldView-4 (WV-4) dataset at full resolution. (a) PAN. (b) Up-sampled MS. (c) GSA \cite{4305344}. (d) BDSD-PC  \cite{8693555} . (e) MTF-GLP-CBD \cite{4305345} . (f) AWLP-H \cite{vivone2019fast} . (g) PWMBF \cite{6951484}. (h) PanNet \cite{8237455}. (i) FusionNet \cite{9240949}. (j) GTP-PNet \cite{ZHANG2021223} . (k) LPPN \cite{JIN2022158}. (l) FAFNet.}
\label{fig_wv4_full}
\end{figure*}

\begin{figure*}[htbp]
\centering
\subfloat[]{\includegraphics[width=1.1in,trim=120 0 120 0,clip]{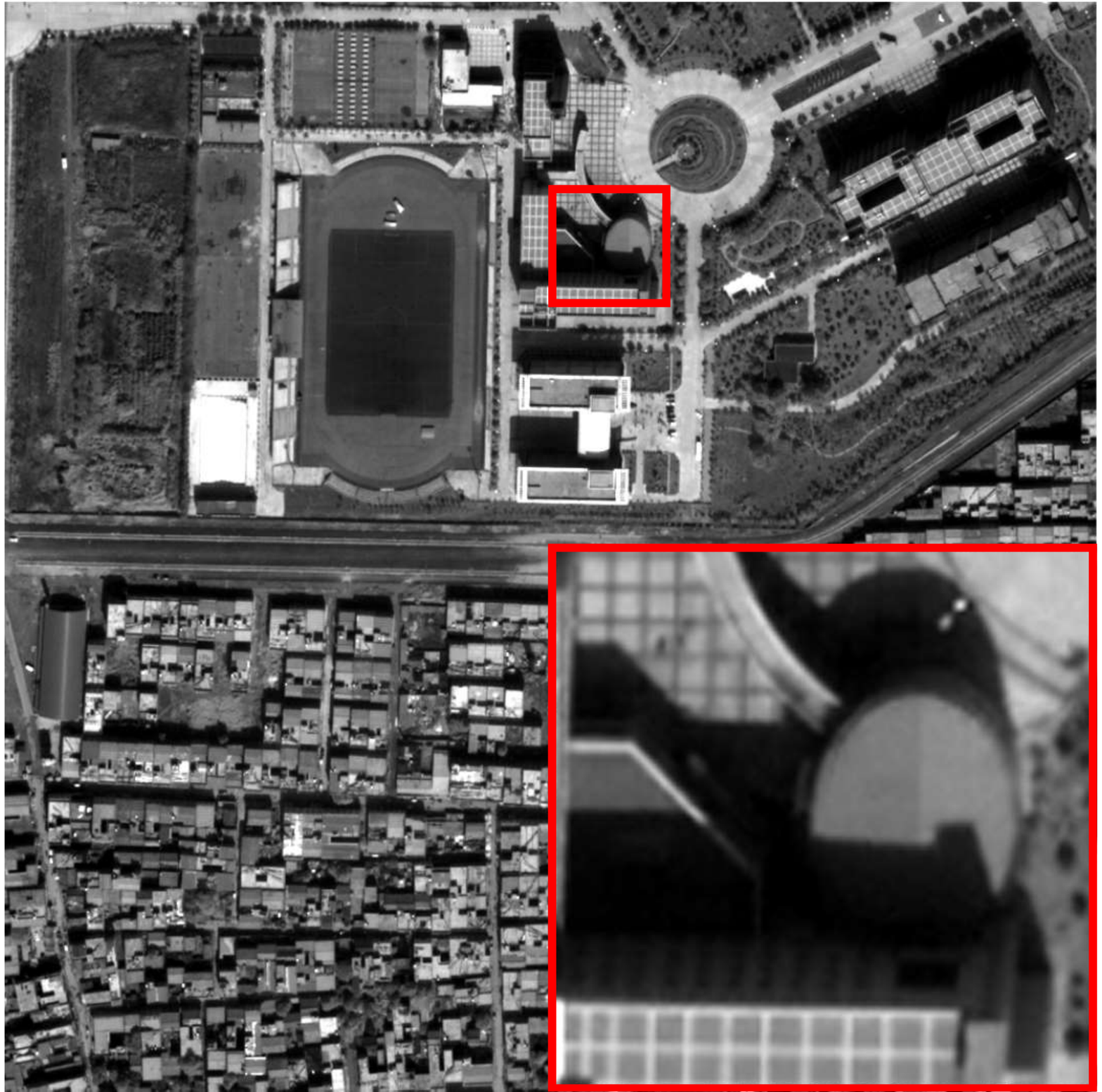}
\label{PAN}}
\subfloat[]{\includegraphics[width=1.1in,trim=120 0 120 0,clip]{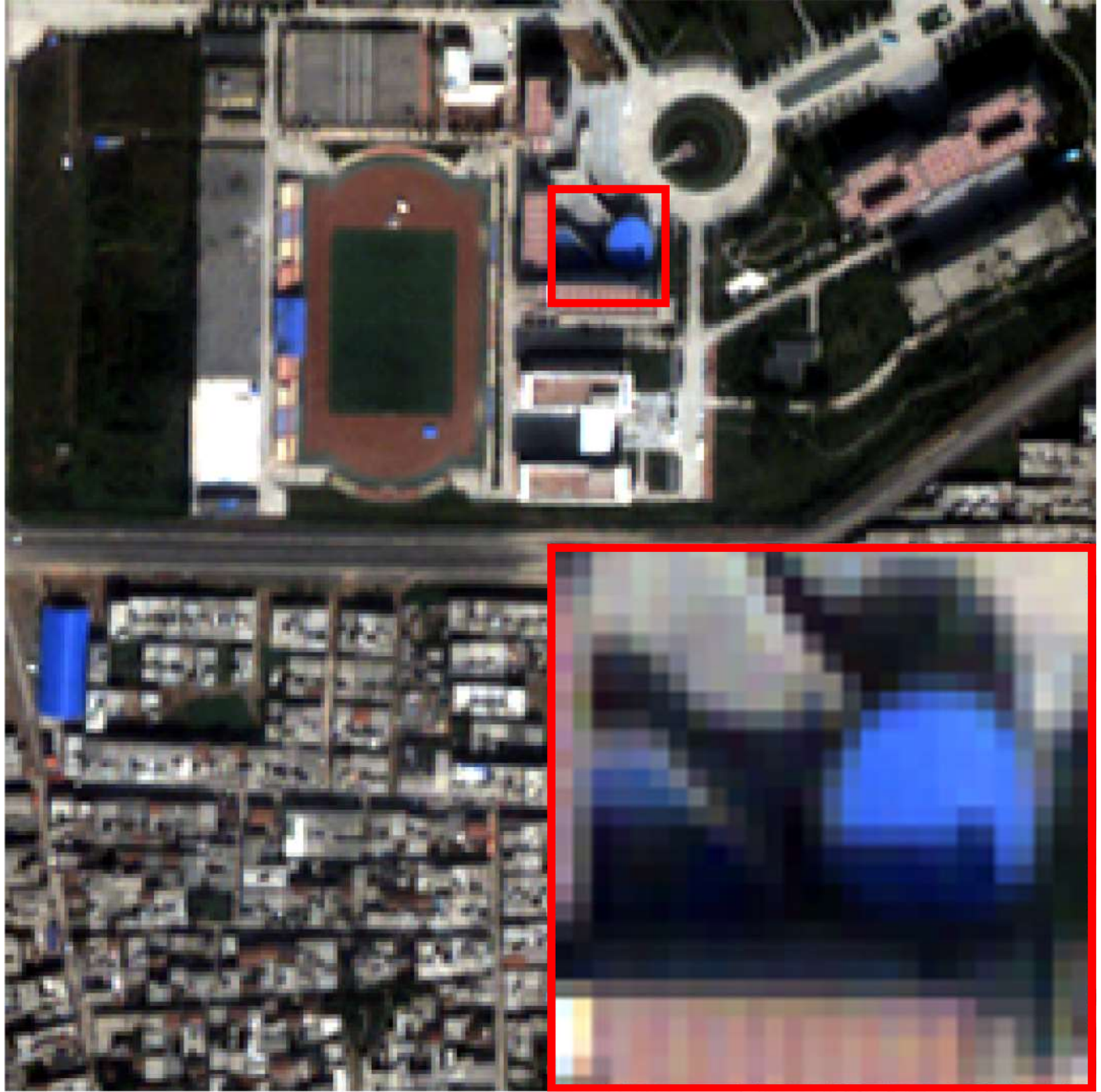}
\label{up-sampled MS}}
\subfloat[]{\includegraphics[width=1.1in,trim=120 0 120 0,clip]{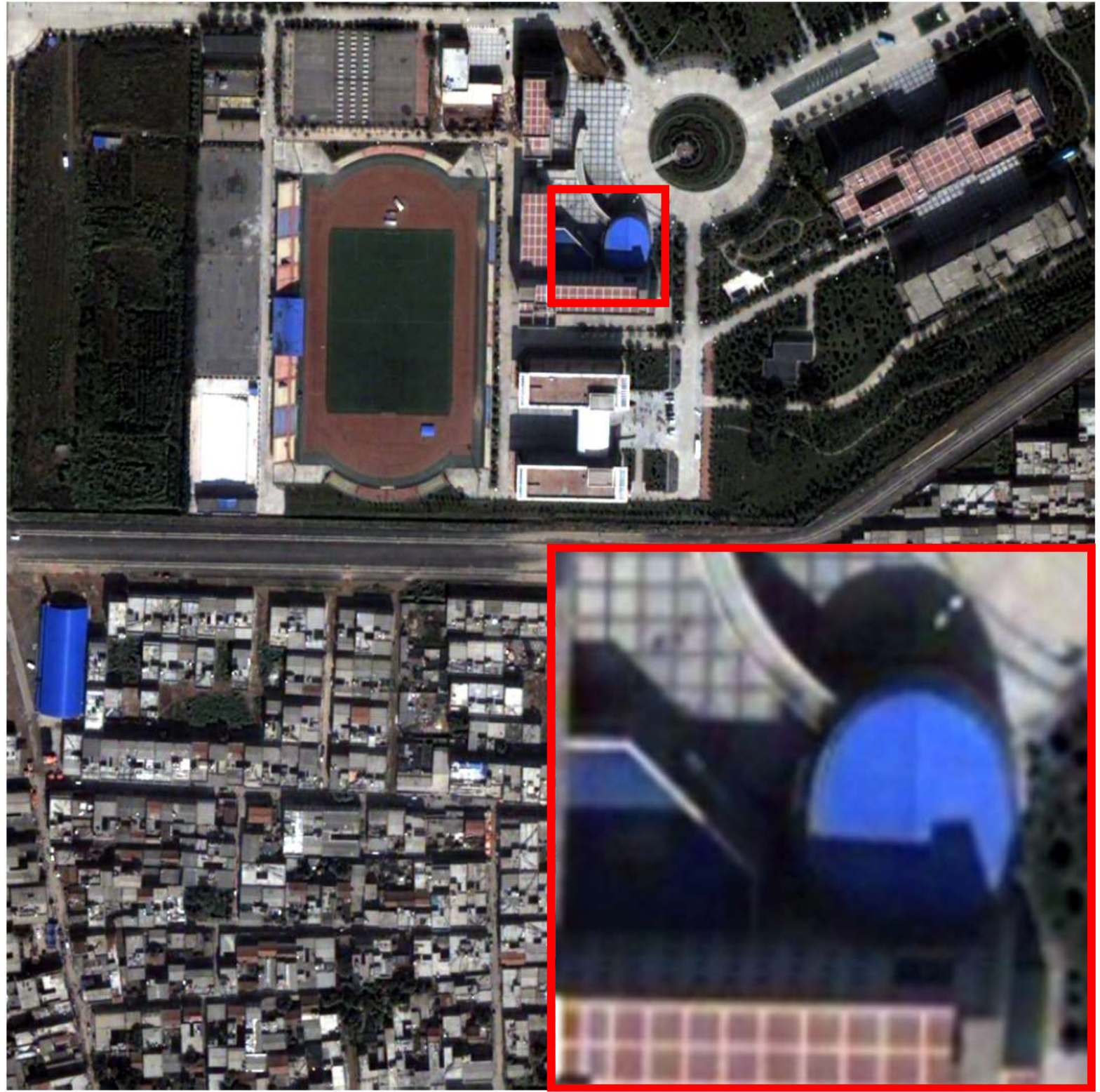}
\label{GSA}}
\subfloat[]{\includegraphics[width=1.1in,trim=120 0 120 0,clip]{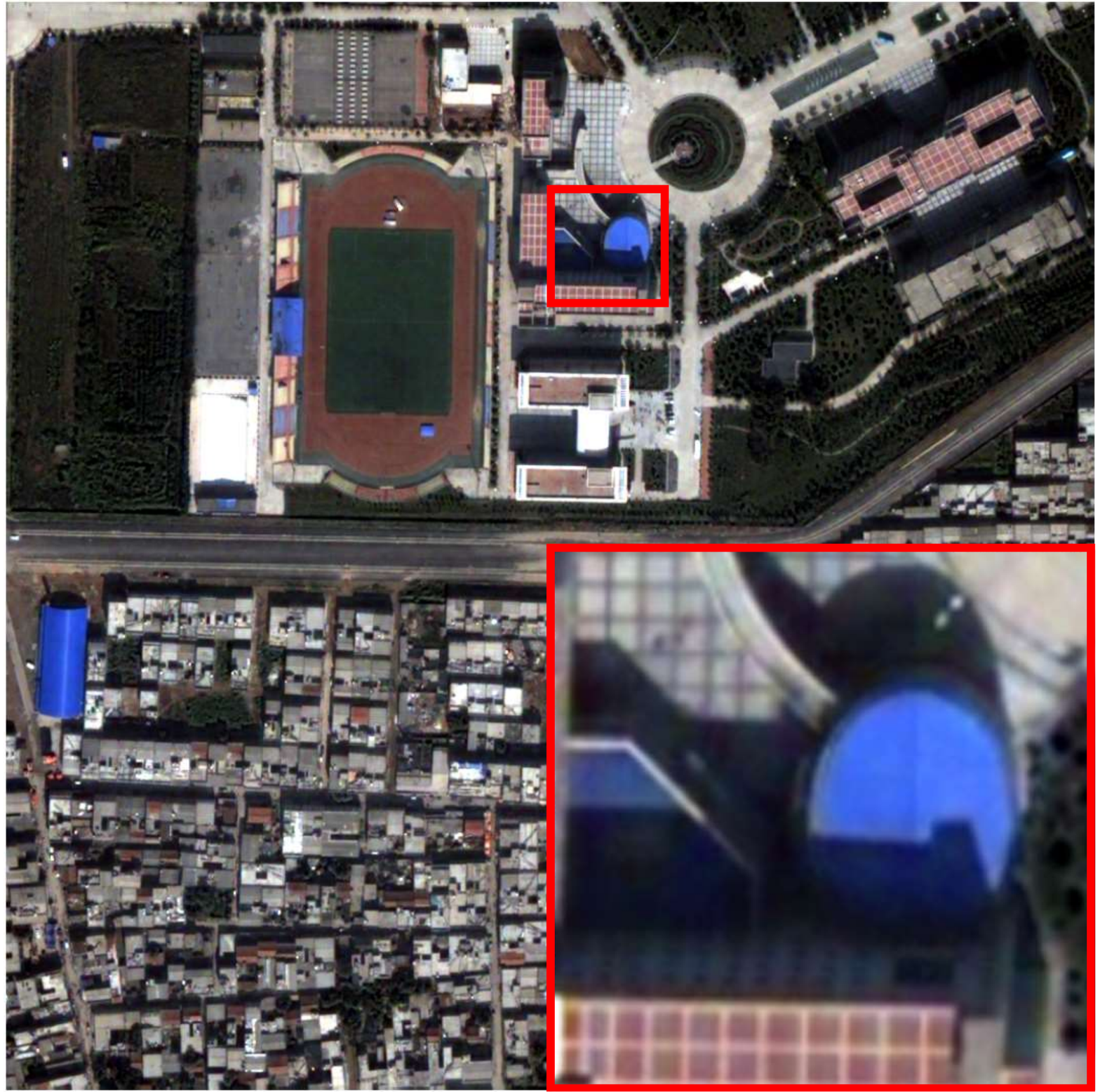}
\label{BDSD-PC}}
\subfloat[]{\includegraphics[width=1.1in,trim=120 0 120 0,clip]{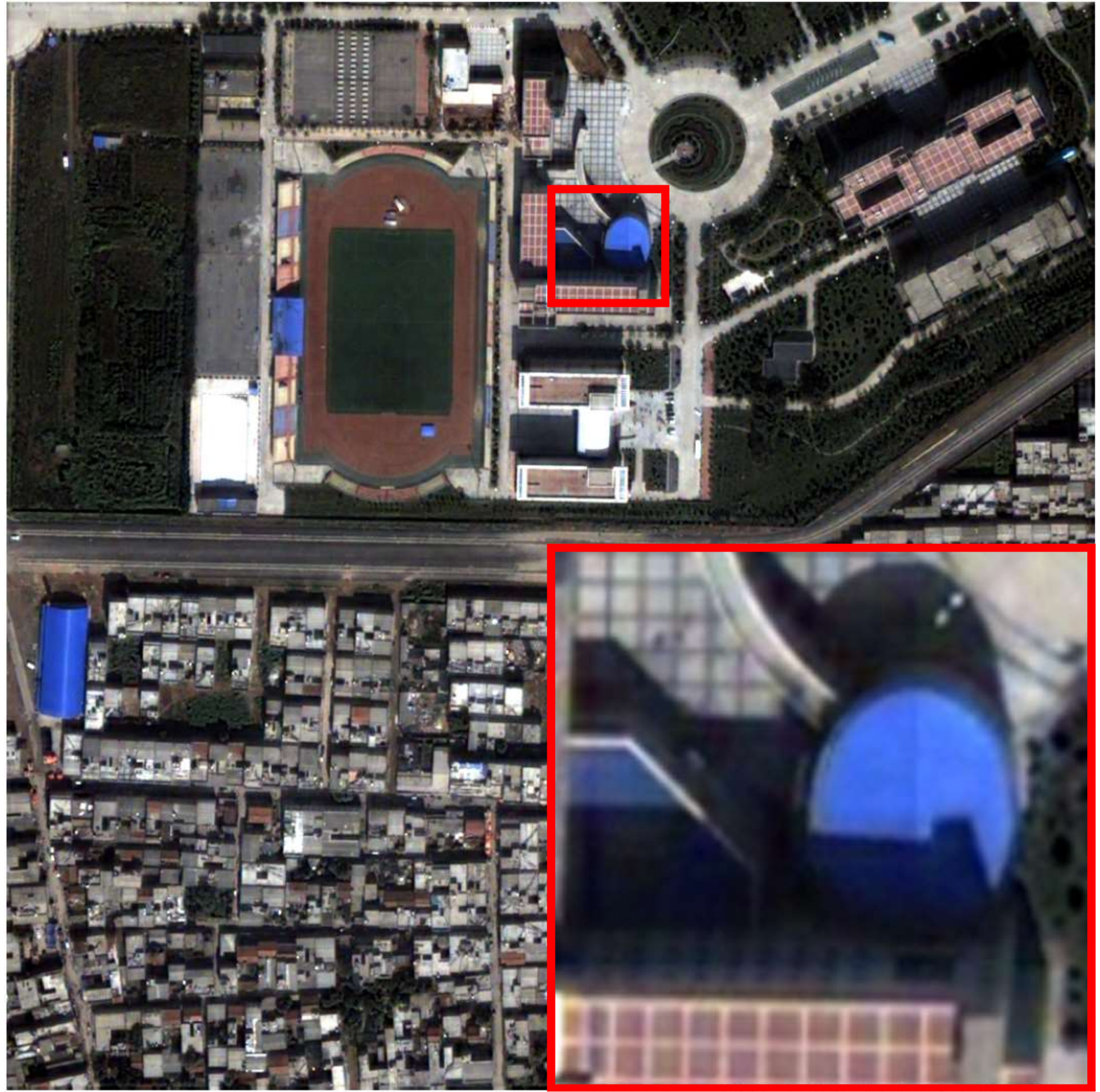}
\label{MTF-GLP-CBD}}
\subfloat[]{\includegraphics[width=1.1in,trim=120 0 120 0,clip]{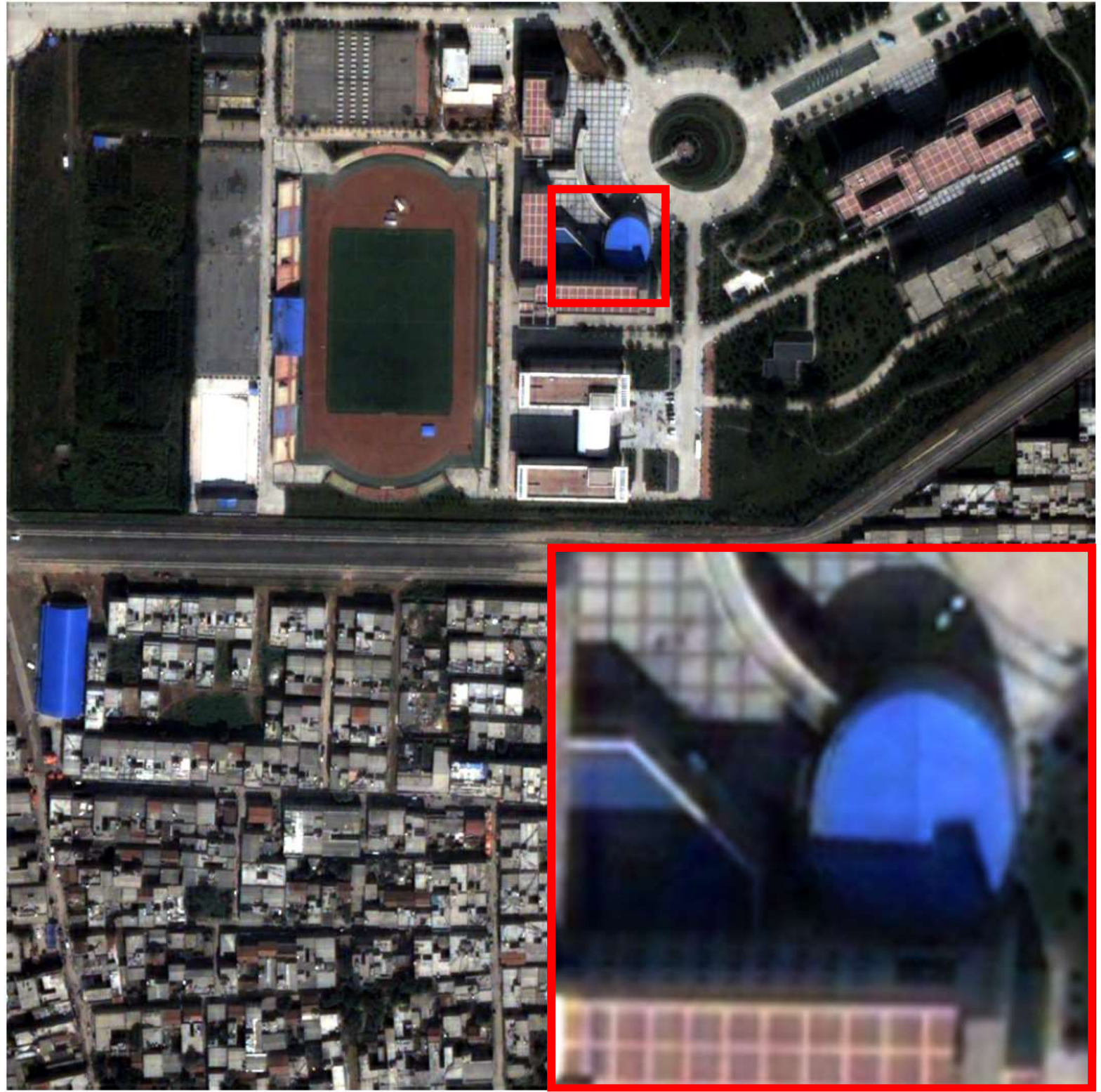}
\label{AWLP-H}}

\vspace{-0.1in}
\subfloat[]{\includegraphics[width=1.1in,trim=120 0 120 0,clip]{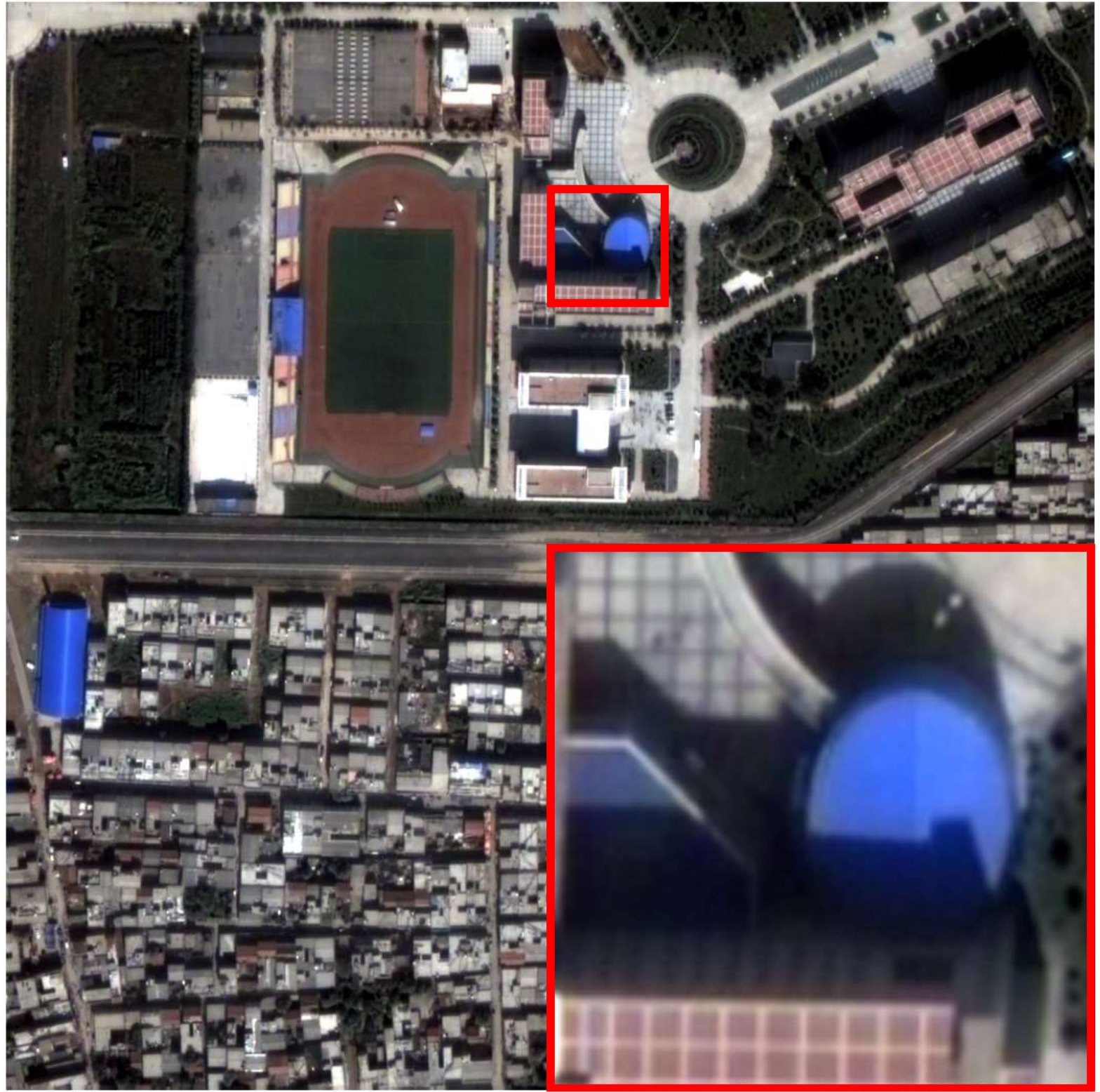}
\label{PWMBF}}
\subfloat[]{\includegraphics[width=1.1in,trim=120 0 120 0,clip]{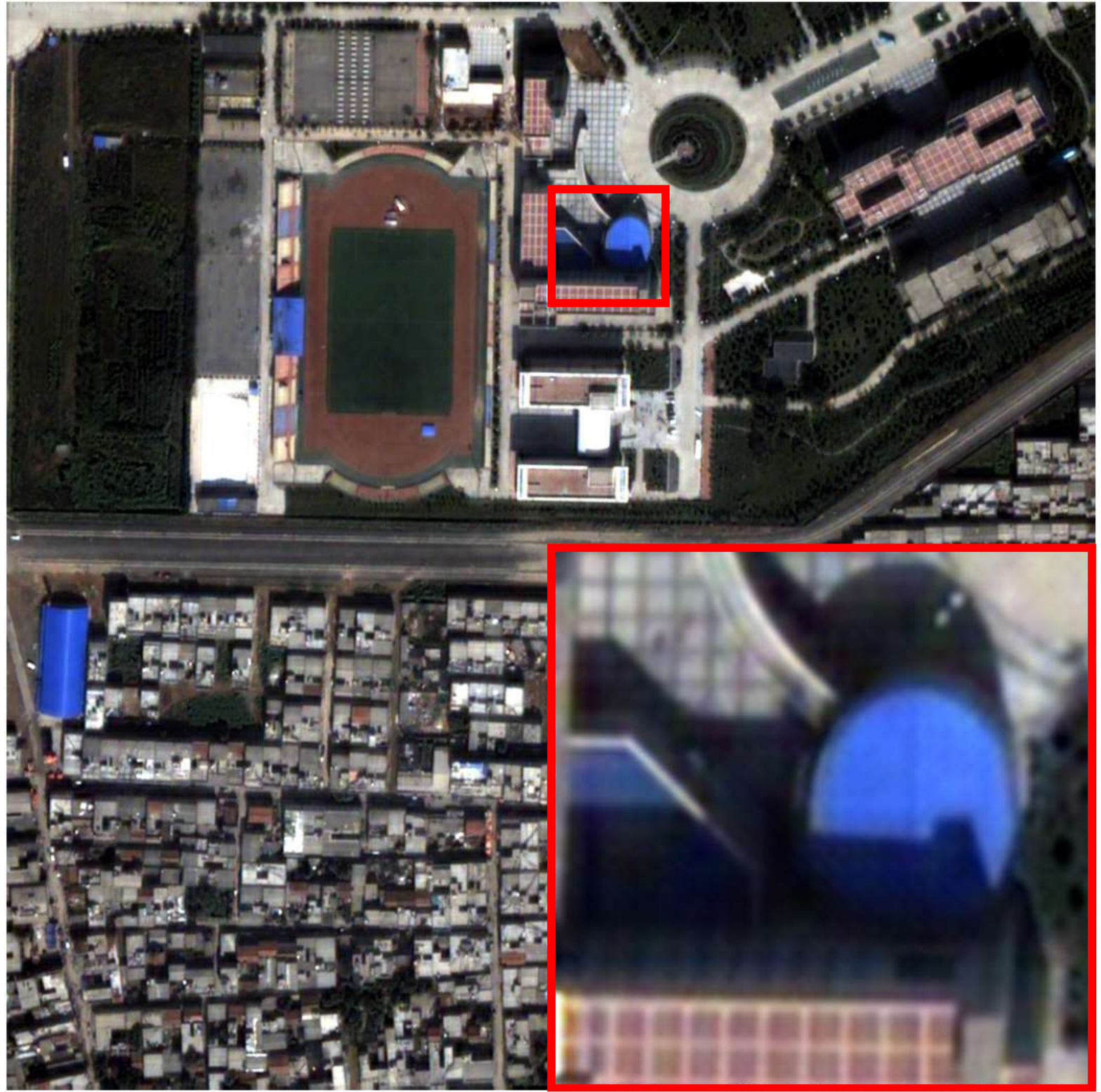}
\label{PanNet}}
\subfloat[]{\includegraphics[width=1.1in,trim=120 0 120 0,clip]{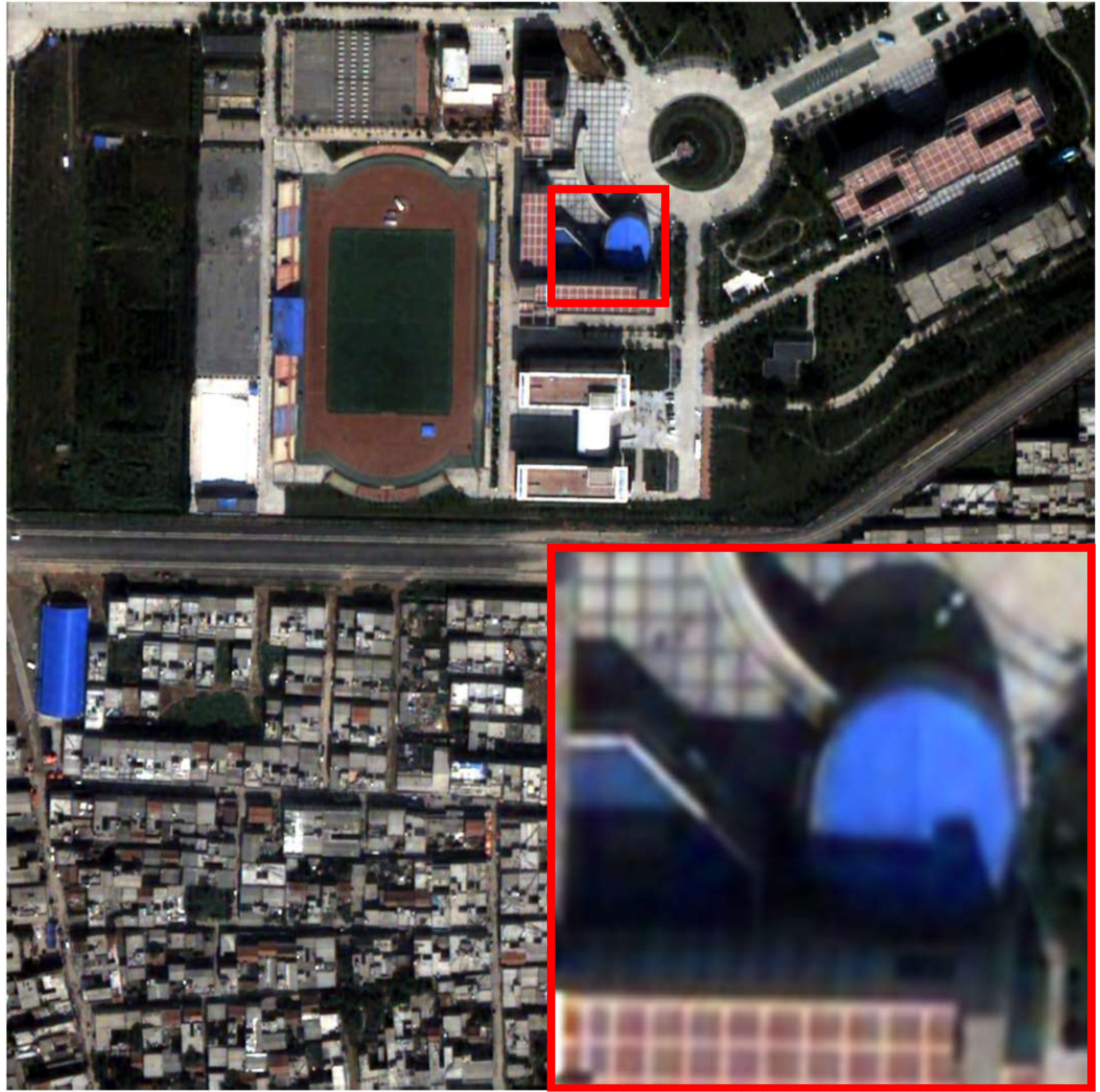}
\label{FusionNet}}
\subfloat[]{\includegraphics[width=1.1in,trim=120 0 120 0,clip]{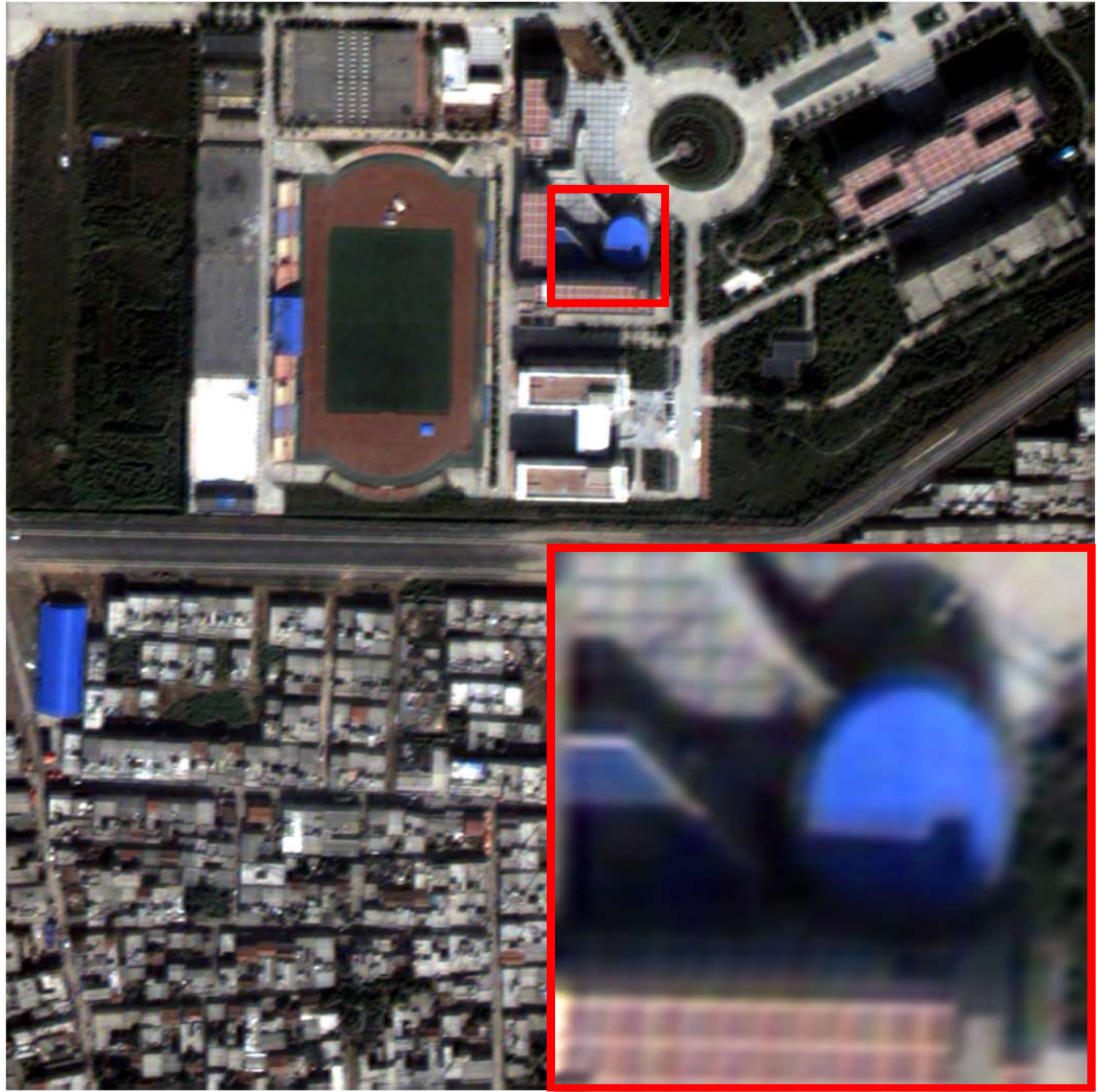}
\label{GTP-PNet}}
\subfloat[]{\includegraphics[width=1.1in,trim=120 0 120 0,clip]{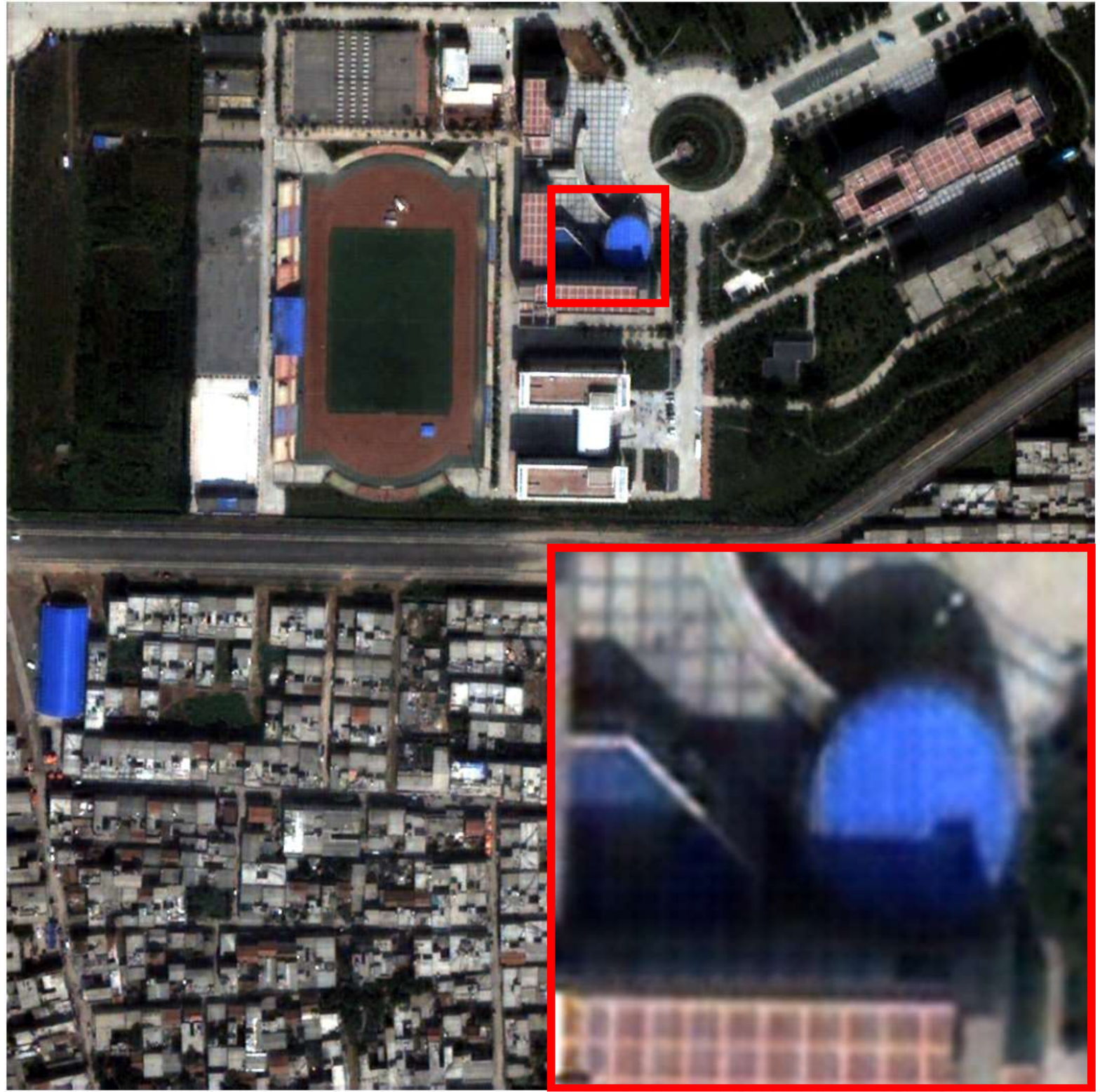}
\label{LPPN}}
\subfloat[]{\includegraphics[width=1.1in,trim=120 0 120 0,clip]{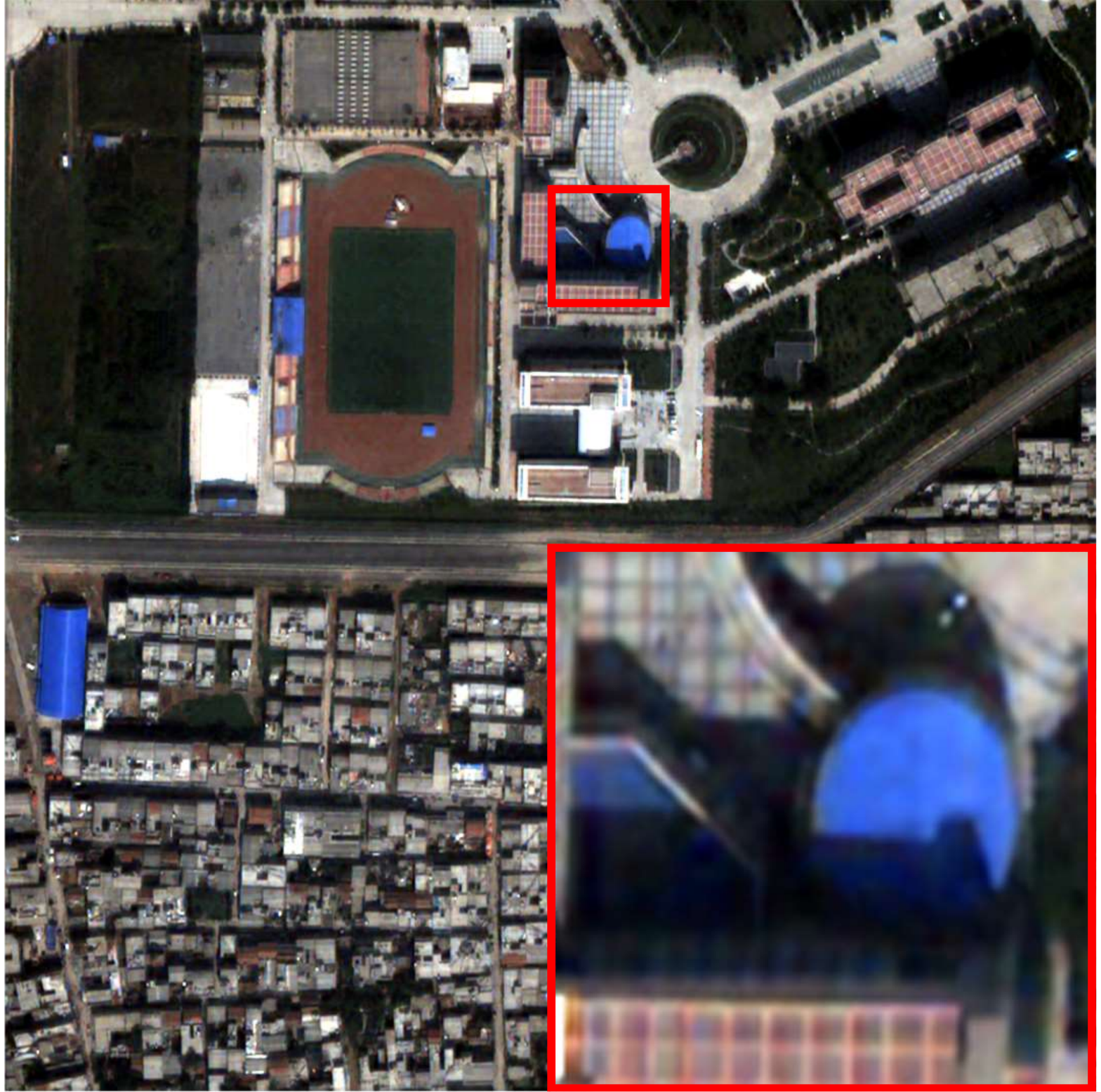}
\label{FAFNet}}
\caption{Visual comparison of different methods on QuickBird (QB) dataset at full resolution. (a) PAN. (b) Up-sampled MS. (c) GSA \cite{4305344}. (d) BDSD-PC  \cite{8693555} . (e) MTF-GLP-CBD \cite{4305345} . (f) AWLP-H \cite{vivone2019fast} . (g) PWMBF \cite{6951484}. (h) PanNet \cite{8237455}. (i) FusionNet \cite{9240949}. (j) GTP-PNet \cite{ZHANG2021223} . (k) LPPN \cite{JIN2022158}. (l) FAFNet. }
\label{fig_qb_full}
\end{figure*}

\begin{figure*}[htbp]
\centering
\subfloat[]{\includegraphics[width=1.1in,trim=120 0 120 0,clip]{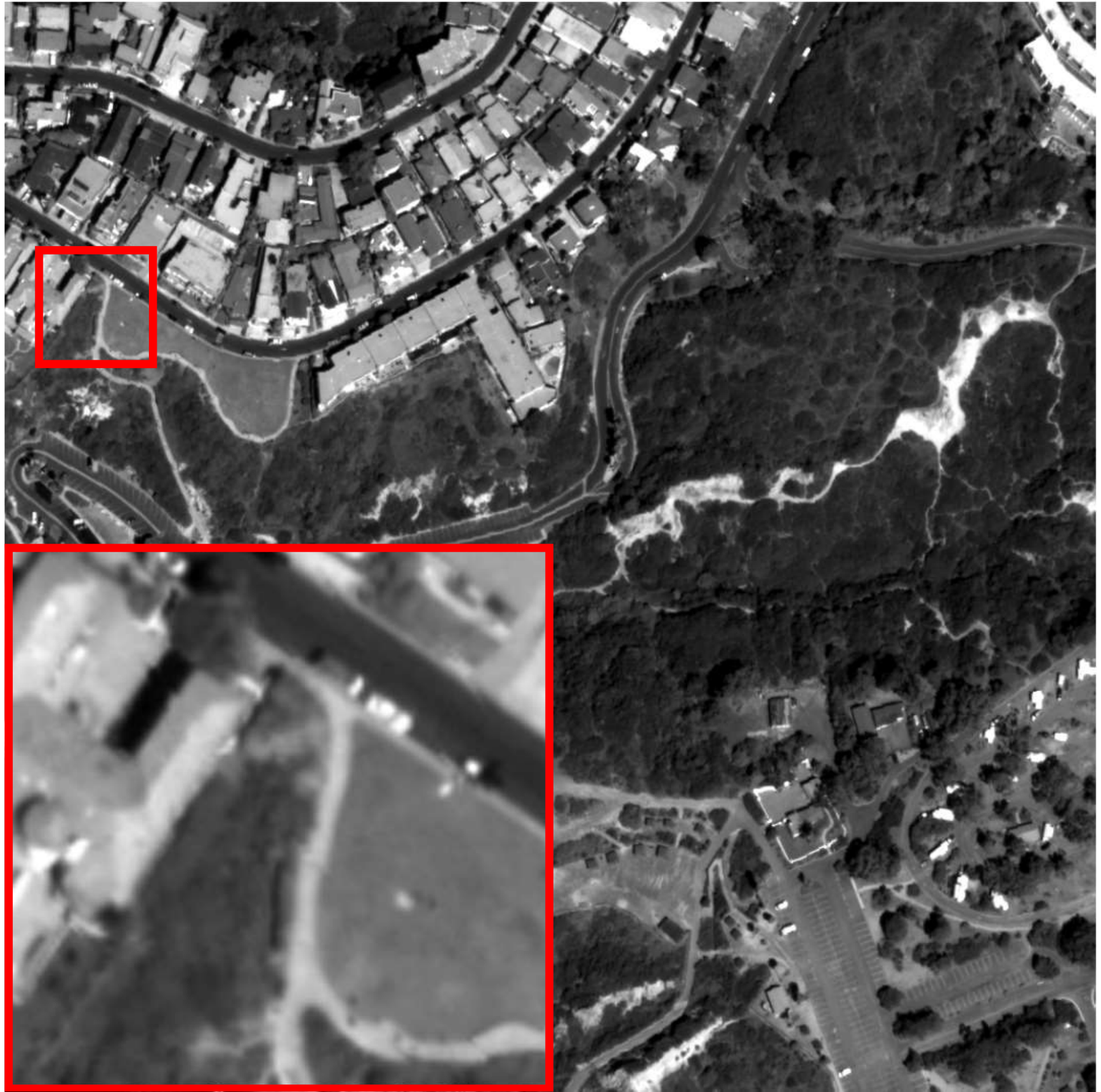}
\label{PAN}}
\subfloat[]{\includegraphics[width=1.1in,trim=120 0 120 0,clip]{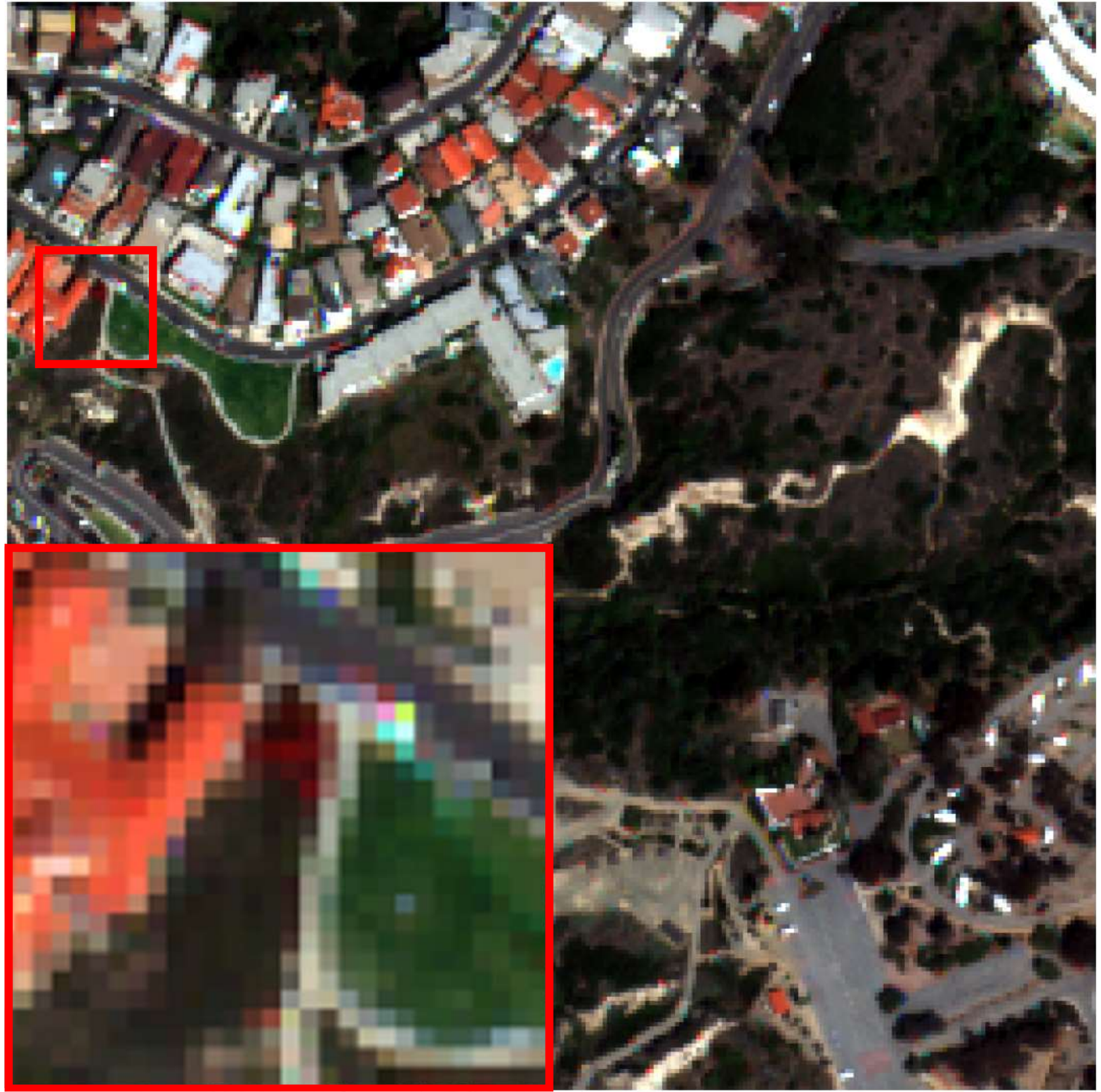}
\label{up-sampled MS}}
\subfloat[]{\includegraphics[width=1.1in,trim=120 0 120 0,clip]{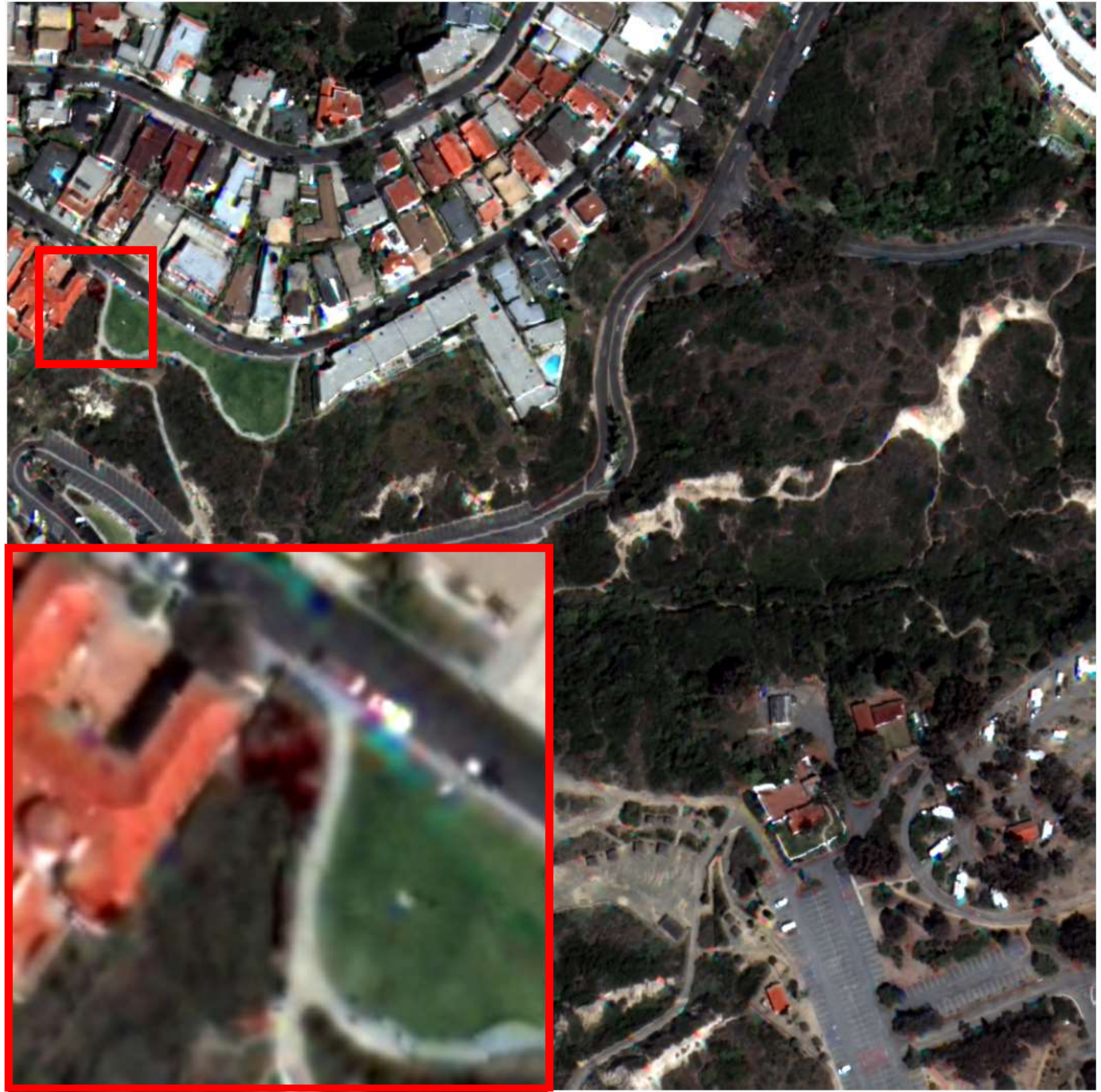}
\label{GSA}}
\subfloat[]{\includegraphics[width=1.1in,trim=120 0 120 0,clip]{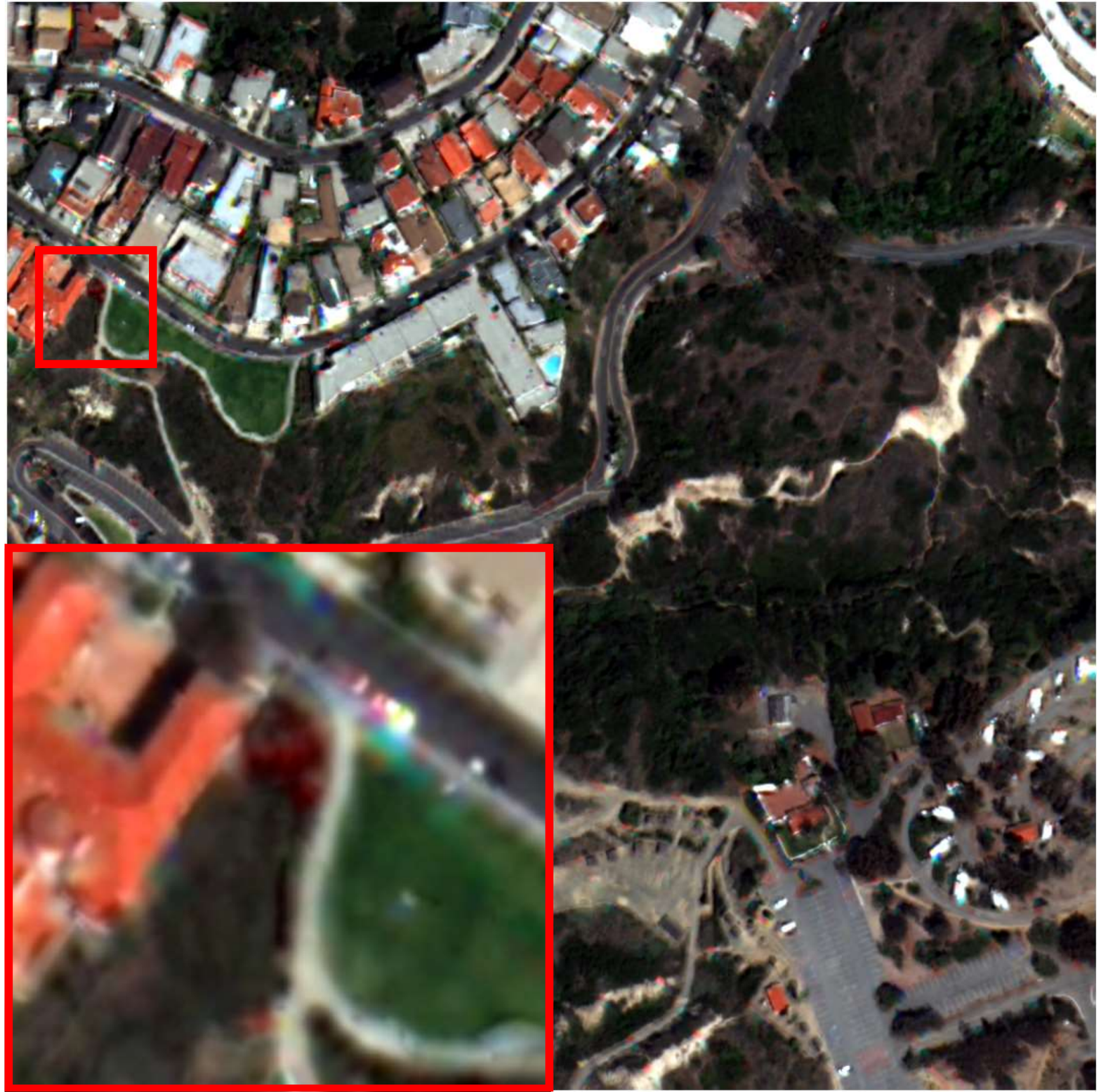}
\label{BDSD-PC}}
\subfloat[]{\includegraphics[width=1.1in,trim=120 0 120 0,clip]{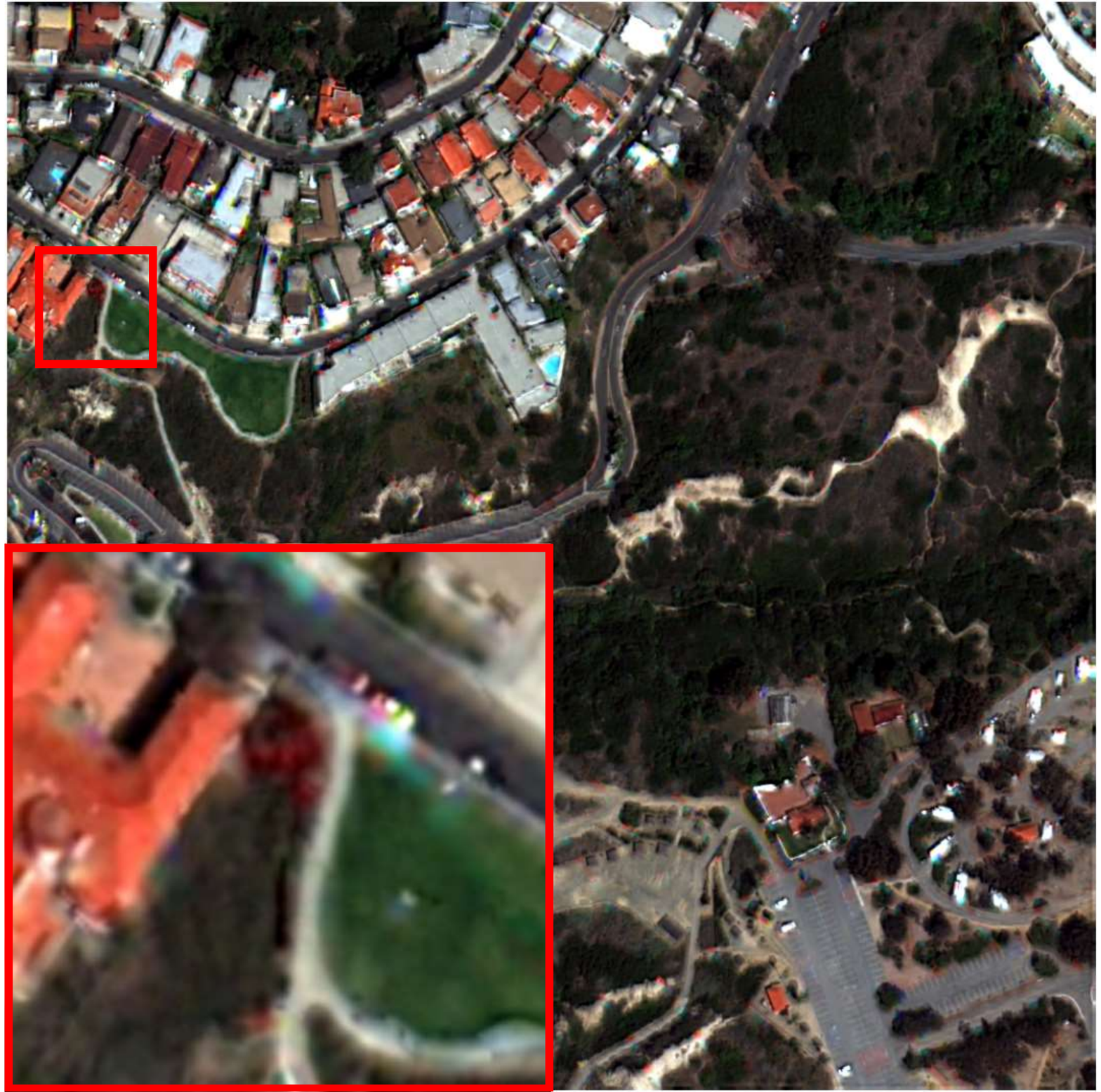}
\label{MTF-GLP-CBD}}
\subfloat[]{\includegraphics[width=1.1in,trim=120 0 120 0,clip]{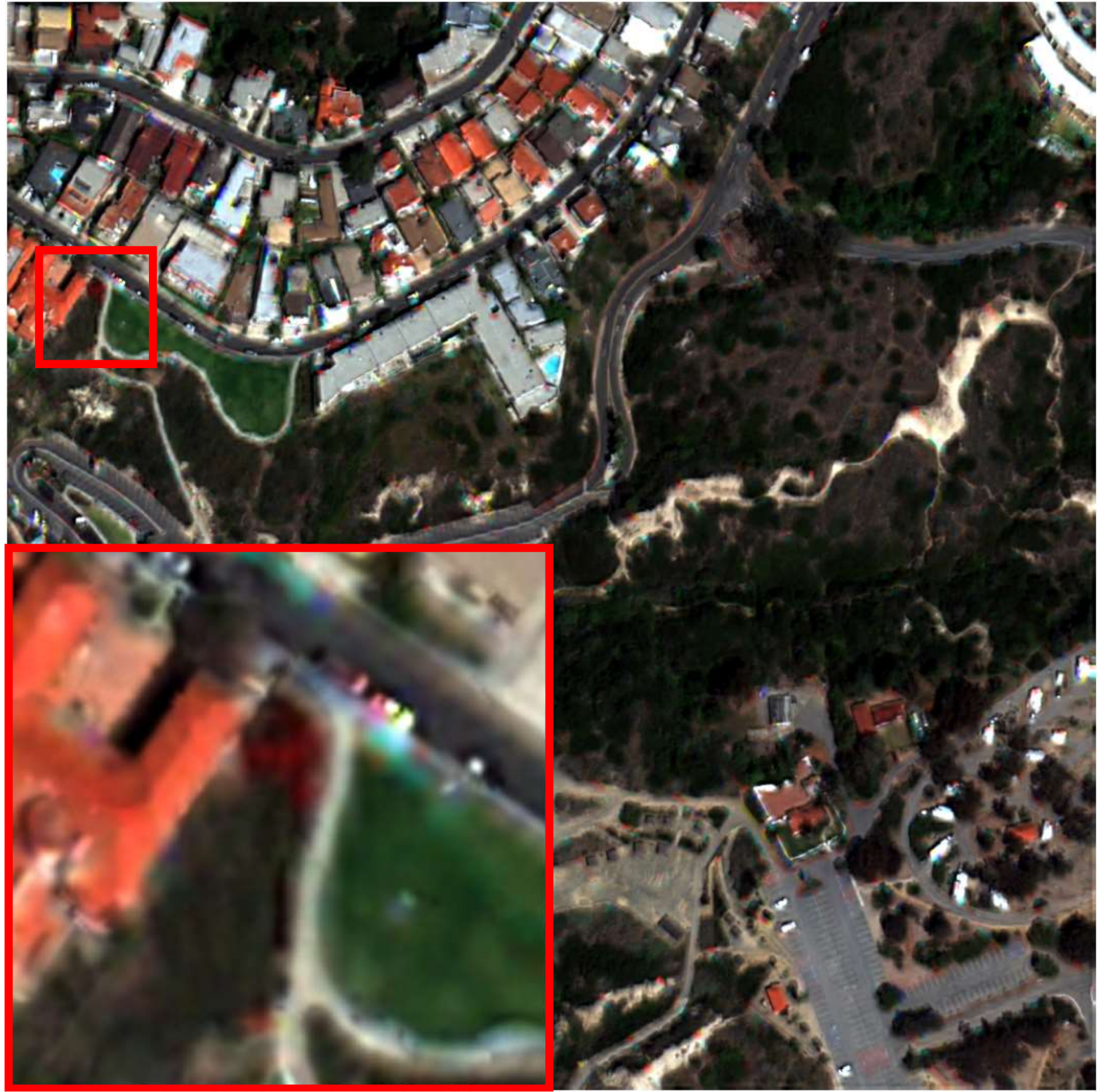}
\label{AWLP-H}}

\vspace{-0.1in}
\subfloat[]{\includegraphics[width=1.1in,trim=120 0 120 0,clip]{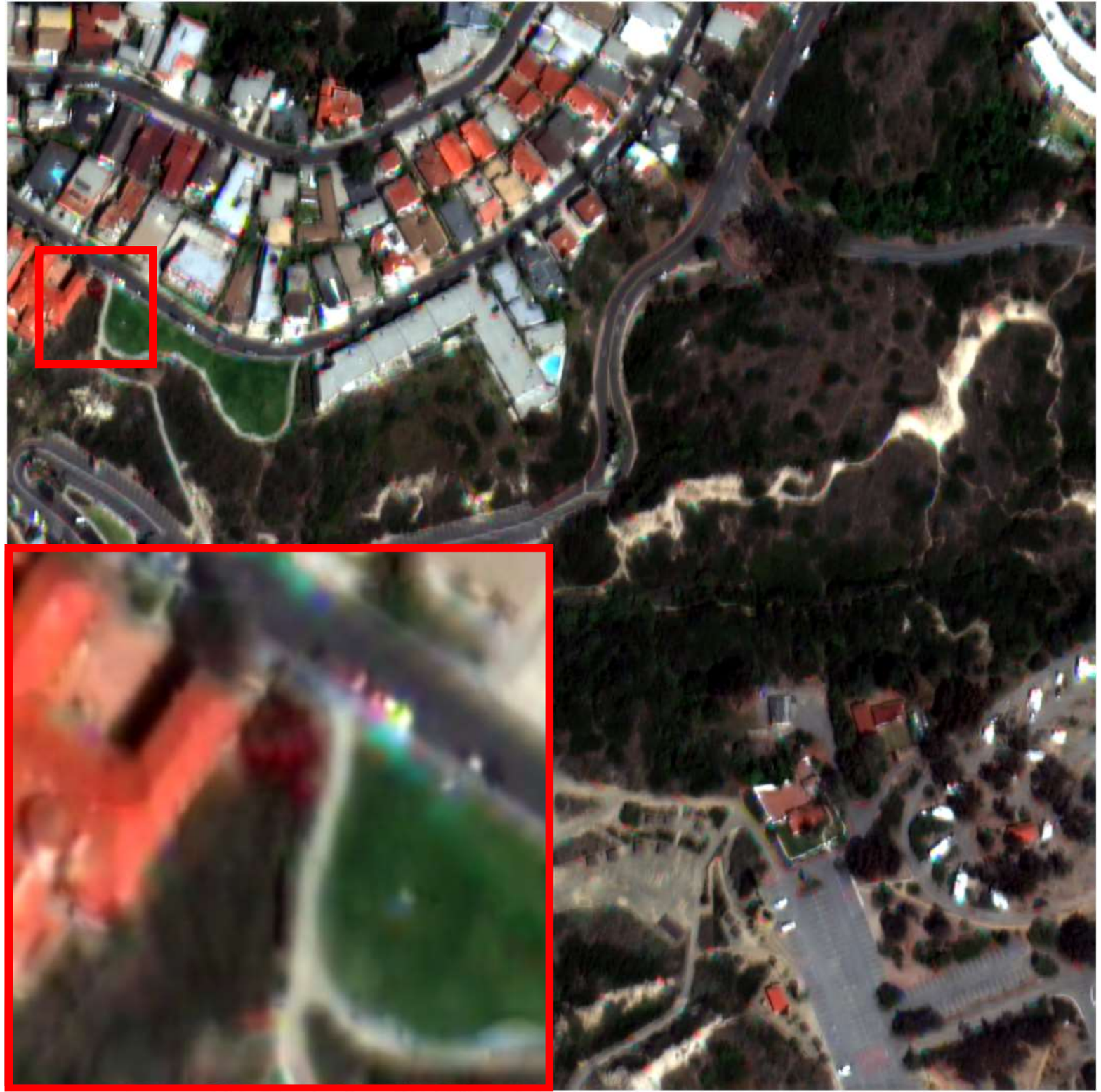}
\label{PWMBF}}
\subfloat[]{\includegraphics[width=1.1in,trim=120 0 120 0,clip]{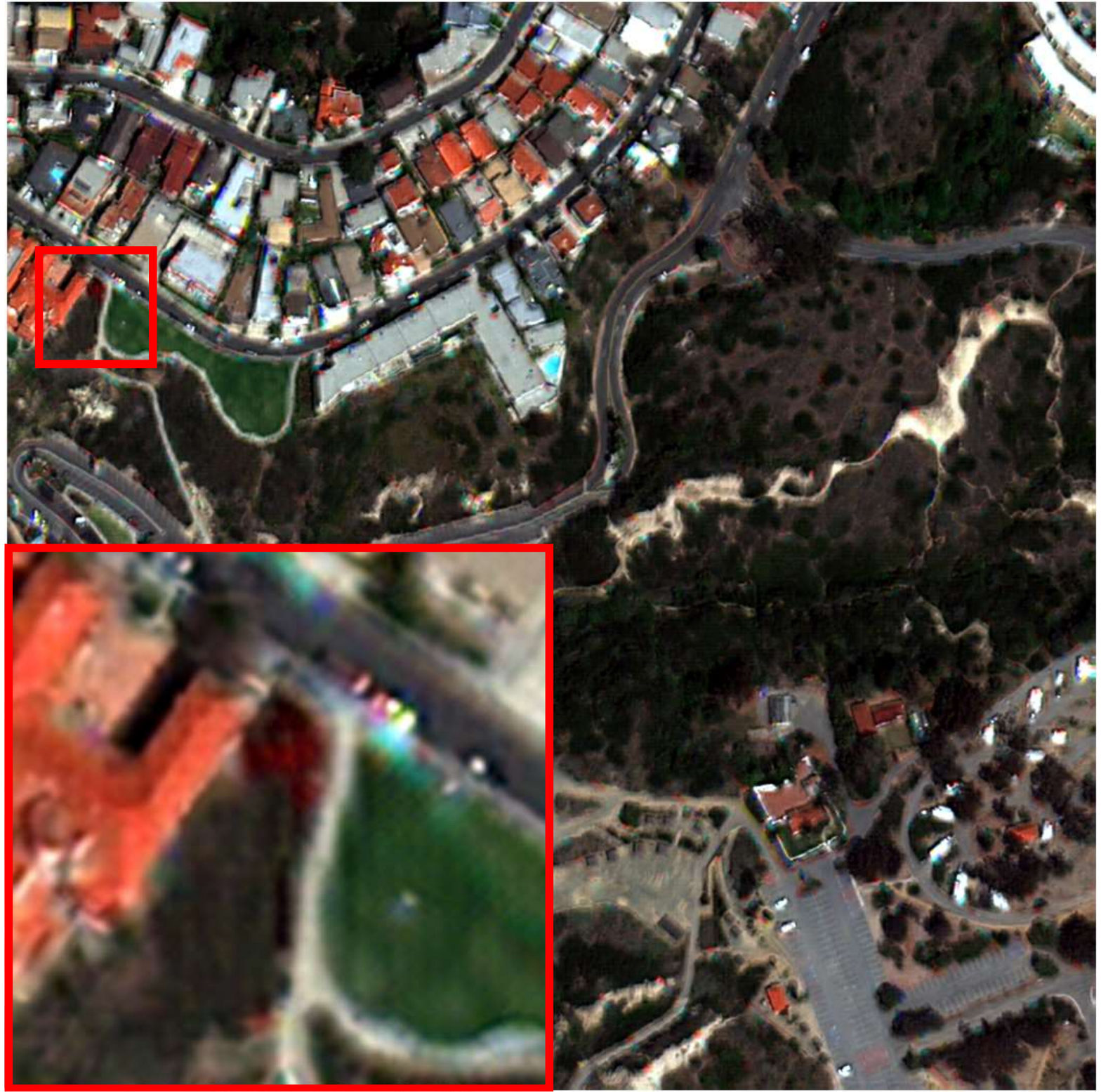}
\label{PanNet}}
\subfloat[]{\includegraphics[width=1.1in,trim=120 0 120 0,clip]{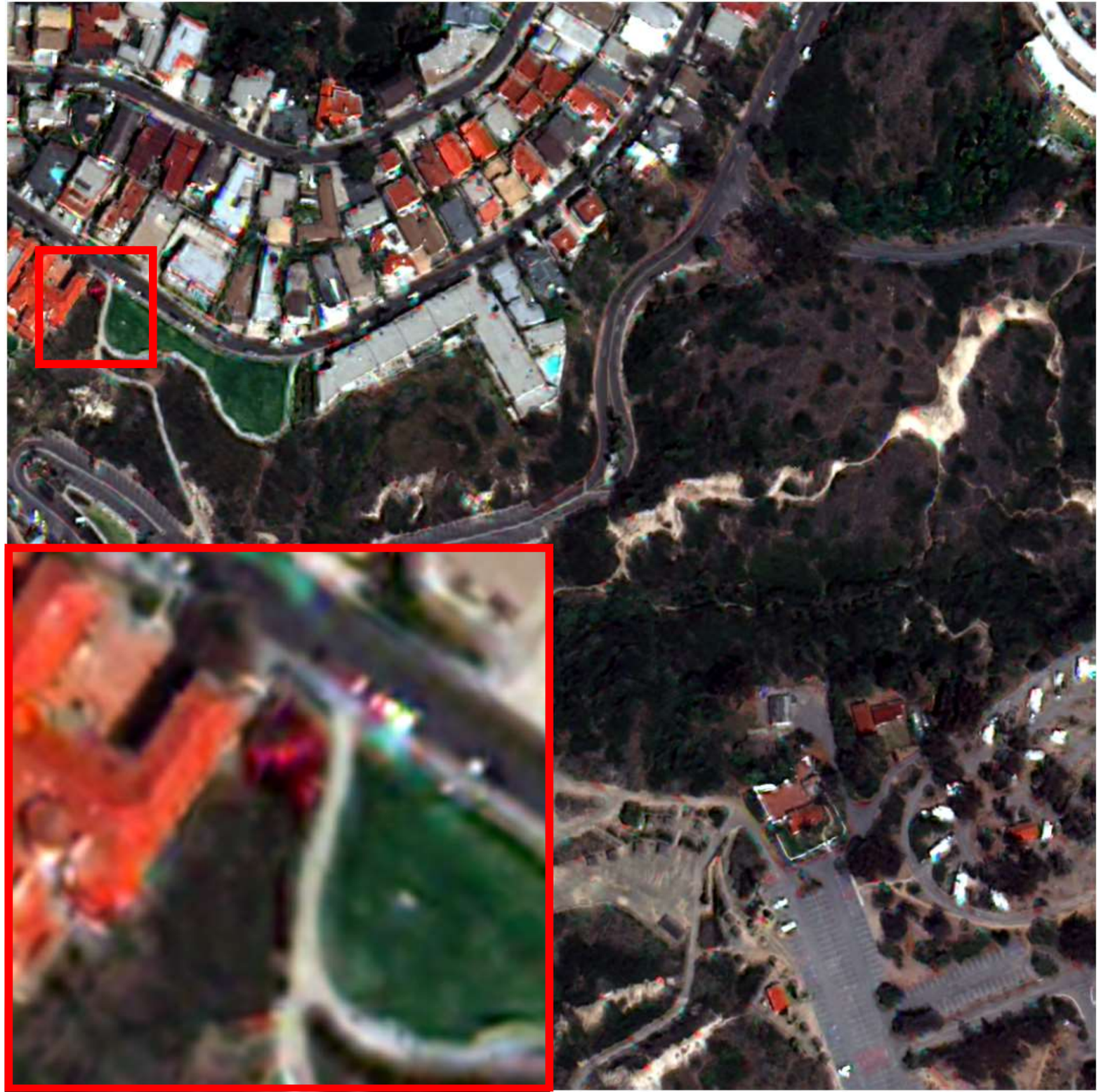}
\label{FusionNet}}
\subfloat[]{\includegraphics[width=1.1in,trim=120 0 120 0,clip]{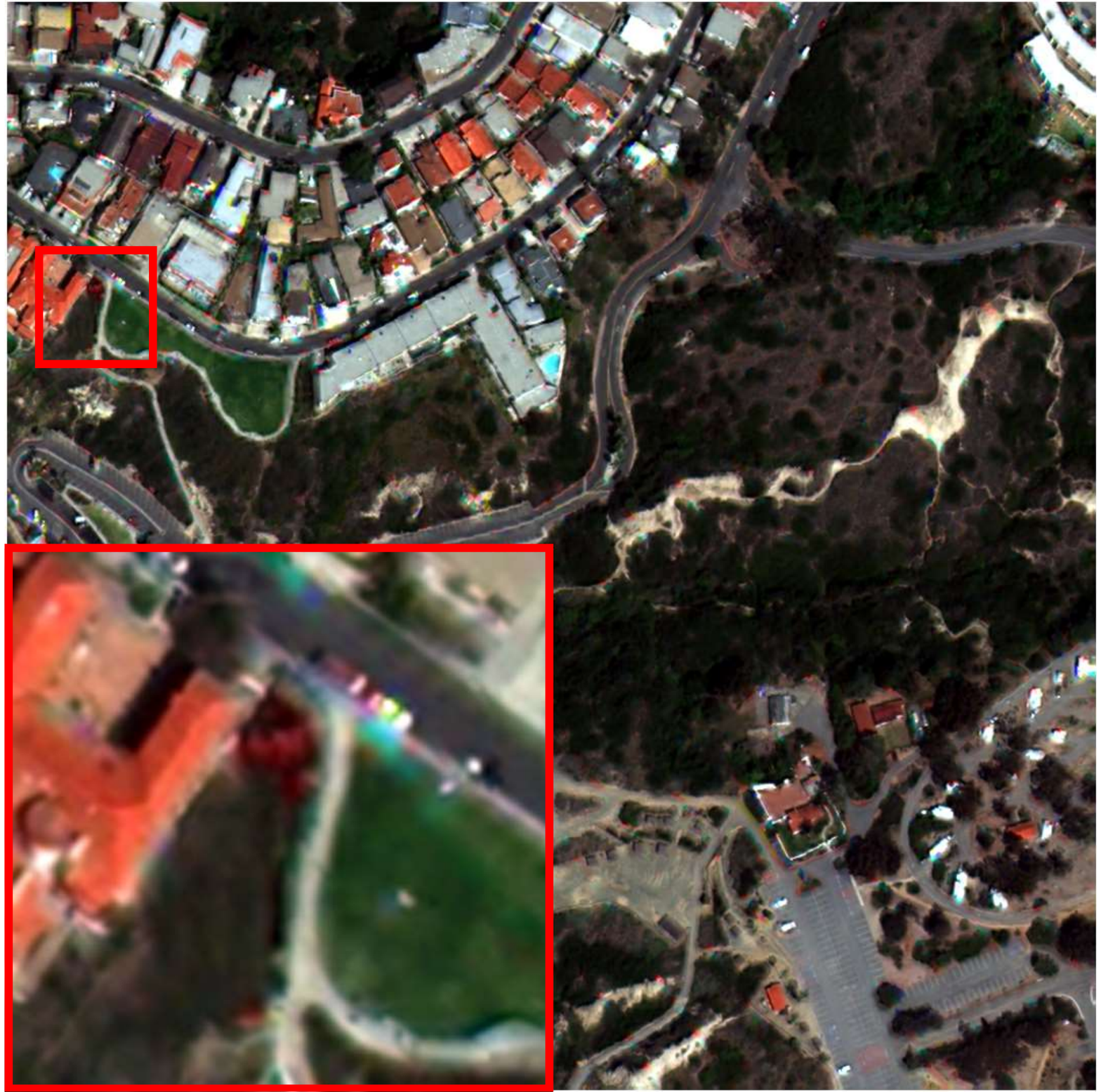}
\label{GTP-PNet}}
\subfloat[]{\includegraphics[width=1.1in,trim=120 0 120 0,clip]{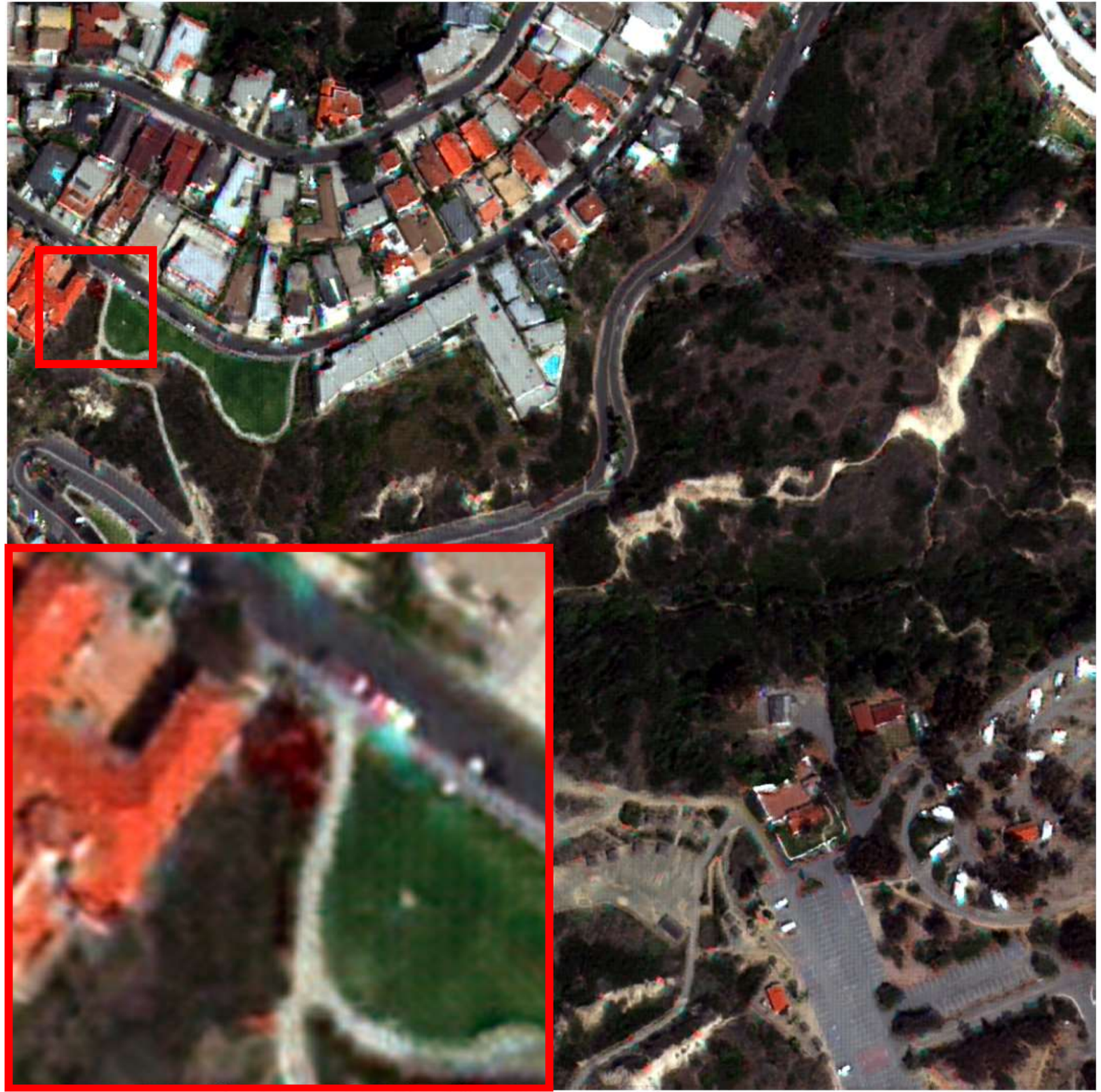}
\label{LPPN}}
\subfloat[]{\includegraphics[width=1.1in,trim=120 0 120 0,clip]{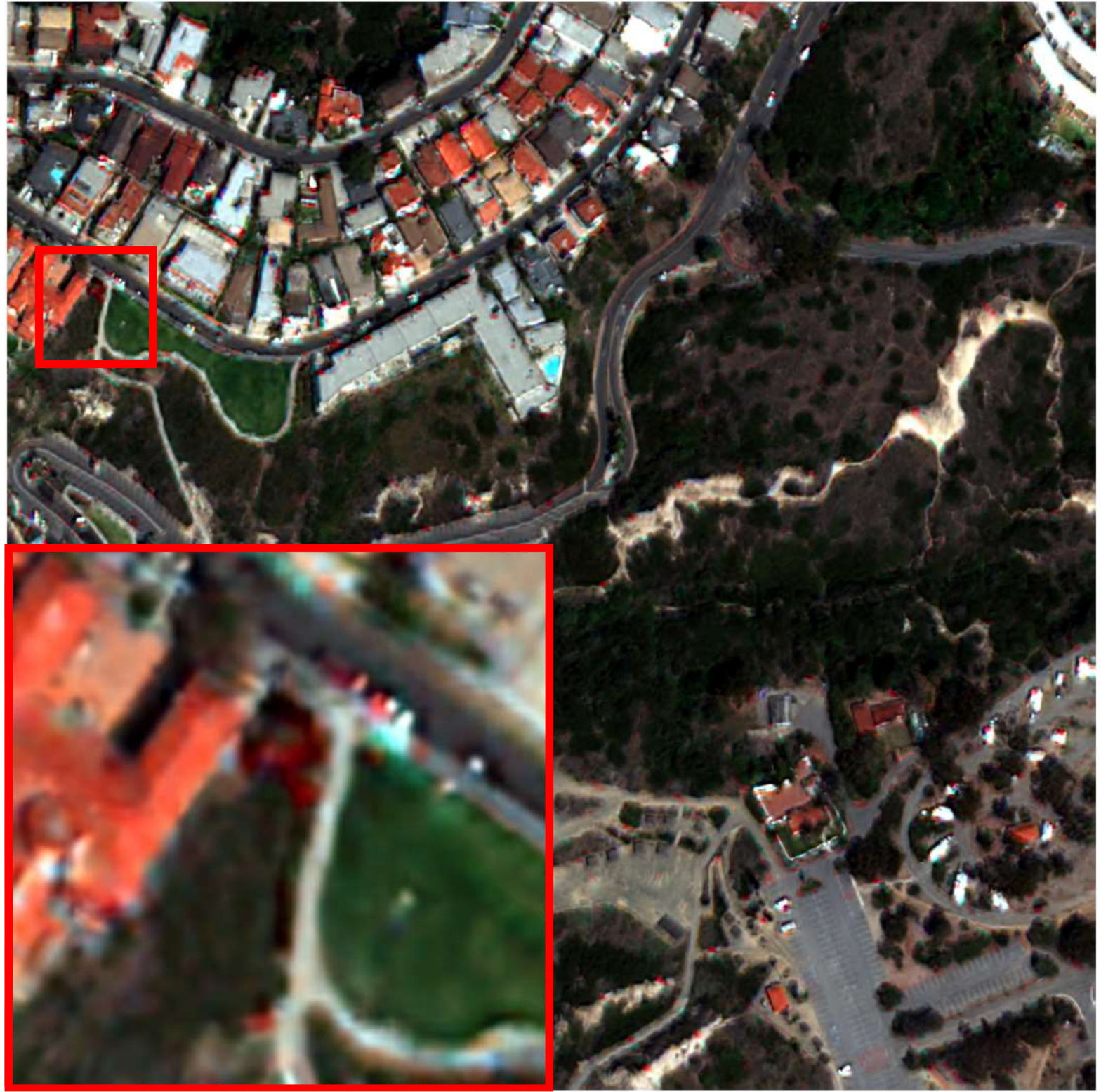}
\label{FAFNet}}
\caption{Visual comparison of different methods on WorldView-2 (WV-2) dataset  at full resolution. (a) PAN. (b) Up-sampled MS. (c) GSA \cite{4305344}. (d) BDSD-PC  \cite{8693555} . (e) MTF-GLP-CBD \cite{4305345} . (f) AWLP-H \cite{vivone2019fast} . (g) PWMBF \cite{6951484}. (h) PanNet \cite{8237455}. (i) FusionNet \cite{9240949}. (j) GTP-PNet \cite{ZHANG2021223} . (k) LPPN \cite{JIN2022158}. (l) FAFNet.}
\label{fig_wv2_full}
\end{figure*}

Due to the fact that the model is trained on simulated training samples obtained through the Wald's protocol \cite{wald1997fusion}, there is a resolution gap between the training samples and the test samples at full resolution, which makes the fusion of full-resolution images more challenging. In this section, we conduct experiments on full-resolution WV-4, QB and WV-2 datasets to verify the effectiveness of proposed model. Similarly, both the visual inspections and the quantitative assessments are used in the comparison. Figs. \ref{fig_wv4_full}-\ref{fig_wv2_full} show the visual comparisons of different methods on WV-4, QB, and WV-2 datasets. Because there are no reference images, original PAN and the up-sampled MS images are shown in these figures to assist the visual inspections.

\begin{table*}[h]
\caption{Average Quantitative Results on 136 Pairs of Test Data from WV-2 Dataset.\label{tab_wv2}}
\centering
\begin{tabular}{c|cccc|ccc}
\toprule
~ & \multicolumn{4}{c|}{\textbf{Reduced resolution}} & \multicolumn{3}{c}{\textbf{Full resolution}} \\		
\midrule
Methods & ERGAS & Q8 & SAM & SCC  & QNR & $D_\lambda$ & $D_s$ \\
\midrule
Ideal value & 0 & 1 & 0 & 1 & 1 & 0 & 0 \\		
\midrule
GSA \cite{4305344} & 4.8655±0.6849 & 0.8017±0.0145 & 6.3368±2.6112 & 0.8594±0.0007   & 0.8322±0.0076 & 0.0695±0.0022 & 0.1084±0.0034 \\
BDSD-PC \cite{8693555}  & 4.8871±0.6243 & 0.8051±0.0149 & 6.3477±2.2189 & 0.8761±0.0002 &  0.9009±0.0136 & \textbf{0.0372±0.0028} & 0.0688±0.0073 \\
MTF-GLP-CBD \cite{4305345} & 4.8155±0.6729 & 0.8021±0.0149 & 6.3446±2.3941 & 0.8607±0.0007  & 0.8706±0.0110 & 0.0657±0.0025 & 0.0721±0.0056 \\
AWLP-H \cite{vivone2019fast} & 4.7503±0.6355 & 0.8084±0.0148 & 5.8756±2.1019 & 0.8912±0.0001 & 0.8313±0.0176 & 0.0948±0.0059 & 0.0893±0.0083\\
PWMBF \cite{6951484} & 5.0787±0.7295 & 0.7782±0.0139 & 7.0062±3.0510 & 0.8560±0.0004 &  0.8075±0.0102 & 0.0822±0.0021 & 0.1240±0.0061 \\
PanNet \cite{8237455} & 4.3835±0.5161 & 0.8220±0.0169 & 5.6708±1.7453 & 0.9029±0.0001  & 0.9067±0.0133 & 0.0475±0.0038 & 0.0528±0.0054 \\
FusionNet \cite{9240949} & \underline{4.1819±0.4734} & \underline{0.8357±0.0174} &  \underline{5.1203±1.2136} &  \underline{0.9151±0.0001} & \underline{0.9109±0.0111} & 0.0475±0.0030 & \textbf{0.0474±0.0047} \\
GTP-PNet \cite{ZHANG2021223} & 4.4662±0.5606 & 0.8165±0.0160 & 5.7287±1.7201 & 0.8951±0.0001 & 0.8833±0.0139 & \underline{0.0413±0.0034} & 0.0835±0.0066 \\
LPPN \cite{JIN2022158} & 4.7112±0.5877 & 0.8070±0.0184 & 5.7579±1.3537 & 0.9014±0.0001  & 0.9097±0.0117 & 0.0462±0.0033 & 0.0503±0.0049 \\
FAFNet &  \textbf{3.9686±0.5140} & \textbf{0.8507±0.0178} & \textbf{4.5525±1.1841}  &  \textbf{0.9241±0.0001}  & \textbf{0.9150±0.0211} & 0.0476±0.0067 & \underline{0.0474±0.0090} \\
\bottomrule
\end{tabular}
\end{table*}

\begin{table*}[h]
\centering
\caption{Ablation Studies on 271 Pairs of Test Data from WV-4 Dataset.}\label{tab_ablation}

\renewcommand{\arraystretch}{1.5}
\begin{tabular}{cc|c|cccc}
\hline
~ &~& ~&\multicolumn{4}{c}{\textbf{Reduced resolution}} \\		
\cline{3-7}
DWT/IDWT layers & HFS loss & Indices & ERGAS & Q4 & SAM & SCC \\		
\cline{3-7}
~&~&Ideal value & 0 & 1 & 0 & 1 \\	
\hline
\checkmark  & \checkmark & \emph{Baseline} (FAFNet) & \textbf{1.1364±0.1101} & \textbf{0.9356±0.0015} & \textbf{1.5026±0.2782} & \textbf{0.9717±0.0001}  \\
-&\checkmark & \emph{Variant 1} & \underline{1.1620±0.1276} & 0.9336±0.0015 & \underline{1.5196±0.2962} & \underline{0.9702±0.0001} \\
\checkmark & - & \emph{Variant 2} & 1.1878±0.1433 & \underline{0.9348±0.0015} & 1.5810±0.3907 & 0.9695±0.0001\\
\hline
\end{tabular}
\end{table*}

Fig. \ref{fig_wv4_full} shows the fusion results on WV-4 dataset. Similarly, PAN and the up-sampled MS are shown in Figs. \ref{fig_wv4_full}(a)-(b), and the results of compared methods are shown in Figs. \ref{fig_wv4_full}(c)-(l). It can be found from the zoomed-in regions that BDSD-PC \cite{8693555}, PWMBF \cite{6951484} and FusionNet \cite{9240949} yield obvious spectral distortion, while the results of GSA \cite{4305344}, MTF-GLP-CBD \cite{4305345}, AWLP-H \cite{vivone2019fast}, PanNet \cite{8237455}, and GTP-PNet \cite{ZHANG2021223} are affected by the blurry effects. Concerning both the fidelity of spectral and spatial information, LPPN \cite{JIN2022158} and FAFNet achieve superior performance, but the fusion product of LPPN \cite{JIN2022158} contains more artifacts. Table \ref{tab_wv4} reports the average quantitative results on 271 pairs of test data from the WV-4 dataset. From Table \ref{tab_wv4}, we can observe that FAFNet obtains the best fusion results compared with other state-of-the-art methods.

Fig. \ref{fig_qb_full} displays a group of pansharpened images produced by different methods on the QB dataset. As shown in Fig. \ref{fig_qb_full}, the GSA \cite{4305344}, BDSD-PC \cite{8693555}, MTF-GLP-CBD \cite{4305345}, AWLP-H \cite{vivone2019fast}, PWMBF \cite{6951484} and FusionNet \cite{9240949} preserve spatial details well, but they yield serious spectral distortion. The result of GTP-PNet \cite{ZHANG2021223} is blurry, such as the building areas. On the contrary, the fusion results of PanNet \cite{8237455} and LPPN \cite{JIN2022158} are affected by severe aliasing artifacts. Our proposed FAFNet produces high-quality pansharpened result with abundant details, clear boundaries and high spectral fidelity. The average quantitative results on 156 pairs of test data are reported in Table \ref{tab_qb}, which indicates that FAFNet obtains the best values in all three metrics. 

The experimental results on the full-resolution data from WV-2 are shown in Fig. \ref{fig_wv2_full}. From the enlarged regions in Figs. \ref{fig_wv2_full}(c)-(l), we can observe that the fused results of BDSD-PC \cite{8693555}, AWLP-H \cite{vivone2019fast} and PWMBF \cite{6951484} have different levels of spatial degradation, especially for the texture of objects, such as buildings and roads, while the results of LPPN \cite{JIN2022158} and PanNet \cite{8237455} suffer from serious checkerboard artifacts. The fusion results of GSA \cite{4305344}, MTF-GLP-CBD \cite{4305345}, FusionNet \cite{9240949}, and GTP-PNet \cite{ZHANG2021223} have different degrees of spectral distortion. Compared with them, the proposed FAFNet has high fidelity in both the spectral and spatial domain. The quantitative evaluations are listed in Table \ref{tab_wv2}, where the average indices of fusion results on 136 pairs of test data from the WV-2 dataset are demonstrated, form which we can conclude that FAFNet performs better on the eight-band WV-2 data compared with other methods.

\begin{table*}[htbp]
\caption{Comparison Of Average Computational Time in Seconds on Different Size.\label{tab_time}}
\centering
\resizebox{\textwidth}{6mm}{
\begin{tabular}{ccccccccccc}
\toprule
PAN Size & GSA \cite{4305344} & BDSD-PC \cite{8693555} & MTF-GLP-CBD \cite{4305345} & AWLP-H \cite{vivone2019fast} & PWMBF \cite{6951484} & PanNet \cite{8237455} & FusionNet \cite{9240949} & GTP-PNet \cite{ZHANG2021223} & LPPN \cite{JIN2022158} & FAFNet\\		
\midrule
256 $\times$ 256 & 0.0545 &  0.0663 & 0.0988 & 0.0536 & 0.3490 & \textbf{0.0027} & 0.1668 & 0.2556 & 0.8215 & \underline{0.0241} \\
1024 $\times$ 1024 & 0.7708 & 0.4418 & 1.3236 & 0.3747 & 3.5229 & \textbf{0.0042} & 2.4387 & 3.9525 & 13.2468 & \underline{0.0708}\\
\bottomrule
\end{tabular}}
\end{table*}

\subsection{Ablation Studies}
To verify the effectiveness of the DWT and IDWT layers as well as the proposed High-frequency Feature Similarity (HFS) loss, we conduct several ablation studies on WV-4 dataset at reduced-resolution. Taking the proposed FAFNet in Fig. \ref{structure_FAN} as \emph{Baseline}, the experimental results are provided in Fig.~\ref{fig_ab} and Table \ref{tab_ablation}. %\textcolor{red}{The visual comparisons among three models are presented in Fig. \ref{fig_ab}, including fusion results and AEMs.}

\subsubsection{Effectiveness of DWT/IDWT layers}
We replace the DWT and IDWT layers with normal Haar DWT and IDWT to evaluate the effects of DWT/IDWT layers. This setting is denoted as \emph{Variant 1}. By comparing quantitative evaluations of \emph{Variant 1} with that of \emph{Baseline} in Table \ref{tab_ablation}, we find that the introduction of DWT and IDWT layers has positive impacts on all evaluation indexes. When comparing fusion results together with AEMs of \emph{Variant 1} and \emph{Baseline} from enlarged region in Fig. \ref{fig_ab}, it is clear that \emph{Baseline} has achieved better performance considering both details enhancement and spectral preservation.

\subsubsection{Effectiveness of HFS loss}
\begin{figure*}[h]
\footnotesize
  \centering
  \scriptsize
  \begin{tabular}{{c}{c}{c}{c}{c}}
  \hspace{-4mm}
  \includegraphics[width=1.1in]{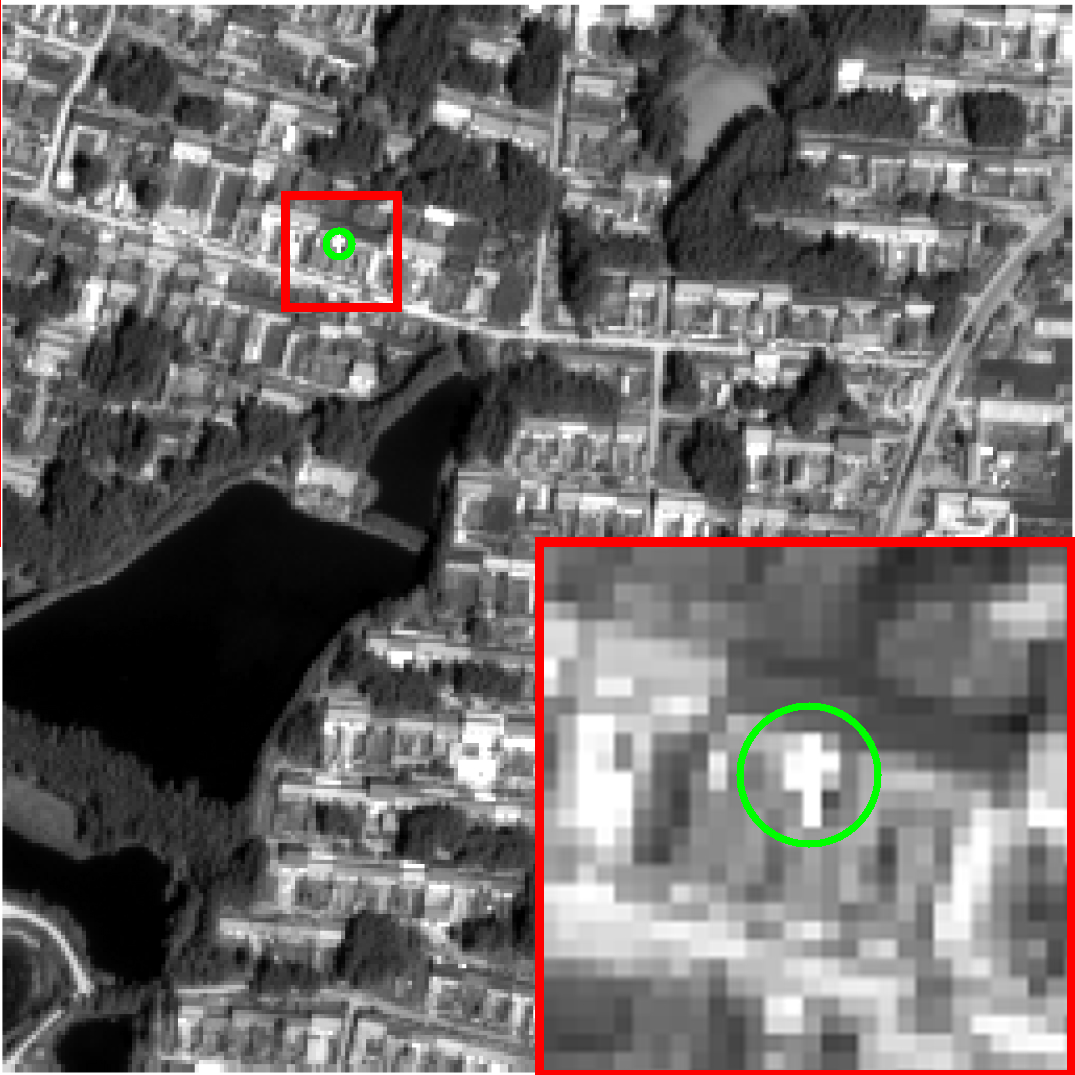} &
  \hspace{-4mm}
  \includegraphics[width=1.1in]{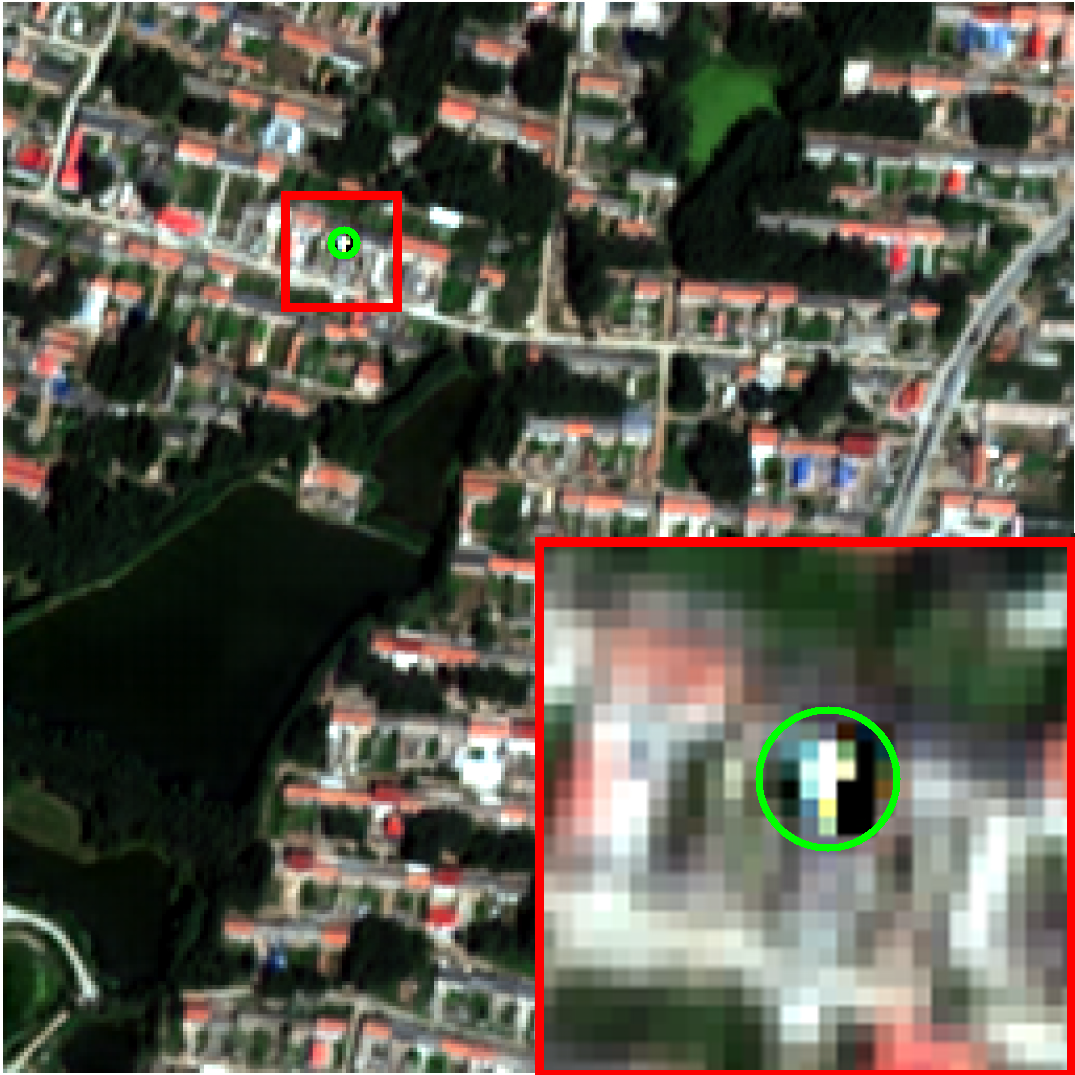} &
  \hspace{-4mm}
  \includegraphics[width=1.1in]{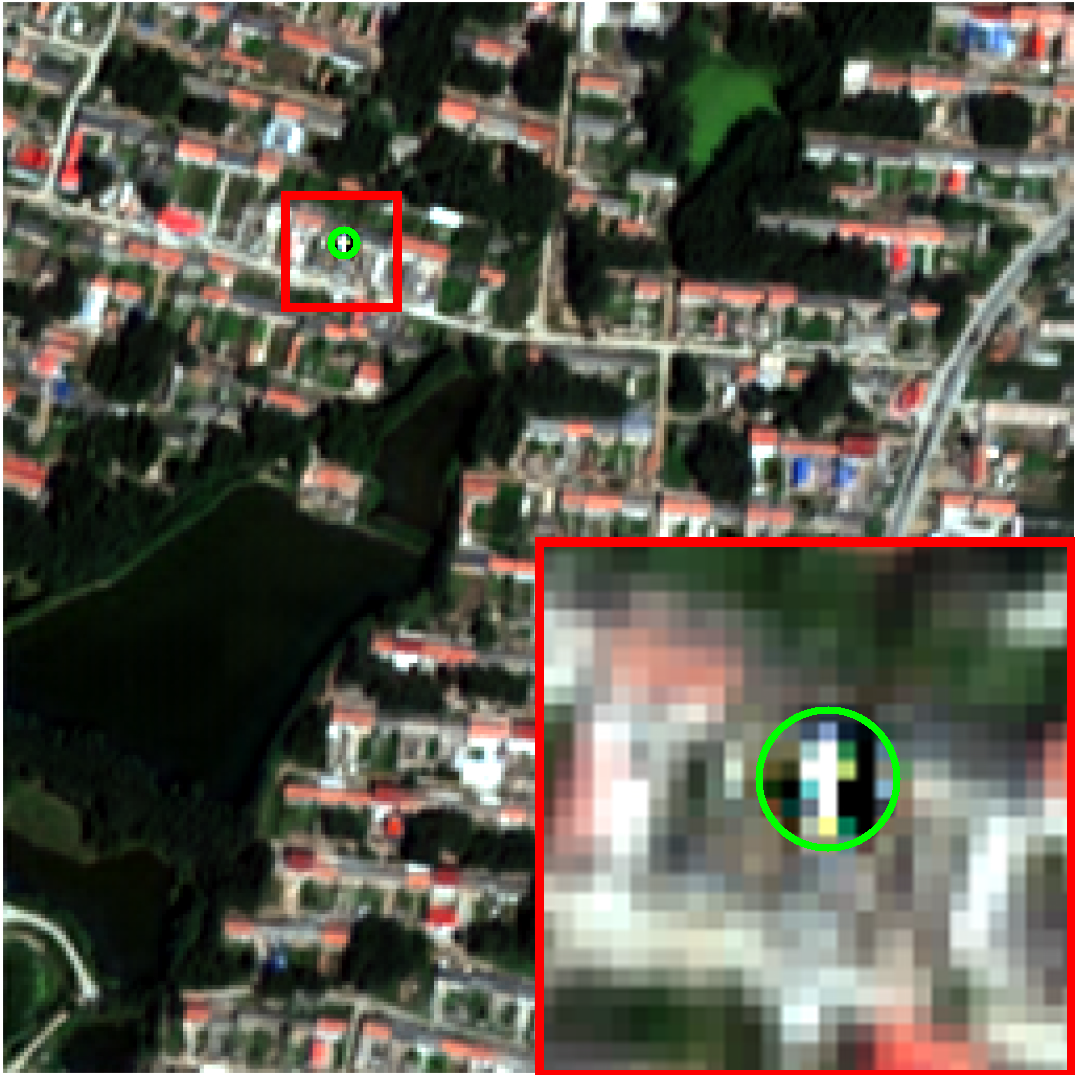} &
  \hspace{-4mm}
  \includegraphics[width=1.1in]{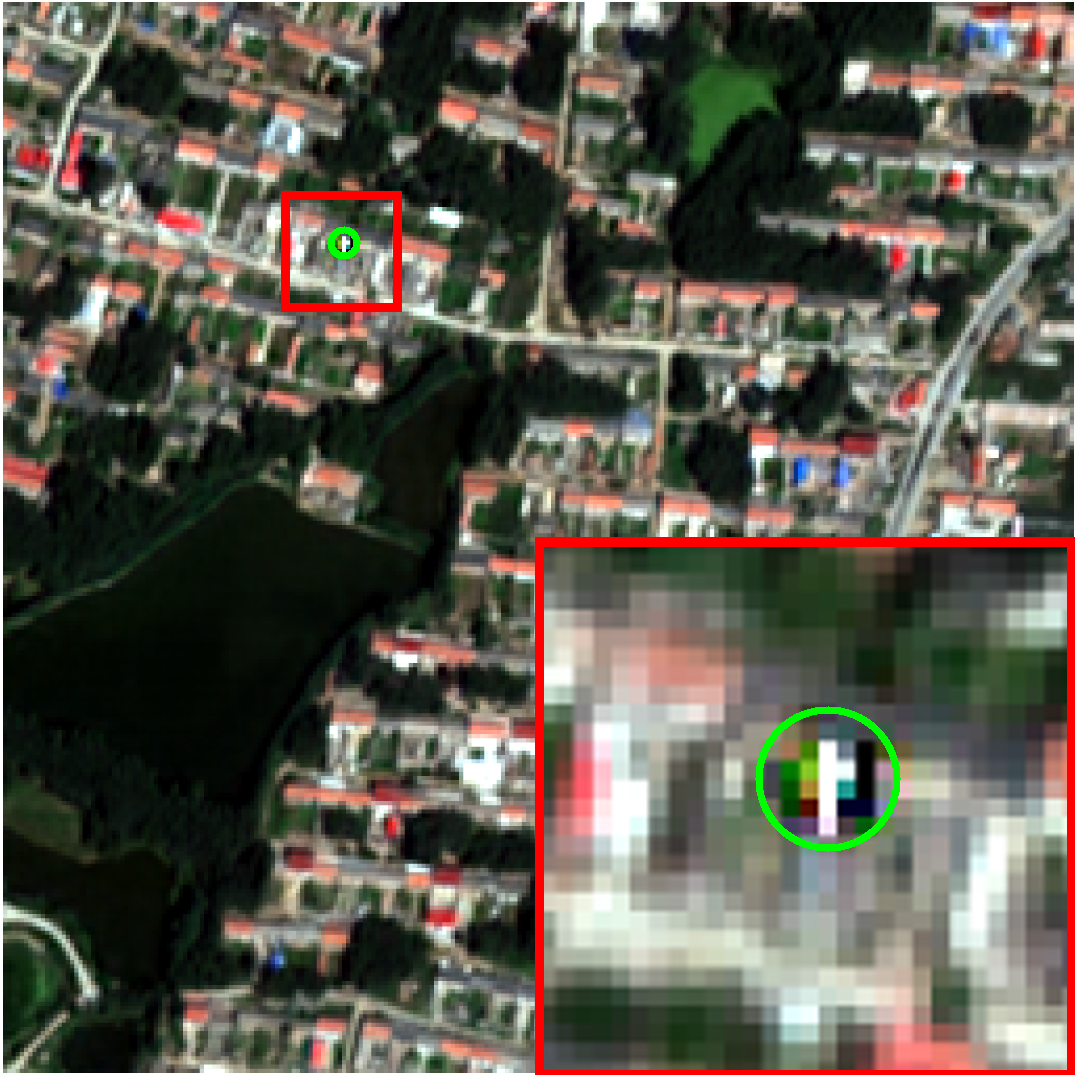} & \hspace{-4mm}
  \includegraphics[width=1.1in]{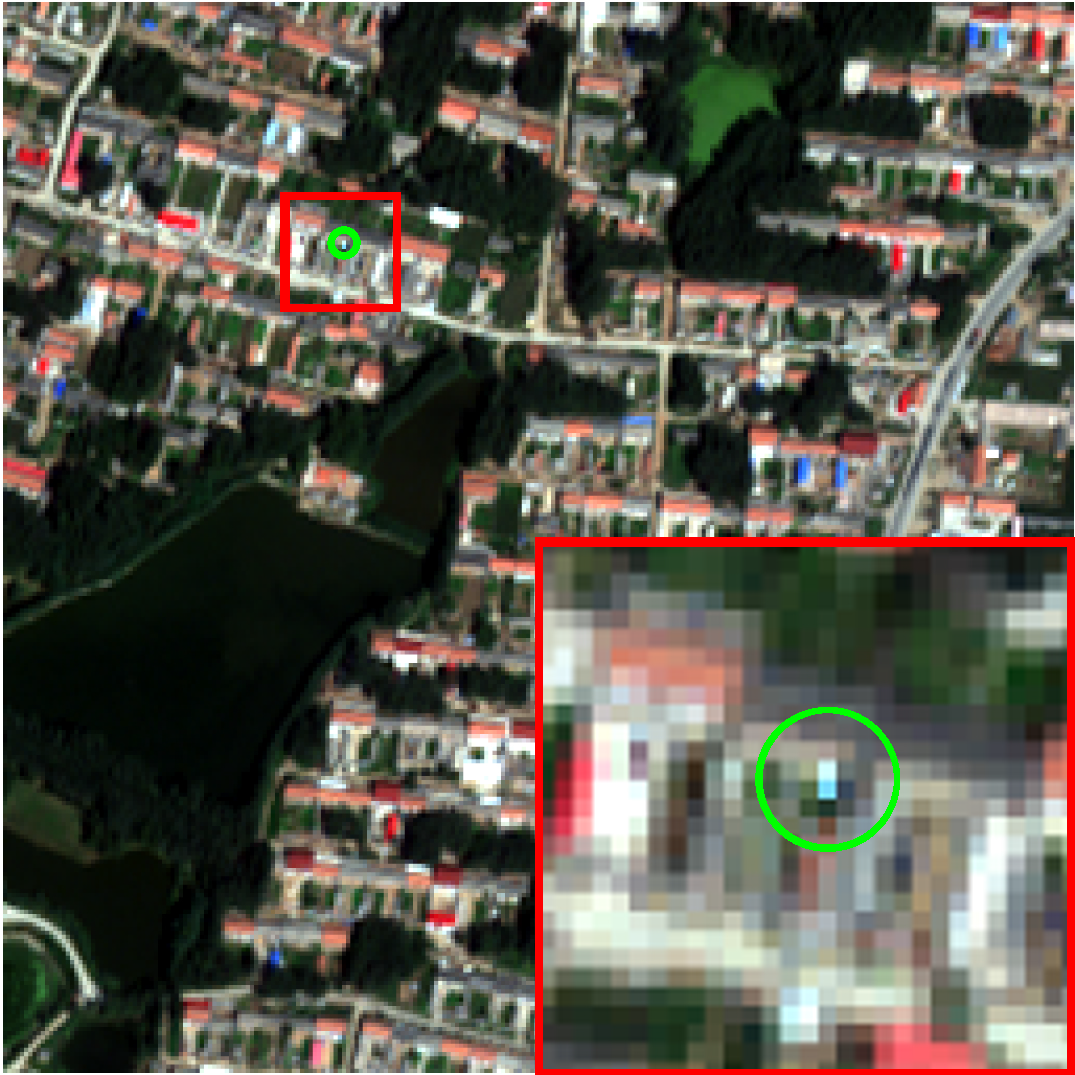} \\

     \hspace{-4mm} PAN &   \hspace{-4mm} \emph{Baseline} (FAFNet) &   \hspace{-4mm}\emph{Variant 1} &   \hspace{-4mm}\emph{Variant 2} &   \hspace{-4mm}Reference \\
   \hspace{-4mm}
  \includegraphics[width=1.1in]{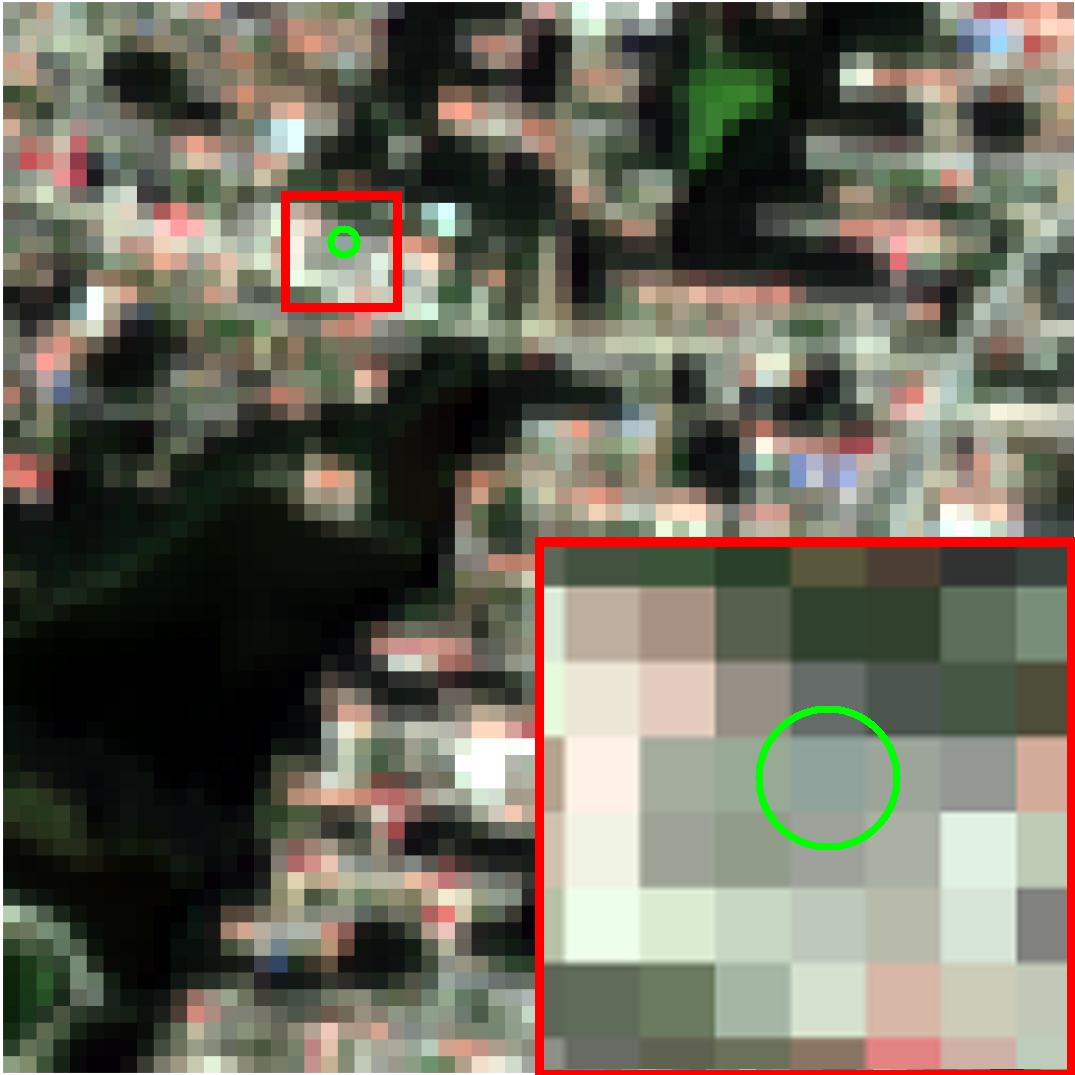} &  
  \hspace{-4mm}
  \includegraphics[width=1.1in]{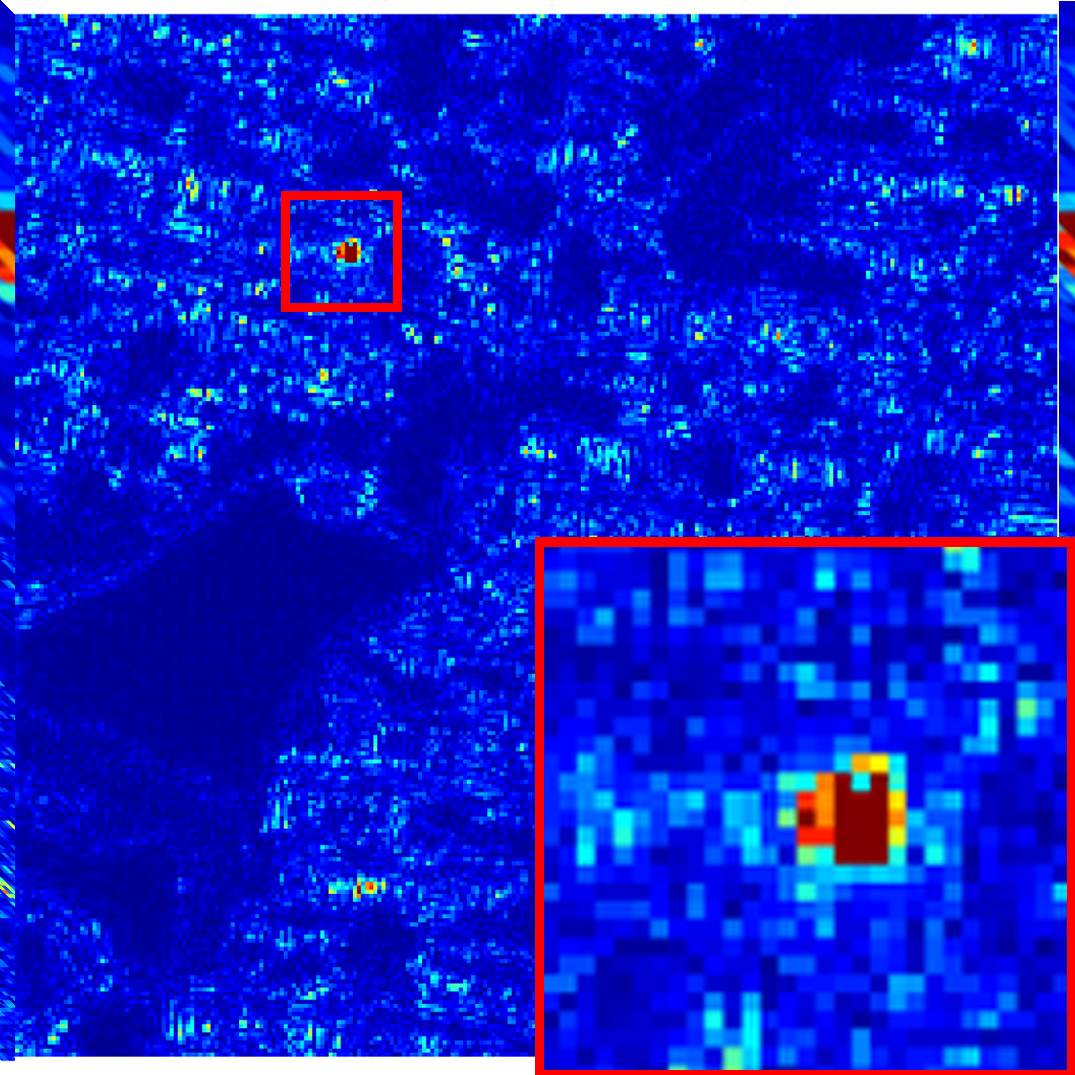} &
  \hspace{-4mm}
  \includegraphics[width=1.1in]{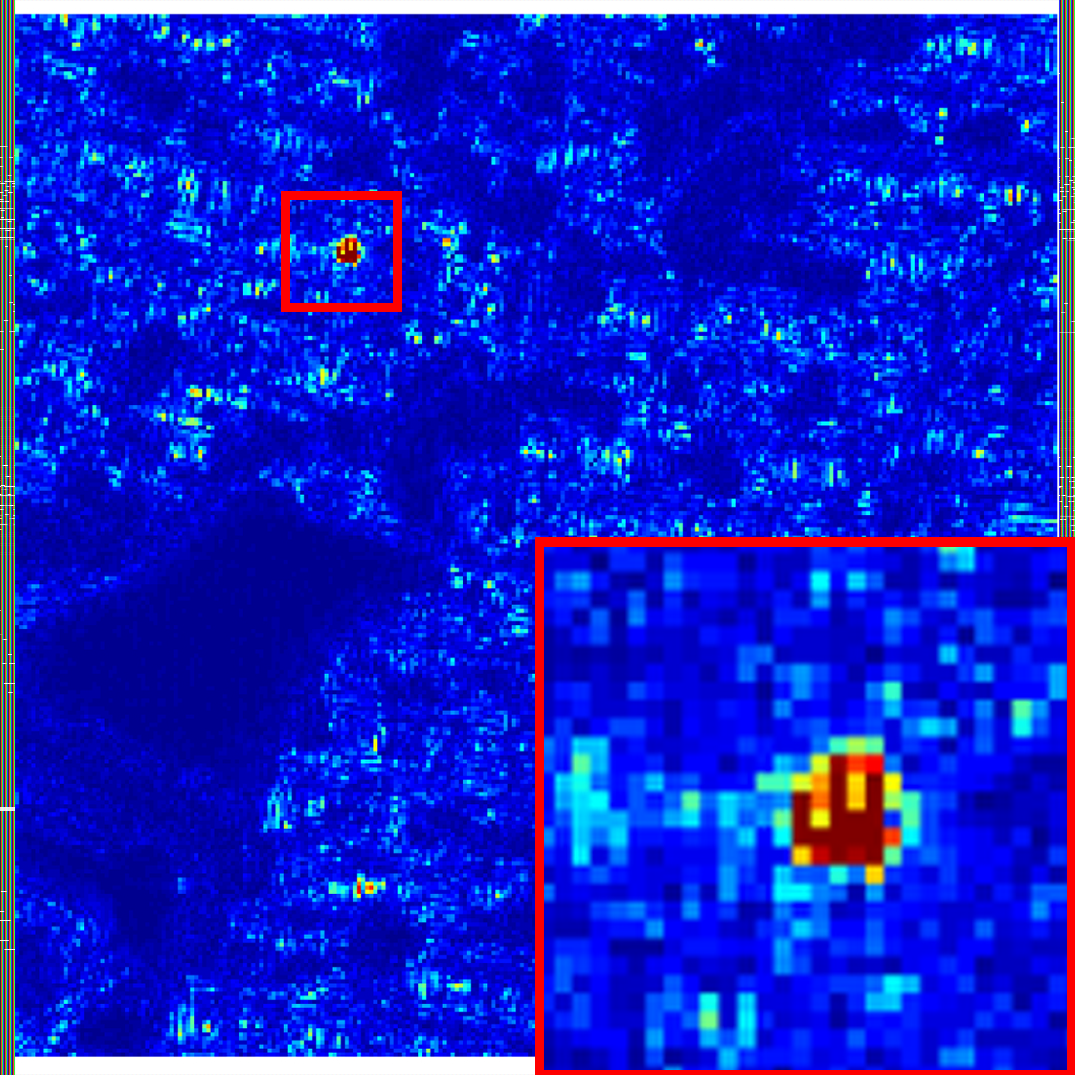} &
  \hspace{-4mm}
  \includegraphics[width=1.1in]{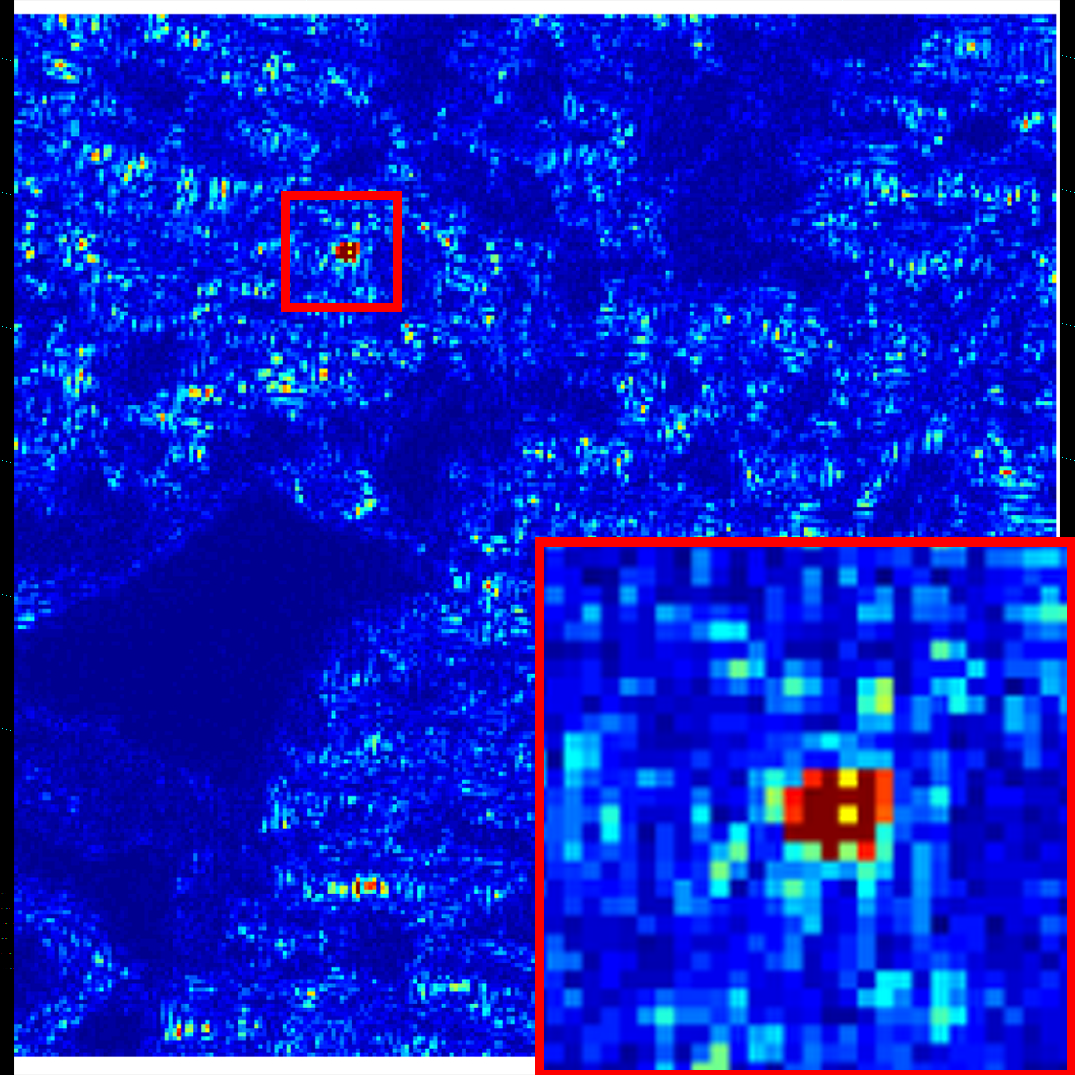} & \includegraphics[height=1.1in,trim=220 0 220 0,clip]{figures/wv4/error/bar.pdf}\\
  
   \hspace{-4mm} MS &   \hspace{-4mm}MAE of \emph{Baseline} (FAFNet) &   \hspace{-4mm}MAE of \emph{Variant 1} &   \hspace{-4mm}MAE of \emph{Variant 2} &   \hspace{-4mm} bar \\
      
  \end{tabular} \normalsize
  \caption{Visual comparisons among \emph{Baseline} (FAFNet), \emph{Variant 1} and  \emph{Variant 2}.}
  \label{fig_ab}
\end{figure*}

To investigate the effects of HFS loss, we neglect the HFS loss from the loss function. This setting is denoted as \emph{Variant 2}. From the comparison between \emph{Variant 2} and \emph{Baseline}, we can find that the evaluation indices are all degraded  especially ERGAS and SAM, which concludes that the HFS loss is of great importance and has the ability of restricting the HF features of PAN as equivalent as possible to the missing HF parts of MS images. As can be seen from Fig. \ref{fig_ab}, both \emph{Variant 2} and \emph{Baseline} have some flaws in their fusion images, especially the color of objects in green circles. By observing the AEMs of \emph{Variant 2} and \emph{Baseline}, we can find that \emph{Baseline} corrects some spectral distortions in \emph{Variant 2}, which confirms the importance of HFS loss.

It should be noted that the configurations of above variants are the same as \emph{Baseline}, except the architectures. From above analyses, we can conclude that the DWT and the IDWT layers, together with the HFS loss are all essential for FAFNet, and each of them can bring specific improvements. By working together, the optimal fusion results both at reduced and full resolution are obtained.

\subsubsection{Impacts of different $\beta$ values}
\begin{figure}[h]
\centering
  \begin{tabular}{{c}{c}}
 \hspace{-4mm}
\includegraphics[width=0.5\linewidth,trim=90 270 90 270,clip]{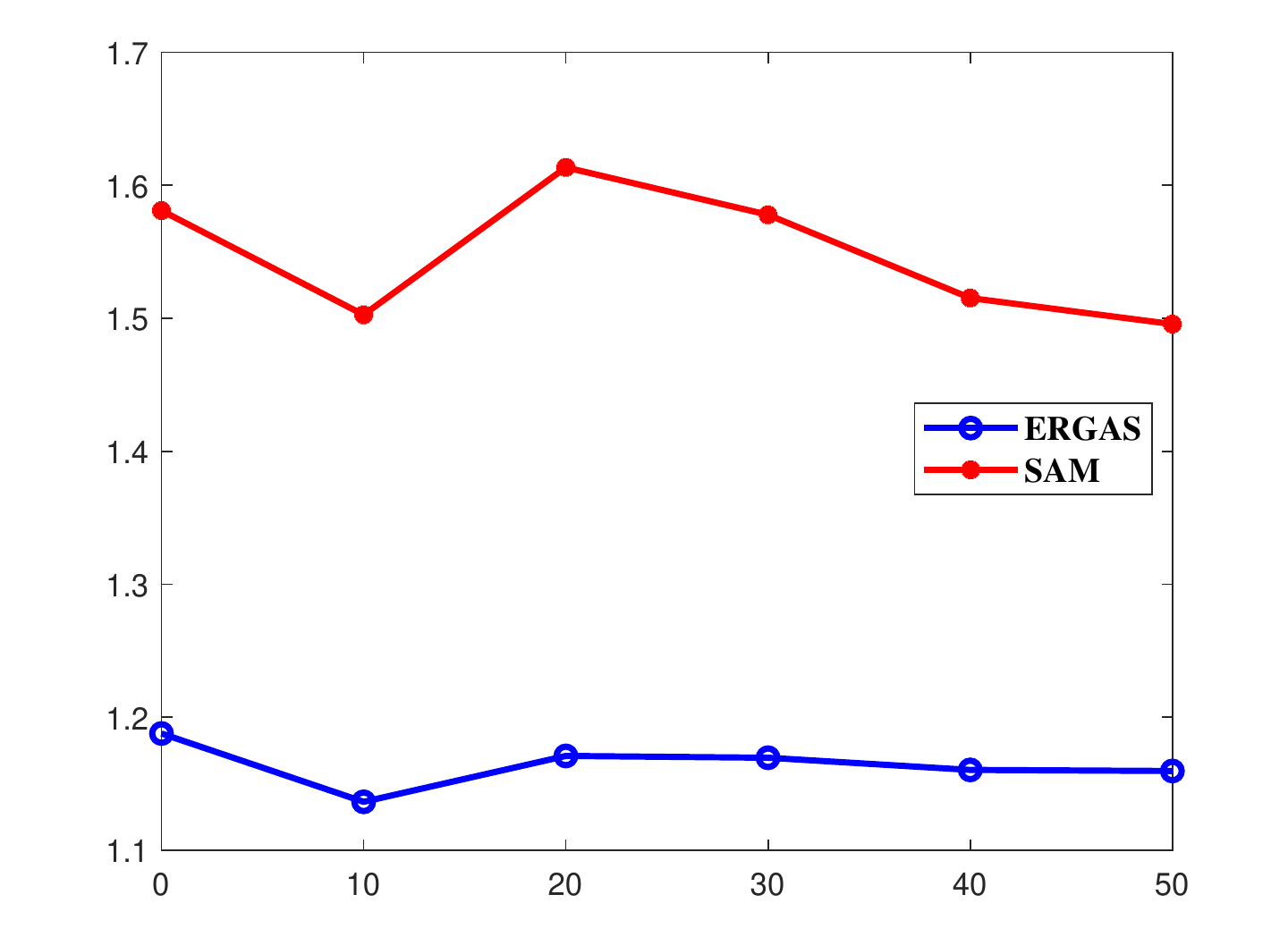}&
\hspace{-4mm}
\includegraphics[width=0.5\linewidth,trim=90 270 90 270,clip]{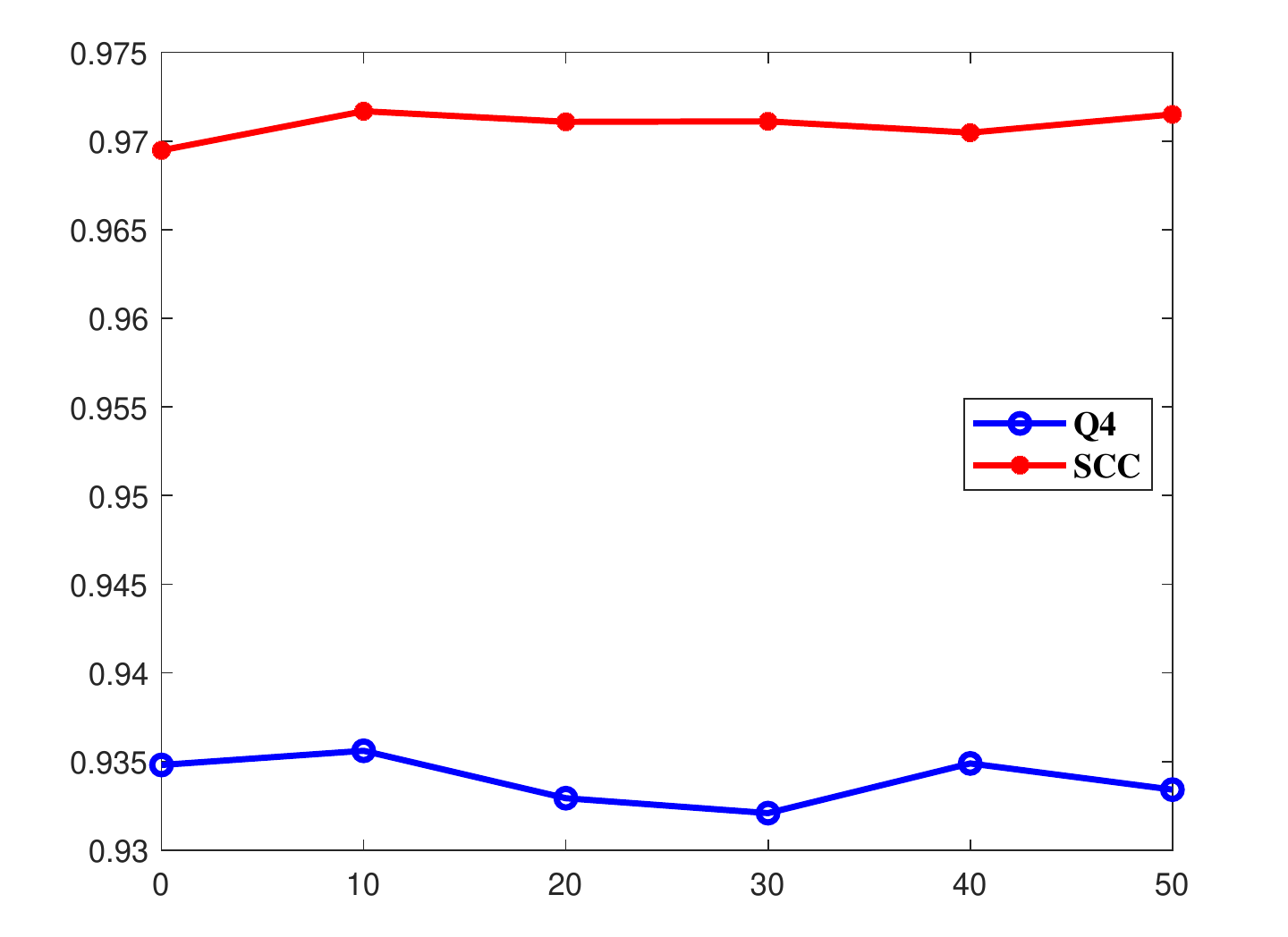}\\
\hspace{-4mm}(a) &   \hspace{-4mm}(b) \\
\end{tabular} \normalsize
\caption{Different $\beta$ settings on WV-4 dataset. (a) Average ERGAS and SAM. (b) Average Q4 and SCC.}
\label{fig_p_ablation}
\end{figure}

In this subsection, we conduct a series of experiments on WV-4 dataset to investigate the impacts of $\beta$ on fusion results. Concretely, we change the values of $\beta$ from 0 to 50 and present the variations of four indexes with respect to different value of $\beta$. The results are shown in Fig. \ref{fig_p_ablation}. We can find that the optimal result is obtained when $\beta$=10. Therefore, we set $\beta$=10 when train FAFNet on WV-4 dataset. Similarly, the configuration of $\beta$=15 is adopted in the experiments on QB dataset, and $\beta$=30 is used for WV-2 dataset.

\subsection{Computational Time}
In this section, we compare the average computational time in seconds for different fusion methods at two scales, i.e. PAN images with the size of 256 $\times$ 256 and 1024 $\times$ 1024. Traditional methods are implemented by MATLAB R2018b on the computer with Intel(R) Core(TM) i5-1135G7 processor. For CNN-based methods, they are tested on the computer with an Nvidia GTX 1080Ti GPU and an Intel(R) Xeon(R) W-2123 CPU. Table \ref{tab_time} lists the average running time in seconds of different methods, from which, we can see that proposed FAFNet model is efficient when compared with most of state-of-the-art methods. This is partially due to the relatively simple architecture of FAFNet.

\section{Conclusion}
In this paper, we propose a frequency-aware fusion network (FAFNet) along with a novel High-frequency Feature Similarity (HFS) loss to learn the correspondence in frequency domain for multispectral pansharpening. In order to enable FAFNet to learn sufficient high-frequency information, DWT/IDWT layers are introduced, and FAFNet works directly in frequency domain to explicitly and adaptively extract and process high-frequency features. To reduce the spectral distortion, HFS loss is further designed to align high-frequency features of PAN and MS. Through the constraint of HFS loss, the high-frequency features of PAN can appropriately complement to that of MS. Extensive experiments on three datasets at both reduced and full resolution demonstrate the advantages of FAFNet. %especially for the full-resolution experiments, which further proves that FAFNet has strong generalization ability in the real scenes. 
FAFNet is an attempt of learning frequency features by CNNs, and we believe that it is a trend for many low-level computer vision tasks.

It should be noted that the training of FAFNet relies on simulated datasets, therefore, due to the scale variations, some undesirable artifacts appear when applied to full-resolution images in real scenes. In the future, we will explore unsupervised pansharpening methods from the aspects of observation modeling and the frequency-aware cross-scale feature learning.

%\section*{Acknowledgments}

%{\appendix[Proof of the Zonklar Equations]
%Use $\backslash${\tt{appendix}} if you have a single appendix:
%Do not use $\backslash${\tt{section}} anymore after $\backslash${\tt{appendix}}, only $\backslash${\tt{section*}}.
%If you have multiple appendixes use $\backslash${\tt{appendices}} then use $\backslash${\tt{section}} to start each appendix.
%You must declare a $\backslash${\tt{section}} before using any $\backslash${\tt{subsection}} or using $\backslash${\tt{label}} ($\backslash${\tt{appendices}} by itself
% starts a section numbered zero.)}


\bibliographystyle{IEEEtran}
\bibliography{references}

\newpage

\vfill

\end{document}